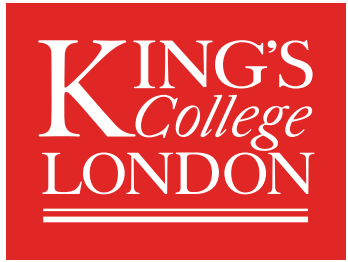

# Whole Brain Network Dynamics of Epileptic Seizures at Single Cell Resolution

## Dominic Robert Wai Chueng Burrows

Supervisors:

Professor Mark Richardson (Primary Supervisor)

Dr Richard Rosch (Secondary Supervisor)

Dr Martin Meyer (Initial primary supervisor)

*A thesis presented for the degree of*

*Doctor of Philosophy*

MRC Centre for Neurodevelopmental Disorders

Institute of Psychiatry, Psychology & Neuroscience

King's College London

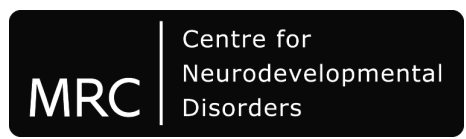

# Contents

# List of Figures

# List of Tables

# Acknowledgements

First and foremost, I would like to thank my supervisor and friend Dr. Richard Rosch. You have given countless hours, weekends and evenings to guiding me on my scientific journey. Your boundless enthusiasm to my work has helped me through dark days, your openness has allowed me to push my research in whatever direction I chose (sometimes regrettably so), and your patience in explaining computational concepts has given me the gift of self-belief, in a field that sometimes terrifies me. Thank you for everything.

I would also like to thank Professor Mark Richardson for his excellent mentorship and guidance through difficult times. Despite your incredibly busy schedule and different expertise, you always found time to provide comforting advice and to shift my perspective towards the clinic, and I am truly grateful for this.

I would like to thank Dr. Martin Meyer for his mentorship in the first few years of the PhD. I am very grateful that you had the openness to take me on as a young researcher, when I had no experience in the lab. You fostered such an exciting and open lab environment which I feel so grateful to have been a part of. These early years laid the foundations for me to grow into the scientist that I am today, so thank you.

Special mention must also go to Dr. Gerald Pao. Thank you for hosting me at the Salk and for all of the science and career chats – you have taught me that it's possible to be an expert experimentalist and a computationalist.

I am immensely grateful to all of my science friends.

To Tom Sainsbury, thank you for committing so much time to teaching me how to code, you've given me a gift that I hope to use for the rest of my life. To Tom Shallcross, thank you for all of the maths lessons, you've helped maths feel less scary to me. To Rachel Williams, you have provided endless emotional support and structural connectivity expertise, so thank you. To Jade Lau, thank you for all of the help and guidance from day 1, from zebrafish breeding to molecular biology.  To Kostas Lagogiannis, thank you for all of the intense science conversations which make me feel like a better scientist. To Giovanni Diana, thank you for being the physicist in my life who would actually explain complex problems to me that no one else could or would. To Catia Fortunato, thank you for our many conversations about manifolds – somehow you make them seem interesting. To Uzma Zahid, thank you for sticking with me till the end.

Lastly, I need to say a huge thank you to my Mum and Dad. You have both sacrificed so much to get me to where I am today. No words on a page can express the gratitude I have for everything you have done for me. I hope that this thesis makes you proud.

'If you don't know where you are going, any road can take you there.'
*The Chesire Cat, Lewis Carrol*

'If the thought of something scares you, then you should probably do it'
*Dominic Burrows*

# Abstract

Epileptic seizures are characterised by abnormal brain dynamics at multiple scales, engaging single neurons, neuronal ensembles and coarse brain regions. At the single neuron level, seizure dynamics are highly heterogeneous, giving rise to complex non-linear behaviours at higher scales, such as high frequency oscillations, bifurcations and deterministic chaos. Key to understanding the cause of such emergent population dynamics, is capturing the collective behaviour of neuronal activity at the microscale. However, linking microscale neuronal dynamics with emergent macroscale dynamics using conventional approaches is challenging – macroscale recordings coarse grain the underlying microscale activity, while typical microscale recordings subsample the global network dynamics. In this thesis I make use of the larval zebrafish to capture single cell neuronal activity across the whole brain during epileptic seizures. Firstly, I make use of statistical physics methods to quantify the collective behaviour of single neuron dynamics during epileptic seizures. Here, I demonstrate a population mechanism through which single neuron dynamics organise into seizures – brain dynamics deviate from a phase transition. Secondly, I make use of single neuron network models to identify the synaptic mechanisms that actually cause this shift to occur. Here, I show that the density of neuronal connections in the network is key for driving generalised seizure dynamics. Interestingly, such changes also disrupt network response properties and flexible dynamics in brain networks, thus linking microscale neuronal changes with emergent brain dysfunction during seizures. Thirdly, I make use of non-linear causal inference methods to study the nature of the underlying neuronal interactions that enable seizures to occur. Here I show that seizures are driven by high synchrony but also by highly non-linear interactions between neurons. Interestingly, these non-linear signatures are filtered out at the macroscale, and therefore may represent a neuronal signature that could be used for microscale interventional strategies. This thesis demonstrates the utility of studying multi-scale dynamics in the larval zebrafish, to link neuronal activity at the microscale with emergent properties during seizures.

# Chapter 1

## Introduction

The brain consists of around $10^{10}$ neurons each forming up to $10^3$ synapses, giving rise to a complex network of ~$10^{13}$ connections. Each node within this network can send an electrochemical signal to all of its connected neighbours at any given moment. These nodes fire spontaneously, and in response to sensory input or changes in behavioural state. Given the dense, recurrent nature of connections in the brain it is perhaps surprising that neuronal dynamics remain broadly stable. In fact, a common feature of large networks is the propensity to exhibit cascading failures, whereby local events trigger network-wide disruptions in normal dynamics – for example during transport network failures or electrical power grid blackouts. Interestingly, such cascading failures are a common feature of brain biology: seizures occur in 10% of people, and across the animal kingdom. Individuals who experience recurring seizures, in particular those diagnosed with epilepsy, live with debilitating symptoms. Historically, approaches to describe seizures have relied on methods which obscure the dynamics of each node to approximate an average effect over the network. Conversely, more recent methods support the measurement of single node dynamics in a sub-region of the network. However, seizures emerge out of the interactions between individual nodes in the network, specifically neurons, and cascading failures may influence the network in its entirety. Approaches to characterise and ultimately intervene in such cascading failures in the brain should therefore utilise the dynamics of neurons across the entire network. The goal of this thesis is to use the larval zebrafish, a system which is amenable to the recording of single cell dynamics across the whole brain, to understand how cascading failures occur as seizures. The daunting complexity of the brain lends itself to computational modelling tools which can reduce the dimensionality of the data and test network mechanisms. This chapter will first provide historical context for our current understanding of why seizures occur. It will then explore current approaches to modelling cascading failures in large networks before providing background to the specific experimental and theoretical tools that I will use to approach this problem. This chapter uses text and figures from some of my previously published work (Burrows et al., 2020; Rosch et al., 2019).

## 1.1   Dynamic Signatures of Epileptic Seizures Across Brain Scales

### 1.1.1   Epilepsy and Seizures

Epilepsy is a collection of neurological syndromes primarily characterised by the recurrence of seizures (Fisher et al., 2014). Seizures are themselves defined as transient changes to behavioural state, caused by excessive or hyper-synchronous brain activity (Devinsky et al., 2018). Seizures are highly diverse – such changes to behavioural state can include muscle jerking, changes to muscle tension, rapid blinking, head nodding, a loss of awareness, changes to sensory experience, the experience of déjà vu, and confusion, among other phenomena (Stafstrom & Carmant, 2015). Epilepsy is further characterised by the cognitive, psychological, physiological and social impairments that occur in most syndromes. In particular, epilepsy patients experience cognitive impairments including lower IQ (Farwell et al., 1985), cognitive regression (Thompson & Duncan, 2005), and impaired memory (Butler & Zeman, 2008) but also psychiatric comorbidities such as anxiety, depression and psychosis (Weatherburn et al., 2017). This underscores both the debilitating and diverse nature of the symptomatology that broadly defines what we call epilepsy.

Epilepsy is a global phenomenon, affecting 65 million people and representing the third greatest societal burden across all neurological disorders (Vos et al., 2016). It has an estimated prevalence of 0.6% and an incidence of 70 diagnoses per 100,000 people per year (Fiest et al., 2017; Hauser et al., 1991). The burden for patients with epilepsy is staggering – in particular, quality of life is significantly reduced, with a loss of independence, increased anxiety, and the avoidance of social and physical activity due to seizures (Harden et al., 2007; Krumholz et al., 2016). Patients also have significantly reduced income (Lindsten et al., 2002) and lower employment rates (Kobau et al., 2008). Furthermore, patients exhibit a high risk of comorbidities with other psychiatric conditions – 24% of patients are diagnosed with a mental health condition (Tellez-Zenteno et al., 2007). There is also a higher risk of premature mortality in epilepsy patients, with a 2.3–2.6 increased rate of death from all causes (Levira et al., 2017; Thurman et al., 2017) and sudden unexpected death in epilepsy occurring in 0.1% of patients (Sveinsson et al., 2017).

Importantly, seizures are a major contributor towards the overall morbidity of epilepsy. Patients whose seizures remain refractory to treatment experience greater comorbidity (Gaitatzis et al., 2004), mortality (Mohanraj et al., 2006) and cognitive decline (McCagh et al., 2009). Partially due to the failure to treat seizures, epilepsy carries high costs to health care providers, associated with clinical assessment, drug treatment and surgery (Allers et al., 2015). In the United States of America alone, the direct costs of epilepsy are estimated at $28 billion annually (Begley & Durgin, 2015). Given the direct and indirect effect of seizures on quality of life, preventing them is key to reducing the burden of epilepsy for patients (Berg et al., 2019).

Current treatment options for epilepsy are typically geared towards reducing seizure propensity. Anti-seizure drugs are the primary treatment for epilepsy, with over 20 drugs approved for use globally. Although anti-seizure drugs can reduce seizure frequency, at present, a third of patients remain refractory to treatment (Kwan & Brodie, 2000). Even for those who initially respond to treatment, the risk of breakthrough seizures is high, occurring in 34% of patients (Bonnett et al., 2017). Furthermore, current anti-seizure drugs cause highly undesirable side effects such as impaired memory and attention (Meador, 2002), due to their systemic targeting of entire receptor subclasses in the brain (Rogawski et al., 2016). Such suboptimal treatment options are reflective of our limited understanding of the specific cellular and synaptic mechanisms driving seizures in each patient. One alternative approach for patients who are unresponsive to anti-seizure drugs, is neuro-surgical intervention. This involves the removal of putative seizure-causing tissue, which leaves 52% of patients seizure free 5 years after surgery, and 47% after 10 years (de Tisi et al., 2011). Nonetheless, given the invasiveness of epilepsy surgery, the high risk of seizure recurrence post-operatively (Zhang et al., 2018), and the low proportion of patients who are eligible for surgery (Wiebe et al., 2001), improved therapeutic avenues are clearly required. To this end, improving our understanding of seizure genesis, will ultimately improve our ability to intervene.

In order to fully appreciate the pathogenesis of epileptic seizures, it is useful to first explore their epidemiology and symptomatology. Seizures are remarkably common, with 10% of people experiencing a seizure in their lifetime (Hauser & Beghi, 2008). In fact, seizures are a phenomenon that occurs in brains throughout the animal kingdom (Buckmaster et al., 2014; Jirsa et al., 2014; Podell, 1996). Seizures can also be triggered by a diversity of conditions, including sleep deprivation, stress, hypoxia, physical trauma, and chemical and electrical stimulation (Jett, 2012; Luttges & McGaugh, 1967; Nakken et al., 2005). This indicates that seizures are a common endpoint for disruptions in the homeostatic regulation of neuronal processes caused by developmental or acute perturbations.

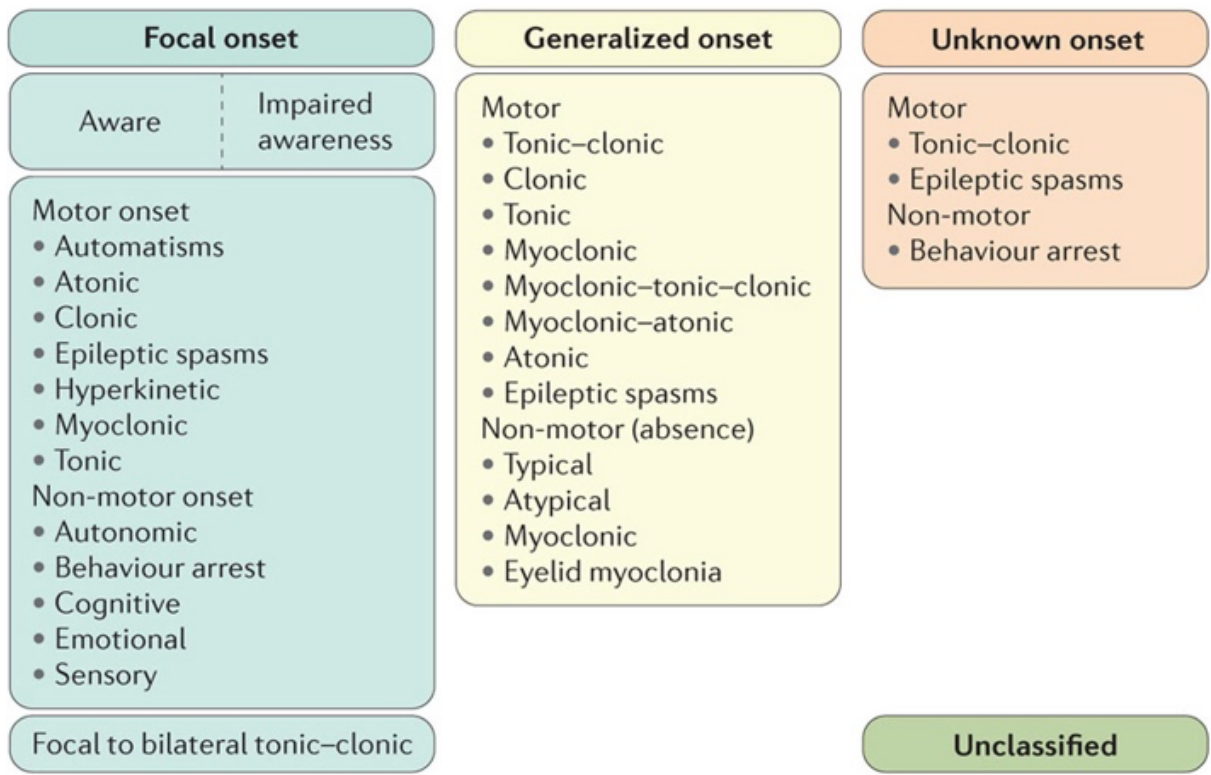

**Figure 1. 1  Seizure classification.**

Seizure classification scheme, according to the International League Against Epilepsy 2017 classification. Seizures are first classified by region of onset (teal = focal, yellow = generalised, orange = unknown). Seizures are further subdivided by presence or absence of awareness (see focal onset). Finally, seizures can be divided by the presence of motor or non-motor symptoms during the seizure. Image source: Devinsky et al., 2018.

Box 1.1 | **Seizure types**

**Absence seizures:** a sudden loss of awareness, often with blank staring (~10s). May be associated with changes to muscle tone (atypical, 5-30s), stiffening contractions in arms, neck and back (myoclonic, 10-60s) or myoclonic eyelid jerks (eyelid myoclonia, <6s).

**Automatism:** repetitive, coordinated movements that resemble voluntary motion, with or without awareness.

**Atonic:** a sudden loss of muscle tone, often resulting in falling or head dropping.

**Behaviour arrest:** reduced amplitude of ongoing motor activity, often causing a pause in activity.

**Clonic:** repeated stiffening and relaxing, appearing as rhythmic jerking (<60s).

**Cognitive:** altered cognition, including disrupted speech, understanding of language and impaired memory.

**Emotional:** changes to mood that were not present before seizure onset.

**Epileptic Spasms:** sudden muscle tension, lasting longer than myoclonic but shorter than tonic seizures (1-2s).

**Focal to bilateral tonic-clonic:** seizures that start focally, before spreading to both hemispheres as a tonic-clonic seizure.

**Hyperkinetic:** proximal limbs engage in irregular ballistic motion such as pedalling, thrusting and jumping.

**Myoclonic:** brief muscle jerks with rapid contractions (1-2s).

**Sensory:** alterate experience of touch, taste, hearing, vision or smell.

**Tonic:** stiffness in muscles, particularly in arms, legs or chest which often happen during sleep (~20s).

**Tonic-Clonic:** stiffness in muscles (tonic) followed by jerking movements in arms and legs (clonic), with a loss of awareness (60-120s).

Although seizures may represent a common abnormality emerging from distinct molecular mechanisms, the expression of seizure symptoms in epilepsy patients is highly diverse, as outlined in Box 1.1. In particular, seizures show variable i) region of onset, ii) degree of awareness, and iii) motor versus non-motor components (Sarmast et al., 2020), and are classified according to the expression of these three features (Figure 1.1). Regarding region of onset, seizures can be of i) focal onset, where seizure activity emerges in one or more brain region or a single hemisphere, ii) generalised onset, where seizure activity emerges across multiple brain regions in both hemispheres concurrently, or iii) unknown onset (Devinsky et al., 2018). Such symptom diversity is likely a reflection of the exact localisation of the abnormal neural activity during the seizure. For example, occipital cortical seizures often can cause visual disturbances, motor cortical seizures can cause motor onset seizures, and somatosensory cortical seizures can disrupt sensory experience (Stafstrom & Carmant, 2015). Given that seizures are common to brain networks exposed to distinct environmental and developmental perturbations, their emergence may depend on convergent neuronal network mechanisms, whose exact symptomatology depends on the brain regions involved. Clearly then, investigations into the neural and network basis of seizures are key to understanding and preventing them.

### 1.1.2 Epileptic Seizure Dynamics at the Macroscale

Epileptic seizures have long been conceptualised as a condition of aberrant brain dynamics. Central to this conceptualisation has been the use of electroencephalography (EEG), which measures voltages over macroscopic brain areas via electrodes typically placed on the scalp. EEG was invented in 1929 by German physicist Hans Berger, and 6 years later was used to record abnormal spikes in electrical potentials during epileptic seizures (Gibbs et al., 1935). Interestingly, Gibbs reported that clinical seizure presentation was preceded some 20 seconds by altered EEG waveforms. This demonstrated for the first time that atypical neuronal dynamics were central to the generation of seizure symptoms.

While EEG data remains central to descriptions of seizure dynamics and their underlying biological mechanisms (Zijlmans et al., 2012), identifying biological causes of EEG features is challenging. This is because EEG is an indirect measure of neuronal activity. EEG measures the build-up of electric potential in the extracellular space, caused by the flow of ions into and out of cells, driven by so-called current sinks and sources (Cohen, 2017). For example, the flow of positive ions into the cell across ion channels (current sinks) during excitatory postsynaptic potentials (Kirschstein & Köhling, 2009), leads to the build-up of negative charge outside of the cell and the formation of a dipole in the extracellular space (Baillet et al., 2001; Hallez et al., 2007). Furthermore, the parallel alignment of apical dendrites across neighbouring pyramidal cells favours the summation of these electric potentials in the extracellular space during synchronised postsynaptic depolarisation, which is sufficient to be seen on EEG (Cohen, 2017; Kirschstein & Köhling, 2009; Mitzdorf, 1985). Therefore, EEG coarse grains over individual neuron dynamics to capture averaged activity over superficial cortical areas. Conventional EEG can broadly be split into two types: i) scalp EEG, where electrical activity is recorded from the scalp, and ii) intracranial EEG, recorded from electrodes placed directly on the surface of the cortex, or implanted inside brain tissue. Despite the macroscopic nature of EEG recordings, such datasets have provided valuable insight into the biological causes of seizures (Jobst et al., 2020).

EEG studies have proven particularly useful in identifying specific dynamic signatures associated with epilepsy syndromes. Certain electrographic abnormalities are indicative of specific syndromes, such as generalised polyspikes in juvenile myoclonic epilepsy or generalised spike wave complexes in absence seizures (Noachtar & Rémi, 2009). Epileptiform EEG patterns with specific regions of onset can also indicate specific syndromes, such as frontal onset seizures in dorsolateral frontal lobe epilepsy (Adcock & Panayiotopoulos, 2012; H. Chen & Koubeissi, 2019; Foldvary et al., 2001). Such differing EEG presentations indicate divergent neuronal pathology. Specific epileptiform EEG dynamics therefore can help to guide hypotheses about underlying neuronal mechanisms, which can for example be tested using mechanistic models (Marten et al., 2009; Rodrigues et al., 2006).

EEG has also been useful for classifying seizures and identifying their general organising principles. Interestingly, despite the diversity of seizure types and epilepsy syndromes, seizures can generally be decomposed into three sequential features: i) an onset, ii) periods of abnormal spikes, and spike and wave events, and iii) an offset (Jirsa et al., 2014). Furthermore, evidence from focal seizure dynamics suggests only 7 distinct seizure onset patterns, defined by variable waveform amplitude and spike frequency (Perucca et al., 2014). Although such attempts to qualitatively classify seizures by EEG recordings alone in small samples are unlikely to capture the full diversity of seizure dynamics, such studies nonetheless underscore the surprising stereotypy of macroscopic seizure features. Interestingly, such stereotyped seizure dynamics may be shared across brain scales in different species, strongly suggesting convergent network mechanisms (Figure 1.2). However, given the coarse nature of EEG dynamics, to understand whether stereotyped macroscale dynamics are caused by similarly stereotyped neuronal activity (F. H. Lopes da Silva et al., 2003), we require access to single neuron dynamics.

Furthermore, EEG studies have been central in the conceptualisation of epileptic seizures as a network phenomenon – emerging from the collective behaviour of multiple brain areas (Richardson, 2012). While the spatial resolution of EEG is limited, its ability to cover distant brain areas with multiple electrodes enables the concurrent recording of multiple macroscale brain regions during seizures. Such approaches have helped to uncover key network properties at the macroscale – intracranial EEG recordings indicate that seizure onset occurs as a reduction in correlation between brain areas, suggesting reduced coupling (Kramer et al., 2008; Schindler et al., 2007). Conversely, macroscale networks are highly coupled during seizure offset (Guye al., 2006; Schindler et al., 2007). Interestingly, both cortical and subcortical macroscopic structures may engage in correlated behaviour during seizures (Norden & Blumenfeld, 2002), such as thalamic and temporal lobe areas in temporal lobe epilepsy (Guye et al., 2006). EEG evidence therefore implicates macroscale network pathology in seizure propagation, rather than a regional, focal core. Such approaches have also been key in formalising seizures as a dynamic network phenomenon, with changes to macroscale connectivity evolving over time, giving rise to variable subnetworks during seizure propagation (Schroeder et al., 2020). In this way, EEG has helped to reframe questions relating to seizure emergence in terms of collective behaviours in interconnected systems, which opens the door to rich theoretical frameworks for the analysis of emergent properties (Kelso et al., 1988).

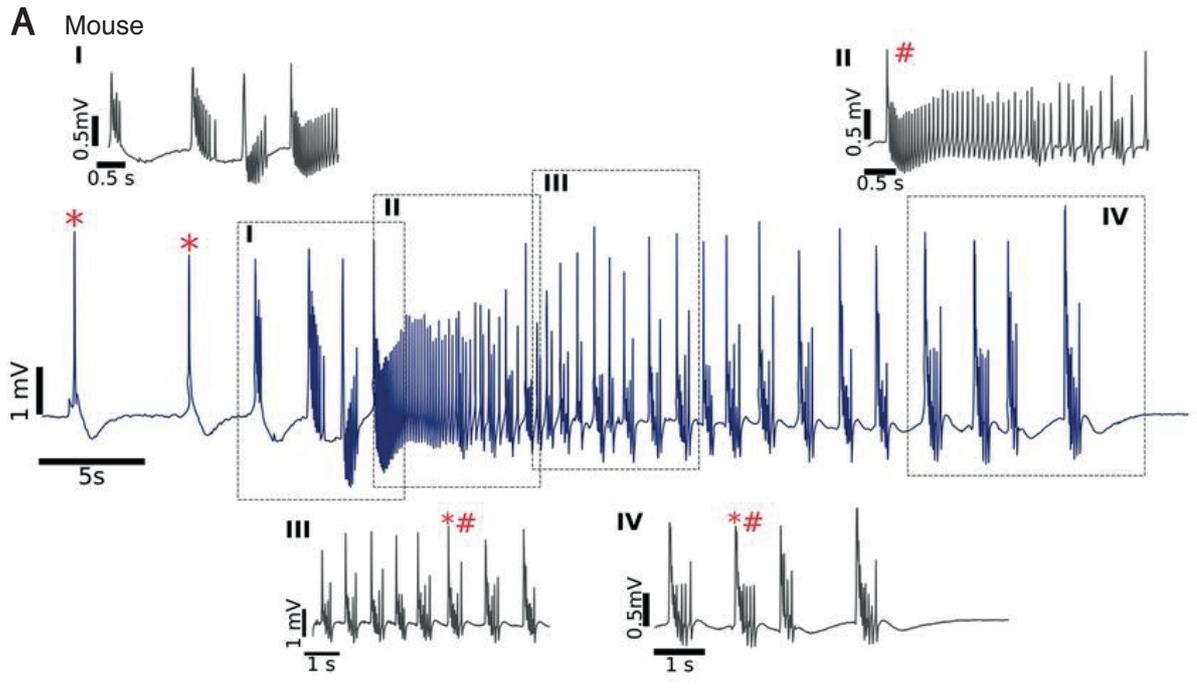

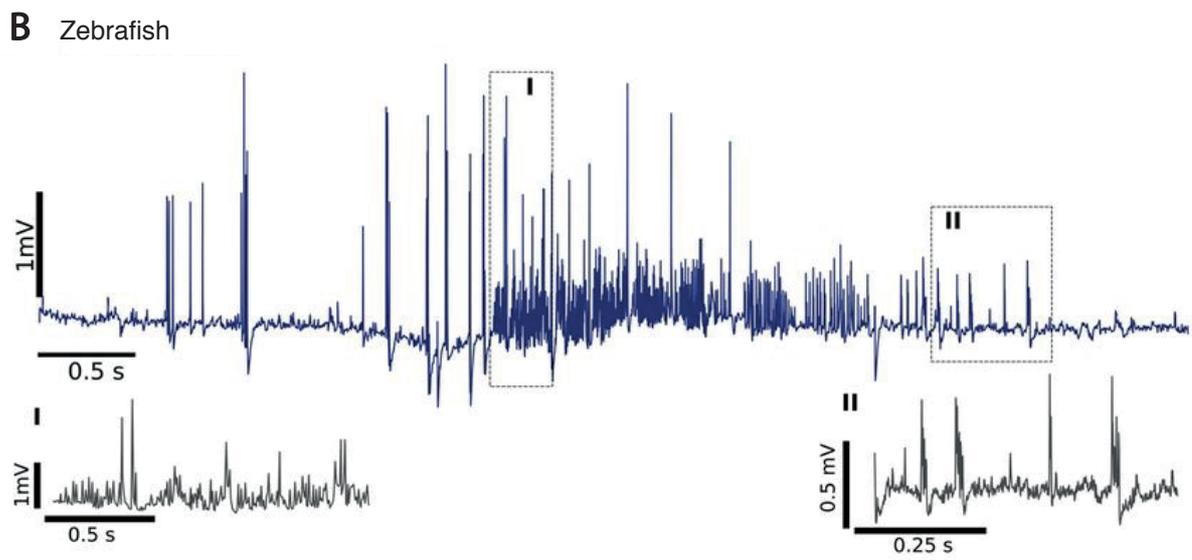

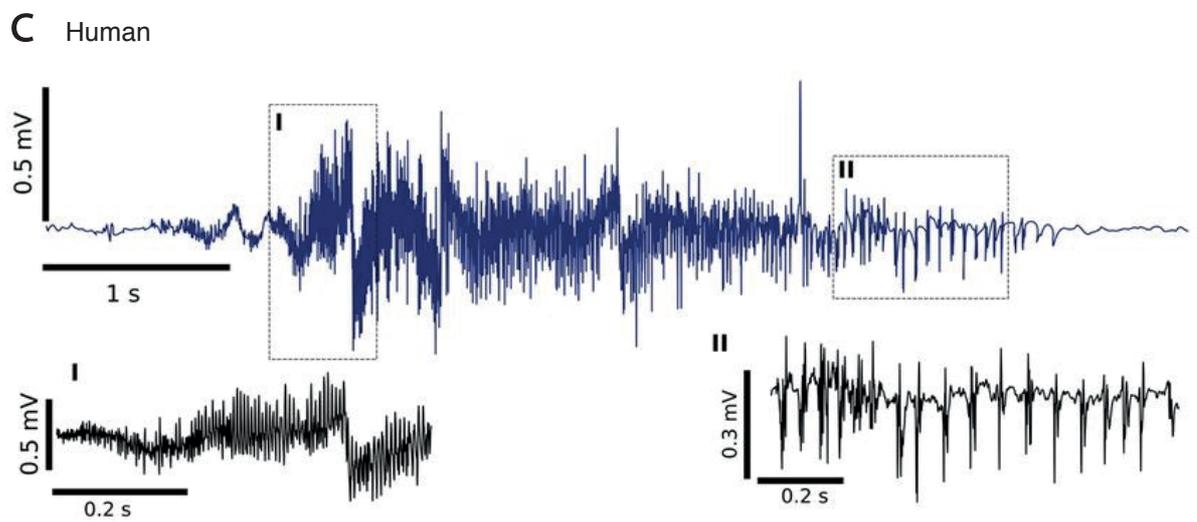

**Figure 1. 2   Seizures patterns may be conserved across species.**

(A) Mouse hippocampal local field potential recordings during chemically-induced seizures over entire recording period (blue) and with zoomed in periods shown (black). Two generic patterns are distinguished during the seizure: high frequency spikes (#) and spike and wave events (*). (B) Zebrafish forebrain local field potential recordings during hyperthermia-induced seizures, exhibiting high frequency spikes (I) and spike and wave events (II). (C) Spontaneous seizures recorded on intracranial EEG in an epilepsy patient, exhibiting high frequency spikes (I) and spike and wave events. Note that all species exhibit slowing down of electrographic patterns during seizure offset. Image source: Jirsa et al., 2014.

While such network approaches are indeed useful for understanding seizures, macroscale recording techniques lack the spatial resolution required to access the distinct functional units and connections through which signal transmission occurs in brain networks – individual neurons and their synapses. Therefore, EEG may obscure the relevant network changes driving seizure genesis (Butts, 2009). For example, EEG can only detect approximately more than 6 cm$^2$ of correlated cortical activity (Kirschstein & Köhling, 2009). Given that seizures could emerge due to small instabilities in the network (C. Wang et al., 2015), and that electric potentials decay with distance (Buzsáki et al., 2012), EEG likely misses key neuronal activity patterns driving seizure genesis. In fact, evidence from 2-photon imaging of evoked seizures in mice indicates the presence of synchronous populations of neurons at the microscale driving seizure onset, whose activity does not register with EEG (Figure 1.3C) (Wenzel, Hamm, et al., 2019). Therefore, access to individual neuron recordings would grant the spatial resolution necessary to localise the neuronal populations that drive the network from normal to seizure states.

Importantly, EEG will also obscure the underlying heterogeneity of neuronal activity during seizure genesis and propagation. In particular, while EEG data indicates a high correlation state during seizures, evidence from microscale recordings suggests heterogeneous firing patterns. For example, a mouse model of absence seizures exhibits asynchronous dynamics, with a large subpopulation reducing activity during the seizure (Figure 1.3A-B) (Meyer et al., 2018). This is by no means a quirk of animal models, with highly heterogeneous firing patterns observed in human

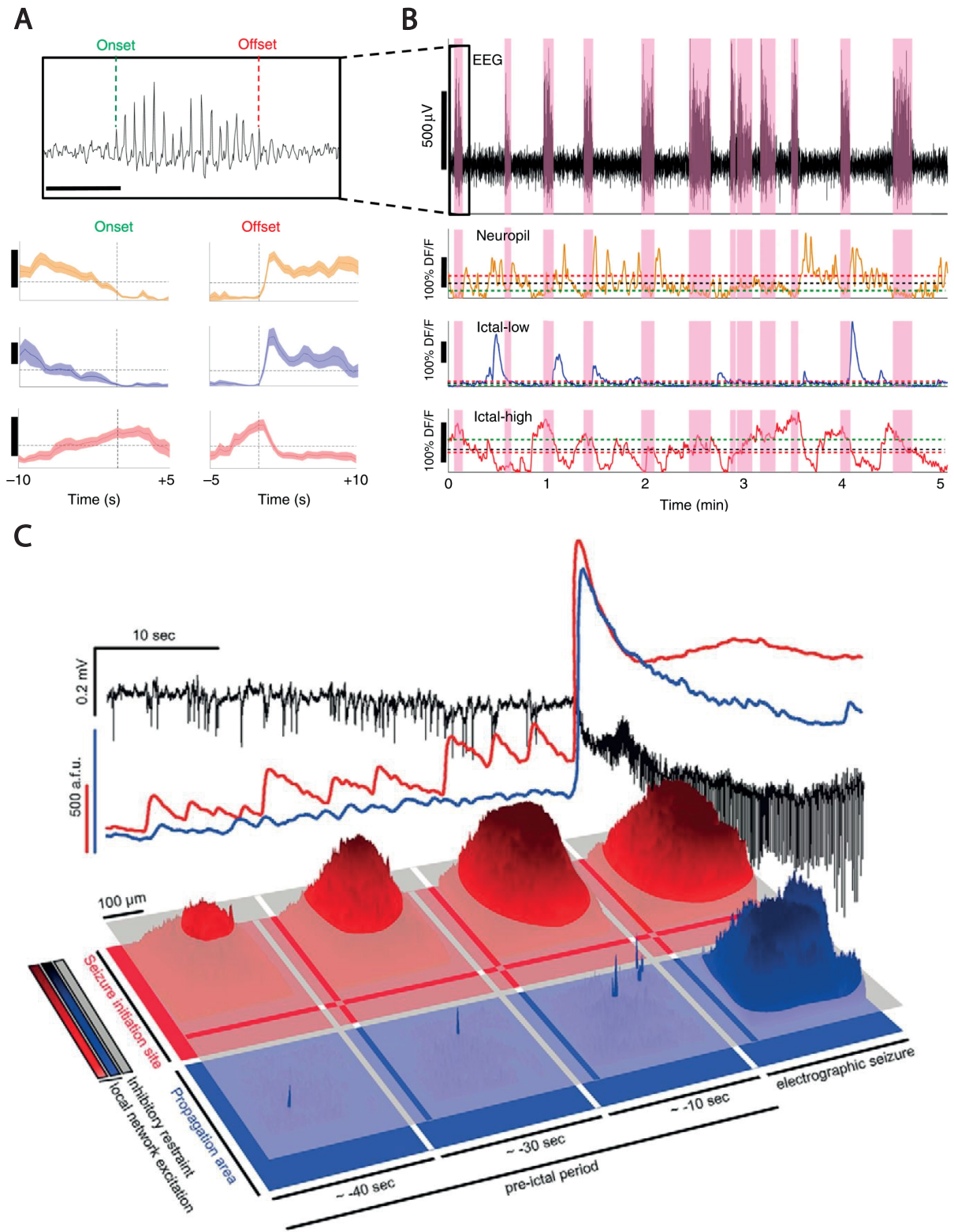

**Figure 1. 3    Single neuron seizure dynamics not captured by macroscale recording techniques.**

(A-B) Concurrent single neuron 2-photon calcium imaging and EEG in visual cortex of a mouse model of absence seizures. Images source: Meyer et al., 2018. (A, top) EEG recordings during seizures, with onset and offset defined as first and last spike of the seizures. (A, bottom) Mean calcium fluorescence (mean ± standard error) during seizure onset and offset for an exemplar neuropil region (orange), ictal-low neuron which reduces firing during the seizure (blue), and ictal-high neuron which increases firing during the seizure (red). (B, top) EEG recording with seizure periods highlighted in pink. (B, bottom) Single trial calcium fluorescence in an exemplar area of neuropil, and an ictal-low and ictal-high region. Notice how ictal-low dynamics are not evident on EEG. (C) Seizure progression model in a mouse model of focal seizures. Image source: Wenzel et al., 2019. (C, top) 40 second pre-ictal local-field potential (LFP) recording (black) followed by a 20 second period of seizure, denoted by epileptiform downward peaks. Corresponding 2-photon calcium recordings from ictal-core (red) and surrounding regions (blue). Notice how ictal-core seizure peaks precede LFP abnormalities by up to 20 seconds. (C, bottom) Population averages in calcium fluorescence in ictal core (red) and surrounding regions (blue), with grey layer representing local inhibitory restraint. Notice how LFP signal of the seizure occurs as single neuron activity propagates to surrounding areas.

recordings as well (Truccolo et al., 2011). This indicates the differential involvement of distinct neuronal subpopulations in seizures, perhaps driven by cell subtype identity (Devinsky et al., 2018). Given that EEG is biased towards correlated depolarisation of pyramidal neurons, such distinctions would not be captured with EEG yet are crucial to understanding the cooperative effect of neuronal subpopulations in the emergence of seizures (Figure 1.3B).

Finally, neuronal dynamics are fundamentally non-linear, but coarse grained measures such as EEG can linearise the underlying dynamics (Nozari et al., 2020). In particular, neuronal activity scales as a non-linear function of its inputs, due to the presence of voltage gated ion channels, fluctuations in membrane potential and neuromodulatory effects, among others. Given that seizure propagation likely relies on such non-linear interactions (Lehnertz, 2008), EEG dynamics cannot account for the underlying input-output relationships between neurons that enable seizure propagation in the network. For example, evidence from a mouse model of focal epilepsy demonstrates the state-dependent involvement of interneurons in seizure dynamics, expressing either anti or pro-ictal contributions (Magloire et al., 2019). Such non-linear dependencies would be smoothed over at the macroscale, thus obscuring key neuronal interactions driving seizure dynamics. Instead having access to single neuron activity can provide the spatial resolution, cell-type identity and non-linear dynamics necessary to understand the network mechanisms that drive seizures.

### 1.1.3 Epileptic Seizure Dynamics at the Microscale

The development of microscale recording techniques applied to animal models and human patients has provided insight into the biological mechanisms driving epileptic seizures. Such approaches range from invasive microelectrode arrays placed into the cortex of epilepsy patients to record spikes, to 2-photon imaging of intracellular calcium transients in animal models of seizures (Devinsky et al., 2018). These approaches conventionally record single cell dynamics or single spike activity, in restricted populations of neurons. A key benefit of such microscale approaches is the ability to bridge disparate scales, by recording EEG and cellular dynamics concurrently. For example, the presence of high frequency oscillations on EEG might indicate highly ictogenic regions (Allen et al., 1992; Fisher et al., 1992; Jacobs et al., 2010; Jirsch et al., 2006), but their exact pathophysiology has been unclear from EEG recordings (Zijlmans et al., 2012). Concurrent microelectrode array recordings have demonstrated that these oscillations are driven by burst spiking in neuronal populations whose activity is synchronised across millimetres of cortex, resulting in oscillations in the 80-150 Hz range (Weiss et al., 2013). Interestingly, these bursts occurred in a rhythmic pattern, phase-locked to lower frequency ictal EEG rhythms. Therefore, the concurrent use of microscale and macroscale recording techniques can help to bridge the gap between network pathology at different scales – here abnormal excessive neuronal coupling driving burst firing in neuronal networks entrains neuronal populations at larger scales to oscillate in phase. In this way microscale network pathology can drive and influence macroscale network pathology during seizures. Therefore, studies with access to single cell activity can provide sufficient spatial resolution to resolve the multi-scale neuronal dynamics underlying seizures.

Cellular recording techniques have also been useful in uncovering the differential contribution of cell subtypes in seizure genesis. For example, evidence from microelectrode array recordings indicate a role for GABA interneurons in limiting seizure spread, exhibiting phase locked spikes to the excitatory ictal rhythm (Schevon et al., 2012; Trevelyan & Schevon, 2013). Furthermore, microscale approaches in animal models can utilise molecular and optogenetic tools, which have been used to uncover the roles of specific receptors and cell subtypes in seizure genesis. Such approaches have demonstrated the paradoxically pro-ictal effect of GABA interneurons – increased interneuron activity precedes seizure onset in mouse hippocampal interneurons (Miri et al., 2018), while optogenetic activation of interneurons in somatosensory cortex causes seizure genesis through post-inhibitory rebound excitation (Chang et al., 2018). Furthermore, over-expression of the K+/Cl-cotransporter KCC2 in excitatory neurons prevents the pro-ictal effect GABA, thus implicating excessive intracellular Cl- and consequential depolarisation of the GABA reversal potential in seizures (Magloire et al., 2019). Similar efforts have also demonstrated that extracellular GABA concentration may synchronise interneuron network firing via tonic GABA conductance, which can further perpetuate epileptiform bursts (Magloire et al., 2022). Therefore, such molecular tools and microscale recording techniques afford sufficient neuronal specificity to test causal hypotheses relating to cellular and synaptic mechanisms of seizure genesis. Of further interest, 2-photon imaging of larval zebrafish during induced-generalised seizures, has demonstrated a role for astrocytic extracellular glutamate release and gap-junction communication in driving seizures (Diaz Verdugo et al., 2019). Taken together, such microscale recording techniques can disentangle the differential involvement of distinct cell subtypes in seizure genesis, which is not possible with EEG.

Microscale studies have also helped to uncover neuronal mechanisms relating to seizure propagation. For example, microelectrode array recordings have shown that the migrating edge of the seizure causes travelling excitatory waves into contiguous areas supporting recruitment (Smith et al., 2016). This indicates the local recruitment of neighbouring neurons to the seizure, which has implications for targeted seizure termination. However, mesoscale recordings of neuronal activity suggest that seizures also propagate to non-contiguous locations which share a high degree of functional and, presumably structural connectivity (Rossi et al., 2017). This indicates that seizure propagation respects the structural bounds imposed on it by synaptic connectivity. In fact, evidence from microscale neuronal recordings suggests that inhibitory mechanisms can support both contiguous and non-contiguous propagation, with a loss of local inhibition to the migrating edge of the seizure, or a loss of inhibition in connected but distant brain regions, respectively (Liou et al., 2018). However, this underscores a major limitation of conventional microscale studies. Due to the presence of long-range connections in the brain, a neuron's dynamics will be influenced by distantly connected neurons (Knösche & Tittgemeyer, 2011). Given the limited spatial coverage of microscale recordings techniques, many of these inputs will be outside the recorded region of interest. Therefore, to fully appreciate the destabilising influence of incoming activity onto network components during a seizure we require access to single neuron dynamics across the entire network.

As previously discussed, epileptic seizures are a network phenomenon, involving interactions between distributed brain areas (Figure 1.4A) (Bartolomei et al., 2017; Richardson, 2012). While most microscale studies record from the epileptogenic focus, there is evidence from macroscale recordings which suggest that changes to correlation that precede seizure onset may emerge away from the epileptogenic zone (Le Van Quyen et al., 2005). Even in patients with focal lesions, distant brain areas can engage in seizure activity, as shown via increased propensity to generate rapid discharges at seizure onset on EEG (Figure 1.4B-C) (Aubert et al., 2009; Sevy et al., 2014; Talairach & Bancaud, 1966). Similarly, evidence from structural imaging studies suggest that focal onset seizures can actually exhibit widespread structural macroscopic network changes beyond the focus, as in focal to bilateral tonic-clonic seizures (Sinha et al., 2021). This strongly indicates that seizures, even those caused by localised structural abnormalities, can emerge out of the interactions between macroscopic brain regions throughout the entire brain network (Davis et al., 2021; Spencer, 2002). Importantly, generalised seizures by definition emerge out of macroscale brain networks, covering areas far greater than conventional cellular recording approaches will support (Tangwiriyasakul et al., 2018). Therefore, conventional microscale recordings subsample the epileptogenic network, which can give rise to false inferences about network properties and dynamics (Priesemann et al., 2014). In this way, to accurately identify the general network principles which support the onset of epileptiform activity, we require access to whole brain networks. In fact, evidence also implicates widespread structural and functional network abnormalities in epilepsy patients which correlate with cognitive impairments,

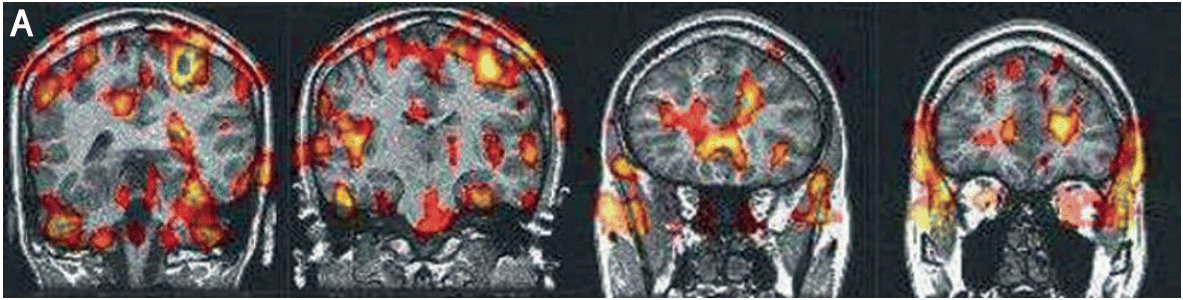

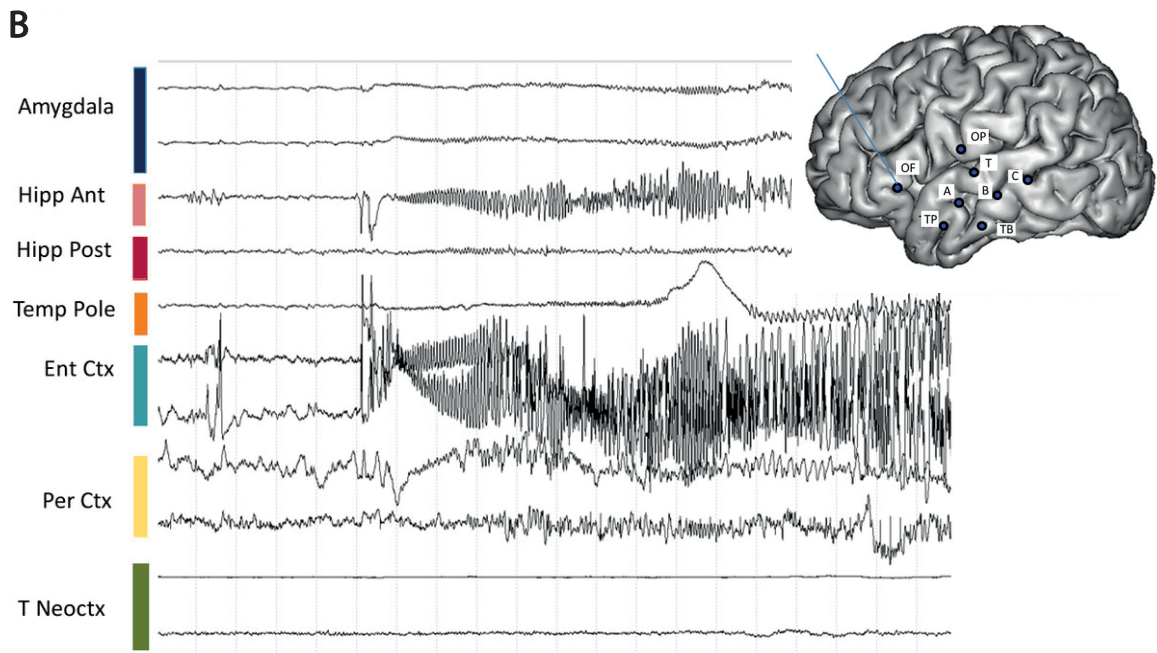

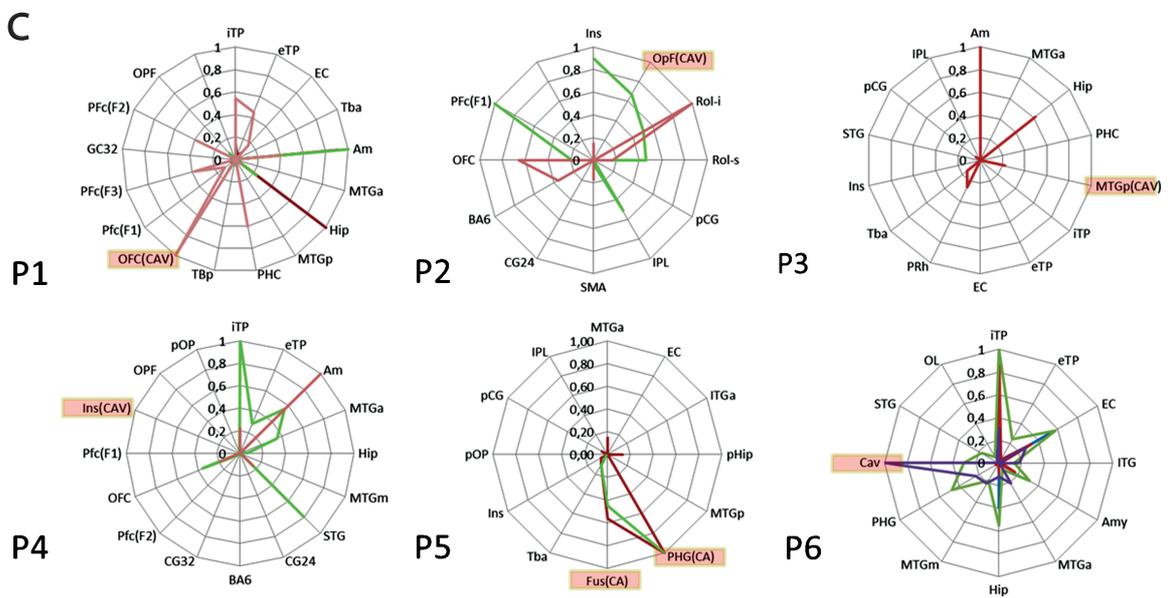

**Figure 1. 4  Seizures emerge across distributed brain areas as a network-wide phenomenon.**

(A) Ictal single-photon emission computed tomography image showing areas of hyper-perfusion during seizure activity. This patient exhibited widespread hyper-perfusion across multiple brain areas belonging to the superior parietal and medial frontal networks. Image source: Spencer, 2002. (B) Intracranial EEG recordings in a patient with temporal lobe epilepsy, showing exact location of electrodes over the brain (top right). Intracranial EEG recordings for each electrode shown during a temporal lobe seizure. The seizure starts with rapid discharges affecting the amygdala (electrode A), the anterior hippocampus (Hipp Ant, electrode B) and posterior hippocampus (Hipp Post, electrode C), and the entorhinal cortex (Ent Ctx, electrode TB). Note that seizure activity appears concurrently across different macroscopic brain areas. Image source Bartolomei et al., 2017. (C) Epileptogenicity index values, a measure of contribution to the seizure ranging from 0 to 1, shown for six patients (P1-6) across different brain areas. Acronyms correspond to localisation of depth electrodes. Acronyms highlighted in pink are areas of focal lesions causing epileptic seizures. Note how there is limited concordance between epileptogenicity and location of focal lesion, suggesting widespread network pathology. Image source: Sevy et al., 2014.

suggesting chronic network alterations in patients (O'Muircheartaigh et al., 2011; Vollmar et al., 2011). This suggests that whole brain network alterations may support both the propensity to express seizures and the impaired cognitive function that co-occurs with epilepsy. Therefore, viewing seizures as the expression of dynamic pathology in structurally altered whole brain networks may provide novel insight into the mechanisms driving seizure genesis and cognitive dysfunction.

Traditional microscale recording techniques offer insight into the cellular and circuit mechanisms of seizures, providing great spatial resolution at the expense of network coverage. Conversely, conventional macroscale recording techniques grant insight into the coarse network dynamics unfolding during seizures, providing wide spatial coverage at the expense of spatial resolution. With this in mind, a model system which can grant access to single cell dynamics across the entire brain network is necessary to understand the emergence of seizures in the brain. Such a model system could help to bridge the gap between microscale and macroscale networks in seizures. In particular, conventional network studies of seizures have traditionally been limited to a single scale, largely due to the technical difficulty of recording multi-scale network dynamics. Therefore, a model system which can access networks at multiple scales, would begin to provide an understanding of the relationship between network dynamics at different scales, which may be central for driving epileptic seizures.

*1.1.4   The Larval Zebrafish – a Tool for Whole Brain Cellular Resolution Imaging*

In the last decade, the larval zebrafish has emerged as a leading vertebrate model to investigate brain dynamics in health (Ahrens et al., 2013) and disease (Diaz Verdugo et al., 2019a; Liao et al., 2019; Marquez-Legorreta et al., 2022). The larval zebrafish has a complex, yet optically accessible central nervous system which has played a key role in investigating the pathophysiology of many human neurodevelopmental disorders ranging from autism to epilepsy (Baraban et al., 2013; Gan-Or et al., 2016), in part due to its high genetic homology to mammals (Shehwana & Konu, 2019). Importantly, its external fertilisation makes the zebrafish embryo suitable for observations of pathological processes from the earliest stages of development (Laughlin et al., 2008; Lawson & Weinstein, 2002). At one week post fertilisation the zebrafish brain contains only ~100,000 neurons (about one millionth of the human brain) and is < 1 mm$^3$ in volume, yet displays canonical vertebrate brain anatomical divisions with high functional homology across brain regions to mammalian counterparts (Figure 1.5A) (Kalueff et al., 2014; Mathuru & Jesuthasan, 2013). Importantly, the larval form of the zebrafish is optically transparent, enabling direct visualisation of the whole brain with single-cell resolution (Ahrens et al., 2013; Antinucci & Hindges, 2016; Kibat et al., 2016). Furthermore, fluorescent transgenic calcium reporters such as GCaMP and RGECO (Akerboom et al., 2012; Dreosti & Lagnado, 2011; Walker et al., 2013) now enable live imaging of single neuron function during behaviour or at rest (Bianco & Engert, 2015; Cong et al., 2017). Such imaging approaches can capture concurrent single neuron dynamics across the whole brain (Figure 1.5B) (Ponce-Alvarez et al., 2018). Thus, optical recordings of neuronal function in larval zebrafish allow for unprecedented spatial resolution (compared to EEG recordings), and brain coverage (compared to mammalian imaging techniques) – ideal for the investigation of seizure dynamics, where single cell interactions across entire brain networks drive epileptic seizures.

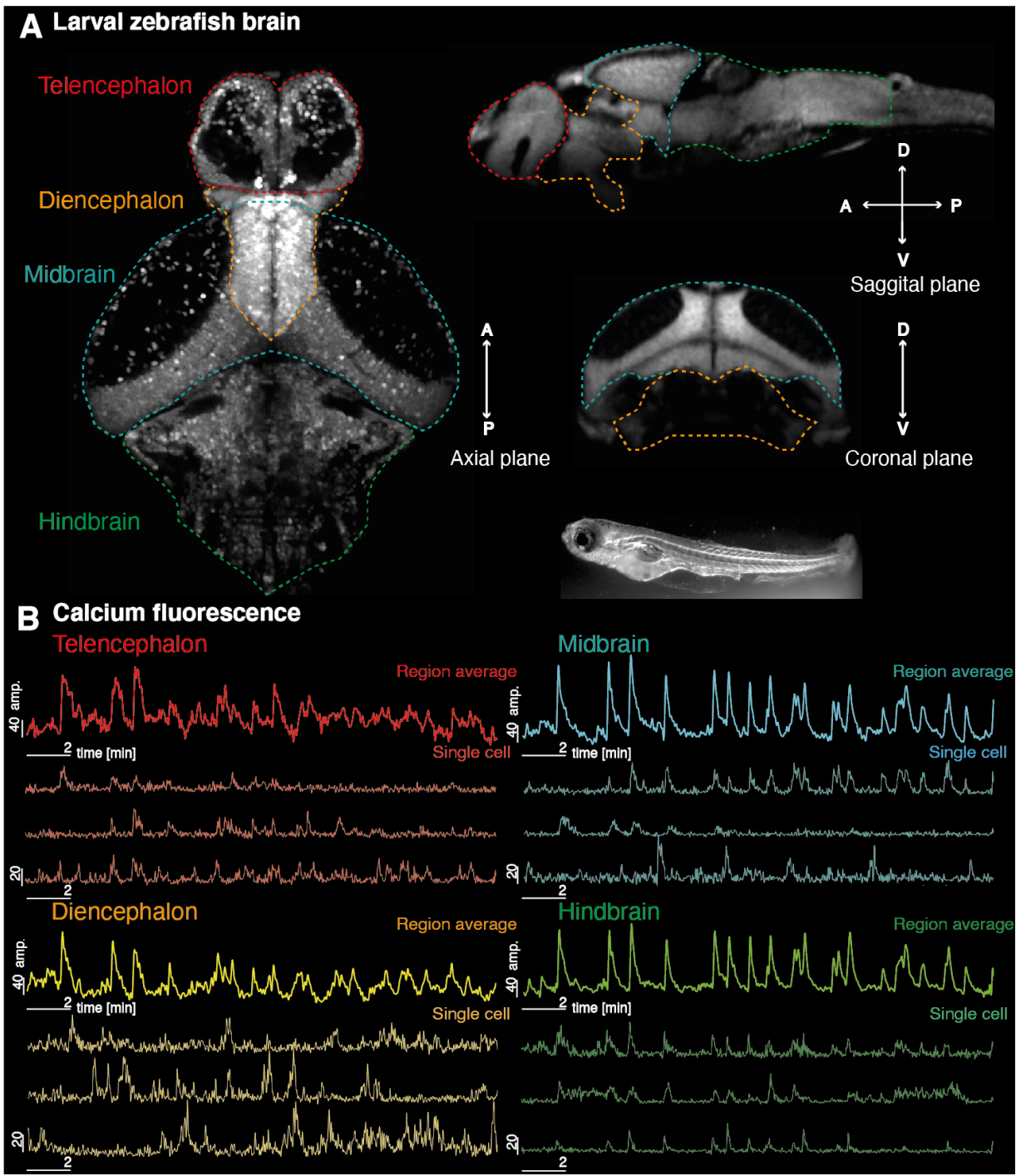

**Figure 1. 5   Whole brain calcium imaging of the larval zebrafish at cellular resolution**

(A) A single axial plane from in vivo 2 photon imaging of a larval zebrafish brain, shown at 7 days post fertilisation on the right. Sagittal and coronal views of a reference image [http://www.zbbrowser.com] are shown to illustrate the 3D location of different brain areas. A zebrafish larva (not to scale) is shown at the bottom of the panel. (B) Example time series data are shown for regional averages of gross anatomical brain regions, as well as example cells identified in each area. Image source: Burrows et al., 2020.

Since an initial publication by Baraban et al., in 2005, interest in zebrafish seizure models have grown substantially (Baraban et al., 2005). This study established protocols to monitor seizure behaviours and electrographic activity in larval zebrafish exposed to pentylenetetrazole (PTZ), a common pro-ictal agent. Multi-channel EEG recordings of adult zebrafish during PTZ-induced seizures have shown qualitative homologies to human EEG recordings, with ictal discharge frequency broadly comparable to scalp EEG in patients (Cho et al., 2017). In electrophysiological recordings obtained from the forebrain or midbrain of agarose-immobilized larvae, clear examples of recurrent brief inter-ictal-like and long duration multi-spike ictal-like discharges are observed with these acute drug manipulations or even hyperthermia (Hunt et al., 2012). Interestingly, calcium imaging approaches have also demonstrated that induced-seizure dynamics in larval zebrafish exhibit a fundamental feature of epileptic seizures – the emergence of sudden state transitions from normal to abnormal brain states (Figure 1.6) (Diaz Verdugo et al., 2019a).  Furthermore, other studies have demonstrated the robustness of epileptic seizure-like activity in larval and adult zebrafish in response to a multitude of different pro-ictal drugs (Alfaro et al., 2011; Mussulini et al., 2013; Winter et al., 2017).

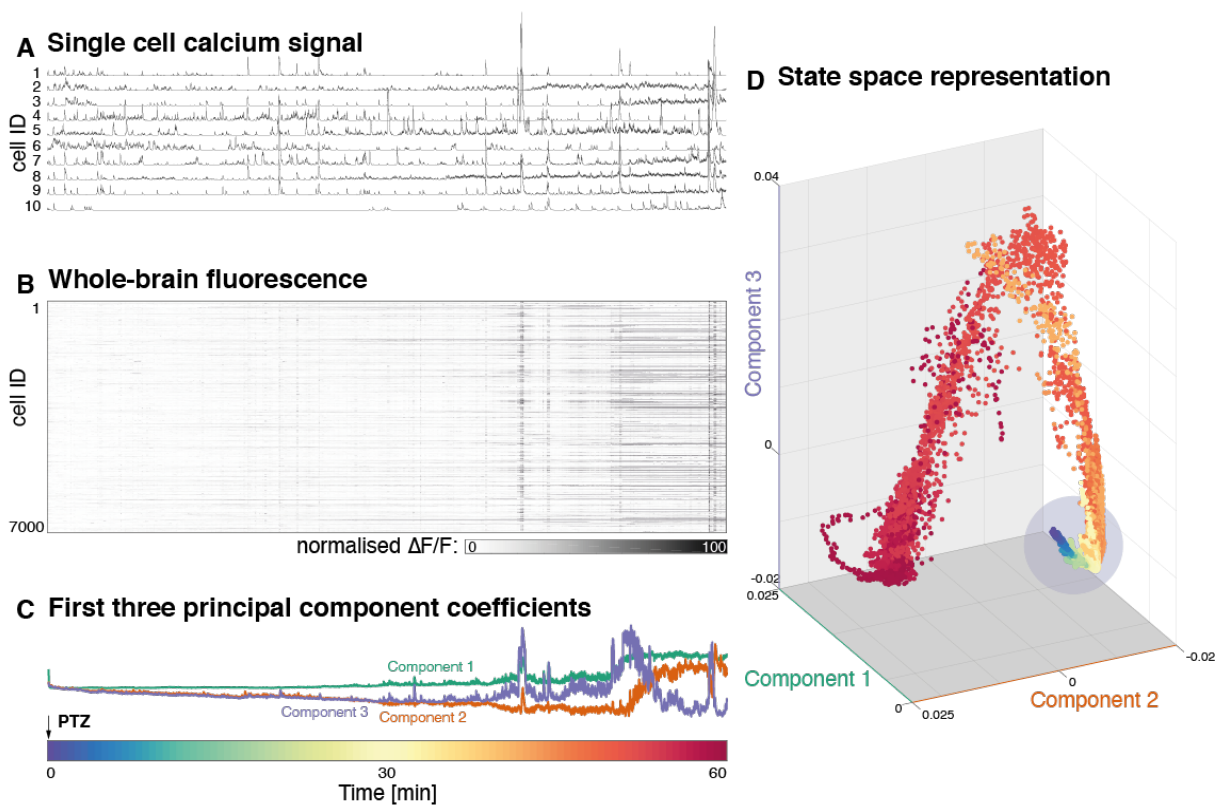

**Figure 1. 6   Whole brain state transition after PTZ exposure.**

(A) Example normalised fluorescence traces are shown for individual neurons for 1 h after addition of PTZ to the bath at time point 0 showing an increase of amplitude and frequency of neuronal firing events. (B) Firing of all >7000 active cells captured in this recording. (C) First three principal components over time varying fluorescence matrix shown in (B). These indicate both a persistent drift in components 2, and 3, as well as drastic changes in the loading of all components towards the end of the recording. (D) The same data is shown as a state space plot. Whilst most of the data points exist in a restricted region/state (indicated by blue circle), the late seizure is characterised by very different activity distribution readily apparent in this low-dimensional projection as points outside of the earlier range. (Time scale for all figures shown as colour bar at the bottom of the figure). Image source: Burrows et al., 2020.

At a behavioural level, bath exposure to pro-ictal drugs leads to a general increase in swim activity and, at higher drug concentrations, whole-body convulsive movements that are rarely observed in untreated wild-type zebrafish. Importantly, drug-induced seizures in larval zebrafish are sensitive to anti-seizure drugs such as benzodiazepines and valproate (Afrikanova et al., 2013; Berghmans et al., 2007) matching their long-established effectiveness in treating drug-induced seizures in rodents (Armand et al., 1998; Georgiev et al., 1991). As such, zebrafish models can recapitulate several key behavioural, electrophysiological and pharmacological features of seizures, thus offering a robust model system for investigating seizure dynamics.

The unique combination of optical accessibility, small brain size, genetic and neuroanatomical homology to mammals, and face validity as a model of seizures, makes the larval zebrafish a key model system for studying whole brain seizure dynamics at single cell resolution. The concurrent emergence of cutting edge imaging techniques supporting the acquisition of such datasets (Ahrens et al., 2013), means that systems neuroscientists now have the tools to unpick the single neuron mechanisms driving brain wide seizure dynamics. Key to making sense of such high dimensional datasets is the use of network modelling approaches, which use mathematical tools to study emergent properties in interconnected systems (Bassett et al., 2018). In fact, several studies have already begun to apply network modelling to make sense of cellular resolution whole brain dynamics in health (Betzel, 2020) and disease (Marquez-Legorreta et al., 2022).

Such approaches can help to link alterations in single neuron interactions with changes to global dynamics, such as a seizure (Figure 1.7). For example, given that whole brain imaging approaches have demonstrated the emergence of paroxysmal global state transitions following PTZ exposure (Figure 1.6C,D), network representations can be used to identify the cellular interactions that drive seizure transitions. Evidence from network representations of brain wide calcium imaging suggest that global seizure transitions are characterised by an increase in long-range pairwise neuronal connectivity, suggesting a loss of the spatial constraints that normally bias connectivity towards nearby neurons (Figure 1.7) (Diaz Verdugo et al., 2019a; J. Liu & Baraban, 2019). Such network descriptions of single neuron activity can be complemented by generative modelling approaches (Hadjiabadi et al., 2021) and cellular resolution structural connectivity atlases (Kunst et al., 2019) to uncover the structural and functional network properties that render the system prone to seizure transitions. For example, some evidence suggests that specific cellular resolution network motifs increase the likelihood of seizure transitions, such as super-hub neurons, single neurons that are both highly interconnected and whose neighbourhood contains many feedforward connections (Hadjiabadi et al., 2021). In this way, network models of single cell, whole brain dynamics may help to bridge the gap between microscale network interactions and macroscale transitions during seizures.

Cellular resolution investigations into global dynamics can also help to shed light on the microscale mechanisms that enable brain-wide seizure propagation, such as during a generalised seizure. Key to these approaches is the use of network models which grant full access to system parameters allowing the testing of specific network mechanisms – these can be used to study the interaction of neuronal parameters such as excitability, connectivity and synapse strength, to identify key parameter regimes that support global seizure propagation. Furthermore, one can also utilise data-driven modelling approaches to uncover the directional flow of activity through whole brain networks (Sugihara et al., 2012; Takahashi et al., 2021). Important to

## A Calcium traces during PTZ-induced seizures

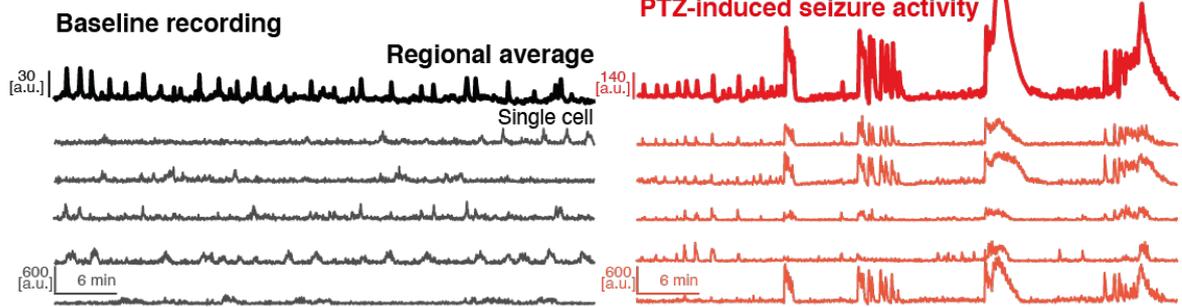

## B Single cell map of whole-brain activity

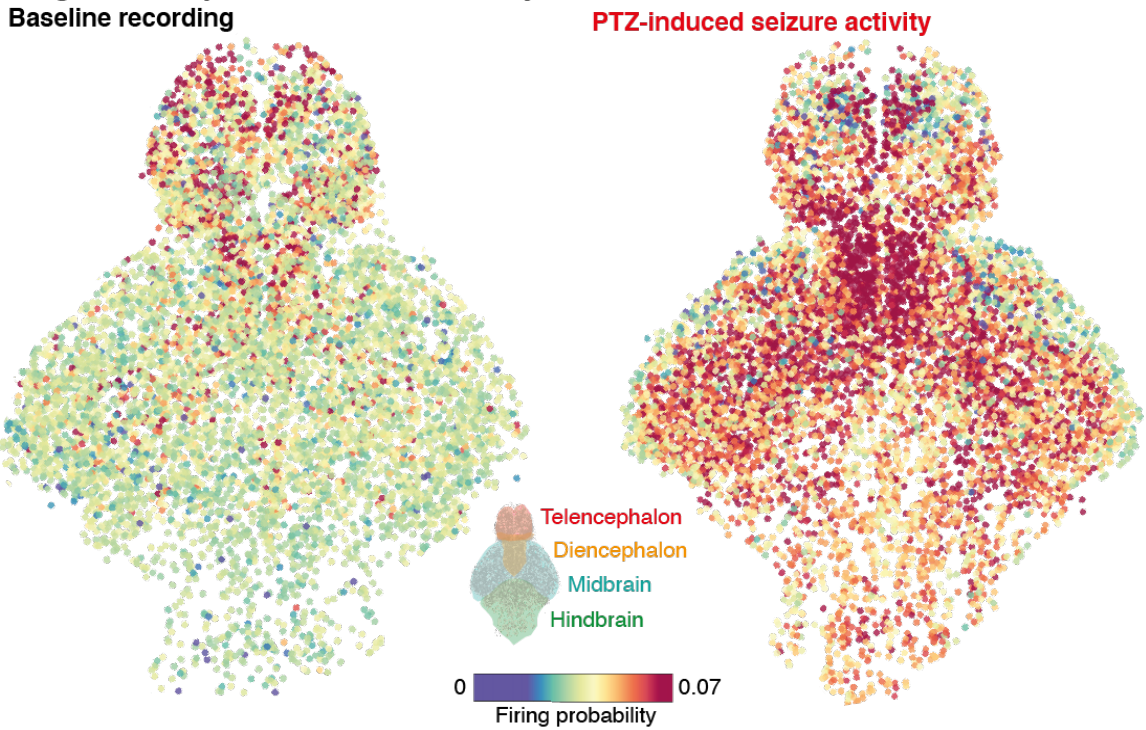

## C Functional connectivity

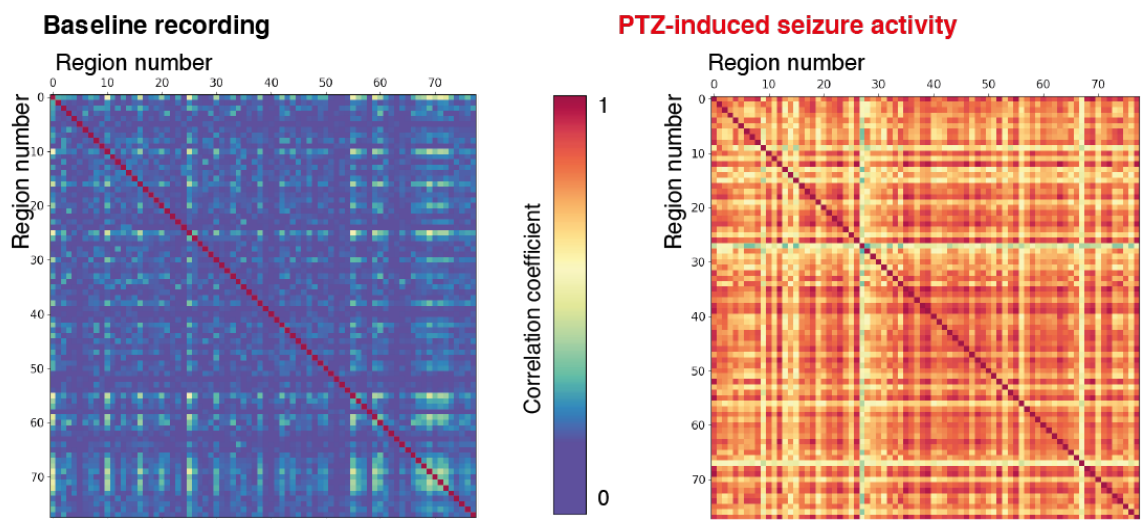

**Figure 1. 7   Increases in whole-brain neuronal synchrony during PTZ-induced seizures.**

(A) Example normalised fluorescence traces are shown as whole brain mean traces and examples of individual neurons at baseline (black) and PTZ-induced seizure (red) conditions (note the difference in amplitude scale). (B) Map of individual cell firing probabilities across the two conditions demonstrates an increase in firing probability in the PTZ-induced seizure condition, with regional heterogeneity across the larval fish brain. (C) Between-region functional connectivity shown as a correlation matrix. Segmented cells were registered to a zebrafish brain atlas, clustered by coarse brain regions (~70 areas) and their traces were averaged together, to measure correlation between major brain regions across baseline and PTZ conditions. Image source: Burrows et al., 2020.

these approaches is the emergence of larval zebrafish brain atlases that enable the identification of neuron subtypes by brain area or gene expression patterns (Gupta et al., 2018; Marquart et al., 2017), and cell specific fluorescent reporters, which together support the mapping of single neuron propagation pathways by cell type. For example, dual genetic labelling of excitatory and inhibitory neurons in the larval zebrafish brain during PTZ-induced seizures, suggest that seizures propagate from areas of high excitatory-inhibitory ratios to areas with balanced ratios, suggesting a loss of inhibitory function in seizure propagation areas (Niemeyer et al., 2022). In this way, network models of cellular resolution brain wide seizure dynamics can help to link single neuron patterns of activity during seizures with whole brain dynamical states, in seizure genesis and propagation.

Studies which leverage brain-wide cellular resolution imaging to understand seizures are limited to a few studies (Diaz Verdugo et al., 2019a; Hadjiabadi et al., 2021; Niemeyer et al., 2022). While they demonstrate the utility of the larval zebrafish for linking microscale alterations to global brain states, some key questions remain unanswered. In particular, it is currently unclear how healthy whole brain networks remain stable – they must balance the need to be sufficiently sensitive to inputs, allowing activity to propagate through the network and activate required downstream targets, but insufficiently sensitive for inputs to destabilise network dynamics as a seizure (Beggs & Plenz, 2003). In this context, I use the term stability to mean the ability of a network to maintain its dynamics within a functional operating range.

Identifying dynamical regimes and neuronal parameters that regulate network stability in whole brain networks may therefore be crucial for understanding why instability occurs as a seizure, and how it spreads throughout the network. Key to answering these questions, is the use of network models which provide complete access to system parameters and components which cannot be afforded in real systems (Bassett et al., 2018).

## 1.2    Modelling Whole Brain Network Dynamics of Seizures

### 1.2.1   Network Models

Neurons communicate with one another via neurotransmitter release across synapses. A neuron's propensity to spike at any given moment is largely defined by its spike history, the incoming presynaptic signals and its corresponding synaptic weights. Therefore, the propagation of neuronal activity through the network, often referred to as cascades (Beggs & Plenz, 2003), will be constrained by the architecture of axonal and dendritic connections between neurons in the brain (Ju et al., 2020). This pattern of interconnected cells gives rise to a complex network whose dynamics, stability and function can be viewed as emergent properties (Bassett & Sporns, 2017), that arise from its individual components and connections – neurons and synapses (Butts, 2009).

In this way, viewing epileptic seizures as an emergent property of neuronal networks has helped to reframe questions relating to seizure genesis in the context of complex systems theory, including dynamical systems and graph theory, which provides mathematical frameworks for quantifying collective behaviour in large networks (Bullmore & Sporns, 2009; Siegenfeld & Bar-Yam, 2020). In particular, network models can be constructed from empirical data, which represent individual components of the system as functional units called nodes, and their interactions as connections called edges (Figure 1.8A). Once constructed, these models may be used to quantify the presence certain network features that influence cascades, information transfer and synchrony, or may be combined with differential equations

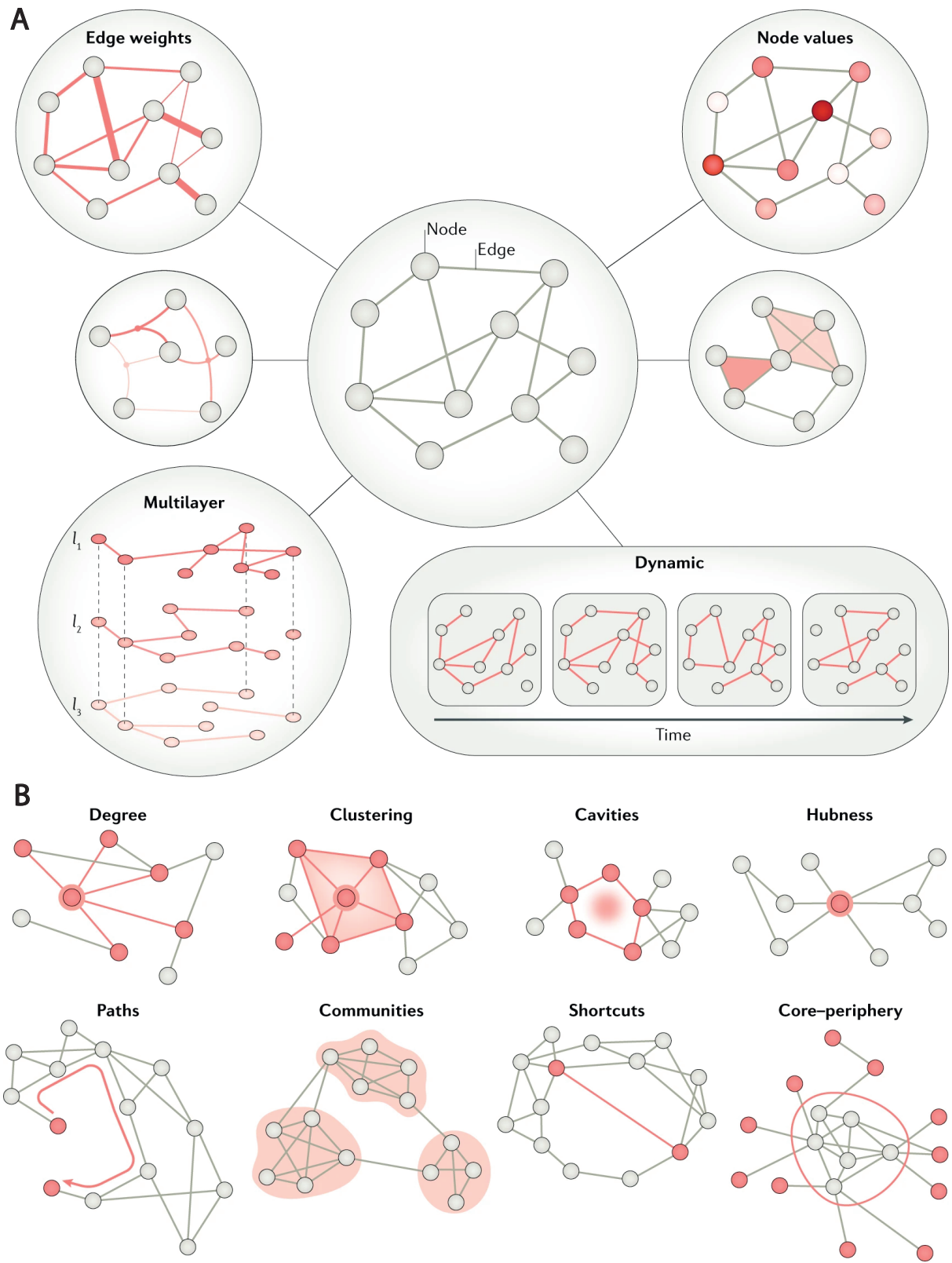

**Figure 1. 8   Network models.**

(A) Basic network models consist of functional units (nodes) and their connections (edges). More sophisticated network models can add strengths to each edge (edge weights) or node (node values), and explicit rules defining the dynamics of node/edge strengths over time.

Multilayer networks may be used to capture different sets of networks, that share some nodes and edges. (B) Common graph theory measures may be estimated based on the combination of nodes and edges. Degree = the number edges that a given node forms in the network. Clustering = the propensity for a nodes' neighbours to connect with each other. Cavities = the absence of edges. Hubness = a node with a high degree relative to other nodes in the network. Paths = a sequence of edges which join one node to another, capturing the potential for information transfer. Communities = groups of densely interconnected nodes. Shortcuts = an edge which connects distant regions of the graph directly, a marker of global efficiency for information transfer. Core periphery = a network with dense connections in the core and sparse connection in the periphery, facilitating information sharing to distributed areas. Image source: Bassett & Sporns, 2017.

that define node or edge evolution to generate network dynamics *in silico* (Figure 1.8) (Bassett et al., 2018). Such approaches have been applied extensively to macroscale seizure data, for example to generate personalised models of seizure propagation (Y. Wang et al., 2019) and to predict surgical outcome (Goodfellow et al., 2016; Y. Wang et al., 2020). Network models may therefore be exploited to study the single cell network properties that regulate stability and drive instability in whole brain networks.

### 1.2.2  Building an Appropriate Model of Whole Brain Seizure Dynamics

Mathematical descriptions of networks act as a conceptual bridge linking observed phenomena with their underlying mechanisms. For example, time-varying seizure phenomena such as high frequency oscillations or spike-wave-discharges can be described using coupled differential equations that capture the evolution of a system's state in time (F. Lopes da Silva et al., 2003). Model based descriptions such as these, offer a range of benefits not afforded to purely experimental approaches, such as the ability to i) provide record all system parameters of interest, ii) experimentally manipulate any system parameter, iii) make experimentally testable predictions about the system, and iv) test the validity of theoretical principles in generating empirical data. The specifics of these models can range from biophysically plausible models representing synaptic physiology (Case & Soltesz, 2011; Papadopoulou et al., 2017), to more abstract network models restricted to capturing key interactions between macroscopic brain areas (Goodfellow et al., 2016; Khambhati et al., 2016). Given the importance of studying seizure dynamics across the whole brain (Richardson, 2012), and the multi-scale nature of the brain in which microscale collective behaviours influence global network dynamics (Deco et al., 2014), what is the appropriate level of detail required to model whole brain seizure activity?

Modelling aims to represent a phenomenon which occurs in real systems, in terms of some of its quantifiable features, in order to gain insight into how it occurs. Key to building models is the process of abstraction, that is simplifying or ignoring details of the real system at some scale in order to identify general principles that emerge at other scales (Bassett et al., 2018). Abstraction is partly a pragmatic approach to building models as gaining access to a systems' dynamics across all scales is impractical and usually impossible. Perhaps more importantly, abstraction is necessary, as a hypothetical all-inclusive model would be too complex to provide sufficient explanatory power to illustrate how a phenomenon emerges (Richardson, 2012). How we choose to abstract in our model is therefore key to the insight we can glean from it. When choosing how to abstract, one must decide what spatiotemporal scales and data features are of interest. To illustrate the relevant levels of abstraction for building whole brain models, we use a general framework for considering abstractions in network models, constructed by Bassett & Sporns.

Bassett & Sporns define a model space consisting of three dimensions. Each dimension captures a continuum of model types extending from one conceptual pole to its opposite, such that the position of a model in space corresponds to a specific combination of each of the three dimensions. Each dimension captures the extent to which a model is i) data versus theory driven, ii) approximating structural versus functional relationships, or iii) built using coarse versus elementary approximations of system units and their connections (Figure 1.9). By visualising models in this way, we can start to understand how a certain level of abstraction will define the relationship a model has with the underlying phenomenon it is trying to model (Bassett et al., 2018).

In the first dimension, data-driven models represent the observed data in a network, without making any assumptions regarding how the data was generated. For example, structural brain networks simply represent the density of white matter pathways (edges) between brain areas (nodes) in a graph, as measured with diffusion tensor imaging (Luat & Chugani, 2008) (Figure 1.9B). Conversely, theory-driven models draw from theoretical principles to define node or edge values, for example by explicitly defining differential equations governing the evolution of oscillatory nodes (Breakspear et al., 2010). In the second dimension, structural models represent physically defined components of a system, where nodes and edges have exact physical counterparts in the real system (Figure 1.9C). For example, nodes and edges may represent neurons and their synaptic connections, such as in the C. elegans structural connectome (White et al., 1986). Conversely, functional models represent features which have no exact physical analogue, such that nodes and edges capture abstract components and relationships. For example, nodes may represent concepts, while edges may capture functional relationships of synchrony or coherence between time series (Schroeder et al., 2020). In the third dimension, elementary models set out to represent discrete, non-overlapping units as network nodes, and their irreducible relationships as edges (Figure 1.9D) (Butts, 2009). Key to defining elementary models is to understand the scale at which interactions may occur – for example while a neuronal network model connected by synapses may be further granularized into proteins and nucleotides, the physical sequestering of these molecules within each neuron may preclude the necessity to further granularize beyond neurons and synapses to capture emergent network dynamics. Conversely, coarse models focus on simplified descriptions of ensembles of elementary units, representing nodes and edges as averages of smaller components. Coarse-grained models may coarse grain explicitly, such as by defining mean field models of neuronal populations (Rosch et al., 2018), or implicitly by representing connectivity between EEG channels over brain areas.

Given the diversity of scales at which collective behaviours might conceivably influence whole brain seizure dynamics, choosing a level of abstraction will depend on the specific question at hand and the datasets available. The vast majority of models of whole brain seizure dynamics represent nodes and edges as coarse-grained approximations. In the next two sections, I will discuss the pros and cons of different levels of abstraction for tackling the question of network stability in brain wide cellular resolution networks, to suggest a move towards elementary models.

**A**

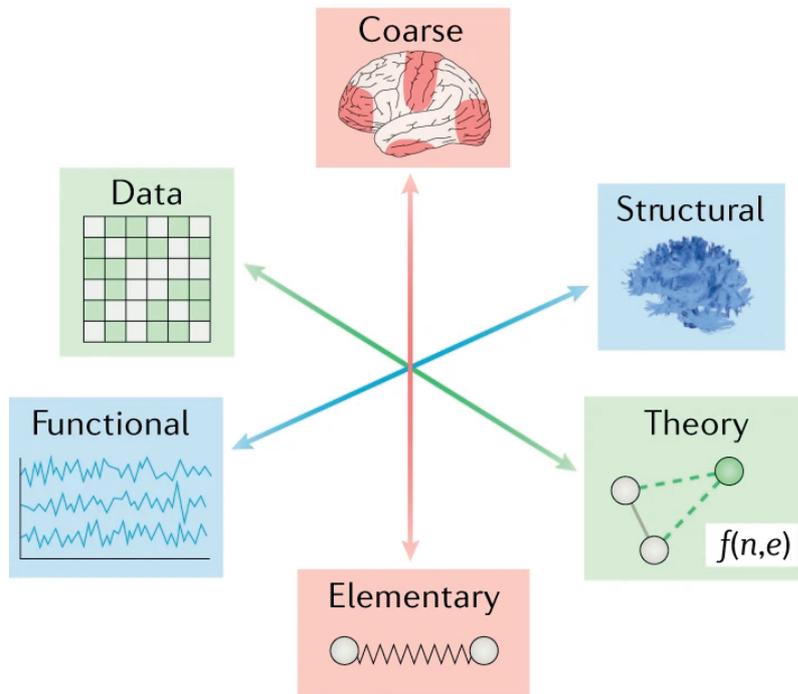

**B**

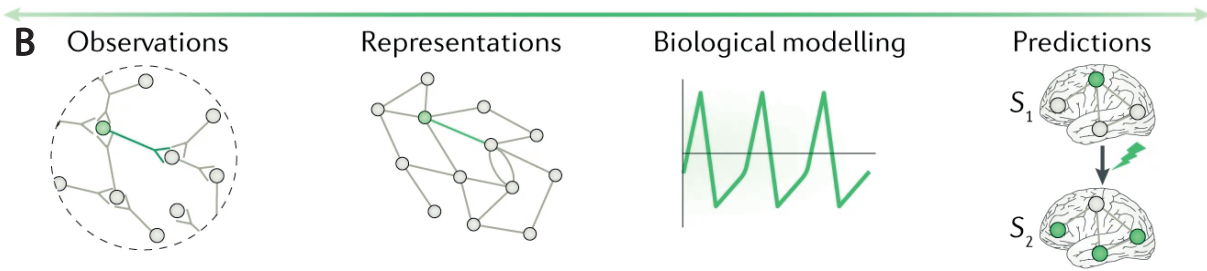

Observations    Representations    Biological modelling    Predictions

**C**

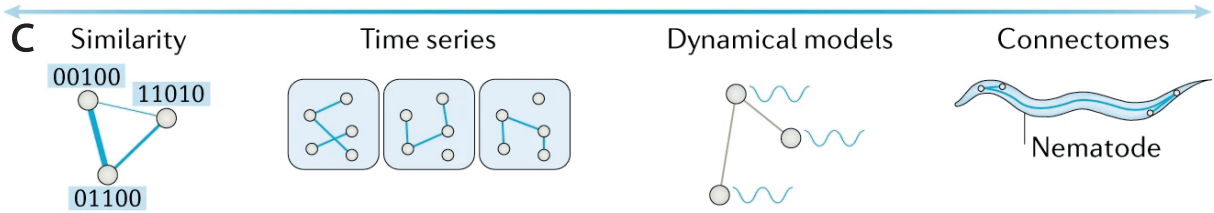

Similarity    Time series    Dynamical models    Connectomes

**D**

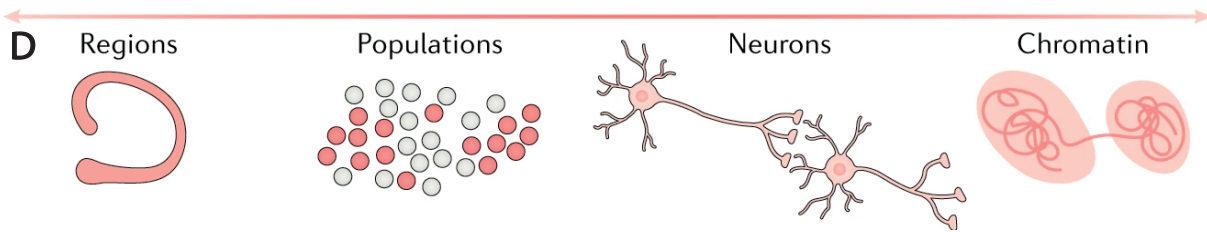

Regions    Populations    Neurons    Chromatin

**Figure 1. 9   Three dimensions defining network model types.**

(A) Bassett & Sporns present a conceptual framework for defining levels of abstraction in network model types, proposing three key dimensions. (B) Data to theory driven models: extending from empirical observations to theoretical assumptions which define nodes and edges. (C) Structural to functional models: extending from nodes and edges with exact physical counterparts, to those with more abstract counterparts. (D) Coarse to elementary models: extending from averaged ensembles of areas or connections, to irreducible, distinct forms. Image source: Bassett & Sporns, 2017.

*1.2.3   Coarse-Grained Models of Whole Brain Seizure Dynamics*

The majority of models of whole brain seizure dynamics live at one end of Bassett and Sporns' model space – coarse-grained approximations. While inferences from such models are limited to questions relating to coarse relationships, these approaches have been invaluable at uncovering the macroscale network principles underlying seizures. Coarse-grained whole brain seizure models include data-driven network models, which construct graph representations from magnetic resonance imaging (MRI) and EEG data modalities (Bartolomei et al., 2017; Richardson, 2012). In such approaches, nodes are coarsely defined as macroscopic brain areas which contain millions of neurons. Similarly, edges capture either functional coherence between coarse-grained nodes (Harding et al., 2012), or macroscopic white matter tracts which represent averaged trajectories over millions of axons (Luat & Chugani, 2008).

Such models have been particularly useful in identifying the macroscale network principles that support seizure propagation. For example, coarse-grained functional models of the epileptic brain have been constructed, in which nodes represent coarse areas and edges represent coherence between functional MRI signals. These approaches have implicated the thalamus as a key hub node (Figure 1.8B) important for the generalisation of seizures – patients with focal to bilateral tonic-clonic seizures exhibit high thalamo-cortical connectivity (He et al., 2020), while the thalamus as a node exhibits high degree and betweenness centrality in the network (Caciagli et al., 2020). Such coarse-functional network representations can therefore uncover altered cortical and subcortical network interactions which might destabilise whole brain dynamics in specific patient populations. However, for these properties to be captured by data-driven coarse models, they must be invariant to spatial averaging and sufficiently stationary over the recording period. Data-driven approaches have also been combined with theoretical principles, for example from network control theory, to predict the destabilising influence of each node on the network (Khambhati et al., 2016), and to identify optimal intervention strategies (Scheid et al., 2021). Given that current interventions to re-stabilise epileptic networks target coarse brain areas, such as neuromodulation or surgical intervention (Ryvlin et al., 2021), coarse-grained models may provide predictive power regarding the epileptogenicity of broad areas at the appropriate scale for seizure intervention. However, this does not preclude the possibility that targeted microscale interventions could offer higher precision and efficacy in re-stabilising network dynamics (Hadjiabadi et al., 2021).

Another major class of coarse-grained whole brain seizure models are theory-driven network models – these apply theoretical assumptions about node dynamics in the form of differential equations, and theory or data-driven edge values to generate network dynamics *in silico*. In such models, nodes conventionally represent averaged neuronal populations known as neural masses (Jirsa et al., 2017; Papadopoulou et al., 2017), while edges typically capture averaged synaptic strengths. One such approach that has been highly influential are mean field models, which emphasize global properties of ensembles of neurons with explicitly defined dynamics. Such models capture the evolution of variables relating to neuronal activity averaged over many neurons (Buice & Chow, 2013), allowing parameters relating to every neuron to be replaced by a lower dimensional set of population variables (Bick et al., 2020). An example of such a model is the Wilson-Cowan model which considers homogeneous populations of recurrently connected excitatory and inhibitory neurons, whose dynamics are defined by mean refractory periods, mean spike thresholds and mean synaptic weights across the network (Wilson & Cowan, 1972) (Figure 1.10A). Similar approaches include the classical Kuramoto model, in which neuronal populations are represented as a network of coupled oscillators, where the macroscopic state of the system is defined by the mean coherence across all oscillators (Bick et al., 2020; Cabral et al., 2011).

Theory-driven coarse and structural models of whole brain seizure dynamics have been particularly useful in identifying population mechanisms driving the onset of seizure state transitions. In particular, neural masses connected via ensemble synapses give rise to a key feature of seizures – the emergence of a qualitative change in system dynamics, known as a bifurcation (Figure 1.6) (Breakspear et al., 2006; Karoly et al., 2018; Y. Wang et al., 2012). For example, Wilson-Cowan models can elegantly capture the emergence of bifurcations when changing internal system parameters – coupled ensemble dynamics can predict transitions from a stable to an oscillatory regime, or the emergence of a bistable regime in which periodic and stable solutions both exist (Figure 1.10). Importantly, such bifurcations can be mapped onto underlying system parameters which control the state transitions. This has linked seizure onset to alterations in excitatory input and population recovery properties (Shusterman & Troy, 2008). In this way, coarse-structural models can help to identify ensemble neuronal parameter changes which may drive the emergence of seizures. However, linking such population parameter changes back to their underlying synaptic and microcircuit properties in heterogeneous neuronal populations is non-trivial.

Coarse-structural models have also been used to identify the dynamical and ensemble neuronal mechanisms that give rise to distinct EEG dynamics for different seizure types. In particular, mean-fields of coupled cortical, thalamo-cortical and thalamic reticular populations have been constructed to identify parameter regimes in which specific seizure bifurcations might arise (Robinson et al., 2002). Such approaches have shown that absence seizure EEG patterns can be captured by a Hopf bifurcation, in which a stable fixed point is lost and a periodic solution emerges – such periodic dynamics are caused by periodic switches between positive and negative feedback in thalamo-cortical loops (Breakspear et al., 2006). Interestingly, tonic-clonic EEG dynamics may emerge from different dynamical mechanisms, with the emergence of a bistable regime due to increases in cortico-thalamic coupling which results in large amplitude oscillations (Breakspear et al., 2006). Similar models, which also include inhibition over a range of timescales have implicated GABAb-mediated inhibitory ensemble potentials in supporting transitions between stable and periodic oscillations in such bistable regimes (da Silva et al., 2003; F. Lopes da Silva et al., 2003). In this way, mean field models have identified both parameter regimes that cause network instability and population interactions that drive emergent seizure dynamics. Such approaches may be further extended, by combining the non-linear dynamics of neural masses enabling different bifurcation types, with personalised structural connectivity data – this can allow for seizure intervention predictions for patients (Jirsa et al., 2017).

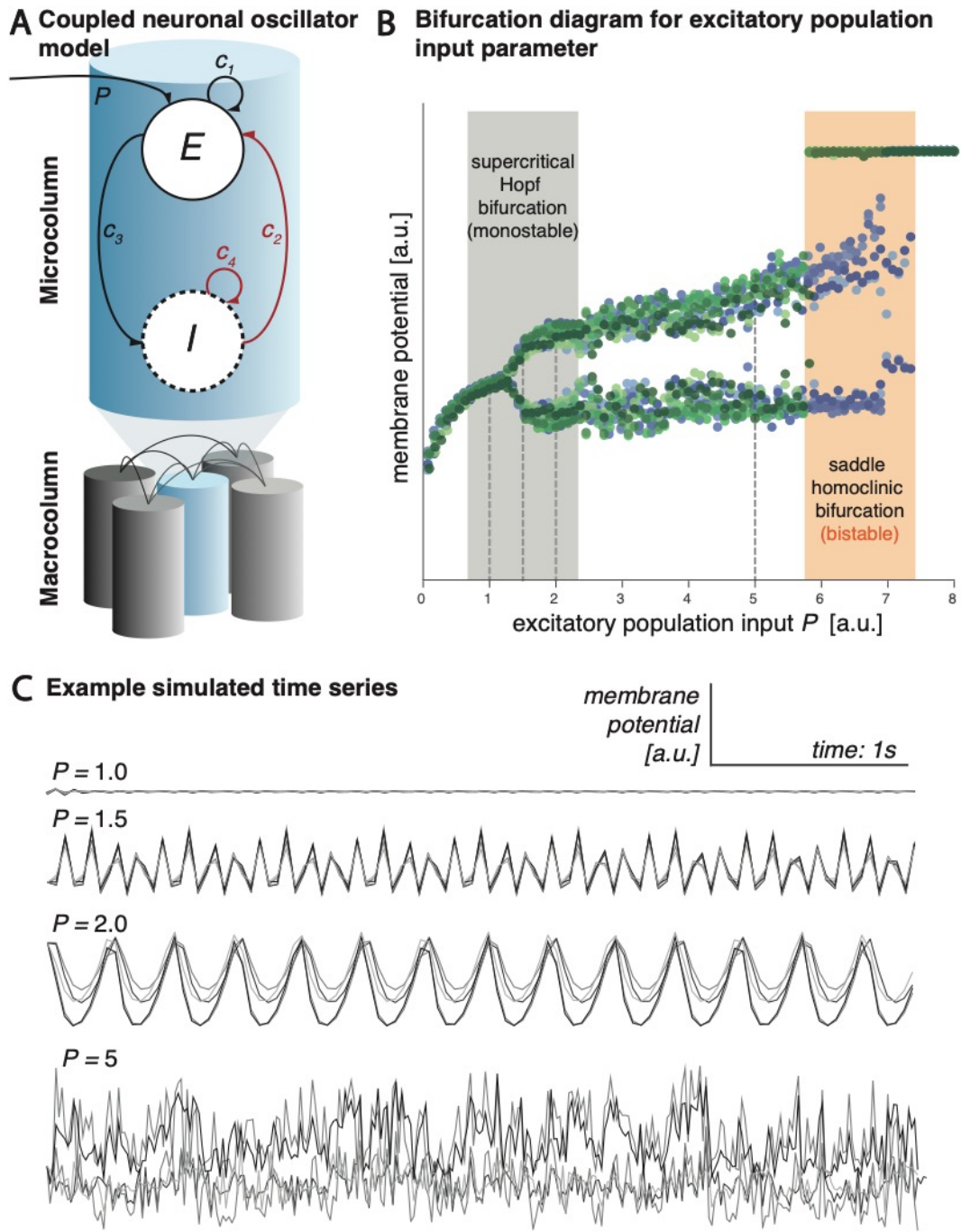

**Figure 1. 10 Bifurcation behaviour using coarse grained coupled oscillators.**

(A) Multiple Wilson-Cowan type neural mass models of excitatory (E) and inhibitory (I) populations were heterogeneously coupled and simulated at different values of excitatory population input P. $c_1$ = excitatory to excitatory synaptic weights. $c_2$ = inhibitory to excitatory synaptic weights. $c_3$ = excitatory to inhibitory synaptic weights. $c_4$ = inhibitory to inhibitory synaptic weights. Synaptic weights capture the average number of synapses per cell, which is homogeneous across the population. (B) This graph shows the membrane potential for five coupled microcolumns at steady state for different values of excitatory population input P, with dots of different shades representing a single microcolumn. Stepwise changes in P cause transitions in dynamics from fixed point steady states (shown as closely overlaid dots per given value P) to oscillatory states (shown as upper and lower sections of the graph for each value of P, for P > 1.3) and back to fixed point (for P > 7.4). Simulations were run in small increasing (blue), and decreasing (green) steps, revealing bistability in the offset of the oscillation (note that outside of this bistability blue and green are largely overlapping). Even in this simplistic model many different state transition phenomena can be modelled, as sudden switches between high amplitude oscillations and fixed points at the bistable offset bifurcation, where two possible steady states co-exist. (C) Time series examples are shown for increasing values of the parameter P for illustration of different dynamic regimes associated with changes in just the single parameter P. Image source: Burrows et al., 2020.

Theory driven coarse and functional models have also been used to infer the macroscale functional network properties that drive network instability and seizure propagation. One such approach is dynamic causal modelling, which estimates unknown model parameters given the data using Bayesian inversion (Friston et al., 2003). Dynamic causal models are generative neural mass models, which describe the dynamics of discrete but functionally connected cortical columns, where each column captures a canonical cortical microcircuit (Kiebel et al., 2008). Such causal models have been applied to EEG traces to characterise the directional progression of seizures (Cooray et al., 2016; Hamandi et al., 2008; Kiebel et al., 2009; Murta et al., 2012; Vaudano et al., 2009). In particular, one study has piloted the application of dynamic causal modelling to whole-brain zebrafish calcium imaging of seizure dynamics, where neural masses were fit to average regional calcium dynamics (Rosch et al., 2018). This approach identified key macroscale reorganisations to functional connectivity that drive seizure activity – increased forward connections onto the midbrain and reductions in synaptic time constants drove seizure propagation. In this way, coarse-functional models can also help to identify the macroscale principles facilitating network instability.

While coarse-grained models have clearly been influential in guiding our understanding of macroscopic network instabilities, such approaches may fail to account for the neuronal heterogeneity that is indicative of seizures at the microscale (Muldoon et al., 2013). Given the emergence of techniques which can capture elementary unit dynamics in whole brain networks, datasets are now available to guide the construction of elementary network models for the study of seizures. Such approaches may well provide novel insight to mechanisms underlying network instability.

### 1.2.4 Towards Elementary Models of Whole Brain Seizure Dynamics

Until very recently, recording techniques which capture neuronal dynamics across the whole brain in living systems have been limited to EEG and functional MRI. Such approaches implicitly coarse-grain underlying dynamics, and therefore modelling efforts were geared towards explaining such macroscale phenomena. Even as model systems and techniques have emerged that facilitate whole brain mapping at the microscale (Ahrens et al., 2013), the dimensionality of such datasets provides a significant barrier to modelling. Coarse models benefit from being simpler to build, as they have fewer interacting components, less computationally expensive, due to fewer system parameters, and more interpretable (Wilson & Cowan, 1972). However, the development of high performance computing has enabled the application of microscale modelling and analytical approaches to high dimensional systems (Takahashi et al., 2021). Furthermore, the growth of unsupervised machine learning algorithms has led to the development of dimensionality reduction techniques which may reduce the parameter space while maintaining key axes of variance not captured by averaging (Cunningham & Yu, 2014). Therefore, with the emergence of high dimensional brain datasets and computational techniques to handle them, computational neuroscientists can begin to build network models which capture the heterogeneity of elementary units in whole brain seizure networks. Such elementary models carry a range of benefits not available to coarse-grained models.

A major assumption of coarse models is that the properties of interest in a population of neurons, can be captured by the population mean. This assumes that the microscale heterogeneity in activity and connectivity of a population may be averaged over to understand its global dynamics (Buice & Chow, 2013). This is problematic when the property of interest is driven by heterogeneous subpopulations and non-trivial connectivity patterns at the microscale – for example the presence of a balanced network state emerges through transient fluctuations of excitatory and inhibitory neuronal dynamics which would not be captured in mean field approaches (Bick et al., 2020; van Vreeswijk & Sompolinsky, 1996; Vreeswijk & Sompolinsky, 1998). Similarly, the emergence of seizures can depend on transient fluctuations in heterogeneous neuronal subpopulations (Muldoon et al., 2013), whose dynamics would be averaged over using mean field models (Wenzel, Hamm, et al., 2019). Therefore, the population mean may not provide a sufficiently detailed description of the networks' dynamics, to understand how seizures emerge. Even if the estimated ensemble mean is sufficient, modelling evidence suggests that neural masses can fail to approximate the true population mean. For example, neural mass models fail to accurately describe mean parameters of simulated spiking networks in a majority of dynamical regimes (Deschle et al., 2021). Therefore, elementary models which capture the heterogeneity of neuronal dynamics may more accurately represent mean properties of the system, and provide a window into properties that would not survive averaging effects, which may be relevant for network instability. Furthermore, given that single neuron fluctuations can drive global brain dynamics (Bonifazi et al., 2009; Brecht et al., 2004; Miles & Wong, 1983), microscale interventions could be used to effectively re-stabilise network dynamics (Hadjiabadi et al., 2021). To this end elementary models can provide intuition into the exact synaptic and neuronal changes that drive global instability, providing greater biological insight than coarse-grained models.

Another major assumption of coarse models is that synaptic coupling between neuronal populations may be captured by mean edge values between ensembles. However, capturing all excitatory to inhibitory connections with a single coupling parameter fails to account for the diversity of local connectivity patterns exhibited by interneurons, many of which show differential synaptic changes in epilepsy (Miri et al., 2018). Therefore, ensemble measures might obscure the microscale interaction

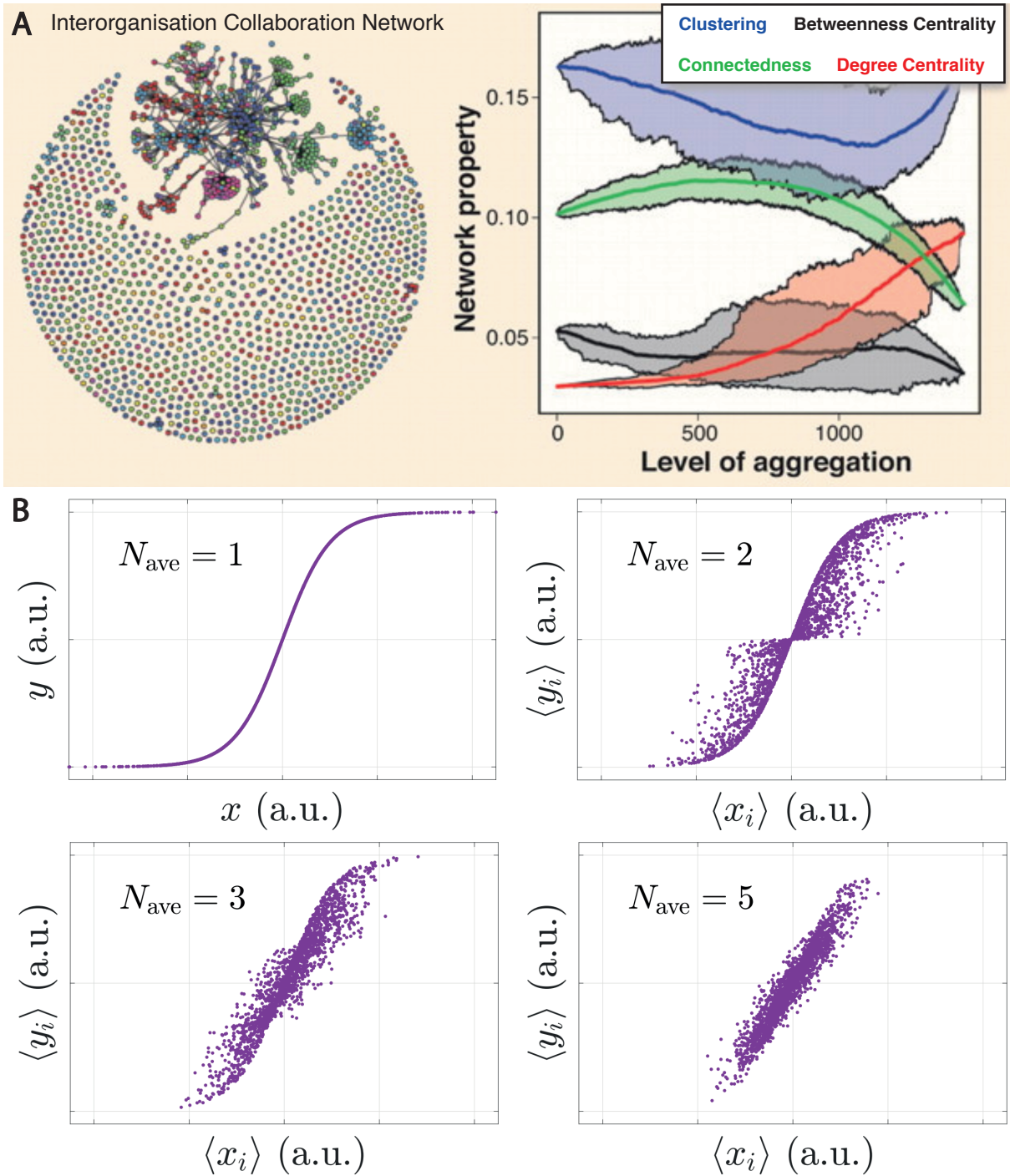

**Figure 1. 11  Coarse graining alters network properties and non-linear relationships.**

(A) A network of interorganisational collaboration during the Hurricane Katrina response. Image source: Butts, 2009. Left image shows the most granular level of representation, where each node represents all subordinates across all organisation. Here each node is coloured by its parent organisation. (Right) As you increase the level of aggregation, by randomly grouping subordinates into parent groups the fundamental structure of the network changes in non-trivial manners. (B) The linearising effect of coarse graining by spatial averaging over nodes. Image source: Nozari et al., 2020. $N_{ave}$ signals, $x_i(t)$, where t = 1, ..., 2000 were randomly sampled where $y_i(t)$ is the non-linear function tanh($x_i(t)$). For each $N_{ave}$ signals, the averages $\langle x_i \rangle$ and $\langle y_i \rangle$ are computed and plotted for increasing $N_{ave}$. Note that despite the non-linear relationship between $x_i$ and $y_i$ as you increase the number of signals that you average over, the mean relationship becomes linear.

and emergent network motifs that drive network instability. Importantly, evidence from graph theory demonstrates that constructing networks by coarse graining over elementary units alters edge distributions and key connectivity patterns (Figure 1.11A) (Butts, 2009). Coarse graining over network nodes may therefore lead to spurious edges and higher order motifs. Similarly, grouping elementary units into larger ensembles can be problematic when the ensembles do not capture clearly defined functional units as is often the case with EEG data (Bassett et al., 2018). This means there is often no clear mapping from abstracted to physical system nodes and edges. In this way, elementary models which use irreducible system units and connections to define network connectivity are less likely to introduce coarse graining artefacts (Butts, 2009). Finally, coarse grained measures of brain dynamics such as functional MRI and EEG filter out non-linear input-output functions (Figure 1.11B) (Nozari et al., 2020), which may be fundamental for seizure propagation (Martinerie et al., 1998). In this way, elementary models which estimate non-linear neuronal relationships and explicitly model microscale connectivity patterns can provide insight into the granular network mechanisms driving global instability.

In the next section, I will discuss how techniques from complex system theory may be combined with single cell network models of whole brain activity to study the emergence of network instability during seizures.

## 1.3    The Emergence of Network Instability

### 1.3.1  Lessons from Cascading Failures in Complex Systems

Viewing epileptic seizures as an emergent property of neuronal networks has helped to reframe questions relating to seizure genesis in the context of complex systems theory, including dynamical systems and graph theory, which provides mathematical frameworks for quantifying collective behaviour in large networks (Bullmore & Sporns, 2009; Siegenfeld & Bar-Yam, 2020). Complex systems theory involves the study of systems containing many interconnected components, and is specifically interested in how the coordinated actions of its nodes gives rise to collective behaviours (Ma'ayan, 2017). A key property of some complex systems is the presence of cascades, which describe the spatiotemporal propagation of activity through the network (Beggs & Plenz, 2003). Importantly, cascading dynamics may be fundamental to network function (Petermann et al., 2009), but if uncontrolled can lead to exponentially growing activity which disrupts network stability (Schäfer et al., 2018). Of particular interest to seizures, complex systems theory has been used to understand how network cascades remain within a stable functional range, how aberrant cascades may ultimately spread through the network, and finally how network function may be restored (Sanhedrai et al., 2022; Siegenfeld & Bar-Yam, 2020). Similar approaches may be leveraged to understand how healthy brain networks allow cascades to propagate through the system without causing seizures, how aberrant cascades spread as a seizure, and how normal function may be restored.

At this stage I should clarify a point of terminology – throughout this chapter I have used the term stability loosely to refer to the ability of a network to maintain dynamics within a functional operating range. However, in the world of dynamical systems, from which much of the analytical approaches applied in later chapters arise, stability has a more formal definition – the propensity for dynamics to converge towards a point in state space. Conversely, instability means that dynamics diverge over time, which is a defining feature of chaos (Babloyantz & Destexhe, 1986). This clarification is important, as seizures invariably are unstable in the functional sense, but not necessarily in the dynamical systems sense (F. Lopes da Silva et al., 2003).

Nonetheless, both definitions of stability are of interest to questions relating to seizures, as instability in either sense could describe seizure emergence and propagation. Going forward I will refer to the functional operating range and dynamical systems definitions of stability as *functional stability* and *dynamical stability,* respectively.

Many complex systems experience environmental perturbations and can exhibit varying degrees of functional stability. For example, damage to a generator in a power grid may lead to a loss of functional stability, or in more resilient networks the continued distribution of electricity to consumers (Pahwa et al., 2014; Schäfer et al., 2018; Witthaut & Timme, 2015). Understanding how such large networks remain functionally stable to perturbations will help to identify key regulating system parameters which might drive the system into instability, such as during a seizure (R. Cohen et al., 2001; J. Gao et al., 2016). For example, statistical physics approaches have been used to hypothesise mechanisms through which a network may self-regulate its own internal parameters in response to perturbations, to maintain its dynamics within a functional range (Zeraati et al., 2021). The identification of such stabilising parameters in neuronal networks may therefore be key to understanding how functional stability fails in the brain, as these parameters may define the transition from normal to abnormal activity characteristic of epileptic seizures (F. Lopes da Silva et al., 2003).

Interestingly, the study of cascading failures in networks, where initial local shocks produce sequences of node failures disrupting global network function, has uncovered key network properties which influence functional network stability. In particular, network models have been used to demonstrate the importance of the topology of the network, specifically the presence of clustered connections and the average path length (Plietzsch et al., 2016; Witthaut & Timme, 2015), and the network size (Pahwa et al., 2014) in regulating the resilience of power grids to cascading failures (Figure 1.12). The identification of such stabilising parameters in network models is therefore key to understanding how functional instability may emerge in real systems. Furthermore, once constructed these models may also be used to predict the implications of altered cascading dynamics on network function

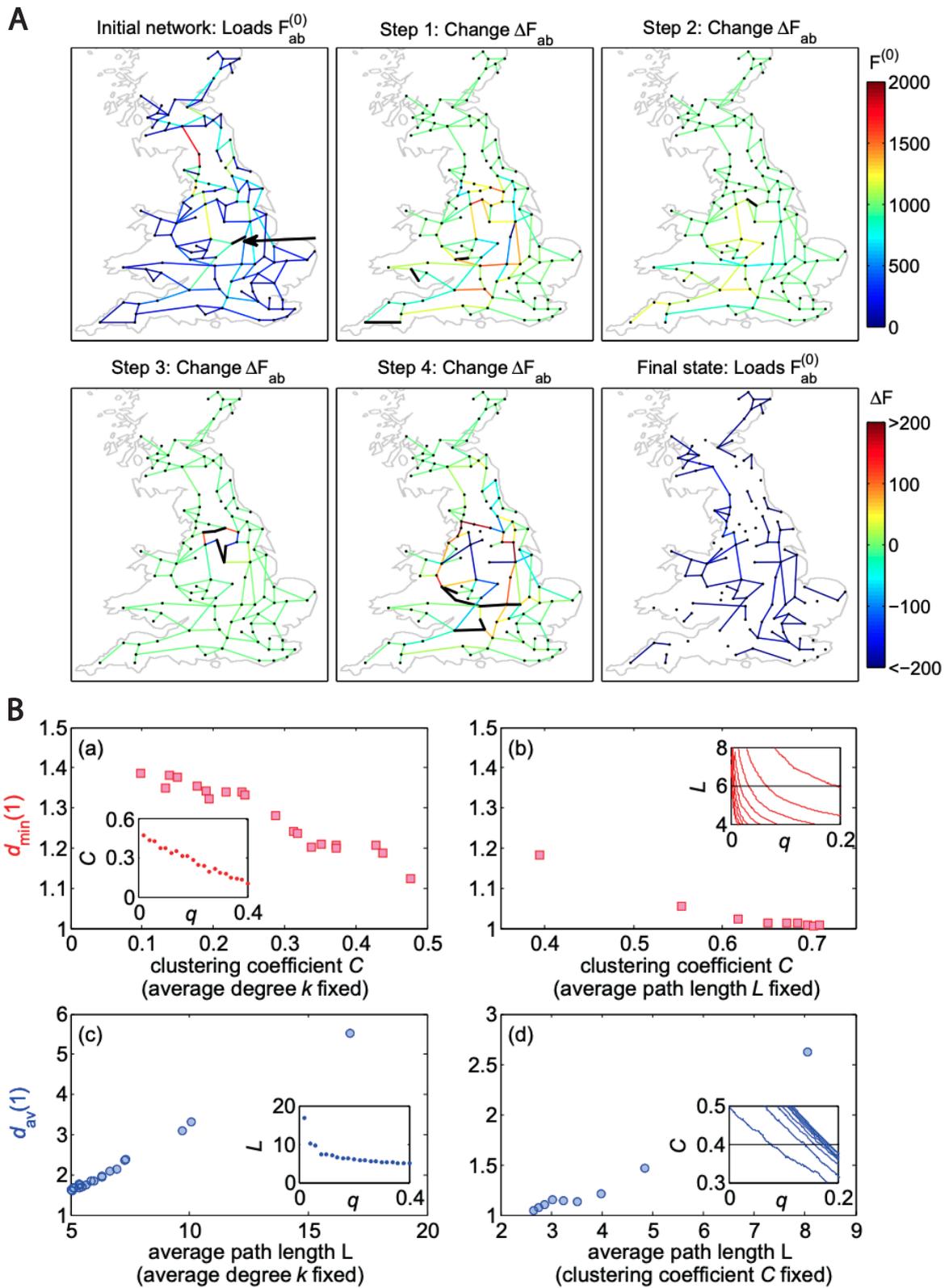

**Figure 1. 12  Identifying control parameters for functional network stability in power grids.**

(A) The propagation of cascading failures in the British power transmission grid. The initial network has loads (F) relatively evenly distributed throughout the network. A single edge (black arrow in initial network) fails, resulting in loads being re-distributed throughout the network, causing edge overload and dropout (thick black lines). In the final network state, the power grid is separated into a series of subnetworks with multiple nodes inoperable, representing disrupted global network function. (B) Network topology as a control parameter influencing network resilience in a Watts-Strogatz model network. (a, b) The mean distance of the nearest overloaded edge to the initial perturbation (dmin(1)) decreases as a function of the clustering coefficient. This indicates that the propensity for cascades to be distributed locally increases higher clustering. (c, d) The mean distance of all overloaded edges (dav(1)) increases as a function of the average path length. This indicates that longer path lengths increase the propensity for global cascading failures. Therefore, global cascading failures are more likely in network with low clustering coefficients and high average path lengths. Panels show parameter changes as a function of the topological randomness q. Image source: Witthaut & Timme, 2015.

– for example to understand the effect of single edge fluctuations on power grid shutdown (Schäfer et al., 2018), or how excessive cascades during seizures gives rise to impaired awareness (Inoue & Mihara, 1998). Importantly, once key regulating parameters and driver nodes are identified, they may be controlled to reach a desirable or restored network state (Jia et al., 2013; Liu et al., 2011; Sanhedrai et al., 2022). In this way, complex systems theory provides analytical tools for the identification of mechanisms driving network instability, and the recovery of normal network dynamics.

Interestingly, elementary network models have identified several key parameters that might regulate the functional and dynamical stability of brain dynamics. Identifying such stabilising parameters in living systems, may therefore be crucial for understanding the emergence of unstable cascades in whole brain cellular resolution networks.

*1.3.2  Single Cell Models of Stable Network Dynamics*

Elementary network models have been used for decades, to relate neuronal firing properties and synaptic interactions to global network states. However, until recently these approaches have not been guided by whole brain cellular resolution datasets. Nonetheless, previous elementary models of large *in silico* networks have helped to identify key microscale mechanisms that could support functionally and dynamically stable network dynamics.

For example, theory-driven elementary models have helped to explain the presence of functionally stable states consisting of irregular spike trains in cortical neurons, which resemble a Poisson process (Vogels et al., 1989). Interestingly, neuronal networks of excitatory and inhibitory neurons randomly and sparsely coupled with strong synaptic inputs, give rise to tightly balanced fluctuations in excitation and inhibition onto a neuron (van Vreeswijk & Sompolinsky, 1996; Vreeswijk & Sompolinsky, 1998). During brief moments this balance is imperfect, releasing a neuron transiently from its inhibition – this gives rise to a balanced state with irregular firing patterns. Therefore, such approaches may be used to map synaptic features onto functionally stable network states relevant for brain function such as balanced dynamics. Furthermore, one can use the predictive power of such models to identify synaptic mechanisms which support a diversity of functionally and dynamically stable regimes – such as the presence of up and down states (Jercog et al., 2017). For example, by altering the distribution of excitatory-inhibitory synaptic strengths, studies have predicted the emergence of a bistable regime in which the system switches between two dynamically stable states consisting of periods of low firing rates and short bursts, respectively (Kriener et al., 2014). In this way, elementary models can identify the synaptic mechanisms that regulate stable global dynamics. Such approaches may be extrapolated to guide our understanding of the microscale features that give rise to network instability.

Of particular relevance to the question of network instability, elementary models have been used to study how brain networks remain sufficiently sensitive to inputs to activate downstream targets, but insufficiently so as to cause cascading failures (Figure 1.13) (Kinouchi & Copelli, 2006). One theory, inspired by statistical physics and supported by *in silico* network simulations (Bak et al., 1988; Bornholdt & Rohlf, 2000), supposes that synaptic plasticity may organise neuronal dynamics near to a phase transition (Beggs, 2008). This dynamical regime, called criticality, is able to balance both high and low activity states (Hesse & Gross, 2014), enabling maximal sensitivity to inputs without causing seizures (Figure 1.13) (Gautam et al., 2015; Haldeman & Beggs, 2005). Interestingly, various theoretical critical regimes separate out distinct states which qualitatively describe hypo and hyper-excitation phenomena (e.g. a seizure) – such as the boundary between high and low activity states (Beggs & Plenz, 2003), dynamical stability and chaos (Dahmen et al., 2019), or stationary and oscillatory dynamics (Cocchi et al., 2017). Therefore, the study of phase transitions may offer a dynamical mechanism for the emergence of seizures, where operating too close to the transition renders the network susceptible to crossing over into states of high activity, synchrony or dynamical instability (Figure 1.13) (Scheffer et al., 2009).

Interestingly, similar approaches have also uncovered how synaptic parameters may be intrinsically altered to adjust the dynamical stability of the network slightly away from criticality, to provide a safety margin between high and low seizure risk regimes according to task requirements (Figure 1.13) (Priesemann et al., 2014). This so-called reverberating regime could account for the presence of both irregular and critical dynamics (Wilting et al., 2018; Wilting & Priesemann, 2019). For example, modelling evidence has implicated both the ratio of excitatory-inhibitory synaptic strengths, and the interplay between adrenergic and cholinergic neuromodulation in regulating near-critical dynamics (Li & Shew, 2020; Munn et al., 2022). In this way, elementary network models can be used to identify dynamical mechanisms which homeostatically regulate network stability, and could be extended *in vivo* or *in silico* to understand how the maladaptive regulation of such stabilising parameters causes instability during seizures (Meisel et al., 2012).

Remarkably, evidence from cellular recordings of whole brain dynamics in the larval zebrafish *in vivo* indicates that neuronal interactions organise global dynamics near to a phase transition (Ponce-Alvarez et al., 2018). Furthermore, theoretical evidence of critical neuronal networks suggests that the disrupted regulation of critical dynamics would cause functionally and dynamically unstable dynamics (Haldeman & Beggs, 2005), reminiscent of seizures. As such, I aim to utilise criticality as a framework to understand the regulation of functionally and dynamically stable dynamics and the emergence of network instability, in cellular resolution whole brain networks.

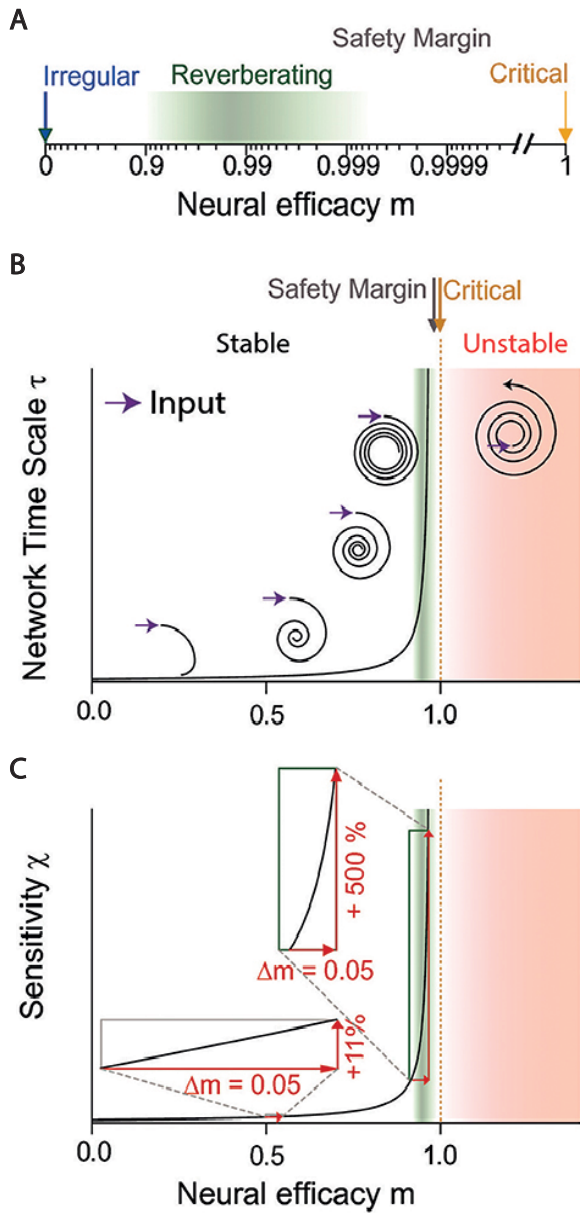

**Figure 1. 13    Dynamically stable and unstable regimes in elementary network models.**

(A) The neural efficacy parameter m defines the sensitivity of the network to inputs. At criticality when m = 1, the network is maximally sensitive to inputs. The irregular state exhibits m ~ 0, and is insensitive to inputs. The reverberating regime (shown in green for m values found in experimental observations) exhibits m ~ 0.9 – 0.99. The safety margin denotes the safe boundary between dynamically stable dynamics of the reverberating/irregular regimes and the phase transition at m = 1, which separates stability and instability. (B) Plot showing the time scale and dynamical stability of dynamics as you get closer to the critical point. Sketch overlaid onto plot visualises the trajectories of perturbations onto the network, assuming the presence of a limit cycle attractor. Note that the time that inputs are maintained in the network exponentially increases as networks approach criticality.

While operating at criticality maximises timescales, operating too close to the phase boundary (yellow dotted line) might render the network at risk of crossing over into dynamical instability. (C) Regimes that can dynamically regulate its dynamics by altering neural efficacy, can support a wide-range tuning of the sensitivity of the network. This is shown by increasing m by 0.05 in the reverberating regime (green) resulting in a 500% increase in sensitivity, compared with 11% increase at lower m. Operating in the stable regime, close to the stable-unstable boundary may provide a flexible, yet safe dynamical regime. Image source: Wilting et al., 2018.

### 1.3.3  Criticality – Whole Brain Dynamics at a Phase Transition

Criticality refers to the behaviour of a system that resides at a phase transition, balancing distinct types of activity (Cocchi et al., 2017). Given that critical systems can optimise various network response properties (Langton, 1990), criticality has emerged as an appealing model of brain function. Interestingly, the cost of residing close to a phase transition may be the propensity to be abruptly perturbed into suboptimal, functionally unstable states (Wilting et al., 2018) (Figure 1.13). In fact, theoretical studies have postulated that deviating from the critical point would lead to seizures (Hsu et al., 2008). However, the extent to which this describes the emergence of network instability *in vivo* is unclear. Furthermore, the single cell network mechanisms stabilising dynamics around criticality, and giving rise to cascading failures once functional stability fails, are currently unexplored. To begin to answer these questions, it is first worth outlining the key theoretical properties of critical systems, as they are central for analyses used in later chapters.

The notion of phase transitions stems from physics, which characterises a system's behaviour into qualitatively distinct phases. For example, the classical states of matter, solid, liquid and gas, describe different emergent properties of microscopic components in the system of interest. To visualise these distinct phases, one can use a phase diagram which shows regions of parameter space for which a system takes on each phase, with each phase separated by a phase boundary (Figure 1.14). A phase transition occurs upon changing some parameter sufficiently such that the system crosses over from one phase to the next – for example, by lowering temperature to transition from liquid water to solid ice (Figure 1.14). Such phase transitions may be classified by their dynamical properties – key to this is constructing a bifurcation diagram (Cocchi et al., 2017) (Figure 1.15). These show the relationship between a global property of interest, known as an order parameter, and parameters which drive the system about the phase transition, a control

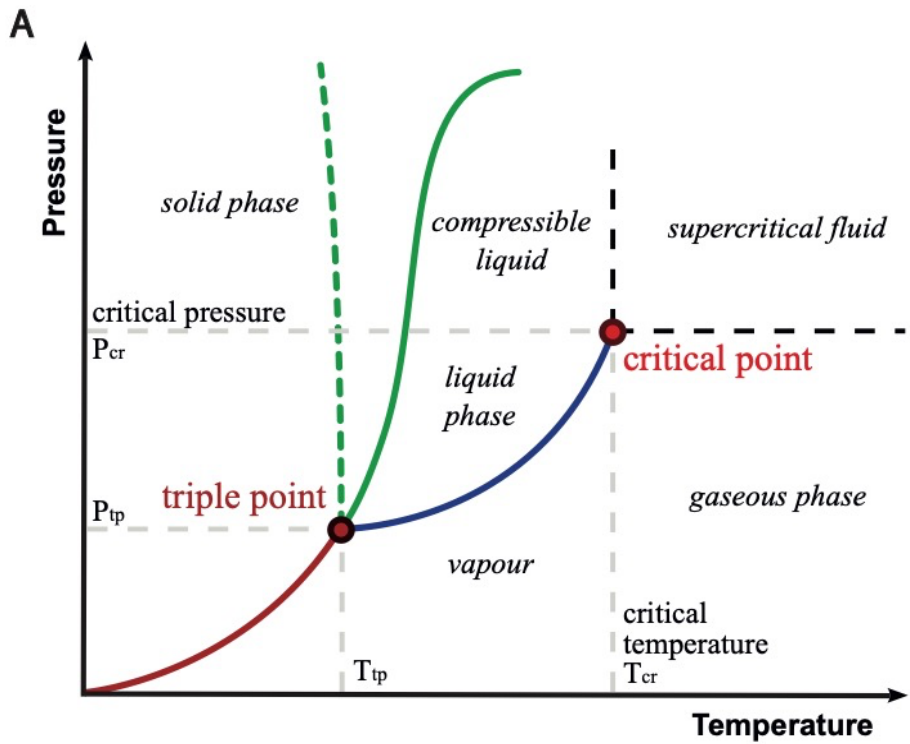

**Figure 1. 14 Phase diagram for pressure and temperature.**

(A) A pressure-temperature phase diagram for a typical material, showing the phase boundaries for solid, liquid and gas phases for different combinations of parameters. Solid lines are phase boundaries – points at which materials change phase, where the free energy changes abruptly. The solid green line shows the boundary between solid and liquid, the red line shows the boundary between solid and gas, and the blue line between liquid and gas. At the triple point, phase boundaries intersect, such that all three phases of matter co-exist. At the critical point, the system is at the critical temperature and pressure meaning that the substance behaves as a gas and a liquid simultaneously – for increasing temperature and pressure, there is no phase boundary as both phases co-exist as a supercritical fluid. Fixing either the critical temperature or pressure and reducing the other gives rise to different properties of matter. Image is licensed from Matthieumarechal, CC BY-SA 3.0, https://commons.wikimedia.org/w/index.php?curid=4623701

parameter. Away from a phase transition, a smooth change in the control parameter gives rise to a smooth change in the order parameter – however, when the system crosses over a phase boundary an abrupt change in the order parameter occurs (Figure 1.15) (Hesse & Gross, 2014). This change can occur either as a discontinuity in the order parameter, known as a first-order phase transition, or by a non-differentiable but continuous change in the order parameter, a second-order phase transition (Kim et al., 1997). This discontinuity in the first-order phase transition results in multi-stable behaviour, in which the systems order parameter stochastically switches between either phase – as in the liquid to gas phase transition for water below its critical temperature (Figure 1.15B) (Santoro et al., 2007). However, it is the second-order phase transition that allows criticality to occur, as here the order parameter can reside exactly between the two states (Cocchi et al., 2017) (Figure 1.15A). This gives rise to a variety of fascinating features, which make criticality an appealing model of brain function.

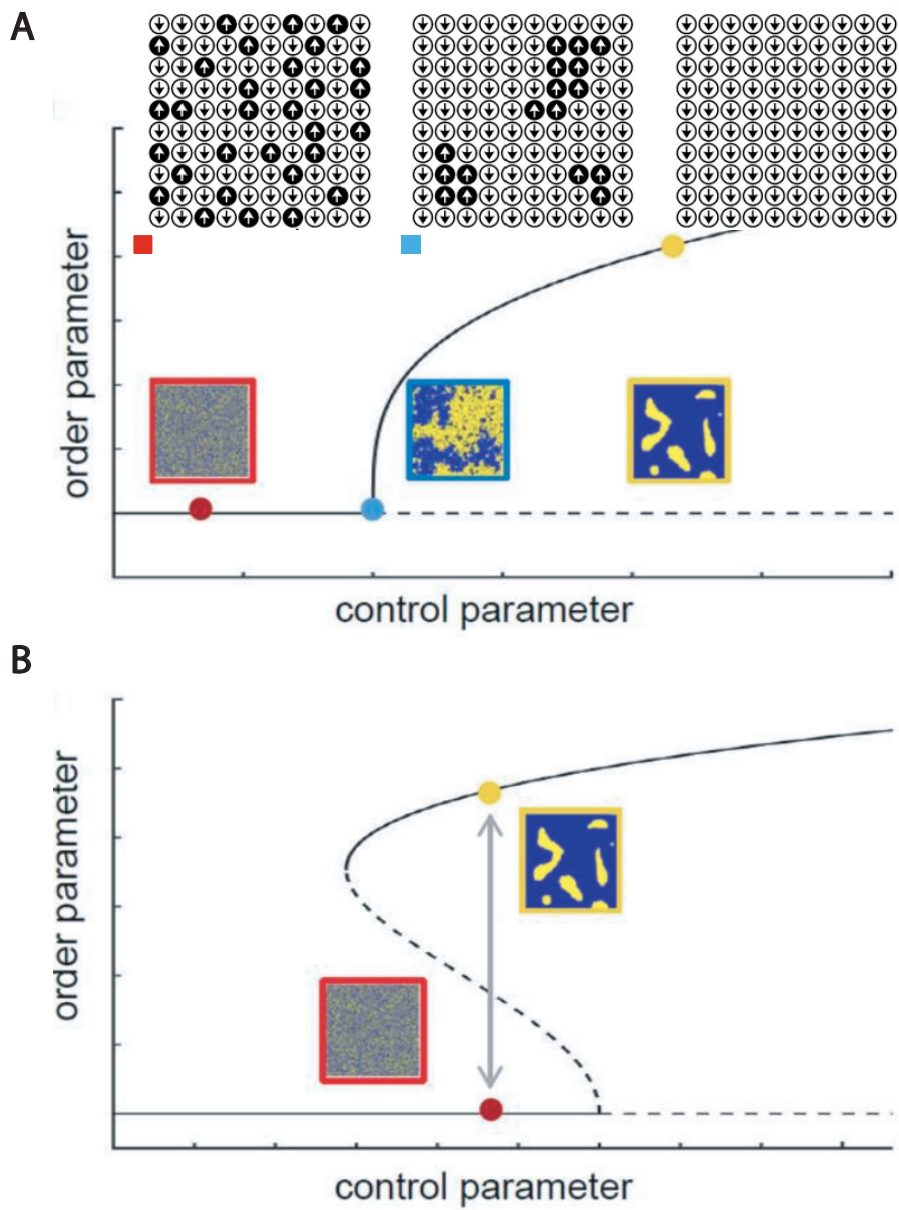

**Figure 1. 15  Bifurcation diagram for the Ising model.**

(A) A second order phase transition. The lattice sites of the Ising model are shown over the entire site (coloured square insets), and for zoomed in sections (above). When the temperature is high (red) the electron spins are not aligned – this means the order parameter is at 0 as there is no magnetism due to the absence of magnetic domains. When the temperature is low (yellow), the electron spins are ordered – therefore the order parameter is high as there are several large magnetic domains covering the ferromagnet, causing magnetism. At the critical point (blue), the electrons spins show clusters of alignment and non-alignment – this gives rise to magnetic domains that can span the entire scale of the system. Note that at the bifurcation the order parameter exhibits a continuous, non-differentiable change such that the system can reside exactly at criticality balancing both regimes. (B) A first order phase transition. Here the system exhibits a discontinuity at the bifurcation, such that the system cannot reside between the two phases but jumps between them – this is an example of multistability.  Images adapted from Cocchi et al., 2017, and Beenhakker, 2019.

To explain some of these properties, I will use a classical physics model for the study of second-order phase transitions – the Ising model. This model captures the behaviour of unpaired electron spins in a ferromagnetic material in a slowly changing magnetic field – in the 2 dimensional model each electron is represented as a site on a square lattice existing in an up or down state (Ising, 1925) (Figure 1.15A). Each electron imposes local fields over its neighbours such that neighbouring electrons tend to share the same spin. In this system, the control parameter is the temperature, while the order parameter is the net magnetisation of the system. To understand the dynamics of the Ising model in different phases, it is useful to think about how groups of electrons can coherently change their spin in response to local field changes – these sequential spin changes across the population are often described as avalanches (Figure 1.16A) (Sethna et al., 2001). Interestingly, when the temperature is below the Curie temperature ($T_C$) the local interactions of neighbouring spins dominate – large portions of the system will flip coherently resulting in large avalanches. This results in most of the lattice being covered by ordered magnetic domains with identical spin, resulting in ferromagnetism (Figure 1.15A) (Ising, 1925). Conversely, if the temperature is increased beyond $T_C$ a phase transition occurs, as thermal fluctuations begin to interfere with electron interactions – here single electrons will flip almost independently as small avalanches. This gives rise to a disordered lattice with small magnetic domains, resulting in paramagnetism (Figure 1.15A). However, at criticality when the temperature = $T_C$, avalanches can span all scales – small magnetic domains regularly flip, some of which can grow into larger domains that span the full scale of the system. At this point the tendency for spin alignment is perfectly balanced with the thermal fluctuations from heat, resulting in a mixture of ordered and disordered phases (Beggs & Timme, 2012).

To consider why this occurs at criticality, it useful to view the propagation of events through a network (i.e. avalanches) as a branching process (Harris, 1963). This is a Markovian process, which describes the propensity for an ancestor, such as an activated node, to give rise to descendants, in this case a neighbouring node that is subsequently activated (Figure 1.16A) (de Carvalho & Prado, 2000). We define $p_i$ as the probability that an ancestor gives rise to $i$ descendants, where $p_i$ $(i = 0, 1, …,\infty)$. We can then define the branching ratio as the weighted average over the probabilities,

$$\sigma = \sum_{i=0}^{\infty} i\, p_i, \quad (1.1)$$

which captures the mean number of descendants. Importantly, branching processes undergo a phase transition at $\sigma = 1$ (Harris, 1963). At the critical point ancestors will on average trigger 1 descendant – one activated node will trigger one of its neighbours on average (Sethna et al., 2001). This means that while all avalanches eventually decay, on average activity will survive in the system, meaning that avalanches of all sizes, from small to large, will eventually occur (Figure 1.16). These dynamics are described as scale-invariant avalanches, and are a vital ingredient of systems at criticality (Beggs, 2008). However, it is worth noting that branching processes are a simplification of physical systems that have defined network topologies, such as the Ising model, and as such they may behave differently around the phase transition (Zierenberg et al., 2020).

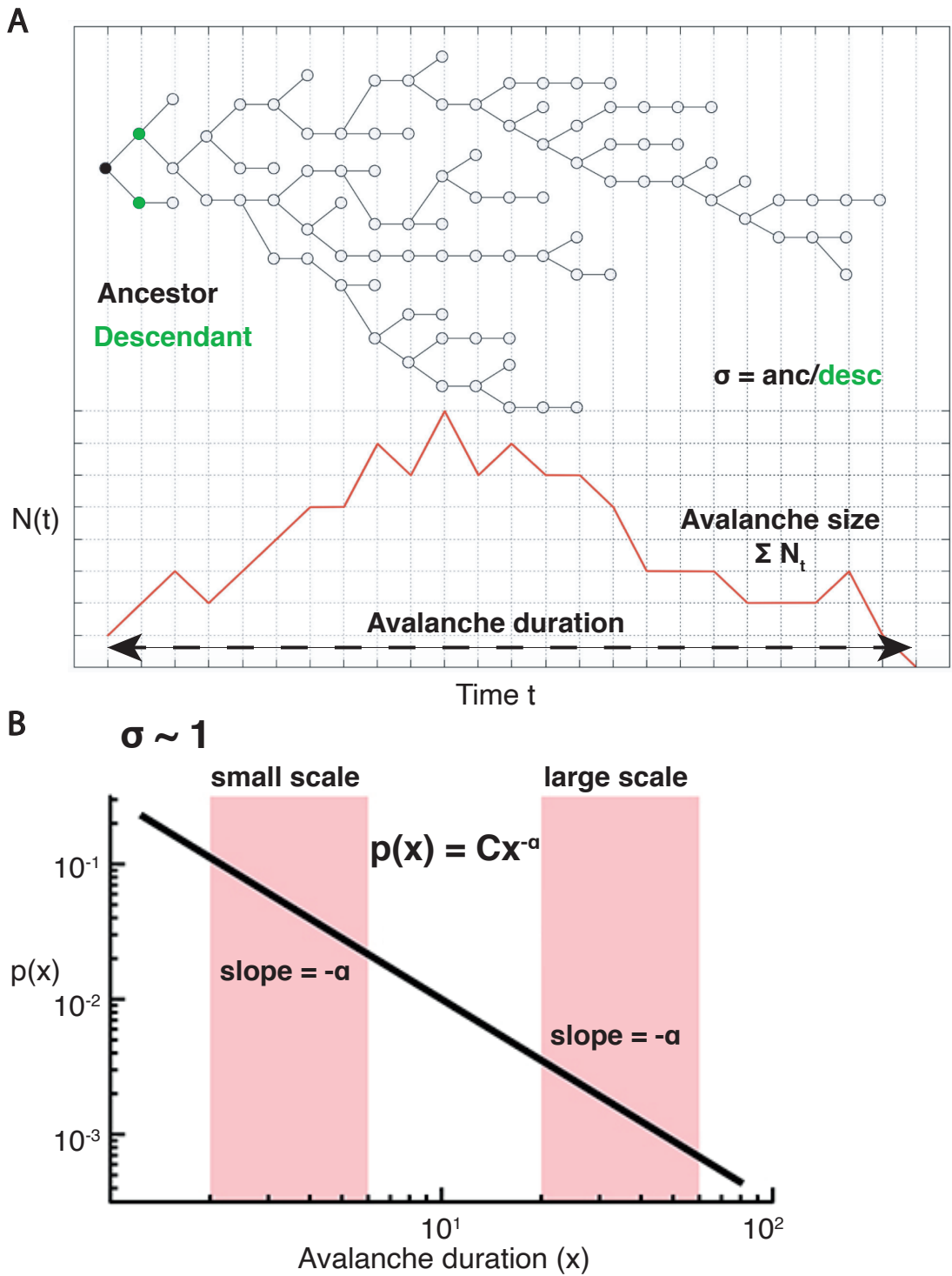

**Figure 1. 16 Critical branching processes and scale invariant avalanches.**

(A) A branching process showing activations of one ancestor node giving rise to descendants at each time step (top). The branching ratio σ captures the average ratio of ancestors to descendants. This branching process can be viewed as the propagation of activity through a network, such as spikes through the brain. Below, shows the propagation of an avalanche through the network over time where N(t) is the number of descendants at time t. Avalanche size is the total number of activations during the avalanche, and duration is the number of time steps for which the avalanche was active. Image is adapted from Corral & Font-Clos, 2012. (B) When $\sigma \sim 1$, the branching process is at a phase transition – on average one ancestor activates one descendant. This give rise to scale-invariant avalanche distributions, which exhibit the same slope regardless of the scale of the avalanches, as shown for small and large scales in pink. Image is adapted from Hesse and Gross, 2014.

At criticality we say the dynamics are scale-invariant because when $\sigma = 1$, the probability distributions for avalanche size and duration are power laws (Figure 1.16B) (Harris, 1963; Levy & Solomon, 1996) – the relationship between the size/duration of an avalanche $x$ and the probability of observing it $p(x)$, is

$$p(x) = Cx^{-\alpha}, \quad (1.2)$$

Power laws are scale invariant functions because no matter how you re-scale the avalanche property of interest, the probability distribution will always retain the same log-linear slope (Figure 1.16B). This means that scaling the input $x$ simply leads to a scaled version of the function $p(x)$, as

$$p(Ax) = A\left(Cx^{-\alpha}\right), \quad (1.3)$$

where the right hand side is just $A(p(x))$. This property is non-trivial, as shown for exponential growth functions,

$$p(x) = b^x$$
$$p(Ax) = b^{Ax}, \quad (1.4)$$

where scaling the input scales the output non-linearly, such that the function takes on different shapes at different scales. Scale invariance means that a system or function is self-similar across time or length scales (Sethna et al., 2001) – if we visualise avalanches at $T_C$ in the Ising model at different spatial scales, the system statistically resembles itself regardless of its size (Figure 1.17). This occurs due to the Markovian nature of our branching process, which means that each avalanche from a given ancestor is statistically identical to subsequent avalanches from its descendants regardless of the size of each generation – this gives rise to self-similarity across different scales and power-law distributions in many different observables (Hesse & Gross, 2014).

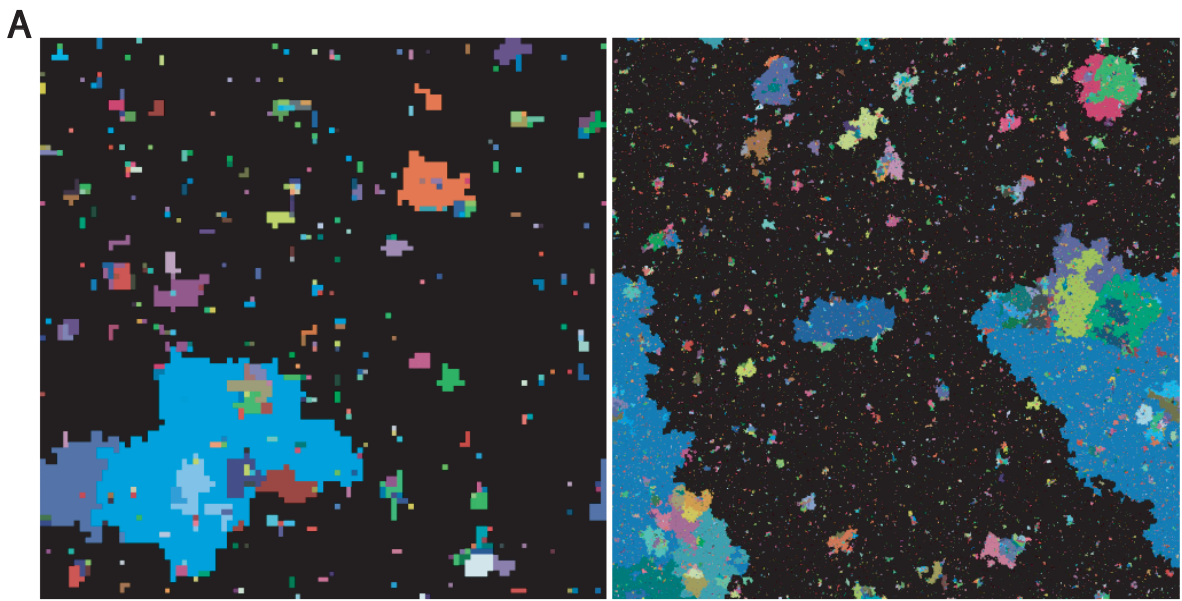

**Figure 1. 17  Self-similar avalanches of the Ising model.**

(A) Cross section of avalanches that occur in the Ising model, where each avalanche is a different colour. Each image taken from a simulation with $100^3$ (left) and $1000^3$ domains (right). Note how despite the difference in scale, when you re-scale the larger system to the area of the smaller system, they look similar. Image source: Sethna et al., 2001.

Importantly, scale invariance imbues the system with a variety of appealing computational properties (Langton, 1990; Packard, 1988). As critical systems usually separate out phases of order and disorder, they can exist at the boundary between dynamical stability and instability, or two different dynamically stable regimes (Dahmen et al., 2019; Haldeman & Beggs, 2005). At criticality the system is weakly stable, as the Lyapunov exponents approach 0 (Haldeman & Beggs, 2005). This means that the systems' dynamics are weakly attracting, such that any perturbations in the system will be retained for long periods, dominating the dynamics. This grants the system maximal memory about inputs, a feature of which is long autocorrelation and high variance (Figure 1.18A). This occurs because at $\sigma = 1$, inputs will stay roughly the same size in the network, whereas at $\sigma < 1$ avalanches will quickly decay, while at $\sigma > 1$, the ongoing activity in the network causes the distance between the input and the representation to grow over time (Hesse & Gross, 2014). Another appealing property at criticality in the Ising model is that the dynamic correlation length between electrons is maximal (Figure 1.18B). Dynamic correlation relies on the ability of electron spins to both correlate and fluctuate, enabling the system to respond to inputs (Beggs & Timme, 2012). Interestingly, at high temperatures electron spins are largely uncorrelated, and at low temperatures electrons are highly correlated but hardly fluctuate (Figure 1.18B). However, at the phase transition, the dynamics are highly variable and yet avalanches can span the full scale of the system, meaning that dynamic correlations emerge between very distant electron spins (Ising, 1925). This means that the system is maximally sensitive to inputs, as even small perturbations have the propensity to flip the entire system (Cocchi et al., 2017). Therefore, criticality endows systems with certain optimal network properties, some of which may be beneficial for brain function.

Remarkably, evidence for critical dynamics is ubiquitous in the physical world – scale invariant avalanche dynamics have been reported in solar flares (Lu et al., 1993), traffic flows (Nagel & Herrmann, 1993), crumpled materials (Abobaker et al., 2015; Sethna et al., 2001), flocking birds (Cavagna et al., 2010), earthquakes (A. Sornette & Sornette, 1989; D. Sornette et al., 1995) and forest fires (Palmieri & Jensen, 2020). While these systems may share little with biological systems such as the brain, evidence from statistical physics supposes that critical systems exhibit universality (Stanley, 1999; Zapperi et al., 1995) – here, the microscale details of collective behaviours at phase transitions become unimportant, such that distinct systems share the same macroscopic properties. Interestingly, a large number of studies have purported to show the presence of critical avalanche dynamics from brain recordings, including in microscale recording from *in vitro* cultures (Figure 1.19) (Beggs, 2008),

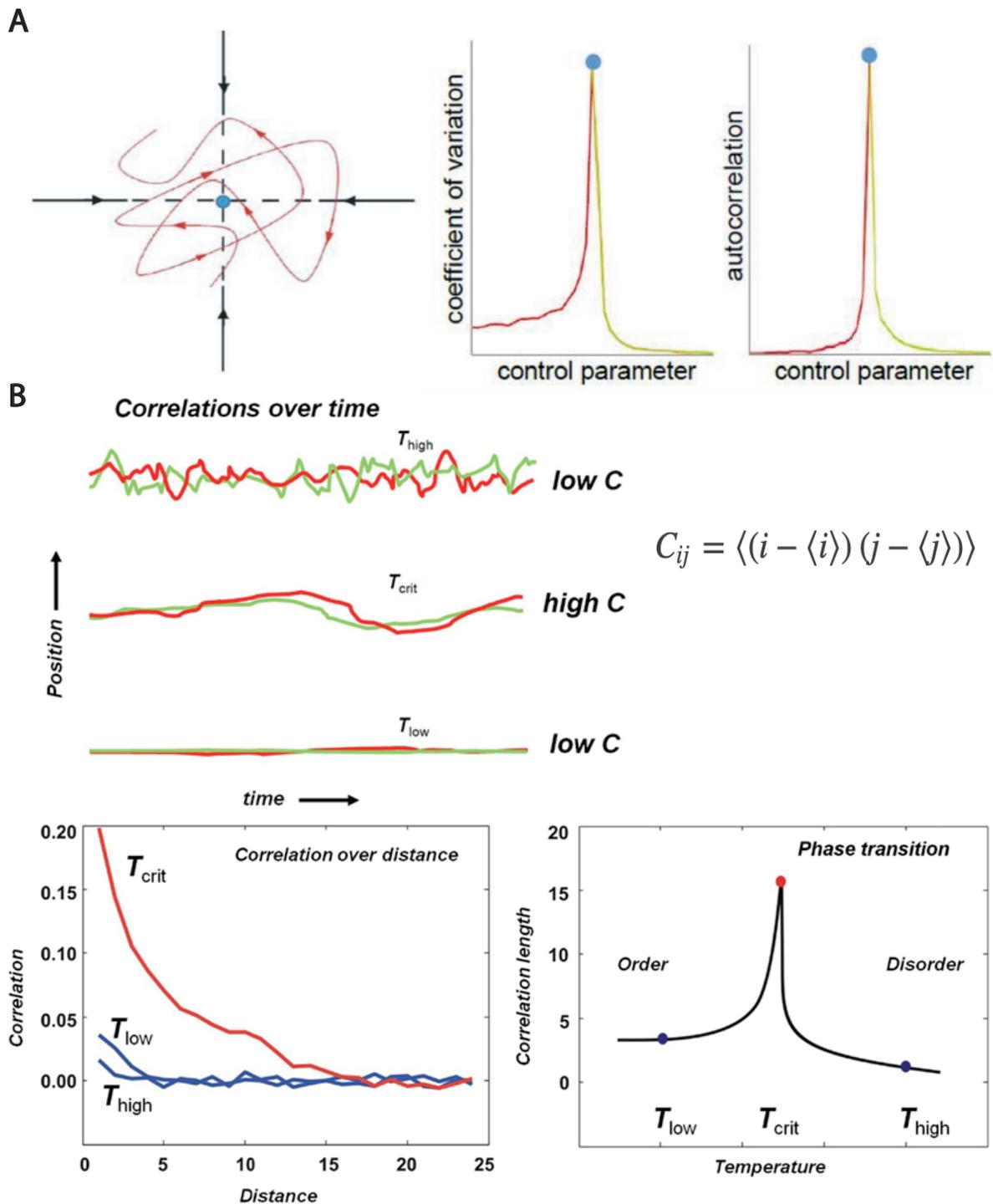

$$C_{ij} = \langle (i - \langle i \rangle)(j - \langle j \rangle) \rangle$$

**Figure 1. 18 Favourable properties of systems at criticality.**

(A) Critical systems are weakly stable. System dynamics are weakly attracted to the critical point (blue), and so fluctuate loosely around criticality enabling long representations of signal inputs (left). This gives rise to high variation and long autocorrelation at the critical point (right). Images adapted from Cocchi et al., 2017. (B) At criticality the dynamic correlation is highest. (top) Dynamic correlation measures the coordinated fluctuations of two time series, shown by the formula on the right for different spin sites i and j. At high temperature, spins fluctuate greatly but independently, giving rise to low dynamic correlation. At low temperatures, while spins are correlated they fluctuate little also giving rise to low dynamic correlation. At criticality, spins both fluctuate and are coordinated. Dynamic correlation is shown a function of distance (bottom left). Correlation length peaks at the phase transition point, shown in red (bottom right).  Images source: Beggs & Timme, 2012.

cortical activity in rodents (Gautam et al., 2015) and non-human primates (Petermann et al., 2009), and even macroscale recordings from humans, in functional MRI (Tagliazucchi et al., 2012) and magneto-encephalography (Shriki et al., 2013). While the idea that such critical dynamics are universal across different systems is highly controversial (Beggs & Timme, 2012; Touboul & Destexhe, 2017), these studies nonetheless provide strong evidence that brain dynamics exhibit some properties of systems at a second-order phase transition. Of particular importance to this thesis, evidence from single unit recordings across the whole brain in larval zebrafish, suggests that brain-wide dynamics exhibit scale invariant avalanches, and a whole host of other features suggestive of critical dynamics (Ponce-Alvarez et al., 2018). Therefore, this indicates that whole brain single neuron dynamics may be organised near to a phase transition – criticality may thus serve as an end point for functional stability in whole brain networks (Ma et al., 2019).

This finding has a series of implications which are central to the question of functional network stability and instability in the brain. Firstly, if single neuron dynamics across the whole brain are critical, how do neuronal networks self-organise to criticality? In the Ising model example, the experimenter manually tunes the temperature to $T_C$. However, for criticality to occur in natural systems, there must be some intrinsic mechanism which allows the system to reside at the phase transition in the face of environmental perturbations.  This is because of fluctuations in external input and in the case of the brain, intrinsic fluctuations to brain dynamics. Interestingly, simple mechanisms can allow systems to self-organise their dynamics to a phase transition, known as self-organising criticality (Bak et al., 1987a, 1988). Therefore, identifying parameters which support self-organising criticality in the brain, may elucidate how

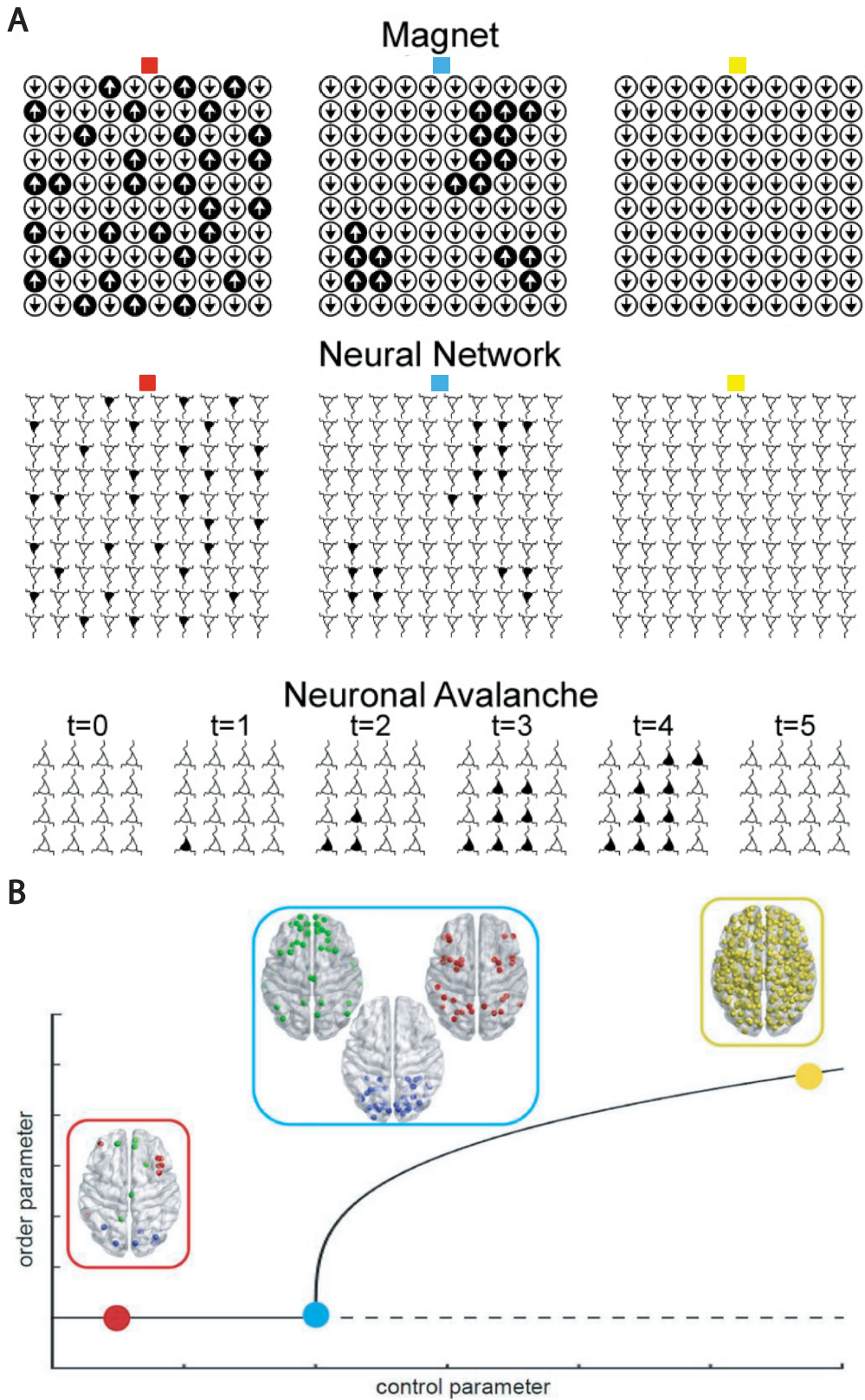

**Figure 1. 19  Scale invariant neuronal avalanches in the brain.**

(A) A ferromagnet may be organised between order (yellow) and disorder (red), at a critical point (blue) exhibiting scale invariant avalanches. Similarly, neuronal networks may be organised to criticality, exhibit avalanche dynamics that span the full scale of the system. Here neuronal avalanches measures the spread of spikes through the network, rather than electron flips through a ferromagnet. Image adapted from Beenhakker, 2019. (B) In the disordered phase (red) brain networks may be uncoupled showing impaired integration, while in the ordered phase (yellow) brain networks may be hyper-synchronous such as during a seizure. At criticality (blue) brain networks may show a balance of segregation and integration. Image source: Cocchi et al., 2017.

neuronal networks maintain functional stability in response to perturbations, and how this may go awry during seizures. Evidence from elementary neuronal network models indicate that synaptic plasticity may serve as a mechanism supporting self-organising criticality (Zeraati et al., 2021) – various models have implicated the role of activity-dependent synaptic depression (Levina et al., 2007), structural homeostatic plasticity (Bornholdt & Rohlf, 2000; Tetzlaff et al., 2010; van Ooyen & Butz-Ostendorf, 2019), Hebbian-like plasticity (Bornholdt & Roehl, 2003; Rubinov et al., 2011), and a combination of Hebbian and homeostatic plasticity (de Arcangelis et al., 2006) in self-organising criticality *in silico*. However, many of these network models employ biologically-inspired but biophysically implausible synaptic parameters and mechanisms, in highly simplified models. Therefore, to understand how the brain might self-organise to criticality as a means of maintaining functional stability *in vivo*, one needs to perturb putative stabilising parameters in living systems.

A second key question is if criticality serves as a stabilising end point for brain dynamics, then does the network instability that arises during seizures emerge as a deviation from the critical point (Figure 1.13)? Some evidence from coarse-grained data indicates that seizure dynamics may emerge as loss of criticality (Arviv et al., 2016), however the statistical indicators used to assess this are weak (Meisel et al., 2012), and coarse-grained dynamics will likely alter critical statistics. Therefore, to accurately describe seizure dynamics in this context, one needs to use robust statistical indicators of criticality in single cell whole brain networks, to confirm theoretical predictions of dynamics in a system driven away from a phase transition. Importantly, given the diversity of computational benefits of operating at criticality (Bertschinger & Natschläger, 2004; Langton, 1990), a loss of critical dynamics may help to explain observed brain dysfunction in epilepsy and during seizures. To this end, data-driven network models of seizure dynamics can be used to map microscale cellular alterations onto global critical dynamics, to explain the emergence of network instability and functional impairments in epilepsy.

Taken together, critical dynamics can provide a unifying framework to understand functionally stable and unstable dynamics in whole brain networks at single cell resolution.

*1.3.4   The Non-Linear Neuronal Interactions Driving Network Instability*
Key to understanding the cause of network instability is the identification of destabilising nodes, whose dynamics and connections may be pathologically altered to propagate excessive activity through the network. The identification of such destabilising nodes can guide interventional strategies to prevent network instability from recurring (Goodfellow et al., 2016) or to restore normal network dynamics (Jia et al., 2013; Y.-Y. Liu et al., 2011; Sanhedrai et al., 2022). To identify such destabilising nodes, one approach is to capture the input-output relationships across all network nodes, to identify the aberrant node-node interactions from which abnormal network dynamics emerge (Hadjiabadi et al., 2021). Interestingly, a neurons' dynamics, as defined by input-output relationships between pre and post-synaptic partners alongside intrinsic excitability changes, are highly non-linear (Stam, 2006). This is because a neuron's membrane potential scales as a non-linear function of its inputs, due to the presence of voltage gated ion channels, intrinsic fluctuations in membrane potential and dynamic changes to synaptic strengths through neuromodulation (Nadim & Bucher, 2014). Estimating neuronal interdependencies using detailed biophysical characterisations across entire brain networks *in vivo* is neither experimentally nor computationally tractable. To get around this problem, many approaches infer neuronal dependencies using analytical techniques applied to empirically measurable signals (Niedermeyer & Lopes da Silva, 2005). By far the most common approach here is correlation, which can capture the extent of linear interactions between two time series. However, such approaches fail to capture the non-linear input-output relationships that are fundamental to driving network dynamics (Stam, 2006). As such, approaches which fail to account for non-linear interactions are likely to miss key node-node interdependencies, particularly in the presence of state-dependent dynamics where correlational structures can change (Sugihara et al., 2012). Therefore, to accurately identify destabilising nodes in brain networks, it is important to use analytical techniques which can account for both linear and non-linear interactions between neurons.

Evidence from EEG studies in epilepsy indicates the presence of non-linear interactions between brain areas during seizures (Le Van Quyen et al., 1998). In fact, accounting for such non-linear interactions has demonstrated utility in improving the localisation of destabilising nodes responsible for seizures (Andrzejak et al., 2001; Lehnertz et al., 2001). This demonstrates the importance of accounting for non-linear interdependencies for studying network instability in epilepsy. However, all previous approaches to study non-linear interactions in epilepsy, have been applied to coarse-grained data (Lehnertz, 2008). As previously discussed, EEG signals represent coarse grained neuronal dynamics, that is averaged fluctuations in postsynaptic potentials which can remove non-linear components of signals. In fact, coarse graining measures tend to filter out non-linear input-output functions (Figure 1.11B) (Nozari et al., 2020). Therefore, to accurately reconstruct the network interdependencies driving network instability, one needs microscale recordings to which non-linear analytical techniques may be applied.

These findings have a series of implications for our understanding of epilepsy, and the development of interventional strategies for targeting network instability in the brain. In particular, if seizures are in fact non-linear phenomena, what is the cellular and dynamical makeup of such non-linear behaviour? A characterisation of non-linear seizure dynamics at the microscale would help to provide a broader description of the cellular dynamics underlying seizures. Furthermore, if coarse graining removes such non-linearities, are macroscale measuring techniques filtering out information that might be useful for guiding interventional strategies? Specifically, can we combine microscale dynamics which retain non-linear information, and effective non-linear reconstruction techniques, to more effectively guide intervention or seizure prevention? To this end, elementary network models which can reconstruct non-linear interdependencies across single neurons brain-wide, may be used to understand the importance of microscale, non-linear dynamics in the emergence and prevention of seizures.

## 1.4   Thesis Aims

This thesis broadly aims to understand the single cell mechanisms that drive network instability in whole brain networks during seizures. To this end, I perform 2-photon imaging of the larval zebrafish in acute and genetic seizure models, to record from functionally stable and unstable brain dynamics at the microscale. I then use tools from complex systems theory, in particular network models and dynamical systems, to describe and quantify the emergence of network instability. My specific aims are as follows:

1. *Confirm criticality as an organising principle for whole brain dynamics (Chapter 3).*
   I measure the collective behaviour of single neurons via avalanche dynamics (Hesse & Gross, 2014). Using this approach, I test for the presence of criticality in whole brain spontaneous activity, to identify an organising principle of healthy population dynamics. I also employ null models to confirm that the presence of critical statistics in fact occurs due to the structure in the underlying data.

2. *Identify regulating parameters which stabilise dynamics around criticality (Chapter 3).*
   Using pharmacological and genetic perturbations to excitation-inhibition balance, I use techniques developed in Aim 1 to test whether the ratio between excitation and inhibition regulates criticality. In doing so I aim to identify a regulating parameter for critical dynamics, to uncover the microscale mechanisms maintaining functionally stable dynamics *in vivo*.

3. *Test whether epileptic seizure and EI-imbalanced dynamics emerge as a loss of criticality (Chapter 3).*

   I apply techniques developed in Aim 1 to epileptic seizure dynamics, to test for the presence of criticality. A loss of critical statistics in single cell brain-wide seizure recordings, would demonstrate a population mechanism through which neuronal activity collectively organises to drive whole network dynamics into functionally unstable states. I also test whether genetic models of excitation-inhibition balance also cause a loss of critical statistics, to identify a novel mechanism underlying network dysfunction in excitation-inhibition balance disorders.

4. *Identify single cell network mechanisms driving network instability in seizures (Chapter 4).*

   I build spiking network models of single cell whole brain seizure dynamics, applied to experimental recordings of epileptic seizures. I use this network modelling approach to test the role of different microscale neuronal parameters in driving network instability and a loss of criticality. In doing so I aim to identify synaptic mechanisms that drive global network instability in brain wide networks.

5. *Identify the effects of single cell network changes on brain function (Chapter 4).*

   Using identified synaptic mechanisms and network models from Aim 4, alongside dynamical systems approaches, I test the effect of synaptic alterations driving seizures on brain function. In doing so, I am to relate microscale synaptic changes with emergent brain dysfunction in epilepsy.

6. *Understand the importance of non-linear interactions in network instability and seizure prediction (Chapter 5).*

   Using techniques from dynamical systems I employ causal inference procedures to study the role of non-linear neuronal interdependencies on driving network instability during seizures. I also test the extent to which such non-linear interactions are useful for seizure prediction.

# Chapter 2

## Methods & Materials

This chapter will focus on the methods shared across results chapters 3, 4 and 5. All methods that are specific to a single chapter are discussed in detail within their respective results chapters. This chapter contains text, data and figures from some of my work currently in preprint (Burrows et al., 2021).

## 2.1   Experimental Model Details

### 2.1.1  Larval Zebrafish Husbandry

Zebrafish larvae were raised at 28°C in Danieau solution on a day and night cycle of 12:12 hours. Adult fish used for breeding were cared for and raised at the Guy's Campus Zebrafish facility, King's College London. All animal work was approved by the local Animal Care and Use Committee (Kings College London), and was carried out in accordance with the Animals (Experimental Procedures) Act, 1986, under license from the United Kingdom Home Office.

Transgenic zebrafish larvae, Tg(HuC:H2B-GCaMP6s), expressing a nuclear localised genetically-encoded calcium sensor were used for wildtype functional imaging experiments (gift from Misha Ahrens, Janelia Research Campus) (T.-W. Chen et al., 2013). To maximise optical transparency, Tg(HuC:H2B-GCaMP6s) larvae were crossed with melanophore deficient (-/-) roy;nacre mitfa mutants (Lister et al., 1999). All imaging experiments were performed at 6 days post fertilisation (dpf).

The creation of foxg1a mutants was carried out in collaboration with the Houart lab, with Hannah Bruce performing gene editing experiments (Bruce, 2022). The foxg1a mutation used in this study was achieved by a TALEN deletion of 5 base pairs (bp) near the start codon of the foxg1a coding region; this deletion results in a frameshift and the induction of an early stop codon. As the foxg1a homozygous mutant fish dies 8-10 around dpf, adult fish stocks were maintained as heterozygous carriers. Progeny from heterozygous incrosses were genotyped by PCR amplification and subsequent sanger sequencing, genotype was determined by the deletion of 5bp. For foxg1 imaging experiments, Tg(HuC:H2B-GCaMP6s) x mitfa mutants were outcrossed with heterozygous foxg1 carriers. This line was then incrossed to create a mixture of homozygous, heterozygous and wildtype Foxg1 TALEN x Tg(HuC:H2BGCaMP6s) x casper larvae.

## 2.2   Data Collection & Pre-processing

### 2.2.1  Genotyping

The genotyping of foxg1a mutants was carried out in collaboration with the Houart lab, with Hannah Bruce performing genotyping experiments (Bruce, 2022). Zebrafish larvae were genotyped by alkaline lysis DNA extraction in 45µl of solution (25mM NaOH, 0.2mM EDTA), heated to 95C for one hour; 50µl of neutralisation buffer (40mM TrisHCL pH8.0) was then added and 1µl was then used as DNA template in a PCR reaction.

### 2.2.2  2-photon Calcium Imaging

For imaging experiments, non-anaesthetised larvae at 6 dpf were immobilised in 2% low-melting point agarose (Sigma-Aldrich) and mounted dorsal side up on a raised glass platform that was placed in a custom-made Danieau-filled chamber. All imaging was performed on a custom built 2-photon microscope (Independent NeuroScience Services, INSS), which utilises a Mai Tai HP ultrafast Ti:Sapphire laser (Spectraphysics) tuned to 940nm. Objective laser power was at 15mW for all experiments. Emitted light was collected by a 16x, 1NA water immersion objective (Nikon) and detected via a gallium arsenide phosphide detector (ThorLabs).

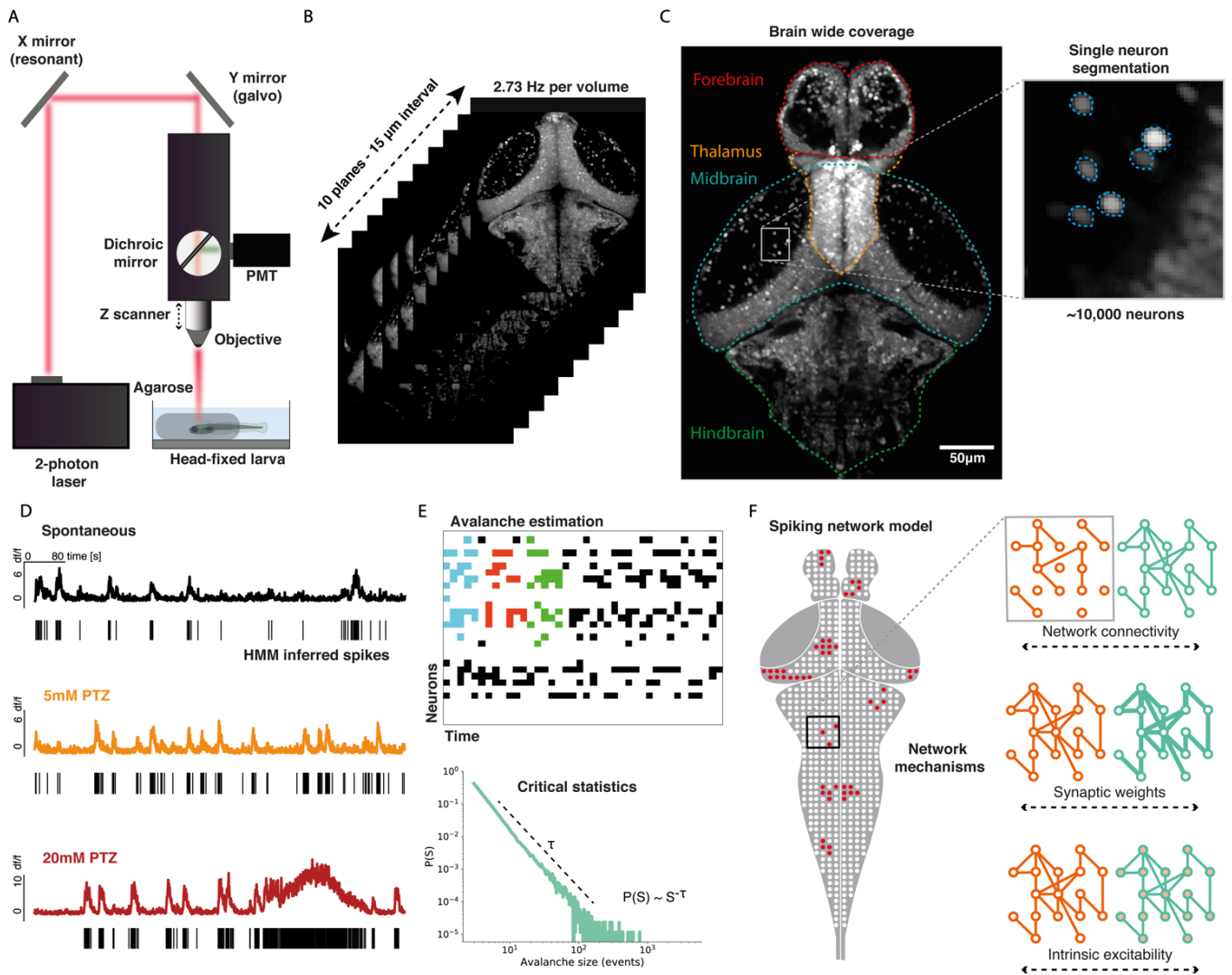

**Figure 2. 1   2-photon imaging setup, data processing and analysis.**

(A) In vivo 2-photon imaging setup with head-fixed larval zebrafish. (B) Imaging was captured across 10 planes with 15 um spacing at an imaging rate of 2.73 Hz per volume. (C) Max projection across zebrafish volume taken with 2-photon microscope, demonstrating coverage of major brain regions. Nuclear localised GCaMP fluorescence allows the segmentation of ~10,000 neurons per brain (right). (D) Single cell traces shown from representative neurons for a single fish, showing normalised calcium fluorescence over time for *spontaneous* (black), *5mM PTZ* (orange) and *20mM PTZ* (red) conditions. A hidden Markov model (HMM) was used to infer spike times (black bars). (E) The spatio-temporal propagation of activity through the network was quantified as avalanches, as shown for 3 example avalanches (coloured by avalanche) for an example raster plot (top). Avalanche statistics were calculated to assess critical dynamics (bottom). (F) A network model of the larval zebrafish brain was constructed, which was used to test the role of specific network mechanisms in driving empirical avalanche dynamics.

Scanning was performed by a resonance scanner (x-axis) and galvo-mirror (y-axis), with a piezo lens holder (Physik Intrumente) adjusting the z-plane (Figure 2.1A). The field of view of the objective in x and y dimensions was sufficient to cover the entirety of forebrain, thalamic, midbrain and cerebellar regions while covering the rostral portion of the hindbrain (Figure 2.1C). Volumetric data was collected across 10 focal planes at 15µm intervals in the z dimension, resulting in a frame rate of 2.73 Hz per volume (Figure 2.1B). Images were acquired at a resolution of 512x512 pixels, with a pixel size of 1.05 x 1.03 µm. Larvae were left for 30 minutes in the light to allow the agarose to set and the fish to habituate to its environment.

### 2.2.3  Pharmacological Induction of Seizures

To induce focal and generalised seizures in wildtype fish I used the GABAa receptor antagonist pentylenetetrazole (PTZ), which elicits clonus-like convulsions and epileptiform discharges, which are removed with conventional anti-seizure medication (Baraban et al., 2005). To capture spontaneous and seizure dynamics I recorded 3 x 30 minute consecutive imaging blocks for each fish: 1) *spontaneous* activity representing normal brain dynamics at rest, 2) *5mM PTZ* giving rise to focal seizures, and 3) *20mM PTZ* causing generalised seizures (Figure 2.2) (Diaz Verdugo et al., 2019a). *Spontaneous* dynamics consisted of low correlation, local ensemble activity and rare, global activity which recruits large parts of the brain (Figure 2.2A). Seizure dynamics consisted of localised, high correlation ensemble activity in the *5mM PTZ* condition (Figure 2.2B) and long-lasting, high-amplitude, synchronous activity recruiting most of the brain in the *20mM PTZ* condition (Figure 2.2C). Data was collected and analysed in 10 fish.

To study spontaneous dynamics and seizure thresholds in foxg1a mutant fish, I recorded 2 x 60 minutes consecutive imaging blocks for each fish: 1) *spontaneous*

**A** Spontaneous

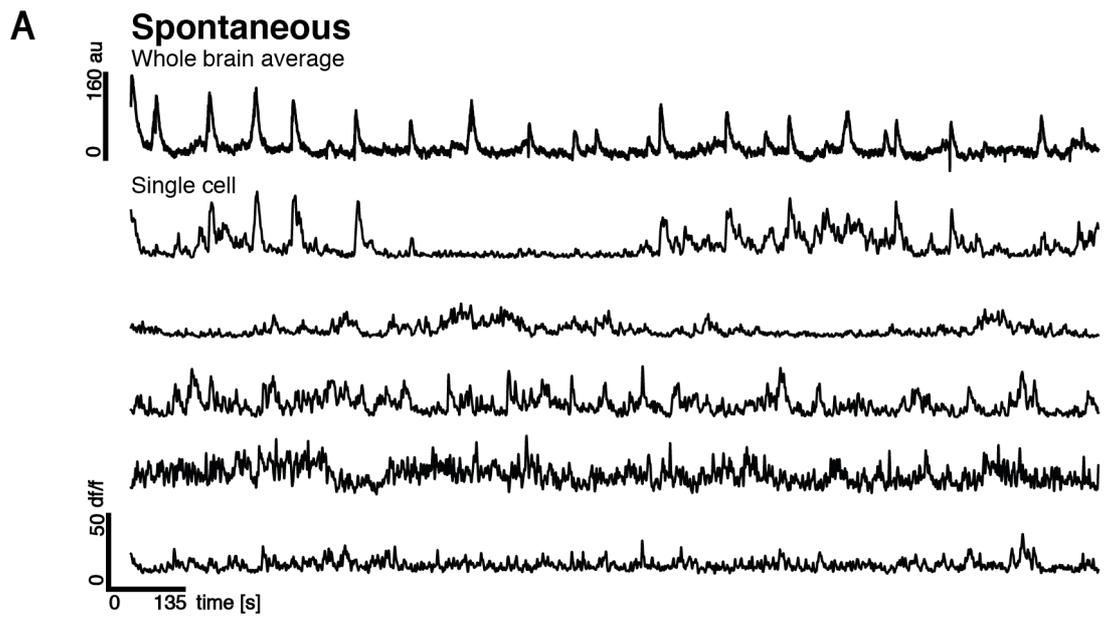

**B**

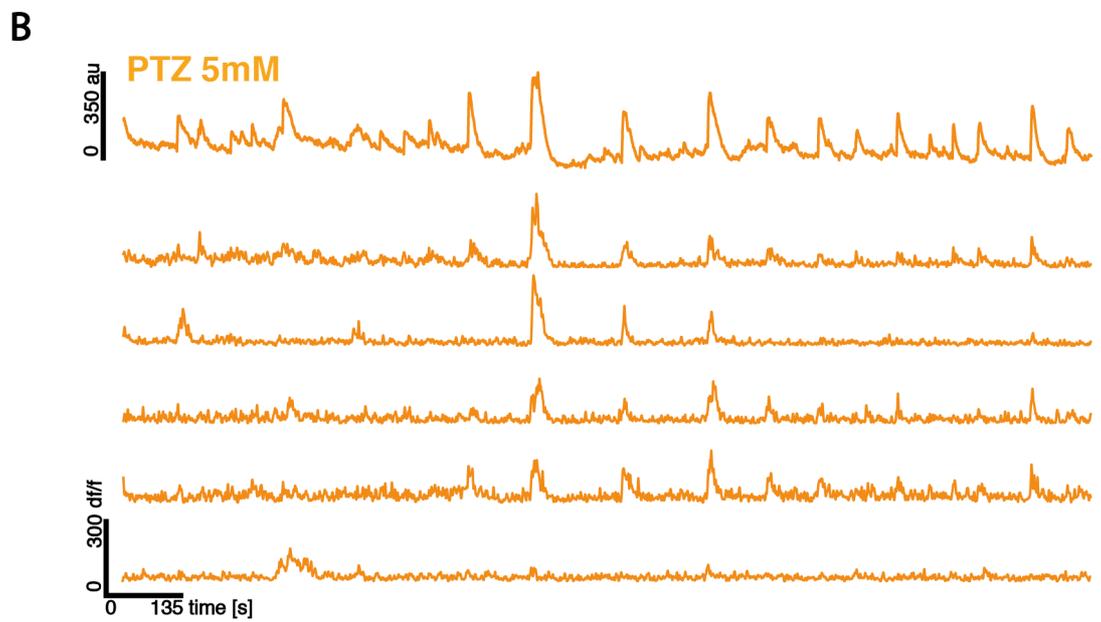

**C**

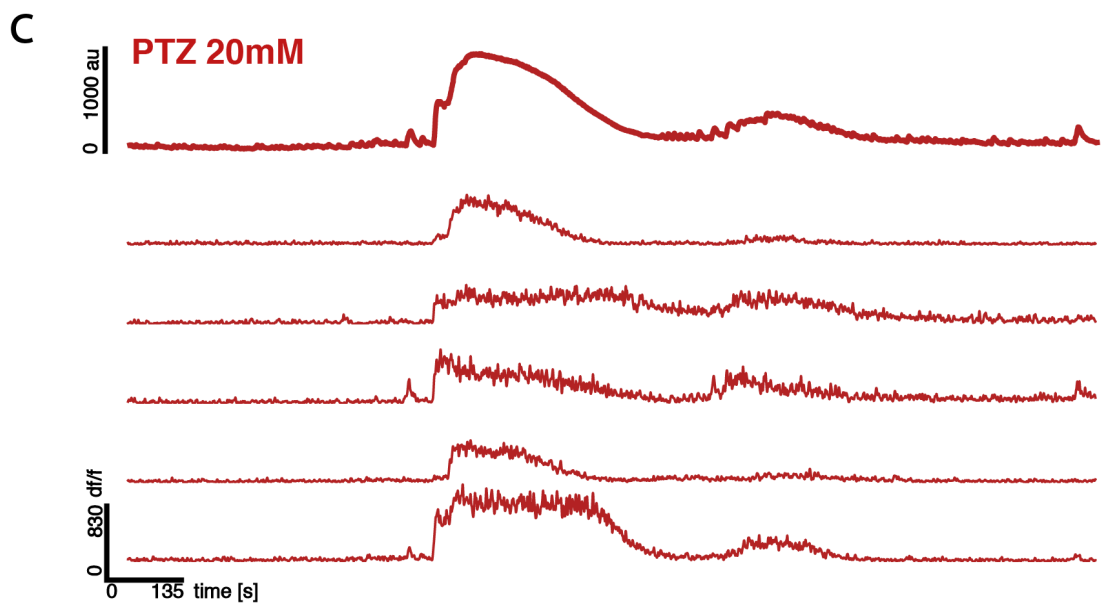

**Figure 2. 2  Single cell and whole brain dynamics during baseline and seizure activity.**

(A-C) Whole brain average (top trace) calcium fluorescence and representative single cell traces (bottom 4 traces) for spontaneous (black), 5mM PTZ (orange) and 20mM PTZ (red) conditions for the entire 30 minute recording period for a representative fish. Spontaneous dynamics (A) showed low correlation interspersed with brief high correlation events. 5mM PTZ dynamics showed regular bursts of correlation (B), while 20mM PTZ showed sustained, high amplitude synchronous cascades (C).

activity representing normal brain dynamics at rest, 2) *2mM PTZ* which is insufficient to cause seizures in wildtype fish but may cause seizures in brain networks that sit closer to the seizure threshold, such as in foxg1a mutants. Data was collected and analysed in 5 foxg1+/+, 8 foxg1+/- and 5 foxg1-/- fish.

For all imaging experiments, PTZ was applied to the bath immediately after each session via a 1ml shot of PTZ suspended in Danieau, after which point imaging was restarted.

### 2.2.4  Calcium Image Registration

The use of nuclear localised GCaMP enables the segmentation of individual neurons according to somatic fluorescence. First recordings were visually inspected for the presence of z-drift, with datasets removed accordingly. Recordings were then corrected for drift in x and y dimensions by using the rigid registration algorithm in Suite2p (Pachitariu et al., 2017). First, a reference image is generated by taking 200 random frames, taking the mean of the top 20 most correlated frames and subsequently aligning the random frames to this reference. The final reference image is then computed as the mean of most aligned frames of the previous step. The registration step is then estimated using phase correlation, which uses spatial whitening to compute the cross-correlation peaks between frames via fast Fourier transforms (Alba et al., 2015).

*2.2.5 Cell Segmentation*

Individual cells were identified using the segmentation algorithm in Suite2p (Pachitariu et al., 2017). Initially, the software models the neuropil signal within each plane to account for signal contamination from the extracellular space due to elongated point spread function in the z-axis in 2-photon microscopes. This estimated neuropil signal is modelled as a series of spatially-localised raised cosine functions, which allows the signal to vary slowly in space. By incorporating this information into a generative model, the software can subtract an estimated neuropil signal from the overall cell signal. Subsequently, Suite2p uses an expectation-maximisation algorithm which estimates the spatial distribution of fluorescence in the recording and the temporal patterning of spikes in fluorescence. During the expectation step the algorithm optimises parameters representing the cellular and neuropil activity by defining cluster membership and scaling for each signal. The maximisation step then determines cellular and neuropil signals using the optimised parameters, by performing linear regression. These steps are repeated until the procedure converges, after which point detected cells are segmented. This enables the segmentation of ~10,000–20,000 cells throughout the whole zebrafish brain (Figure 2.1C).

*2.2.6 Removal of Noisy Cells*

While the segmentation process is effective at identifying most of the cells in the recording, some segmentations may be false positives due to noise. To remove these non-cells, I designed a false-positive detection algorithm which relies on the assumption that true cells would show fluorescence spikes with a slower decay time than shot noise events, due to the decay time of GCaMP6s. The false-positive detection algorithm works as follows: 1) split the trace for each cell into 9 frame windows, 2) find the minimum fluorescence value across each window, 3) find the single maximum value across all minimum window values for each cell. This creates a distribution of max-of-min fluorescence window values (a measure of the fluorescence decay speed), for all segmented cells across the brain. A threshold was then chosen to remove non-cells. This enabled the accurate segmentation of ~10,000-20,000 neurons per fish.

### 2.2.7  Regional Labelling of Segmented Cells

In order to map segmented cells onto their corresponding brain regions, a further registration step was performed. I used the Tg(HuC:H2B-RFP) image stack from the Zebrafish Brain Browser repository as a reference template (Tabor et al., 2019). Non-linear registration was then performed to map a mean image of each 2-photon image plane onto the reference. Specifically, we used the *SyNRA* function in ANTSPy to perform Symmetric normalisation: consisting of rigid, affine and deformable transformations (Klein et al., 2009). Custom python code was used to apply the above registration, mapping an image plane to the reference, to its corresponding segmented cell coordinates, to map the cells onto the reference. Each cell was labelled according to its x-y-z position in the reference space, according to the Gupta atlas which consists of genetic labels from 210 transgenic lines (Gupta et al., 2018).

### 2.2.8  Calcium Signal Normalisation

In order to calculate the normalised fluorescence values for each cell, I first estimated the baseline of the recording. This was done with the underlying assumption that the fluorescence values in each cell are composed of a relatively static baseline and fast calcium events which underlie neuronal activation. For each neuron, the baseline was estimated by calculating the lowest 5th of percentile of the fluorescence for the entire recording. From this baseline, normalised fluorescence values (ΔF/F) were calculated for each time frame by subtracting the raw fluorescence from the baseline, and then dividing by baseline (Figure 2.1D).

### 2.2.9  Detection of Calcium Events

In order to estimate calcium events from the raw fluorescence I applied a Hidden Markov Model, developed by our lab in a previous paper (Diana et al., 2019). Briefly, the method models a neuron as a Bernoulli process with a latent variable $s_t$ representing its hidden on/off states. The fluorescence trace is decomposed into the sum of calcium transients $c_t$ with fixed decay constant $\lambda$ and onset defined by $s_t$, baseline activity $b_t$ and Gaussian noise. The algorithm iteratively maximises the probability of the latent state at time $t$, given the state of $t$ - 1 and the empirical observation of fluorescence value $x_t$. This enables the robust estimation of the onset of exponentially decaying calcium transients. The model requires the selection of parameter $q$, defining the event probability. This was chosen by visual inspection of which $q$ gave the most accurate representation of calcium traces in our data ($q =$ 0.59) (Figure 2.1D).

### 2.2.10   Detection of Seizure State Transitions

For analysis of wildtype fish, data was compared across entire 30 minute imaging blocks (n=10) and shorter 400 frame periods, to examine seizure state transitions (n=9). Generalised seizure transitions were defined as the abrupt appearance of whole-brain, synchronous cascades (Figure 2.3A). In order to define the onset of a state transition (in PTZ 20mm imaging blocks) we required a clear separation between local ictal and globally synchronous ictal events: any brain recordings where the maximum mean brain fluorescence was at least 4x the minimum fluorescence were included in state transition analysis. This removed 1 dataset in which ongoing, oscillatory activity occurred throughout the recording, making it difficult to identify specific generalised state transitions (Figure 2.3B). For the remaining 9 datasets, to identify the beginning of generalised state transitions we used a 30 frame sliding window over the mean whole-brain fluorescence trace, calculating the maximum change from the start of the window and the subsequent 29 frames. The window with the highest difference was defined as the start of the generalised state transition.

In order to compare dynamics across state transitions, we compared activity from 400 frames just before the generalised seizure (*pre-ictal*), the first 400 frames of the generalised seizure (*ictal-onset*) and 400 frames from a randomly selected segment of spontaneous recording (*baseline*) (Figure 2.3A).

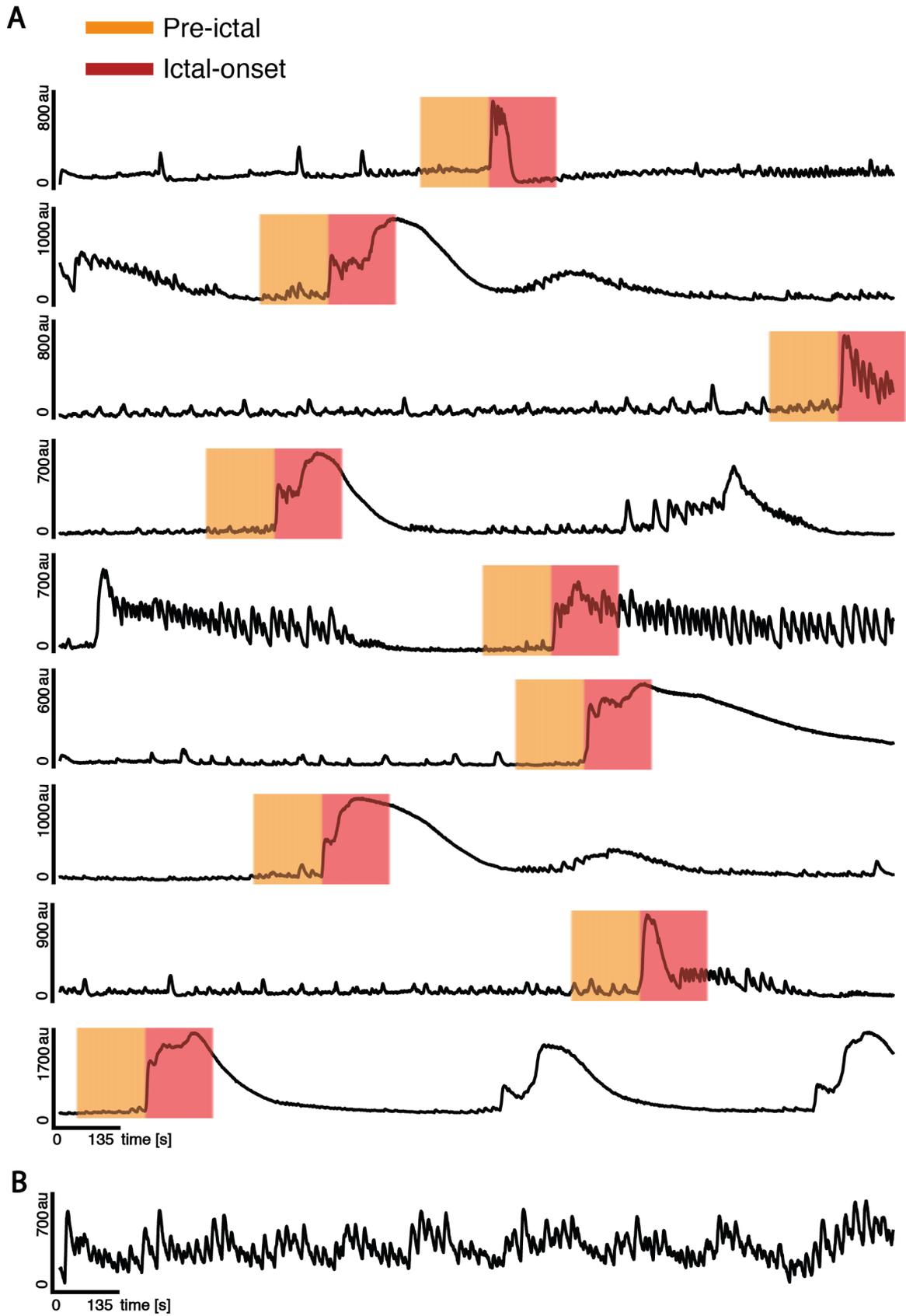

**Figure 2. 3   Identification of seizure state transitions.**

(A) Mean whole brain traces showing 400 frame pre-ictal (orange) and ictal-onset (red) periods used for 9 datasets. Start of state transition was identified using a 30 frame sliding window, with the window with the maximal difference over the preceding 29 frames identified as the start of the ictal-onset period. (B) One dataset without clear state transitions from pre-ictal to ictal-onset state was removed.

## 2.3   Statistical Analysis & Resources

### 2.3.1  Statistical Tests
D'agostino's $K^2$ test was used to test for normality in data distributions ($\alpha = 0.05$). Paired t-tests or Wilcoxon signed-rank tests were used to compare *spontaneous, 5mM PTZ*, *20mM PTZ,* and state transition datasets in cases of normality, and non-normality respectively ($\alpha = 0.05$). Independent t-tests or Mann-Whitney U tests were used to compare different foxg1a genotypes and network models in cases of normality, and non-normality respectively ($\alpha = 0.05$). Bonferroni corrections were used for multiple comparisons.

### 2.3.2  Software
Data was analysed using custom code written in Python. Image registration and cell segmentation was performed using Suite2p (Pachitariu et al., 2017). Neural network simulations were run using Brian2 (Stimberg et al., 2019). Convergent cross mapping was run on Lorenz data using custom written Python code, and on whole brain datasets using kEDM (Takahashi et al., 2021). Statistical hypothesis tests were performed using scipy. Graphs were generated using matplotlib and seaborn.

### 2.3.3  Resources
Details of key reagents and resources are included in Table 1. Further information and requests for resources should be directed to the author (dominic.burrows@kcl.ac.uk).

| Regent or Resource | Source | Identifier |
|---|---|---|
| Tg(elavl3:H2B-GCaMP6s) | (J. Freeman et al., 2014) | ZDB-TGCONSTRCT-141023-1 |
| Foxg1 TALEN x Tg(HuC:H2BGCaMP6s) x casper lines. | (Bruce, 2022) | |
| Scanimage Software - Vidrio Technologies (image acquisition) | | https://docs.scanimage.org/ |
| Suite2p (image registration and segmentation) | (Pachitariu et al., 2017) | https://github.com/MouseLand/suite2p |
| hidden markov model (calcium transient estimation) | (Diana et al., 2019) | https://github.com/giovannidiana |
| Brian2 (Python network modelling software) | (Stimberg et al., 2019) | https://github.com/brian-team/brian2 |
| Kokkos – Empirical Dynamic Modelling | (Takahashi et al., 2021) | https://github.com/keichi/kEDM |

*Table 2. 1*   **Key resources table.**

### 2.3.4  Data Availability

Datasets supporting the thesis are available upon request from the author (*dominic.burrows@kcl.ac.uk).*

### 2.3.5  Code Availability

Custom written Python code can be accessed at:

1. Calcium imaging pre-processing:

   https://github.com/dmnburrows/img_process

2. Seizure state transition detection:

   https://github.com/dmnburrows/seizure_detection

3. Avalanche analysis and calculation of critical statistics:

   https://github.com/dmnburrows/criticality

4. Spiking network modelling of avalanche dynamics and network response properties:

   https://github.com/dmnburrows/avalanche_model

5. Metastability and covariance eigenspectra analyses:

   https://github.com/dmnburrows/seizure_dynamics

6. Convergent cross mapping & lagged coordinate embedding:

   https://github.com/dmnburrows/empirical_dynamic_modelling

# Chapter 3

# Critical Avalanche Dynamics in EI Balanced and Imbalanced Networks

This chapter contains text, data and figures from some of my work currently in preprint (Burrows et al., 2021).

## 3.1   Introduction

A delicate balance between excitation and inhibition supports the diverse repertoire of neural dynamics needed for behaviour. This excitation-inhibition (EI) balance manifests across scales, with invariant ratios of excitatory to inhibitory synapses along dendritic segments (Iascone et al., 2020) and finely balanced proportions of neuronal subtypes in cortical circuits (Cobos et al., 2005). Such synaptic and circuit EI balance gives rise to proportional, phase-locked excitatory-inhibitory responses, as measured *in vivo* during spontaneous activity (Okun & Lampl, 2008), neural oscillations (Atallah & Scanziani, 2009), sensory-evoked activity (B. Liu et al., 2011) and behaviour (Zhou et al., 2014). Balanced dynamics are likely important for healthy brain function, as multiple neurodevelopmental disorders are linked to EI imbalance, including schizophrenia, autism and epilepsy (Fritschy, 2008; R. Gao & Penzes, 2015; Lee et al., 2017; Žiburkus et al., 2013). Our understanding of the role of EI balance in optimising computation *in vivo* comes from small population recordings, suggesting roles in shaping receptive fields (B. Liu et al., 2010), supporting coincidence detection (Wehr & Zador, 2003) and enabling gain control (Bhatia et al., 2019). How EI balance shapes population dynamics and computation across whole-brain neuronal networks remains an open question, due to the technical difficulties of collecting and analysing such high dimensional datasets.

One appealing solution is to borrow concepts from statistical physics, which estimates microscopic variables probabilistically to explain emergent macroscopic properties in complex systems. Using such approaches, it has been claimed that neuronal populations exhibit dynamics analogous to atomic spins in a ferromagnetic lattice whose interaction strengths are tuned to the transition point between order and chaos, known as criticality (Bak et al., 1987, 1988; Sethna et al., 2001). Interestingly, the spontaneous collective behaviour of a system at a phase transition can give rise to optimal computational properties (Langton, 1990; Packard, 1988). Although increasingly evidence suggests the brain may operate in a near-critical state (Priesemann et al., 2014), various signatures of criticality in the brain have been documented empirically: avalanche power-law distributions, exponent relations, universal scaling and critical branching (Harris, 1963; Sethna et al., 2001), have been reported in *in vitro* cultures (Beggs & Plenz, 2003), *in vivo* calcium imaging (Ponce-Alvarez et al., 2018), functional MRI (fMRI) (Tagliazucchi et al., 2012), electroencephalography (EEG) and magnetoencephalography (MEG) recordings (Arviv et al., 2016; Meisel et al., 2012). Importantly, *in silico* evidence suggests that EI balanced networks give rise to scale-invariant avalanches (Poil et al., 2012) – spatio-temporal cascades of neural activity that span the entire scale of the system, a key feature of criticality. Furthermore, perturbing EI balance *in vitro* removes signatures of criticality (Haldeman & Beggs, 2005; Shew et al., 2011), while inhibitory synaptic connectivity can shape critical dynamics (Lombardi et al., 2017; Ma et al., 2019). Therefore, the tendency of the brain to homeostatically maintain EI balance (Pan-Vazquez et al., 2020; van Vreeswijk & Sompolinsky, 1996) could serve to stabilise its dynamics to a phase transition, a phenomenon known as self-organising criticality (Bak et al., 1988). The identification of EI balance as a regulating parameter for critical dynamics *in vivo* would demonstrate a biological mechanism through which brain networks stabilise their dynamics within a critical, optimal range. While the search for control parameters *in silico* points to plasticity mechanisms which may set EI balance (Levina et al., 2007; Ma et al., 2019; Zeraati et al., 2021), the role of EI balance has yet to be tested *in vivo.*

Given the link between EI balance and criticality, it has been theorised that EI imbalance-associated brain dysfunction may emerge as a deviation from the critical point (Zimmern, 2020). In fact, epileptic seizures and epileptiform activity, which emerge due to pathological EI imbalances, are associated with a loss of critical statistics in EEG and MEG in patients with epilepsy (Arviv et al., 2016; Hobbs et al., 2010; Meisel et al., 2012). This has led to the prediction that epileptic seizures manifest as a supercritical state (Beggs & Plenz, 2003), in which inputs exponentially grow in time thus saturating the system (Harris, 1963), and dynamics become more chaotic (Haldeman & Beggs, 2005), resulting in a loss of functional network stability. However, whether the single neuron dynamics driving brain-wide network instability during epileptic seizures (See Section 1.1), collectively give rise to a supercritical state away from criticality, is unknown. Furthermore, the extent to which the abnormal brain dynamics associated with EI imbalance disorders may also occur due to the emergence of supercritical dynamics is currently unexplored (Shen et al., 2019; Wong et al., 2019; Zimmern, 2020). Showing this experimentally, would demonstrate a dynamical mechanism through which a loss of synaptic EI balance gives rise to the unstable network dynamics of whole-brain networks which are characteristic of epilepsy and EI imbalance disorders.

A major roadblock to accurately assessing critical statistics in brain networks is the technical difficulty of recording neural activity at sufficient spatial scale and resolution. Conventional macroscale recording techniques, such as EEG, coarse-grain the underlying dynamics which obscures the heterogeneity of cellular activity and could alter critical statistics (Keller et al., 2010; Meyer et al., 2018; Muldoon et al., 2013). Conversely, conventional microscale recordings subsample the network which can produce spurious critical statistics (Priesemann et al., 2009). Furthermore, microscale approaches also typically record from a single brain region, and given that distinct areas may be differently tuned to criticality this might bias any inferences to that of the specific brain area (Suryadi et al., 2018). Therefore, to make strong claims about the critical nature of entire neuronal systems we require a system enabling single-cell recordings across the whole brain. Here we take advantage of the transparency of the larval zebrafish to perform *in vivo* functional imaging of the whole brain at single cell resolution (Ahrens et al., 2013). A recent study has already demonstrated the presence of critical statistics in the larval zebrafish brain (Ponce-Alvarez et al., 2018). In this chapter, I measure avalanche dynamics in spontaneous and EI-imbalanced dynamics, using both pharmacological and genetic models of EI imbalance, to assess the presence of criticality in cellular resolution whole brain networks.

I test 3 key hypotheses: i) whole brain single cell networks are organised around criticality, ii) EI balance shapes critical dynamics *in vivo*, and iii) EI imbalance due to a loss of inhibition emerges as a supercritical state away from criticality. We can make several predictions based on critical systems theory to test our hypotheses. Firstly, key statistical markers of criticality should be present in spontaneous neuronal dynamics to provide evidence for critical dynamics. Secondly, changing a regulating control parameter for critical dynamics should give rise to a sharp change in critical statistics (Binney et al., 1992). Thirdly, for EI-imbalance disorders to emerge as a supercritical state we would expect the emergence of bimodal avalanche distributions, a loss of critical exponent relationships and increases in branching ratios above 1 (Beggs & Plenz, 2003; Harris, 1963). I demonstrate spontaneous dynamics exhibit key statistical markers of criticality, suggesting that single cell dynamics organise macroscale brain dynamics to a phase transition. Furthermore, graded perturbations to EI balance give rise to abrupt changes in critical statistics, indicating a role for EI balance in shaping critical dynamics *in vivo*. Finally, EI imbalanced networks exhibit some properties expected for a supercritical network driven away from a phase transition, suggesting that EI imbalance dynamics may emerge due to a loss of criticality.

## 3.2   Results

### 3.2.1   The Estimation of Neuronal Avalanches

A key dynamical property of neuronal systems at criticality is the presence of neuronal avalanches – cascades of neural activity that propagate in space and time. Given that the underlying synaptic connectivity is not known, one must infer the flow of activity from multiple time series. Traditionally, studies that measure neuronal avalanches have done so within a limited spatial window and thus have assumed that all neurons can connect with one another (Beggs & Plenz, 2003). However, given that my recordings cover a broad area over most of the zebrafish brain, and that neurons generally exhibit a strong propensity for local connections across species (Kaiser, 2014), I introduce a spatial constraint limiting neuron-neuron activity flow. To do this I adapt methods used in previous studies to estimate avalanche size and duration (Ponce-Alvarez et al., 2018; Tagliazucchi et al., 2012).

Neuronal avalanches are estimated using binarised calcium traces from zebrafish imaging data (See Section 2.2). First, I assume that neuron $u$ can only activate other neurons in the population $P$ that lie within a local neighbourhood $N_{u,v}$ where $v$ represents the closest $k$% of cells to $u$. This introduces a spatial constraint to avalanche propagation such that activity can only flow between putatively connected neurons, preventing disparate cascades combining into one large avalanche (Figure 3.1). $k$ was selected (0.16%) so that the mean neighbourhood radius (15.5um ± 0.78) lay within the neighbourhood range used to calculate critical statistics in previous studies (Ponce-Alvarez et al., 2018).

I adapt the avalanche calculation algorithm defined by Ponce-Alvarez et al., for single cell data. Briefly, an avalanche begins when at least 3 cells within a neighbourhood $N_{u,v}$ are active at $t_x$, that were not part of an avalanche in the previous timestep. Here I label the neurons currently active at $t_x$ which belong to the avalanche as the set $A_x$ = {$a,b,c...z$}. All active neurons of $P$ at $t_x$ connected to any of $A_x$ via a neighbourhood are included into the set $A_x$.

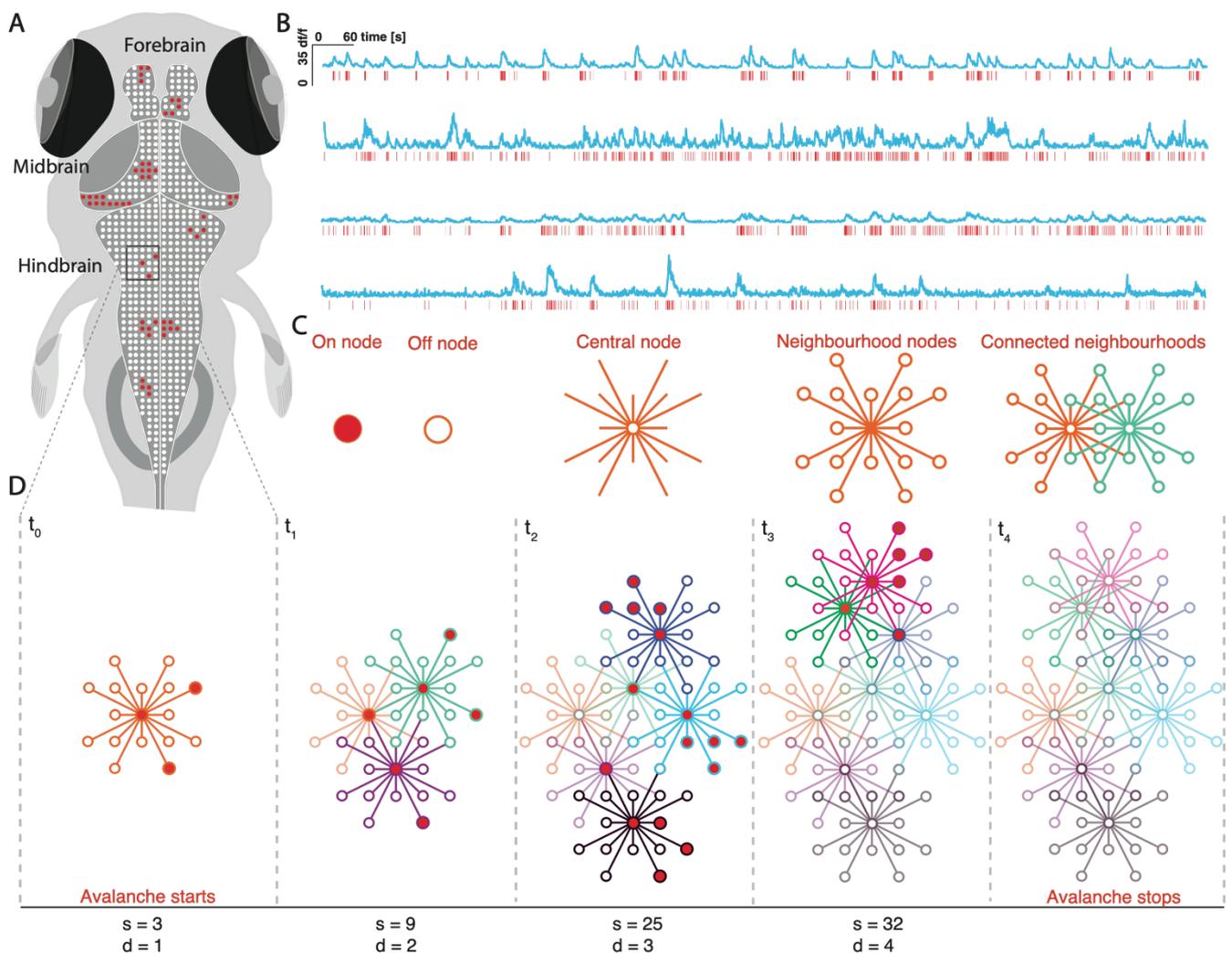

**Figure 3. 1   Neuronal avalanche estimation algorithm.**

(A) Neurons exhibit on (red circles) and off (white circles) dynamics giving rise to ensembles that grow in space and time. (B) Hidden Markov model estimation of calcium transients showing normalised traces (blue), and estimated calcium transients (red). (C) Avalanche calculation legend. (D) For avalanches to begin, at least 3 nodes within the same neighbourhood must be active at t0. To propagate in time, any avalanche node active at t0 must also be active at t1. Once this step is satisfied, any nodes active at t1 that are connected via a neighbourhood to avalanche nodes at t1 become avalanche members. Avalanches terminate when no more nodes are active. Avalanche size (s) is the total number of activations and duration (d) is the number of time steps of the avalanche.

At each subsequent step avalanche propagation iterates as follows (Figure 3.1):

1. If at least one neuron that was part of avalanche $A_x$ is also active at $t_{x+1}$ then the avalanche continues in time, where any active neurons carried forward in time form the set $A_{x+1}$ = {$a,b,c...z$}.
2. Any neurons from $P$ active at $t_{x+1}$ that are connected to any of $A_{x+1}$ via a neighbourhood are included into the set $A_{x+1}$.

Once step 1 is no longer satisfied the avalanche terminates. Avalanches whose active neighbourhoods converge are grouped into a single avalanche. Using this approach, I was able to track the progression of avalanches during each imaging period for each fish. For each avalanche, its size was calculated as the total number of binarised events during the avalanche, and its duration was the number of time steps for which the avalanche was active (Figure 3.1). Next, I constructed probability distributions for avalanche size and duration for each fish, which exhibited long tails on logarithmic axes indicating a wide range of avalanche sizes and durations (Figure 3.2).

The neuronal avalanche framework assumes that activity spreads through the network via action potentials passing between physically connected neighbours – that is, avalanches emerge out of the spatial and temporal structure of network activity. However, purely stochastic dynamics if thresholded can give rise to apparent neuronal avalanches, even in the absence of spatial and temporal structure (Touboul & Destexhe, 2010). Applying avalanche estimation procedures to such datasets would lead to the false notion that neuronal activity was propagating through the network, even when individual units spike independently. Therefore, to confirm that neuronal avalanches indeed occurred in my data, I compared empirical avalanche distributions from resting state datasets (n=10) with a series of null distributions, to understand whether my avalanche distributions could emerge in spatially and temporally random systems. The first null model is the *spatial* null, which tests the hypothesis that empirical avalanche dynamics emerge from local neighbourhoods of neurons putatively connected neurons, which is the central assumption of the neighbourhood approach described above. The *spatial* null was generated by randomly shuffling the locations of neurons such that a neurons' neighbourhood would consist of neurons scattered across the brain rather than nearby neurons. The

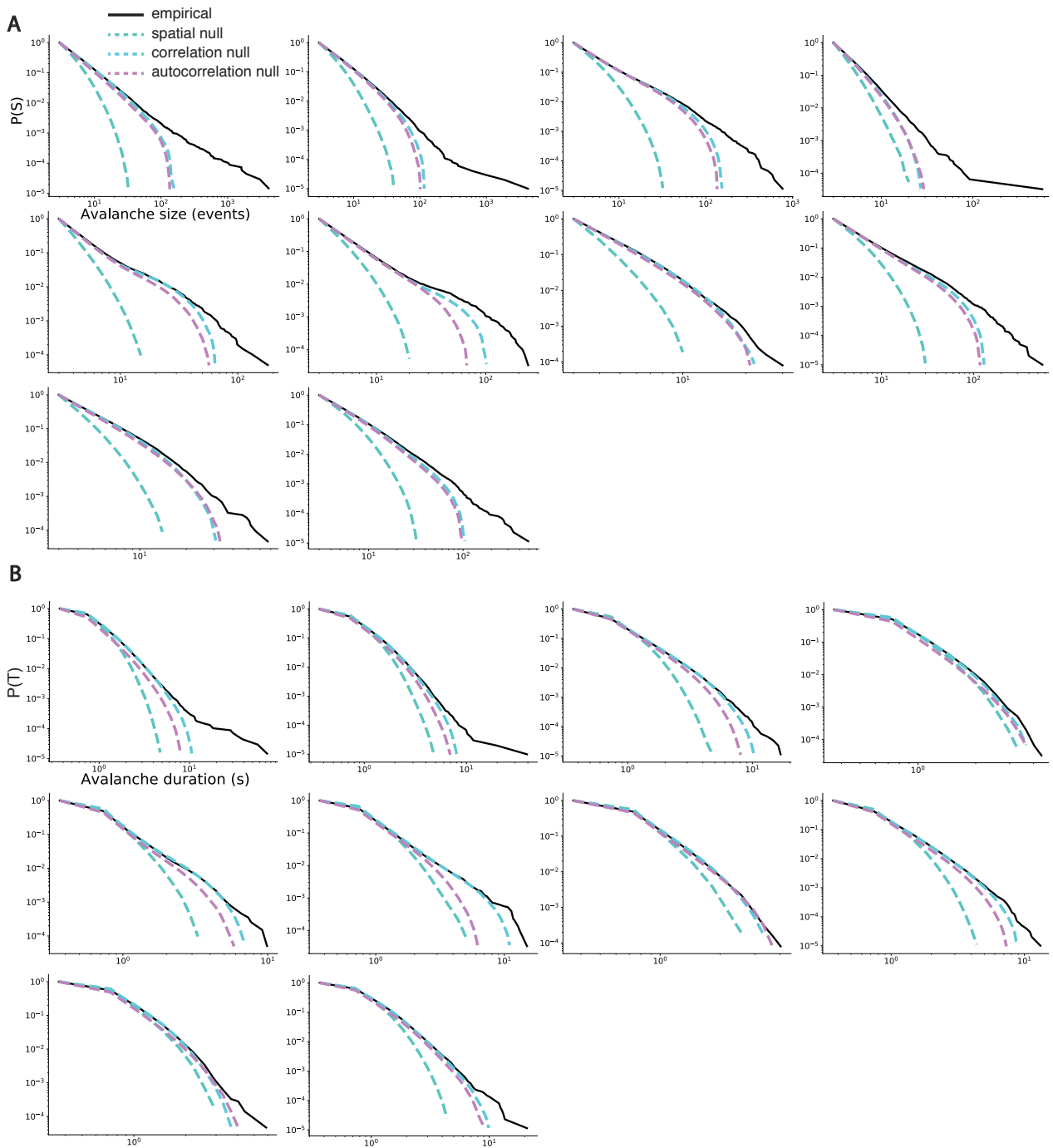

**Figure 3. 2   Null models demonstrate temporal and spatial structure in empirical avalanche distributions.**

(A-B) Complementary cumulative distribution functions for avalanche size (A) and duration (B), comparing empirical (black) with the 3 null models, spatial (teal), correlation (cyan) and autocorrelation nulls (magenta) for each fish. Null models fail to generate sufficient heavy tails for empirical avalanche distributions, suggesting the importance of non-random spatial and temporal structure in underlying data.

second null model is the *correlation* null, which tests the hypothesis that observed avalanche dynamics are generated by the correlated fluctuations of neurons in the network. The correlation null was generated by circularly permuting the time series independently for each neuron, such that the correlations between neurons was lost, while retaining the autocorrelation structure within each timeseries. The third null model is the *autocorrelation null,* which tests the hypothesis that observed avalanche dynamics are generated by the integration of activity over time within neurons, rather than from shot noise (Touboul & Destexhe, 2010). The *autocorrelation* null was generated by randomly shuffling the binarised activity of all neurons in a dependent fashion, such that autocorrelation was lost but the cell-cell correlation retained. In order to compare avalanche distributions from empirical and null datasets, 50 nulls were randomly generated for each fish and for each null type, with avalanches calculated as above.

Using this null modelling approach, I was able to assess whether randomly generated neuronal activity, resulting from shuffling spatial structure, cell-cell correlation and autocorrelation respectively, would generate statistically similar avalanche distributions (Figure 3.2). This would imply that my empirical avalanches emerge in the absence of spatial or temporal structure, and thus the avalanche framework is not appropriate to describe the dynamics. Upon visual inspection I found that, all null model distributions exhibit faster exponential decay at the tail ends of their distributions than empirical distributions (Figure 3.2). This indicates that spatial and temporal structure is necessary to generate long and large avalanches. To statistically compare empirical and null avalanche distributions, I first calculated the area under the curve (AUC) for each distribution. The AUC for a given empirical distribution was then compared with the AUC of each of its 50 null distributions. The mean AUC difference therefore represents the average difference in the shape of an empirical distribution to its null distribution, which is measured in the absence of temporal or spatial structure (Figure 3.3). To test for the significance of this difference, I compared the mean AUC difference between empirical-null distributions, with the expected difference as calculated between null-null distributions (i.e. the within group difference for a given null distribution). Comparing this distribution of mean AUC differences, I found that empirical avalanche distributions were significantly different to null distributions, across all 3 null model types for avalanche size (*spatial*: t = 4.88, p < 0.01, *correlation*: t = 2.96, p < 0.01, *autocorrelation*: t = 6.09, p < 0.001) and duration (*spatial*: t = 11.2, p < 0.0001, *correlation*: t = 2.76, p <0.05, *autocorrelation*: t = 3.56, p < 0.01). This suggests, that empirical avalanche distributions emerge out of the local interactions between contiguous neurons (spatial null), their correlated firing (correlation null) and the long decay of activity within each neuron (autocorrelation null). Of particular note, the *spatial* null distributions exhibit the fastest decay which indicates that locally constrained activity is central to the propagation of empirical neuronal avalanches (Figure 3.2). Therefore, my avalanche algorithm captures key spatial and temporal structure in the underlying data and thus may serve as a useful description of population dynamics.

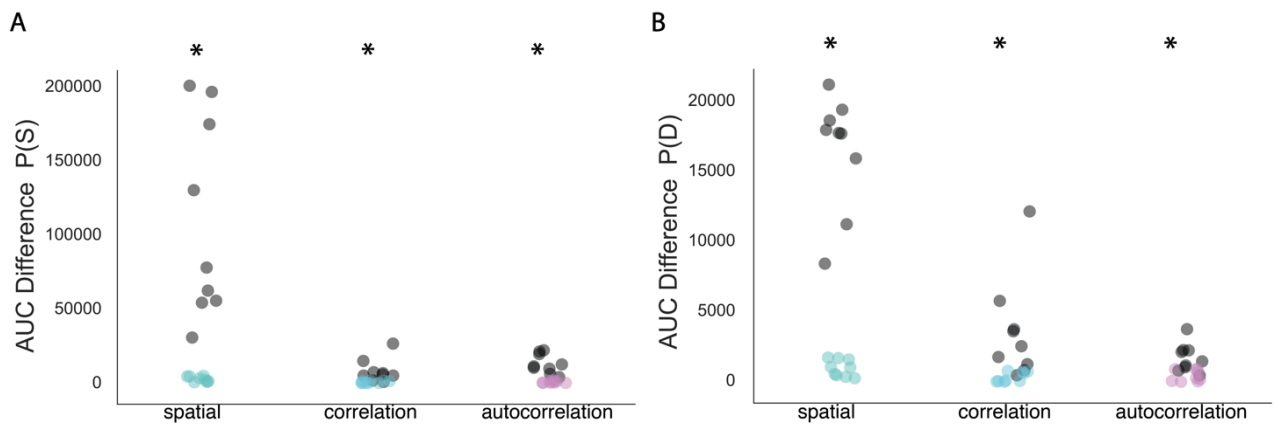

**Figure 3. 3   Null models show spatial and temporal structure is required for the generation of empirical avalanche distributions.**

Mean AUC differences across empirical and null avalanche distributions, for avalanche size (A) and duration (B). AUC differences were computed for each empirical avalanche distribution (black) against each of its 50 null distributions, with an average taken. AUC differences for spatial (turquoise), correlation (cyan) and autocorrelation (magenta) null distributions were also computed across all null distribution pairs within the same group. Empirical distributions are significantly more different from null distributions than each null model, indicating that spatial, correlation and autocorrelation are all required to generate empirical avalanche distributions *: p < 0.05.

To probe the role of EI balance in shaping critical dynamics, I first wanted to measure avalanche dynamics in networks where the balance between excitation and inhibition is within a physiological range. I reasoned that healthy brain networks should exhibit synaptic levels of excitation and inhibition which are appropriately balanced for behaviour – as such, spontaneous avalanche dynamics in wildtype fish should emerge out of synaptic EI balance. Therefore, I used spontaneous brain dynamics as a readout of the EI balanced state.

First, to confirm the presence of critical dynamics in whole brain spontaneous activity, as previously reported (Ponce-Alvarez et al., 2018), I tested for the presence of scale-invariant neuronal avalanches – avalanches of varying size and duration, spanning the entire range available in the system which is a defining feature of criticality. Scale-invariant neuronal avalanches at criticality give rise to power law relationships for avalanche size and duration (See Section 1.1.3). To demonstrate why power laws should emerge in critical systems, I use a scaling argument derived by Sethna et al., which assumes that critical systems are self-similar (Sethna et al., 2001). If we know that a function should be self-similar we can demonstrate that power laws will emerge.

A self-similar function $F(x)$, must equal itself rescaled by some factor, such that

$$F(x) = (1 - \alpha\delta)\, F\big(x(1 - \delta)\big). \quad (3.1)$$

Here, $\delta$ is a small shift in our function input $x$ - $x\delta$ and output $F(x)$ - $\alpha\delta(F(x))$, which represents rescaling our function by some factor. $\alpha$ is a constant that relates a change to the input $\delta$ with the corresponding change in the output.

By rearranging 3.1 one can show that

$$F(x) = (1 - \alpha\delta)\, F(x - x\delta)$$

$$F(x) = F(x - x\delta) -\ \alpha\delta F(x - x\delta)$$

$$F(x) - F(x - x\delta) = -\alpha\delta F(x - x\delta)$$

$$\frac{F(x) - F(x - x\delta)}{\delta} = -\alpha F(x - x\delta). \quad (3.2)$$

Finally, by multiplying the left side of equation 3.2 by $x$ and letting $\delta$ approach 0, we get an expression for the derivative of $F$ in terms of $x$, which can be rearranged to show that

$$\frac{xF(x) - xF(x - x\delta)}{\delta x} = -\alpha F(x - x\delta)$$

$$x\frac{dF}{dx} = -\alpha F(x)$$

$$\int \frac{dF}{F(x)} = -\alpha \int \frac{dx}{x}$$

$$\ln F = -\alpha \ln x$$

$$F(x) = x^{-\alpha}, \quad (3.3)$$

where $F(x)$ describes a power law relationship between $F$ and $x$. Therefore, any self similar system, such as one at criticality, will exhibit power laws in its observable variables.

Specifically, critical systems should exhibit power laws for avalanche size $S$ and duration $T$ with the following form:

$$P(S) \sim S^{-\tau}, \quad (3.4)$$

$$P(T) \sim T^{-\alpha}. \quad (3.5)$$

In order to statistically evaluate the presence of power-laws in my avalanche data, I use an importance sampling approach to compare the likelihood of my avalanche distributions being generated by a power law compared to a candidate distribution, using an algorithm designed by Dr Giovanni Diana (Burrows et al., 2021). Here I used the log normal distribution as the alternative distribution, as it is the most rigorous comparative candidate distribution for power laws tests (Alstott et al., 2014).

For a given power law exponent ε, the likelihood is given by

$$\frac{1}{[z(\varepsilon)]^M} \prod_{k=1}^{M} s_k^{-\varepsilon}, \quad (3.5)$$

 with avalanche size/duration defined from $s_1,\ s_2,\ s_3 \dots s_M$.

The normalisation constant $z(\varepsilon)$ is a function of the exponent

$$z(\varepsilon) = \sum_{s=a}^{b} s^{-\varepsilon}, \quad (3.6)$$

where *a* and *b* are the maximum and minimum avalanche cutoffs (*avalanche size: a* = 3, *b* = maximum observed for each fish, *duration: a* = 2, *b* = maximum observed for each fish).

The log likelihood (log L) is then defined as

$$logL = -M \log(z) - \varepsilon \sum_k \log(s_k). \quad (3.7)$$

Log likelihoods were calculated across a range of sampled ε, which were weighted by the log probability of observing $\varepsilon_i$ from the prior and proposal distributions. To be precise, the weight $w_i$ for exponent $\varepsilon_i$ is given by

$$w_i = \frac{likelihood(data|\varepsilon_i)\ prior(\varepsilon_i)}{proposal(\varepsilon_i)}. \quad (3.8)$$

 Marginal likelihoods (ML) were then estimated as the empirical means of all the weights

$$ML = \frac{1}{N} \sum_k w_k. \quad (3.9)$$

Finally, the likelihood ratio (LR) was calculated as

$$LR = MLpowerlaw - MLlognormal, \quad (3.10)$$

with positive values indicating the presence of power law distributions. I applied this method to my empirical distributions for avalanche size and duration, to statistically assess the presence of power laws. Avalanche distributions for most fish approximated linear distributions on log-log axes (Figure 3.4). Importantly, using log likelihood ratio testing, we find that all datasets are better explained by power law than lognormal distributions, as demonstrated by the fact that all LR values are positive. This indicates that EI balanced activity in spontaneous whole brain networks exhibits scale-invariant avalanche dynamics, as expected at criticality.

Next, I wanted to measure the exact values of avalanche exponents in my data. Various studies have claimed that avalanche dynamics exist at a critical point, belonging to a mean-field-directed percolation universality class ($\tau$ = 1.5, $\alpha$ = 2.0) (Zapperi et al., 1995), or the 3 dimensional random field Ising class ($\tau$ = 2.0, $\alpha$ = 2.8) (Perković et al., 1995). Interestingly, my avalanche exponents are higher than expected for either class, for both avalanche size ($\tau$ = 2.98 ± 0.13) and duration ($\alpha$ = 3.58 ± 0.14) (Figure 3.4). Avalanche exponents for both size and duration were highly variable, suggesting a diversity of brain states captured in the spontaneous recording period. Therefore, this indicates that our resting state neuronal dynamics do not belong to the directed percolation or Ising model universality classes, which have been previously linked to brain activity (Beggs & Plenz, 2003; Ponce-Alvarez et al., 2018). Nonetheless, the presence of scale invariant avalanche dynamics are an important indicator of criticality, while whole brain larval zebrafish dynamics could belong to different universality classes than previously suggested.

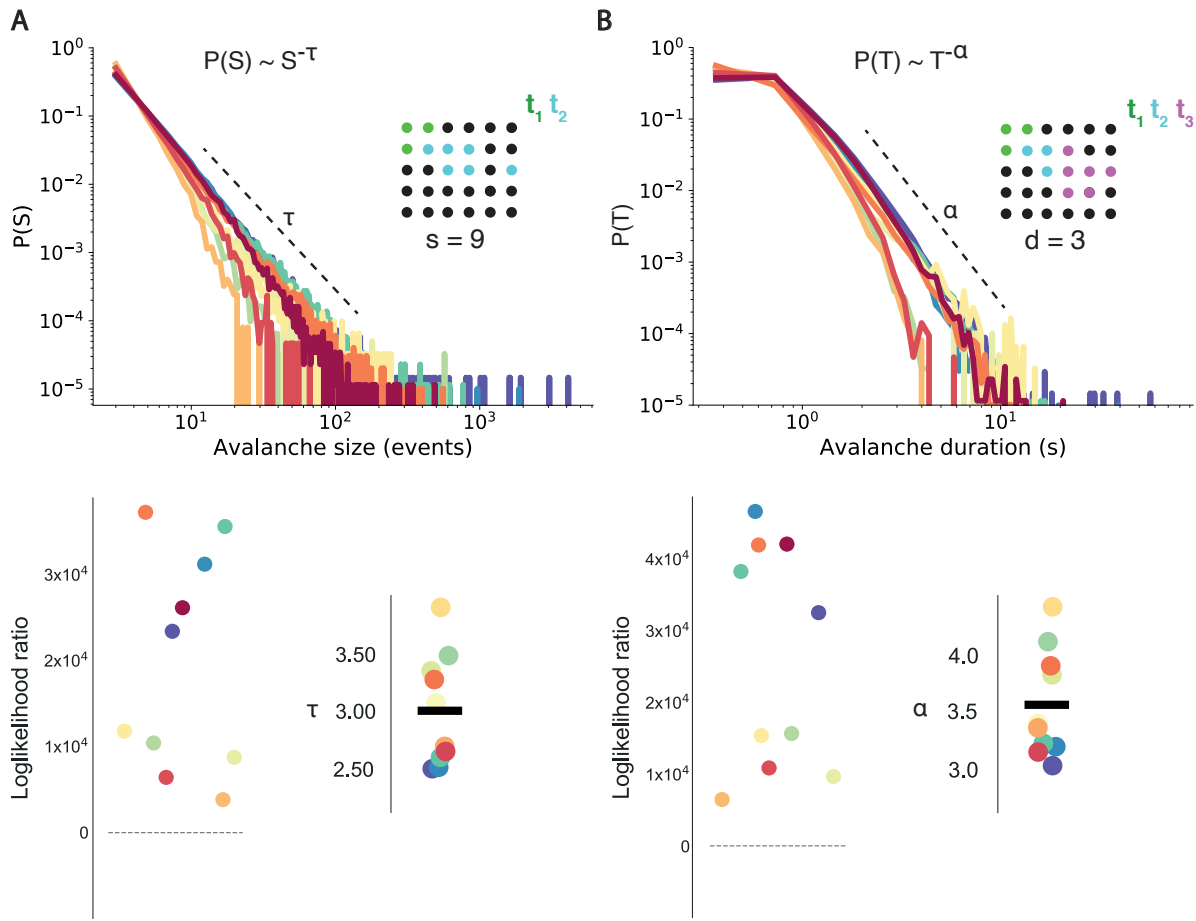

**Figure 3. 4    Power law distributions for avalanche size and duration in spontaneous data.**

Empirical distributions for avalanche size (A) and duration (B) coloured by fish, with exponents estimated from power-law slope (dotted line) (top). (top, right) Avalanche schematic demonstrating calculation of avalanche size (A) and duration (B) for single avalanche events. Black neurons are off. Coloured neurons represent active neurons at time point $t_x$ . Loglikelihood ratio tests for size (A) and duration (B) comparing power-law versus lognormal distributions. Positive values indicate datasets are better explained by power-law than lognormal distributions (bottom left). Power law exponents plotted for avalanche size (A) and duration (B).

### 3.2.3 Scaling Exponents & Exponent Relations in Spontaneous Activity

While power laws are indeed found in systems at criticality, they can emerge in non-critical and stochastic systems, and therefore other markers are needed as further proof of critical dynamics (Miller, 1957). Another key property of criticality, is that if equations (3.4) and (3.5) hold, then there must be a scaling relation between avalanche size $S$ and duration $T$, as

$$\langle S \rangle(T) \sim T^{\frac{1}{\zeta \nu z}}, \quad (3.11)$$

where $\langle S \rangle$ (T) is the average size for a given duration, and $1/\zeta \nu z$ is a combination of critical exponents, where $\zeta$ defines the dependence of the power law cut off, $\nu$ defines the divergence of the size of avalanches, and $\nu z$ defines the divergence of the duration of avalanches, as the critical point is approached (Perković et al., 1995). Put simply, the expected size of an avalanche should scale with its duration as a power law, with exponent $1/\sigma \nu z$. This occurs because many critical systems exhibit self-similarity across space and time scales, such that a single scaling exponent can describe the relationship between size $S$ and $T$ at different scales.

Therefore, to assess the presence of a scaling relationship between $S$ and $T$ in spontaneous activity data, I plotted the mean size of an avalanche $\langle S \rangle$ against its duration and estimated the slope of this curve using linear regression (Figure 3.5A). Interestingly, the relationship (3.11) clearly holds for the majority of my data, with all plots falling onto a power law relationship as expected for a scaling function at criticality (Fontenele et al., 2019). The scaling exponent $1/\sigma \nu z$ for spontaneous activity was highly variable across fish ($1/\sigma \nu z = 1.31 \pm 0.07$) (Figure 3.5C), and somewhat lower than theoretically derived values for the random field Ising model, which estimates $1/\sigma \nu z \sim 1.75$ (Perković et al., 1995). Interestingly, my scaling exponents clearly map onto those observed in critical cortical recordings *in vivo* ($1/\sigma \nu z \sim 1.28$) (Fontenele et al., 2019) and *in vitro* ($1/\sigma \nu z \sim 1.3$) (Friedman et al., 2012). Therefore, while spontaneous avalanche dynamics may not belong to a directed percolation or Ising model universality class (Ponce-Alvarez et al., 2018), they shares key critical exponents with supposedly critical neuronal systems.

However, similar to the presence of power law relationships in avalanche size and duration, scaling exponents can also emerge in non-critical systems (Touboul & Destexhe, 2017). An alternative criterion which may be more robust is the relationship between avalanche exponents, which theory demonstrates should follow the form

$$\frac{1}{\zeta \nu z} = \frac{\alpha - 1}{\tau - 1} . \quad (3.12)$$

This states that at criticality, there is a specific relationship between the scaling

**A**

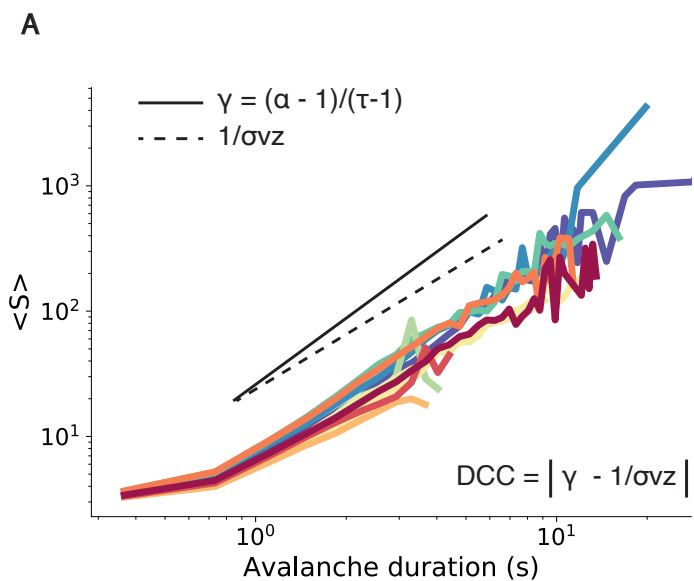

**B**

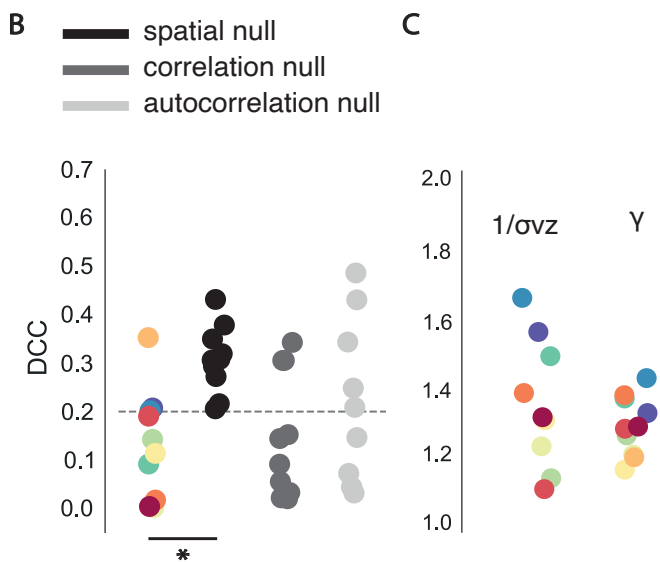

**Figure 3. 5    Scaling exponent and exponent relation in spontaneous activity.**

(A) The scaling exponent $1/\sigma \nu z$ captures the relationship between avalanche size and duration, here visualised as the mean size $\langle S \rangle$ for each duration. Notice how this follows a power law relationship, with the dotted line showing the estimated slope for $1/\sigma \nu z$. Each line is coloured by each fish as in Figure 3.4. Solid line shows the slope for $\gamma$ from a hypothetical dataset, which captures the relationship between size $\tau$ and duration $\alpha$ exponents. DCC is the deviation from criticality coefficient, which measures the absolute difference between $\gamma$ and $1/\sigma \nu z$. (B) DCC values are plotted for each fish against spatial, correlation, and autocorrelation null models (dotted line = critical threshold of DCC < 0.2) (Ma et al., 2019). (C) Values for the scaling exponent 1/σνz and the exponent relation γ shown for each fish. * = p < 0.05.

exponent 1/σνz and the critical exponents for avalanche size and duration.

To demonstrate why this relationship occurs, given that the function ⟨S⟩ (T) in (3.11) is self-similar at criticality, I can use the scaling argument from Sethna et al. to rescale the function by a small amount δ, such that

$$S(T) = (1 + \gamma\delta)\, S\big(T(1 + \delta)\big)$$
$$S(T) = T^{\gamma}. \quad (3.13)$$

This says that the function *S(T)* relates to *T* through an exponent γ which captures the relationship between a change in *T* and *S* when rescaling.

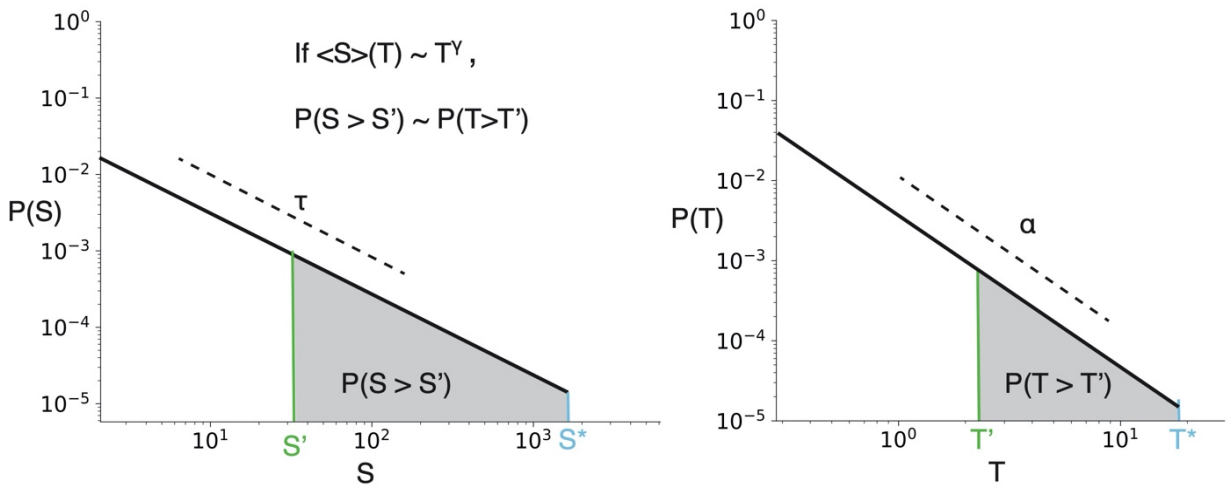

**Figure 3. 6** **Relationship between avalanche size and duration distributions.**
Probability density functions for avalanche size (S, left) and duration (T, right). If the scaling relationship between size and duration holds, then the probability S > S' will approximately equal the probability T > T', for any value T' and its corresponding S'. This occurs due to the power law relationship between S and T, such that changing T changes S in a proportional manner.

From Figure 3.6, it is clear that

$$P(S > S') \sim P(T > T'), \quad (3.14)$$

because due to the scaling law between S and T, moving along *P(T)* will move proportionally along *P(S)*.

Then using (3.4) and (3.5) I can show that

$$P(S > S') = \int_{S'}^{S^*} S^{-\tau}$$

$$P(S > S') = \frac{S^{1-\tau}}{1-\tau} + c \quad (3.15)$$

$$P(T > T') = \int_{T'}^{T^*} T^{-\alpha}$$

$$P(T > T') = \frac{T^{1-\alpha}}{1-\alpha} + c, \quad (3.16)$$

and by using (3.14) and removing the denominators for simplicity, I get

$$S^{1-\tau} \sim T^{1-\alpha}$$

$$(T^{\gamma})^{1-\tau} \sim T^{1-\alpha}$$

$$\gamma(1-\tau) \sim 1-\alpha$$

$$\gamma = \frac{1-\alpha}{1-\tau}$$

$$\gamma = \frac{\alpha-1}{\tau-1}. \quad (3.17)$$

Therefore, if a scaling relationship exists between avalanche size and duration as expected at criticality, then $1/\sigma vz$ should equal $(\alpha - 1)/(\tau - 1)$. In fact, studies which have simulated critical and non-critical systems suggest that expected exponent relationships can separate out truly critical from non-critical systems which generate power-law distributions (Touboul & Destexhe, 2017). However, one should note that even exponent relationships can emerge in random systems (Destexhe & Touboul, 2021). Nonetheless, exponent relations offer an adjunct test for criticality alongside the presence of power laws and scaling relations.

To assess the presence of critical exponent relations in spontaneous activity, I apply the deviation from criticality coefficient (DCC), first developed by Ma et al. (2019). DCC is defined as the absolute difference between $\gamma$ and $1/\sigma vz$ (Figure 3.5), calculated from the equation $(\alpha - 1)/(\tau - 1)$ and the slope of $\langle S \rangle$ (T), respectively. In spontaneous datasets, I find that $\gamma$ values map very closely onto $1/\sigma vz$ ($\gamma = 1.31 \pm 0.03$, $1/\sigma vz = 1.31 \pm 0.07$) (Figure 3.5C). This results in DCC values close to 0 for most datasets (DCC = 0.13 ± 0.04), suggesting the presence of critical exponent relations in spontaneous activity (Figure 3.5B). Importantly, critical exponent relations were significantly less well preserved in spatial null models (DCC = 0.31 ± 0.02, t = -4.80, p < 0.001) compared to empirical data, although not for correlation (DCC = 0.15 ± 0.04, t = -0.48, p = 0.64) and autocorrelation (DCC = 0.21 ± 0.05, t = -1.60, p = 0.15) nulls (see Figure 3.5B). This suggests that empirically observed exponent relations emerge due to the spatial localisation of active neurons during an avalanche, and would not occur in spatially random systems. Therefore, whole brain spontaneous avalanche dynamics exhibit exponent relationships expected at criticality, which provides further evidence for the presence of critical dynamics in EI balanced networks.

### 3.2.4  Branching Ratios in Spontaneous Activity

To further assess the presence of critical dynamics in whole brain spontaneous activity, I estimated the branching parameter. Some evidence suggests that brain activity may reside near to a phase transition that is captured by a branching process (Beggs & Plenz, 2003). Here, the phase transition separates a regime in which avalanches exponentially decay (subcritical), and one in which avalanches exponentially grow (supercritical). At the critical point, avalanches can span all scales, from small to large (Harris, 1963).

A branching process, or a Galton-Watson process, describes how an element can generate other elements, e.g. an active neuron giving rise to other active neurons (Harris, 1963). The process assumes that each element can give rise to a random number of offspring elements $N$, where each $N$ is independent and identically distributed over all ancestor elements with probabilities defined as $P(N=0) = p_0$, $P(N=1) = p_1, \ldots P(N=n) = p_n$, where $n = 0,1,\ldots \infty$. The number of offspring elements in the $t + 1$ generation is defined as

$$Z_{t+1} = \sum_{i=1}^{Z_t} N_i, \quad (3.18)$$

where $N_i$ is the number of offspring elements produced by the ancestor element $i$ of generation $Z_t$. Using equation 3.18, one can show that the expected number of offspring elements at generation $t$ is given by

$$\langle Z_t \rangle = \sigma^t, \quad (3.19)$$

where σ is known as the branching parameter and captures the expected number of offspring elements from a given ancestor, defined as

$$\sigma = \langle N \rangle. \quad (3.20)$$

From equation 3.19, it is clear that when σ < 1, the mean number of elements $Z_t$ exponentially decreases, while when σ > 1, $Z_t$ exponentially increases (Corral & Font-Clos, 2012). In particular, it can be shown that the probability of a branching process becoming extinct $P_{ext}$ is given by

$$P_{ext} = \begin{cases} 1 & if \ \sigma \leq 1 \\ x & if \ \sigma > 1 \end{cases}, \quad (3.21)$$

where $x < 1$. Therefore, avalanches will always eventually become extinct when $\sigma < 1$, whereas avalanches have the possibility for non-extinction when $\sigma > 1$. Thus $\sigma = 1$ separates exponential growth from exponential decay of avalanches – here, the sizes of avalanches generated from such a branching process can span all scales (Corral & Font-Clos, 2012).

In order to test the presence of a critical branching process in spontaneous neuronal activity as previously reported (Beggs & Plenz, 2003), I estimated the branching parameter $\sigma$ in my data. $\sigma$ was estimated as the mean ratio of descendants to ancestors at each time step over all avalanche events defined as

$$\sigma = \frac{1}{t} \sum_{n=1}^{t} \frac{descendants}{ancestors}, \quad (3.22)$$

where $t$ is the number of avalanche propagation timesteps across all avalanche events in a given dataset.

This method allowed me to estimate the branching parameter from my empirical avalanche data to ascertain the presence of critical branching $\sigma = 1$. Interestingly, whole brain spontaneous datasets exhibit $\sigma$ slightly below 1 ($\sigma = 0.93 \pm 0.03$) in spontaneous activity (Figure 3.7B), which indicates the presence of a slightly subcritical state (Priesemann et al., 2014). However, $\sigma$ estimated on a network with a topology such as the brain is likely to be underestimated due to the convergence of activity onto shared descendents. Importantly, I find that empirical $\sigma$ is significantly closer to unity than *spatial* ($\sigma = 0.77 \pm 0.03$, t = 21.8, p <0.0001), and *correlation* ($\sigma = 0.92 \pm 0.03$, t = 4.45, p <0.01), but not *autocorrelation* nulls ($\sigma = 0.93 \pm 0.03$, t = -0.28, p = 0.79) (Figure 3.7B). This suggests that empirically derived $\sigma$ close to 1, is a feature of the underlying spatial structure and correlation in neural activity, and would not emerge in a random system. Therefore, whole brain spontaneous dynamics exhibit σ close to 1, expected for a system in the vicinity of the critical point. That the dynamics are slightly below 1 suggests the network may be slightly subcritical, which may act to prevent unwanted state transitions (Wilting & Priesemann, 2019).

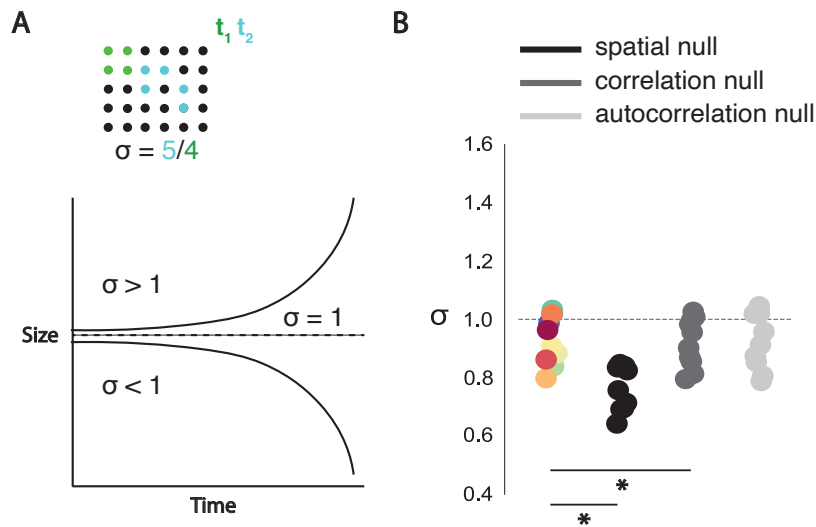

**Figure 3. 7   Branching ratios in spontaneous activity.**

(A, top) Avalanche schematic demonstrating calculation of branching ratio. Black neurons are off. Coloured neurons represent active neurons at time point $t_x$. Branching ratio σ is the number of descendants at a given time step ($t_2$) divided by the number of ancestors ($t_1$)  (A, bottom) Branching ratio schematic. When σ < 1 activity exponentially decays in the system, but when σ > 1 activity grows exponentially. When σ = 1, activity persists in the system without causing exponential growth .(B) Branching ratio relationship σ plotted for each fish against spatial, correlation and autocorrelation null models (dotted line = expected value of σ~1). Empirical data coloured by fish, as in Figure 3.4. * =  p < 0.05.

### 3.2.5  Long Range Correlations in Spontaneous Activity

Long range correlations are another key hallmark of critical systems, as at criticality the correlation length is maximal (Section 1.3). Long-range correlations between distant brain areas give rise to power law scaling of correlation as a function of distance in resting-state activity (Expert et al., 2011; Lombardi et al., 2020; Ponce-Alvarez et al., 2018). Therefore, to further test the presence of critical dynamics in whole brain spontaneous activity, I test for the presence of correlation function power law relationships. Specifically, I define the correlation function

$$r(d) \sim d^{-\eta}, \quad (3.23)$$

which defines the relationship between the correlation between two time series *r* and the distance between the sources of those time series *d*, where *η* captures the slope of the power law relationship (Figure 3.8). To calculate the correlation function for neuron pairs, *r* was assessed as the Pearson's correlation coefficient between neuron time series', while *d* was the Euclidean distance in space. To assess the function *r(d)*, neuron-neuron paired distances were placed in 200 bins, equally spaced across the range of neuron-neuron distances of all neurons in the brain. *r(d)* was then estimated as the mean correlation within each bin. Given that these relationships are not distributions, I cannot perform the likelihood ratio tests described above. Therefore, to assess closeness of fit to power laws I calculated the Euclidean distance between a power-law fitted by linear regression and the empirical function.

Applying the above procedure to whole brain spontaneous activity data, I found that *r(d)* qualitatively follows a power law relationship which is well approximated by a least squares linear regression fit (Figure 3.8). Interestingly, I find a slope *η* of 0.14 ± 0.03, which is slightly lower than previously reported indicating slower reductions in correlations as a function of distance (Ponce-Alvarez et al., 2018). That spontaneous datasets qualitatively follow power lar relationships suggests the presence of both short and long-range correlations in spontaneous activity, which is expected in systems at criticality.

Taken together, EI balanced, spontaneous activity exhibits a variety of statistical properties expected at criticality: i) power-law relationships of neuronal avalanches, (ii) scaling relationships between avalanche exponents, iii) theoretically predicted exponent relationships, (iii) branching ratios near to 1 and (iv) long-range correlation power laws. However of note, several findings expected for certain critical systems were not present: i) universal critical exponents for avalanche size, duration and scaling relationships, and ii) σ = 1. Nonetheless, the majority of expected statistical features of criticality were present in whole brain cellular resolution spontaneous activity datasets, indicating that EI balanced networks may reside near to criticality, but that this critical point may not belong to a critical branching process or other presupposed universality classes.

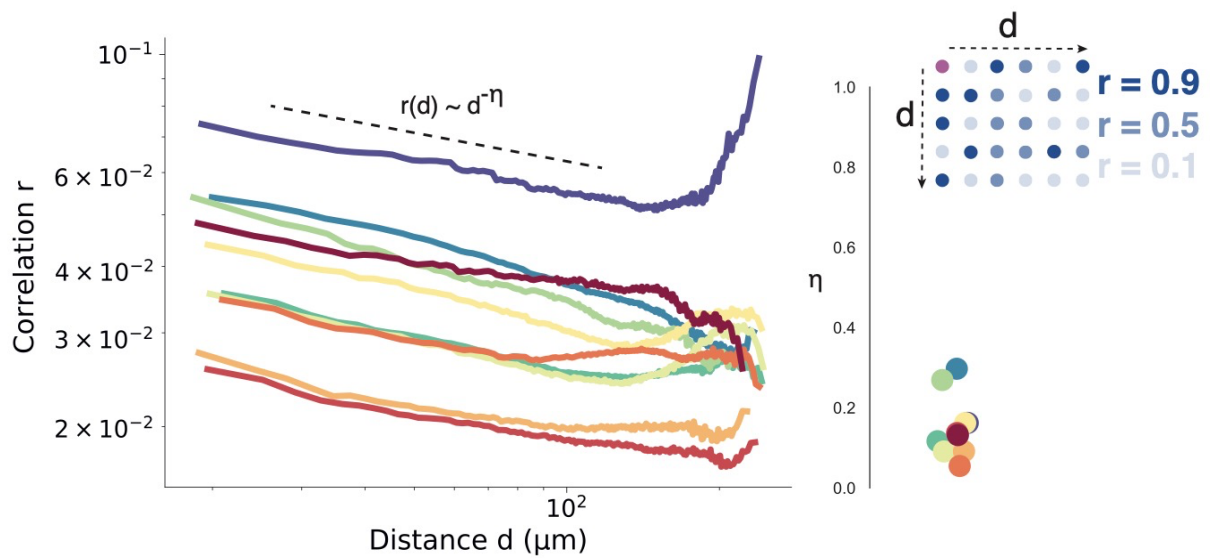

**Figure 3. 8   Long range correlations in spontaneous activity.**

(Top right) Correlation function schematic demonstrating estimation of correlation (r) as a function of distance (d). Magenta neuron = neuron of interest. Other neurons are coloured by their correlation r to neuron of interest. (Left) r(d) follows an approximate power-law (dotted-line) with exponent η. Each r(d) is coloured by fish, as in A. (Bottom right) η for each dataset.

### 3.2.6  The Effect of Avalanche Parameters on Critical Statistics

While the presence of the aforementioned critical statistics are strong indicators for the existence of critical dynamics, it is important to note that these statistics may vary according to avalanche detection parameters. For example, avalanche exponents vary according to temporal binning and avalanche-neighbourhood diameters (Beggs & Plenz, 2003; Ponce-Alvarez et al., 2018; Priesemann et al., 2013). Similarly, σ is affected by temporal binning and subsampling procedures (Priesemann et al., 2009).

Therefore, in order to assess the robustness of extracted statistics in spontaneous activity as a measure of criticality, I performed a sweep over a wide range of parameter values which define my avalanche algorithm (See Section 3.2.1). In particular, I chose to vary the neighbourhood size and the event probability in the hidden Markov model (Figure 3.9), as previous studies indicate that these parameters would influence critical statistics (Ponce-Alvarez et al., 2018).

I found that avalanche exponents are dependent on both parameters, with changes in event probability and neighbourhood size smoothly changing avalanche exponents for size and duration (Figure 3.9A). This demonstrates the danger of using specific exponent values to assess the presence of universality across datasets based on specific exponent values, as these will vary according to different methods. Instead, alternative markers such as power law relationships, and scaling and exponent relations should be used to demonstrate the presence of criticality. Importantly, I find that avalanche power law relationships are retained across all parameter combinations suggesting that spontaneous activity datasets are indeed scale invariant (Figure 3.9B).

I also find that $\sigma$ values vary smoothly according to both event probability and neighbourhood size, with lower event probability and smaller neighbourhood size giving rise to lower $\sigma$ (Figure 3.9C). This indicates that $\sigma$ values may be a relatively weak marker of criticality as they are particularly sensitive to parameter changes. As such, inferences drawn from exact values of $\sigma$ in my datasets should be done with caution, and instead changes to $\sigma$ under different conditions may be used to infer alterations in system excitability.

Interestingly, DCC values varied smoothly as a function of event probability but not neighbourhood size – lower event probability gave rise to greater DCC values (Figure 3.9D). Nonetheless, near-critical values (DCC < 0.25) were retained for event probabilities that accurately matched calcium transients (event probability = 0.57 - 0.61), regardless of neighbourhood size. This indicates that DCC values may be used with avalanche algorithms that accurately capture the spatiotemporal dynamics of brain activity, to assess the presence of criticality.

Taken together, while some critical statistics were sensitive to avalanche parameter changes, several key markers were consistent within reasonable avalanche parameter ranges for our data. Going forward avalanche parameters were selected for further analysis based on previous studies for neighbourhood size and according to which event probability best matched the underlying calcium transients on visual inspection (see Section 2.2.9 and Section 3.2.1).

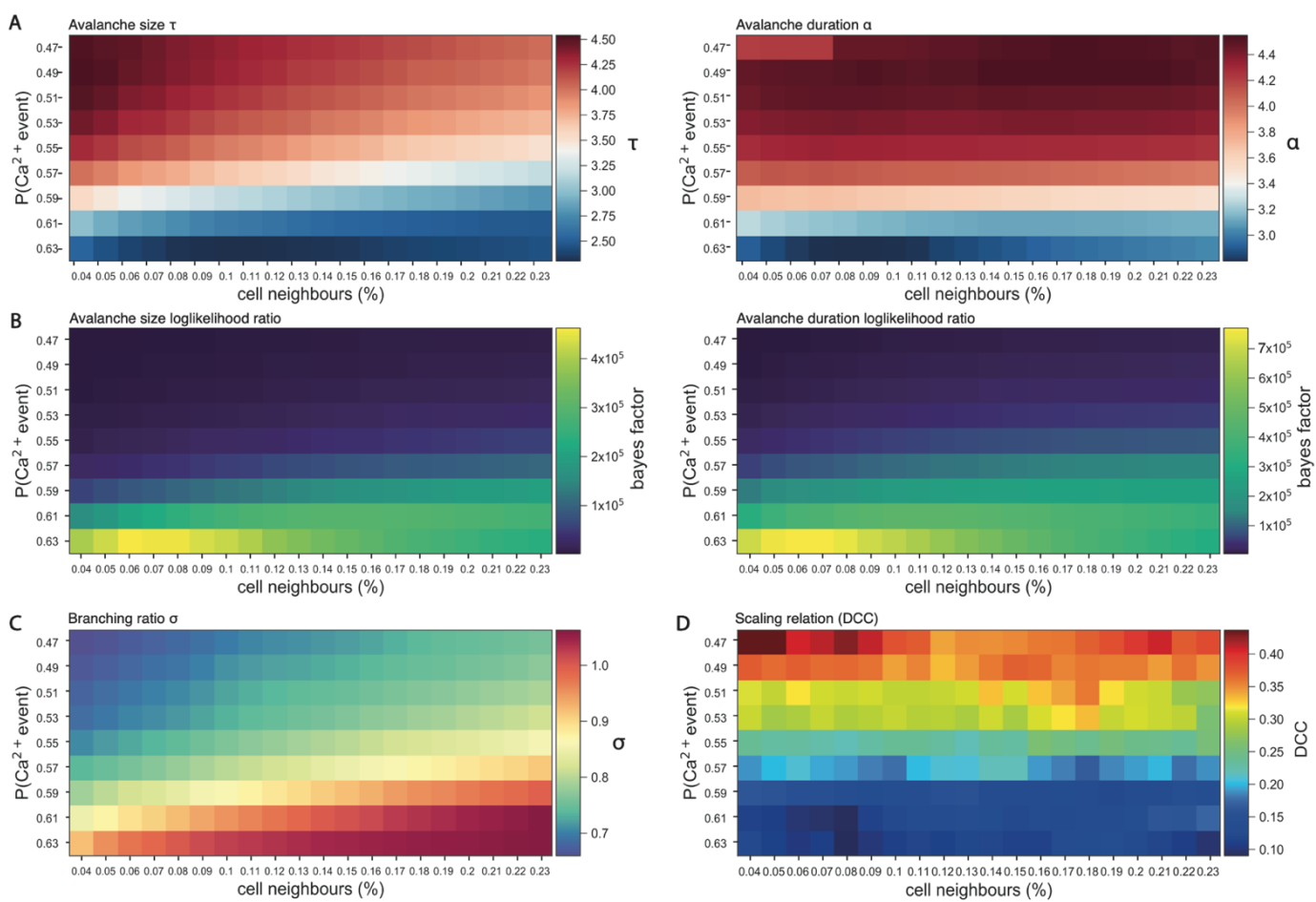

**Figure 3. 9    Critical statistics are parameter dependent.**

Critical statistics vary across different model parameter combinations – the event probability in the hidden Markov model (P(Ca2+ event)) and the % of neurons assigned to a neighbourhood. (A) Exponents for avalanche size (left) and duration (right) vary according to both event probability and neighbourhood size. (B) Avalanche distributions for size (left) and duration (right) are better explained by power law than lognormal across all parameters as shown by positive bayes factor for all parameter combinations. Bayes factors are calculated as loglikelihood ratios for power law vs lognormal distributions. (C) Branching ratio varies according to both event probability and neighbourhood size. (D) Exponent relations as measured by the deviation from criticality coefficient (DCC) varies according to event probability.

### 3.2.7   The Effect of Drug-induced EI Imbalance on Critical Statistics

Having validated key critical statistics and applied these to spontaneous activity data to support the presence of near-critical dynamics in EI balanced networks, I wanted to probe what biological parameters might support the stabilisation of brain dynamics to criticality. One potential biological parameter is the specific ratio of synaptic excitation and inhibition onto single neurons (Iascone et al., 2020) (See Section 3.1). Therefore, I wanted to test whether synaptic EI balance acts as a control parameter for shaping criticality *in vivo*. To test this, I theorised that altering such a control parameter for a critical system, should push the brain away from criticality thus disrupting i) baseline avalanche exponents, ii) scaling relationships, iii) exponent relations, iv) branching ratios and v) long-range correlation power-laws. Furthermore, given that a critical point separates distinct phases, even small changes to a control parameter should give rise to abrupt changes in the behaviour of the system (Binney et al., 1992). Therefore, I disrupted EI homeostasis in a graded manner with differing concentrations of the GABA-inhibitor pentylenetetrazole (PTZ) causing acute, induced epileptic seizures during calcium imaging (see Section 2.2.3). Furthermore, I was also interested in whether EI-imbalance induced seizures would cause a loss of criticality and the emergence of a supercritical regime – to test this I assessed the presence of i) bimodal avalanche distributions, ii) a loss of exponent relations, and iii) increases in branching ratio beyond 1 (Figure 3.7A).

I first compared avalanche exponents, to demonstrate if perturbing EI balance alters cascading dynamics. *20mM PTZ* caused a frequent hypersynchronous, high activity state consistent with generalised epileptic seizure activity, suggestive of a severe disruption of EI balance (Figure 2.2). Avalanche exponents significantly decreased for avalanche size (*spontaneous*: τ = 2.98 ± 0.13, *20mM PTZ*: τ = 2.56 ± 0.05, t = 4.59, p < 0.01), and duration (*spontaneous*: α = 3.58 ± 0.14, *20mM PTZ*: α = 3.30 ± 0.11, t = 4.06, p < 0.01) (Figure 3.9, 3.10). I also compared avalanche exponents following more graded disruptions to EI balance (*5mM PTZ*), resulting in focal seizures that were localised to brain regions. 5mM PTZ also resulted in a significant decrease in avalanche exponents for avalanche size (*5mM PTZ*: τ = 2.65 ± 0.10, w = 0, p < 0.01) and duration (*5mM PTZ*: α = 3.34 ± 0.12, w = 2.0, p < 0.01), compared with spontaneous activity (Figure 3.9, 3.10). This indicates an increased frequency of larger and longer avalanches during generalised and focal seizures, suggesting a role for EI balance in shaping avalanche dynamics, as expected for a control parameter for criticality. Here, it is important to note that some 20mM PTZ datasets (Figure 3.11) may not be well described by a single scaling law, and therefore observed avalanche exponent values should be interpreted with caution.

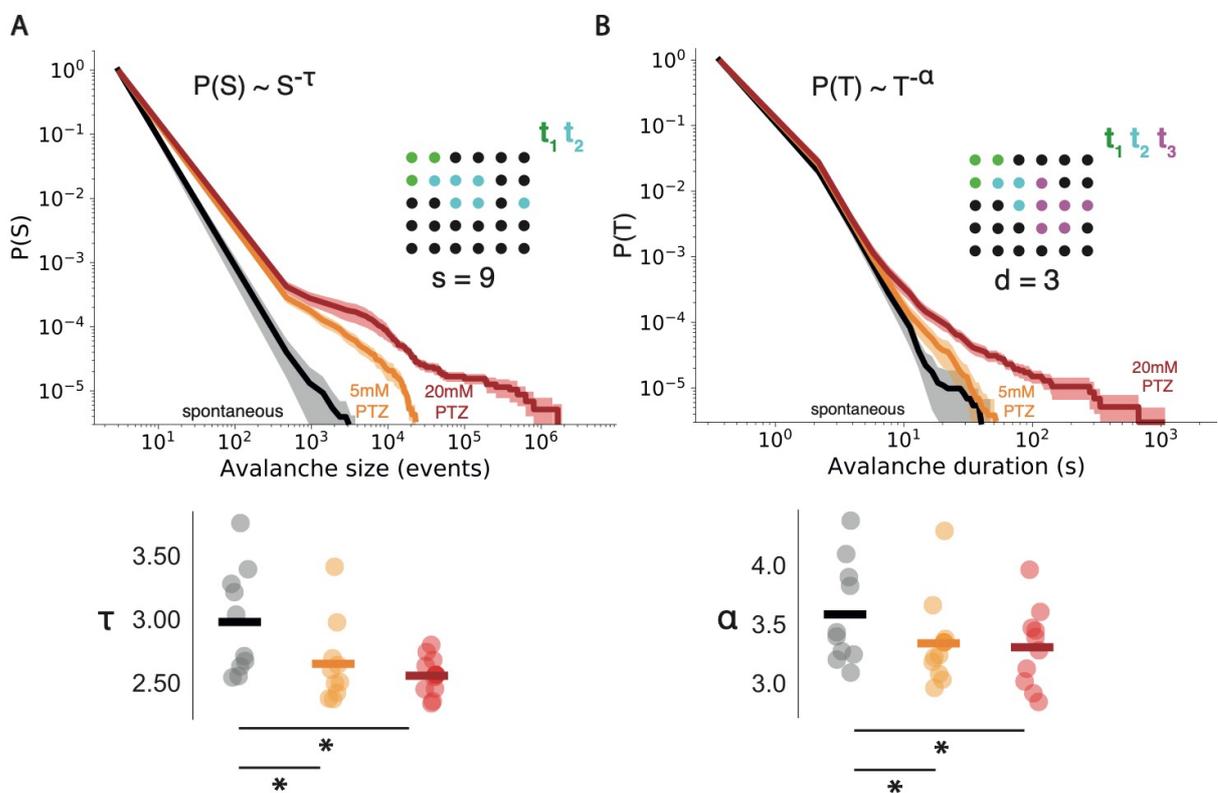

**Figure 3. 10 Drug-induced EI imbalance alters avalanche properties.**

Complementary cumulative distribution functions for avalanche size (A) and duration (B), comparing mean distributions for spontaneous (black, solid), 5mM PTZ (orange, dotted) and 20mM PTZ (red, dotted) datasets. Shaded regions represent the standard deviation. (top, right) Avalanche schematic demonstrating calculation of avalanche size (A) and duration (B) for single avalanche events. Black neurons are off. Coloured neurons represent active neurons at time point $t_x$. Avalanche exponents compared for size ($\tau$, A) and duration ($\alpha$, B) in spontaneous (black), 5mM (orange) and 20mM PTZ (red) conditions. * = p<0.05

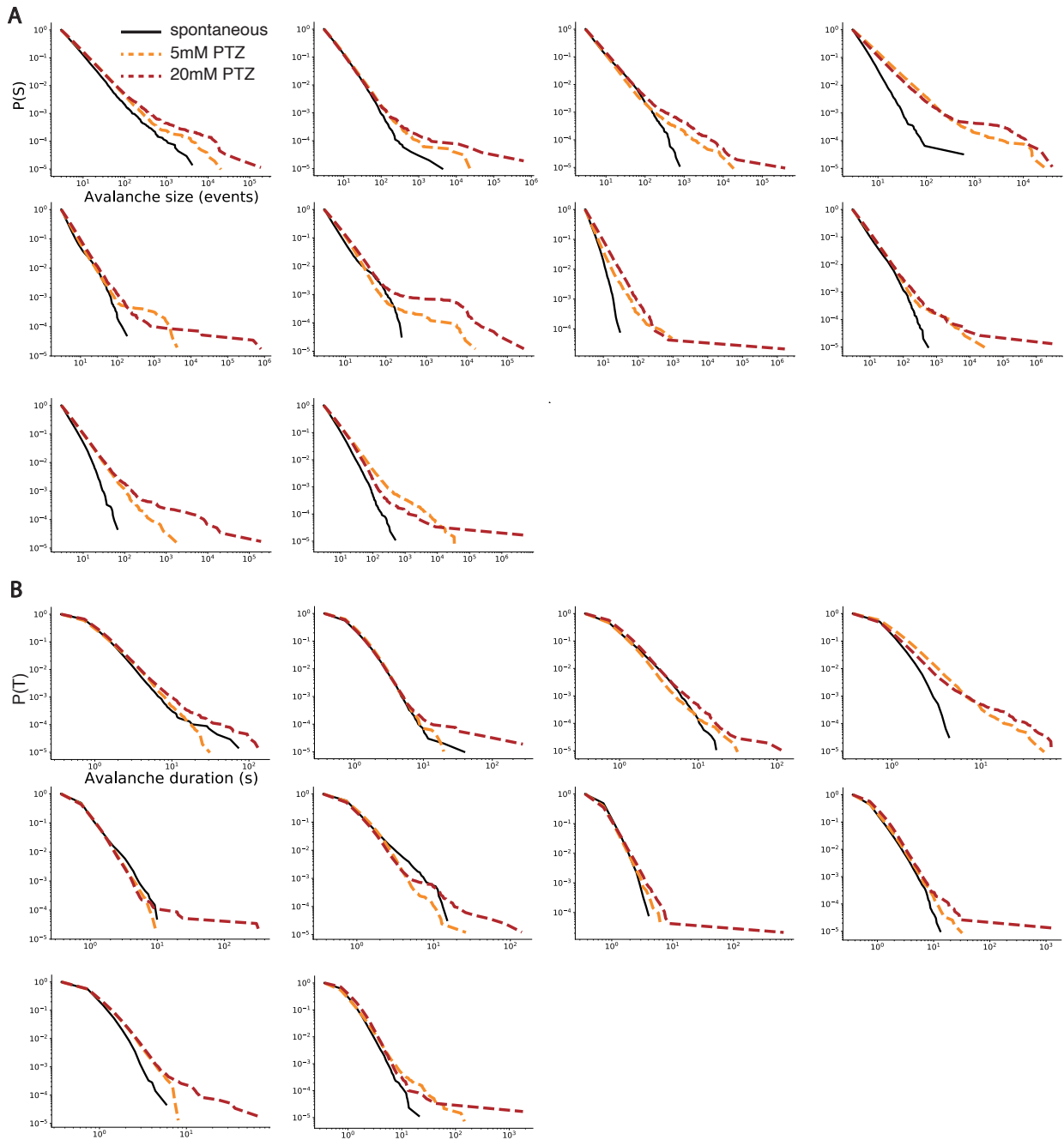

**Figure 3. 11  Avalanche properties during drug-induced EI imbalance for all fish.**

(A-B) Complementary cumulative distribution functions for spontaneous (black, solid line), 5mM PTZ (orange, dotted line) and 20mM PTZ (red, dotted line) periods for avalanche size (A) and duration (B) for each fish.

Upon visual inspection, empirical avalanche distributions (Figure 3.9, 3.10) in both 5mM and 20mM conditions caused a deviation from expected power law relationships. The fact that avalanche size distributions exhibit the emergence of characteristic scale at the tail end of the distribution, is suggestive of bimodal avalanche distributions expected at supercriticality (Beggs & Plenz, 2003). However, for avalanche duration this change manifested more clearly as a change in power law slope without necessarily causing a loss of log linear shape. In order to statistically evaluate the presence of a bimodal distribution instead of a power law, one would need to apply appropriate likelihood ratio tests, but this is beyond the scope of this work. Nonetheless, the emergence of a bump in the tail end of the distribution for avalanche size is suggestive of a loss of criticality and the emergence of supercritical dynamics during epileptic seizures.

To further test the role of EI balance in shaping critical dynamics, I assessed the sensitivity of the scaling exponent $1/\sigma\nu z$ to different PTZ doses. I found that perturbing EI balance causes a significant increase in $1/\sigma\nu z$ in both 5mM ($1/\sigma\nu z$ = 1.96 ± 0.11, t = -7.1, p < 0.001) and 20mM conditions ($1/\sigma\nu z$ = 2.04 ± 0.09, t = -7.0, p < 0.001) (Figure 3.11A,C). That $1/\sigma\nu z$ exponents deviate from those found in cortical recordings at criticality, and my spontaneous EI balanced datasets ($1/\sigma\nu z$ ~1.3), demonstrates a loss of expected scaling exponents in critical systems *in* vivo as a consequence of changing EI imbalance (Fontenele et al., 2019; Friedman et al., 2012). This further suggests a role of EI balance in shaping avalanche dynamics and expected scaling relations expected at criticality.

Interestingly, upon visual inspection, while the relationship between *S* and *T* clearly follows a power law in spontaneous datasets, this relationship appears to be lost during EI imbalance. Specifically, while medium sized avalanches in EI imbalanced conditions appear to follow the scaling relationship in spontaneous activity ($1/\sigma\nu z$ ~ 1.3), as avalanche get larger in size they may occur at a faster rate than expected from the baseline scaling function (Figure 3.11A, Figure 3.12). This indicates the emergence of large seizure cascades that emerge very quickly, resulting in a loss of self-similarity. This strongly suggests a loss of criticality due to EI imbalance, indicating a role for EI balance in shaping critical dynamics *in vivo*.

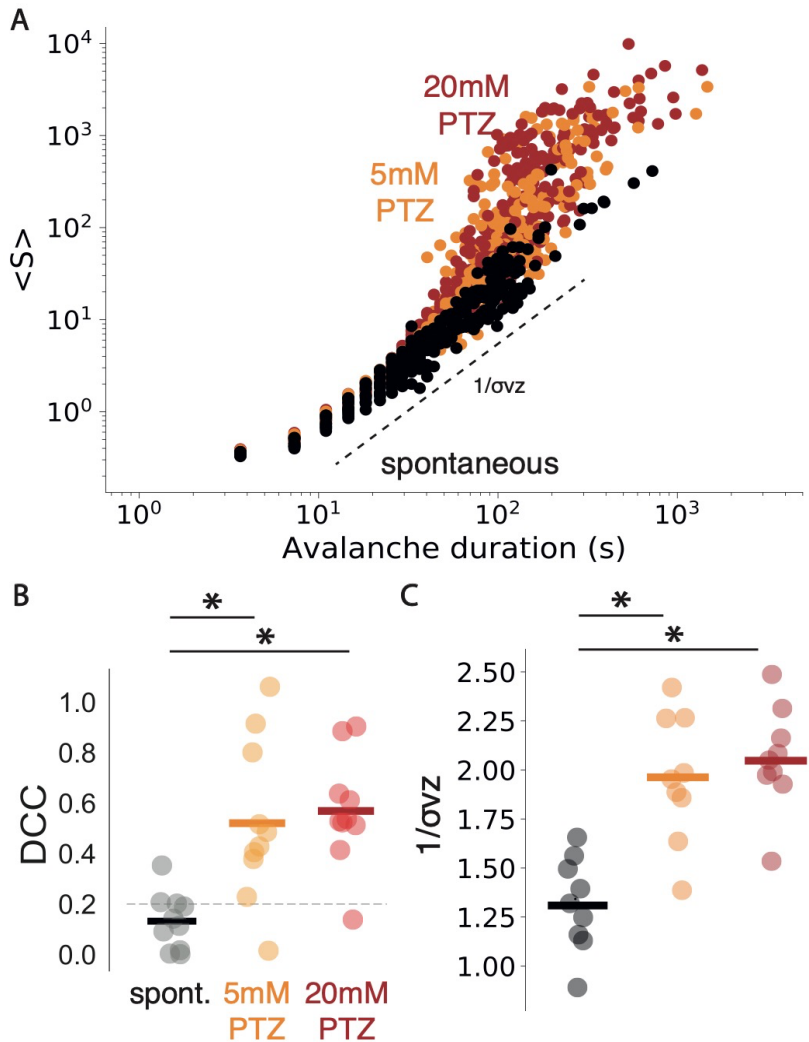

**Figure 3. 12 Drug-induced EI imbalance alters scaling exponents and exponent relations.**

(A) Scaling exponent $1/\sigma \nu z$ captures the relationship between avalanche size and duration, here visualised as the mean size $\langle S \rangle$ for each duration. Notice how this follows a power law relationship, which breaks down due in the 5mM (orange) and 20mM (red) EI imbalanced conditions. Each dot represents an $\langle S \rangle$ (T) relationship for a single fish. DCC (B) and $1/\sigma \nu z$ exponent (C) values are plotted for each fish, comparing spontaneous, 5mM and 20mM conditions. * = p < 0.05.

A loss of self-similarity in avalanche scaling relationships would imply a loss of exponent relations, as outlined in equation (3.12), because the relationship in equation (3.14) would break down. Therefore, to confirm that EI imbalance also disrupts critical exponent relationships, I calculated DCC values in PTZ datasets. I found that perturbing EI balance causes a loss of critical exponent relations, with *20mM PTZ* (DCC = 0.57 ± 0.07, t = -5.70, p < 0.001) and *5mM PTZ* (DCC = 0.52 ± 0.10, t = -3.24, p = 0.01).  resulting in a significant increase in DCCs (DCC = 0.57 ± 0.07, t = -5.70, p < 0.001) (Figure 3.11B). This further supports the notion that scaling functions are no longer self-similar during EI-imbalance induced seizures. Therefore, altering EI balance disrupts exponent relations, an alternative marker of criticality, suggesting a role for EI balance in shaping critical avalanche dynamics. This evidence also points towards the presence of a supercritical state, as both critical scaling exponents and exponent relations should be disrupted in a supercritical state away from criticality.

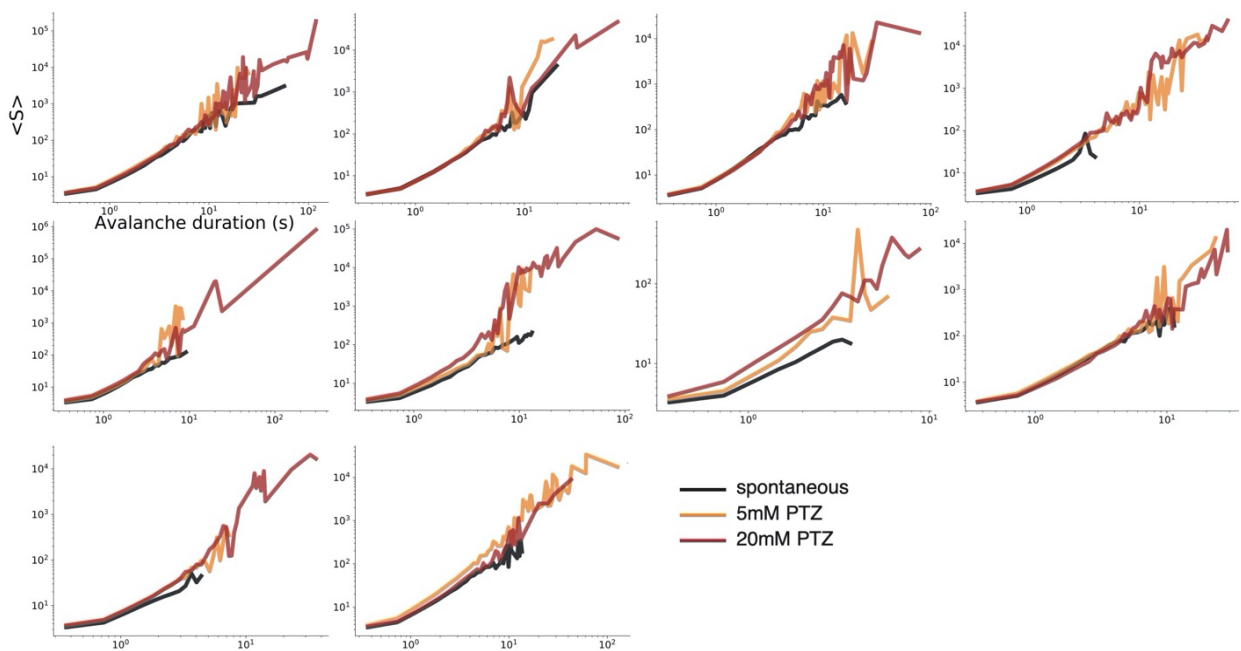

**Figure 3. 13  Avalanche scaling functions during drug-induced EI imbalance for all fish.**

Scaling functions here visualised as the mean size ⟨S⟩ for each duration, are shown for spontaneous (black), 5mM PTZ (orange) and 20mM PTZ (red) conditions.

Next, I assessed the sensitivity of the branching parameter σ to changes in EI balance. If brain activity may be described as a branching process, then any control parameter which shapes dynamics around the critical point should alter σ accordingly. Interestingly, I found a significant increase in both the *20mM PTZ* (σ = 1.01 ± 0.01, t = -3.11, p < 0.017) and *5mM PTZ* conditions (σ = 1.00 ± 0.02, t = -3.79, p <0.01), suggesting a key role of EI balance in shaping branching relationships (Figure 3.12B).

Regarding the presence of supercritical dynamics in EI imbalance-induced seizures, according to the branching process model of criticality, σ values should increase beyond 1 for the presence of a supercritical state (Harris, 1963) (Figure 3.13A). However, my EI imbalanced dynamics get closer to 1 for both 5mM and 20mM conditions (Figure 3.13B). This could point towards the transition from slightly subcritical dynamics to a critical branching regime where σ ~ 1 during seizures. However, I note that the model of branching processes assumes infinitely large networks, whereas brain networks are of finite size. This means that observing σ > 1 in supercritical systems with a topology is unlikely due to the high propensity for multiple activity sources to coalesce onto single nodes thus causing the artificial termination of branching processes (Zierenberg et al., 2020). This means that the propagation of activity will always be bounded by the size of the system in such instances rendering σ > 1 unlikely. An alternative approach to identify supercritical branching is to look over shorter periods, to capture the growing phase of avalanches before they are bounded by the system size. Furthermore, I also note that my longer *20mM PTZ* recordings contain both periods of both generalised ictal activity and post-ictal depression (see Figure 2.2C), and therefore σ closer to unity could arise due to averaging effects rather than critical dynamics. Therefore, I also compared σ during shorter windows of *baseline* and *ictal-onset* periods (see Section 2.2.10). Using this approach, I found that σ shows a greater magnitude increase beyond 1 in periods only containing the onset of generalised seizures (*spontaneous*: σ = 0.94 ± 0.03, *ictal-onset*:  σ = 1.05 ± 0.02, t = -3.41, p <0.01) (Figure 3.12C). This provides evidence for the presence of a supercritical regime during seizures, indicating that population dynamics are driven away from the phase transition.

However, as previously outlined in Section 3.2.6 the interpretation of exact values of σ from empirical data can give misleading representations of system state (Priesemann et al., 2014). Nonetheless, σ can be useful as a comparative measure across the same system under different conditions. That σ significantly increases from baseline values (Figure 3.12A), alongside a deviation from critical scaling and exponent relationships, shows some evidence for the emergence of a supercritical state away from criticality. That perturbing EI balance causes σ to increase beyond near-critical values at rest, indicates a role of EI balance in shaping critical dynamics.

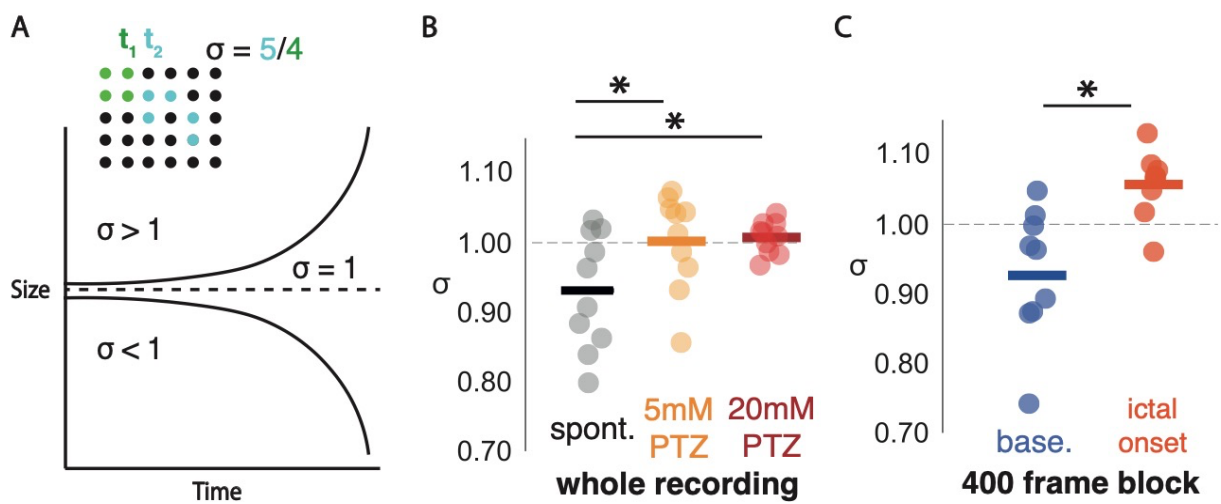

**Figure 3. 14  Drug-induced EI imbalance alters branching ratios.**

(A, top) Avalanche schematic demonstrating calculation of branching ratio. Black neurons are off. Coloured neurons represent active neurons at time point $t_x$. Branching ratio σ is the number of descendants at a given time step ($t_2$) divided by the number of ancestors ($t_1$)  (A, bottom) Branching ratio schematic. When σ < 1 activity exponentially decays in the system, but when σ > 1 activity grows exponentially. When σ = 1, activity persists in the system without causing exponential growth .(B) Branching ratio relationship σ compared across spontaneous (black), 5mM PTZ (orange) and 20mM PTZ (red) conditions. (C) Branching ratio relationships compared across shorter 400 frame periods of baseline (blue) and ictal-onset periods (crimson). * =  p < 0.05.

Finally, I assessed the sensitivity of correlation power law relationships to changes in EI balance. Upon visual inspection I found that the pairwise correlation increases due to EI imbalance across all neuron pairs regardless of distance, suggesting a reorganisation of the normal rules constraining functional connectivity to lower values (Figure 3.14A). I also found a significant increase in the correlation function slope $\eta$, indicating a faster decay of correlation as a function of distance due to EI

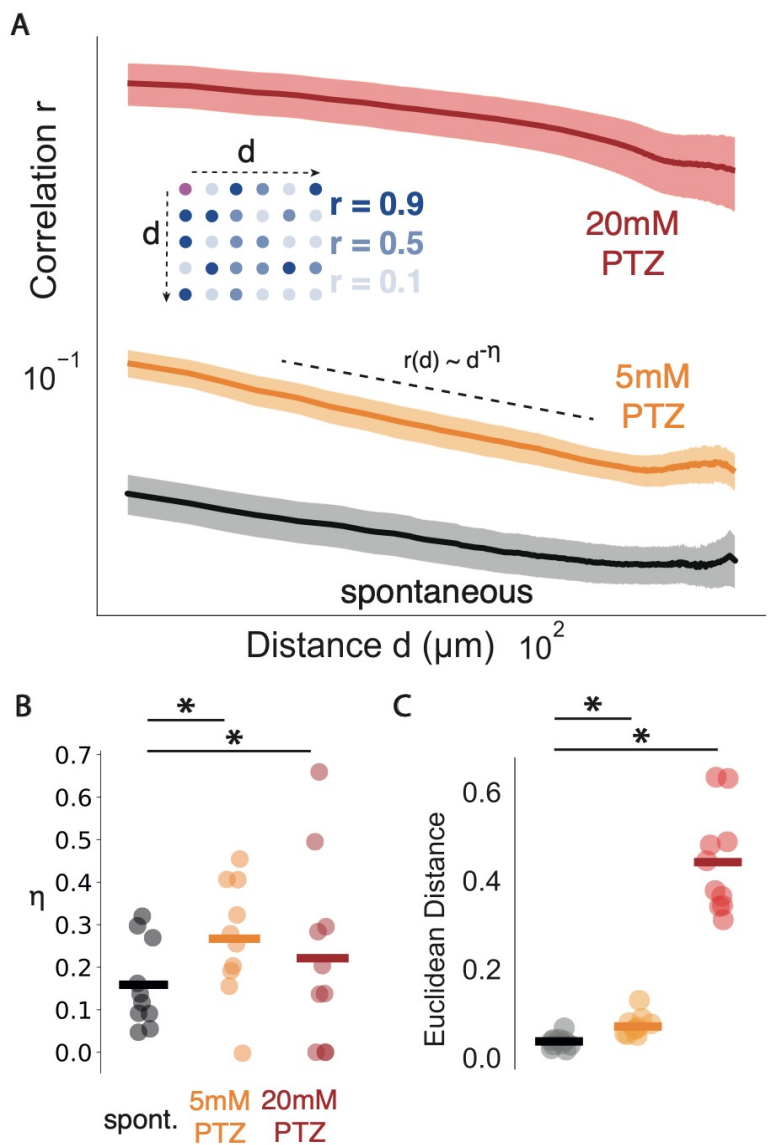

**Figure 3. 15 Drug-induced EI imbalance alters correlation functions.**

(A) Correlation function r(d) compared for spontaneous (black), 5mM PTZ (orange) and 20mM PTZ (red) datasets. Shaded regions represents the standard deviation. (inset) Correlation function schematic demonstrating estimation of correlation (r) as a function of distance (d) (middle). Magenta neuron = neuron of interest. Other neurons are coloured by their correlation r to neuron of interest. (B) η compared across spontaneous (black), 5mM PTZ (orange) and 20mM PTZ (red) conditions. (C) Euclidean distance from fitted power law compares power law fit for spontaneous (black), 5mM PTZ (orange) and 20mM PTZ (red) periods. * = p<0.05.

imbalance, albeit in the presence of higher correlation overall (*5mM PTZ: η* = 0.26 ± 0.04, t = -3.05, p < 0.05*; 20mM PTZ: η* = 0.30 ± 0.06, t= -2.89, p <0.05). Interestingly, correlation functions in the *20mm PTZ* condition are less well fit to power-laws than spontaneous datasets (*spontaneous:* distance from power-law = 0.04 ± 0.00*, 20mM PTZ:* distance from power-law = 0.44 ± 0.04, w= 0.0, p <0.01) (Figure 3.14). Similarly, *5mM PTZ* causes a graded, but significant loss of correlation function power-laws, giving rise to characteristic scale at the tail end of the function (*5mM PTZ:* euclidean distance from fitted power-law = 0.07 ± 0.01, w = 1.0, p <0.01) (Figure 3.15). This suggests that altering EI balance changes the correlation structure of the network, such that correlation no longer scales as a power of distance. Therefore, altering EI balance disrupts long-range correlation power laws that are a hallmark of critical systems, further indicating a role of EI balance in shaping critical network properties.

Taken together, I find that perturbing EI balance causes a loss of critical statistics with i) changes to baseline avalanche exponents, ii) a departure from baseline scaling relations, iii) a loss of theoretical exponent relationships, iii) increases in σ, and iv) a loss of long-range correlation power law relationships. The sensitivity of critical statistics to EI balance changes suggests a direct relationship between EI balance alterations and collective avalanche dynamics around criticality, as expected for a control parameter. Interestingly, the above evidence also serves to demonstrate that microscale neuronal activity collectively drives whole network dynamics away from criticality during seizures. While some evidence suggests that such dynamics are in keeping with a supercritical state, namely i) the appearance of characteristic scale in avalanche size distributions, ii) a loss of scaling and exponent relations, and iii) increase in σ above 1, further evidence is required to quantitatively demonstrate a loss of power law relationships and robust increases in σ above 1 as expected from branching theory.

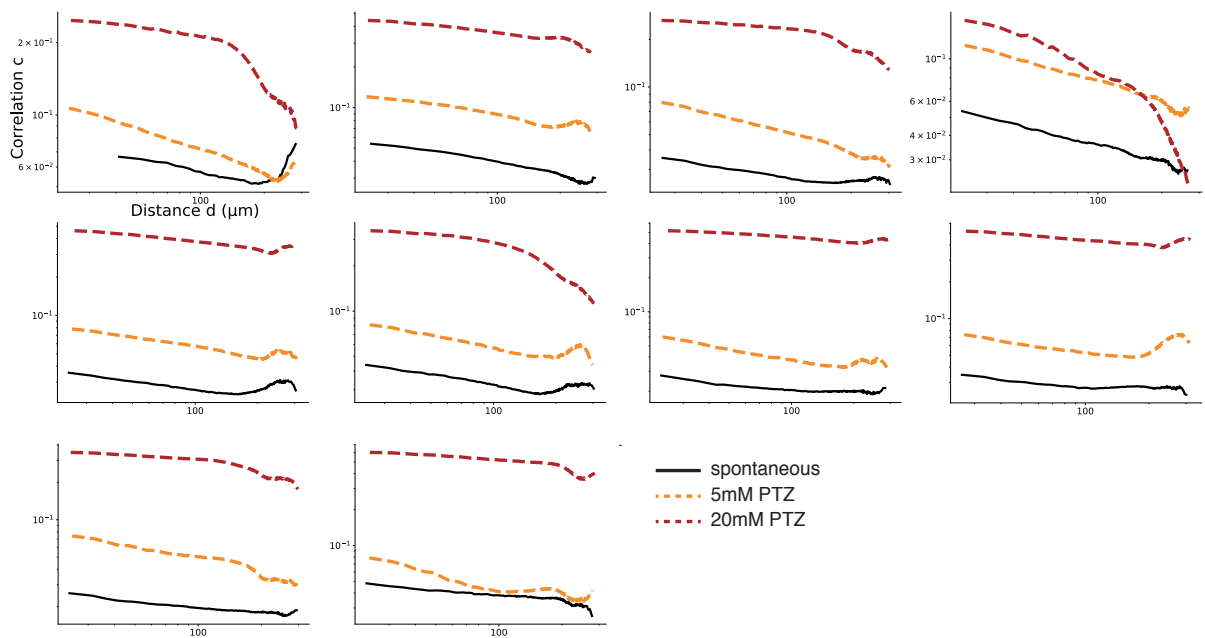

**Figure 3. 16  Correlation functions during drug-induced EI imbalance for all fish.**
Correlation functions for spontaneous (black, solid line), 5mM PTZ (orange, dotted line) and 20mM PTZ (red, dotted line) periods for each fish. 5mM PTZ and 20mM PTZ datasets are less well fit to power laws than spontaneous datasets.

*3.2.8  The Effect of Genetically-Induced EI Imbalance on Critical Statistics*
Systems at criticality can optimise various network properties (Shew et al., 2009), and as such criticality may be a favourable state for brain dynamics. That EI imbalance-induced seizures give rise to some supercritical statistics indicates that aberrant seizure dynamics emerge due to a transition from critical to supercritical states. But what about the abnormal brain dynamics that generally occur in genetic EI imbalance disorders (Sohal & Rubenstein, 2019)? For example, FOXG1 syndrome manifests as aberrant ictal and inter-ictal dynamics (Cellini et al., 2016; Seltzer et al., 2014; Wong et al., 2019), and is linked to interneuron deficits in the developing telencephalon (Shen et al., 2019). Of particular note, Hannah Bruce in her 2022 thesis demonstrated a reduction in the number of interneurons in the larval zebrafish telencephalon, suggesting a loss of EI balance towards excitation. Therefore, could this genetic disruption of EI balance cause brain dynamics to deviate from critical to supercritical dynamics, as we see during pharmacological EI imbalance in seizures? If so, critical dynamics may provide a population framework for understanding pathological emergent network dynamics in brain disorders such as FOXG1 syndrome.

To test whether genetic perturbations to EI balance cause a loss of critical statistics, I use foxg1a homozygous (-/-) and heterozygous (+/-) zebrafish mutants as models for genetically-induced EI imbalance (see Section 2.1.1). While the foxg1a zebrafish mutant has been characterised at the structural level (Bruce, 2022), its neuronal dynamics have yet to be characterised. In order to understand how reductions in the numbers of interneurons influences collective behaviours in the network, I first need to understand the changes to single neuron excitability – single neuron properties are central in driving collective behaviour such as avalanche dynamics and thus can help to relate genetic perturbations to EI balance with emergent properties across populations. As such, I first characterised single neuron firing properties in foxg1a -/- and +/- mutant lines and corresponding wildtype siblings. Given that observations of foxg1 mutant behaviour suggested no obvious seizures, I imaged both spontaneous and low concentration PTZ conditions (*2mM PTZ*), with the aim of capturing resting state activity and activity in seizure prone networks under conditions of reduced inhibition, to capture the full diversity of brain dynamics in genetically perturbed EI imbalanced networks (see Section 2.2.3). Finally, reductions in inhibitory neuron number are only known to occur in the telencephalon (Bruce, 2022) and therefore I restricted my analysis to telencephalic neurons only (see Section 2.2.7).

To probe changes in single neuron excitability, I was first interested in the propensity of neurons to spike (Figure 3.16A). Here, I reasoned that fewer interneurons in the network should give rise to a loss of inhibition and therefore increased spike number. To test this, I calculated the mean spike count for a given fish as

$$\frac{1}{n}\sum_{i=1}^{n} k_i, \quad (3.24)$$

where $n$ is the number of neurons in the network and $k_i$ is the number of spikes over the recording period for neuron $i$. Interestingly, I found a significant reduction in the mean spike count in foxg1a -/- mutants compared with wildtype siblings, with no change in heterozygous foxg1a +/- mutants compared to wildtype (*foxg1a +/+:* 1119.8 ± 131.30*; foxg1a +/-:* 1126.7 ± 121.61, U = 16.0, p = 0.30; *foxg1a -/-:* 179.4 ± 25.4, U = 0.0, p <0.01) (Figure 3.16B). This surprising result indicates reduced excitability of single neurons, due to a loss of inhibitory interneurons. Furthermore, foxg1a -/- fish exhibit heightened sensitivity to *2mM PTZ*, increasing its mean spike count by ~ 1,000 spikes. However, foxg1a -/- and +/- did not exhibit statistically different spike counts compared with wildtype siblings in the presence of PTZ (*foxg1a +/+:* 1270.0 ± 154.64*; foxg1a +/-:* 1362.3 ± 144.60, U = 13.0, p = 0.34; *foxg1a -/-:* 1215.6 ± 209.01, U = 9.0, p =0.45) (Figure 3.16B). The observed large increase in mean spike number in foxg1a -/- fish in the presence of PTZ, that was not observed in wildtype fish indicates that reducing interneuron number in the telencephalon may lead to a heightened sensitivity to GABAa receptor blockade, which might suggest the network is closer to the seizure threshold. However, this conclusion is not in keeping with the general finding of reduced spiking count in the baseline state which is suggestive of reduced excitability.

Next, I was interested in the amplitude of calcium events (Figure 3.16A), reasoning that fewer interneurons in the network might lead to increased activity which would manifest as the accumulation of intracellular calcium in neurons. To test this, I calculated the mean fluorescence amplitude for a given fish as

$$\frac{1}{n}\sum_{i=1}^{n} F_i, \quad (3.25)$$

where $n$ is the number of neurons in the network and $F_i$ is the mean fluorescence over the recording period for neuron $i$. $F_i$ is calculated as

$$F_i = \frac{1}{t} \sum_{x=1}^{t} f_x, \quad (3.26)$$

where $t$ is the number of timesteps and $f_x$ is the normalised fluorescence at time step $x$. In keeping with findings of reduced spike count in the baseline state, I found that mean amplitudes were also significantly decreased in foxg1a -/- mutants compared with wildtype fish, with no change in the heterozygous foxg1a +/- mutants (*foxg1a +/+: 7.13 ± 0.46; foxg1a +/-: 6.93 ± 0.43, U = 18.0, p = 0.41; foxg1a -/-: 4.50 ± 0.27, U = 0.0, p <0.01*) (Figure 3.16C). However, I found that the foxg1 mutants do not exhibit altered firing properties in the presence of PTZ, indicating no changes to activity due to reduced inhibition (*foxg1a +/+: 7.29 ± 0.34; foxg1a +/-: 6.90 ± 0.36, U = 13.0, p = 0.34; foxg1a -/-: 5.88 ± 0.31, U = 1.0, p <0.025*). Overall, this indicates that reductions in interneurons reduces the calcium event amplitude of single neurons further suggesting reduced excitability in the network.

Next, I calculated the average duration of a calcium transient (Figure 3.16A), to characterise the presence of continuous spike bursts, which provides an alternative characterisation of altered firing properties in single neurons. Here I reasoned that fewer interneurons in the network should lead to disinhibition and thus longer, uninterrupted spike trains. I calculated the mean transient duration for a given fish as

$$\frac{1}{n} \sum_{i=1}^{n} D_i, \quad (3.27)$$

where $n$ is the number of neurons and $D_i$ is the mean transient duration over the recording period for neuron $i$. $D_i$ is calculated as

$$D_i = \frac{1}{y} \sum_{x=1}^{y} d_x, \quad (3.28)$$

where $y$ is the number of periods with continuous spiking (including single spikes) and $d_x$ is the duration of transient $x$. I found that while mean transient durations decreased in foxg1a -/- mutants compared with wildtype fish, this change was not significant, with no change in the heterozygous foxg1a +/- mutants (*foxg1a +/+:* 1.31 ± 0.05*; foxg1a +/-:* 1.29 ± 0.04, U = 19.0, p = 0.47; *foxg1a -/-:* 1.21 ± 0.05, U = 7.0, p = 0.15) (Figure 3.16D). Mean transient durations also did not change across genotypes following PTZ exposure (*foxg1a +/+:* 1.29 ± 0.05*; foxg1a +/-:* 1.31 ± 0.04, U = 14.0, p = 0.40; *foxg1a -/-:* 1.26 ± 0.07, U = 8.0, p = 0.36). Therefore, the duration of bursts in single neurons is not altered by the reduction in interneurons.

Finally, I was interested in how the pairwise connectivity in the network was altered in response to a loss of interneurons, reasoning that a loss of interneurons should give rise to increase functional connectivity through the telencephalon. Here I calculated the mean pairwise correlation for a given fish as

$$\frac{1}{n} \sum_{i=1}^{n} R_i, \quad (3.29)$$

where $n$ is the number of neurons and $R_i$ is the mean pairwise Pearson's correlation between neuron $i$ and all other neurons in the network. $R_i$ is then calculated as

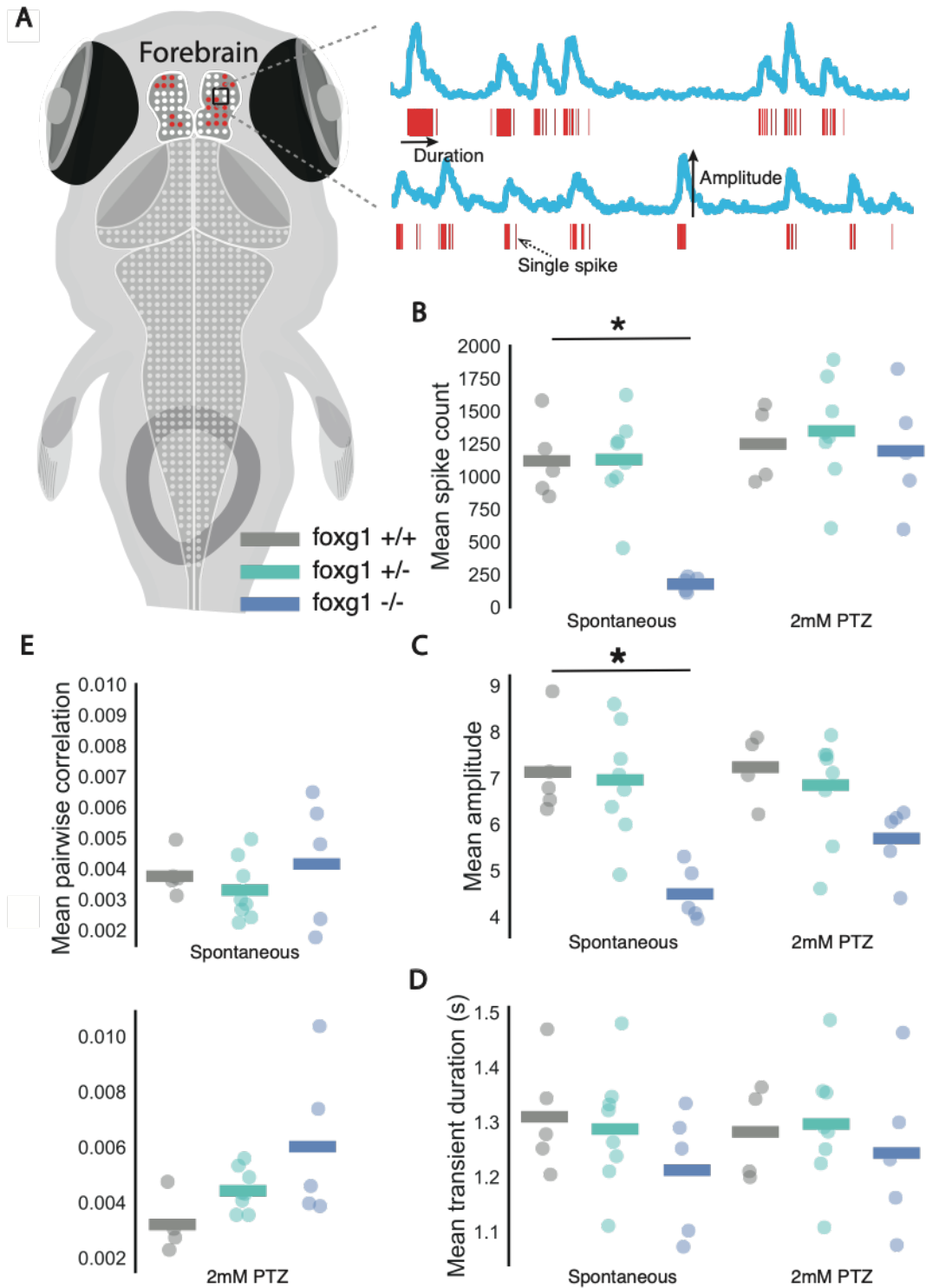

**Figure 3. 17 Genetically-induced EI imbalance reduces single neuron excitability.**

(A) Larval zebrafish brain schematic, with all foxg1a analyses performed across forebrain neurons. Blue traces are example forebrain calcium signals, with underlying estimated spike trains shown below in red. Spike count for each neuron was the total number of spikes throughout the recording. Amplitude for each neuron was the average normalised calcium fluorescence over the whole recording. Transient duration for each neuron was the average duration of all continuous spike trains. Pairwise correlation for each neuron was the average correlation with all other neurons. Mean spike count (B), mean amplitude (C), mean transient duration (D) and mean pairwise correlation (E), averaged over all neurons per fish, compared across foxg1+/+ (grey), foxg1+/- (teal), and foxg1-/- (blue) fish for spontaneous and 2mM PTZ conditions. * = p<0.025.

$$ R_i = \frac{1}{n-1} \sum_{j=1}^{n-1} r_{ij}, \quad (3.30) $$

where *n* is the number of neurons in the population, and $r_{ij}$ is the Pearson's correlation between neuron *i* and *j* given by

$$ r_{ij} = \frac{\sum (i_t - \bar{\imath})(j_t - \bar{\jmath})}{\sigma i \ \sigma j}. \quad (3.31) $$

I found no changes to mean pairwise correlation across different genotypes (*foxg1 +/+:* 0.004 ± 0.00*; foxg1 +/-:* 0.003 ± 0.00, U = 13.0, p = 0.17; *foxg1 -/-:* 0.0042 ± 0.00, U = 11.0, p = 0.41), which suggests no alterations in the functional connectivity across the brain due to a loss of interneurons (Figure 3.16E). Although foxg1-/- and +/- mutants exhibit a small increase in pairwise correlation due to PTZ exposure compared with wildtype (*foxg1 +/+:* 0.0032 ± 0.00*; foxg1 +/-:* 0.0044 ± 0.00, U = 5.0, p = 0.04; *foxg1 -/-:* 0.006 ± 0.00, U = 3.0, p = 0.06), this change does not survive multiple comparison corrections. Overall, this indicates that reducing interneuron number does not alter the pairwise connectivity of the network.

Taken together, I find that increasing the relative number of excitatory to inhibitory neurons, by removing interneurons in the telencephalon, gives rise to a surprising reduction in the excitability of single neurons.Interestingly, the network may also rest closer to seizure threshold as shown by marginally heightened sensitivities to low concentrations of PTZ, although the effect size of this change is small. These alterations in single neuron firing properties may well give rise to altered collective behaviours in avalanches and thus critical dynamics, and as such I extended my analysis to avalanche dynamics.

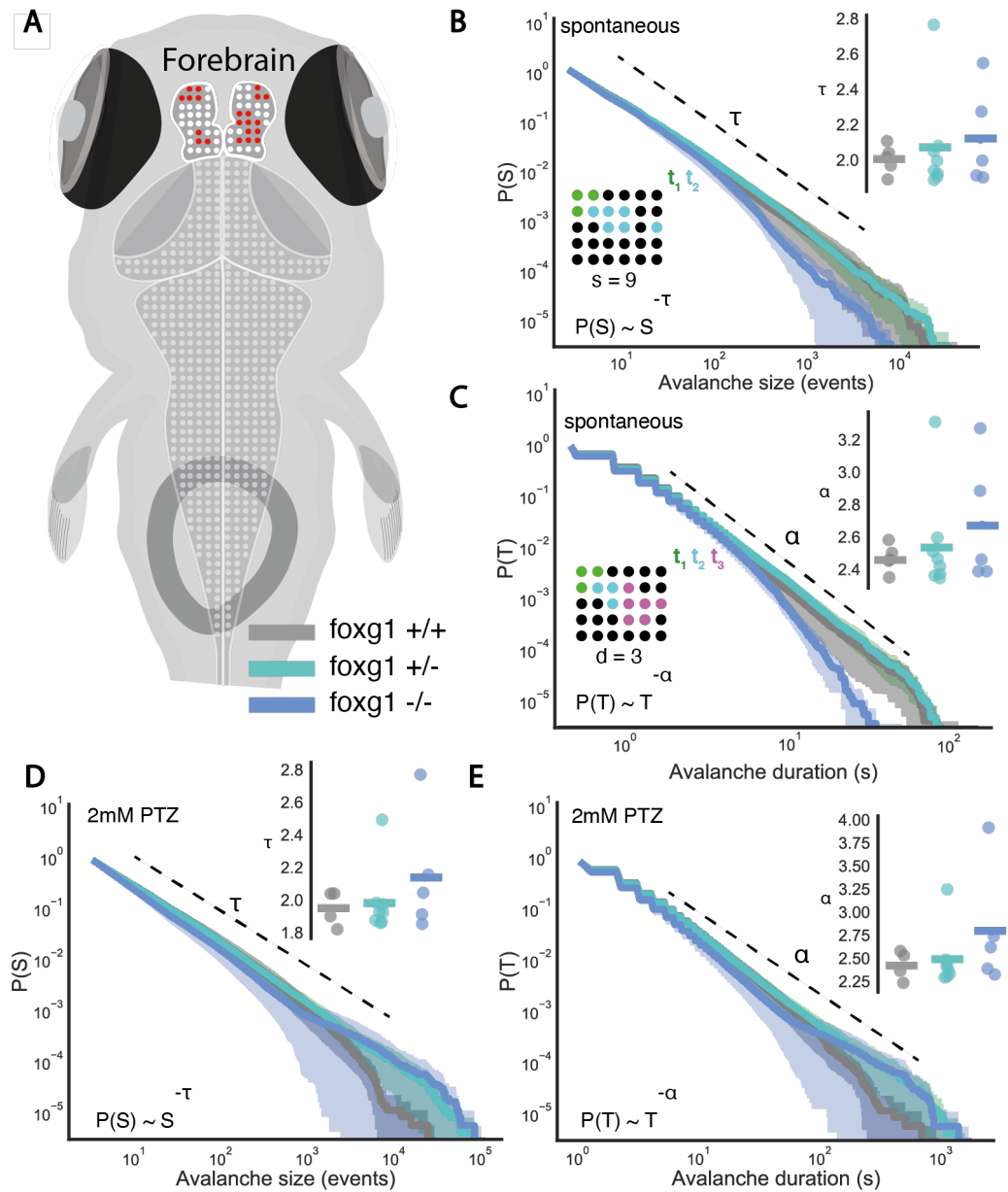

**Figure 3. 18  Genetically-induced EI imbalance effects on avalanche dynamics.**

(A) Larval zebrafish brain schematic, with all foxg1a analyses performed across forebrain neurons. Complementary cumulative distribution functions during spontaneous (B = avalanche size, C = avalanche duration) and 2mM PTZ (D = avalanche size, C = avalanche duration) conditions compared across foxg1+/+ (grey), foxg1+/- (teal), and foxg1-/- (blue) fish. Shaded regions represent the standard deviation. (Insets) Avalanche schematic demonstrating calculation of avalanche size (B) and duration (C) for single avalanche events. Black neurons are off. Coloured neurons represent active neurons at time point $t_x$. (Outsets) Avalanche exponents compared for size (τ) and duration (α) in spontaneous and 2mM PTZ conditions.

First, I calculated avalanche dynamics and constructed empirical distributions for avalanche size and duration across datasets. Interestingly, all datasets follow power law distributions and are better explained by power law than lognormal distributions (Figure 3.17). This indicates that avalanche dynamics are scale invariant even with a reduction in interneurons. Qualitatively, I noticed a slight increase in avalanche slopes for avalanche size (*foxg1a +/+:* 2.00 ± 0.04*; foxg1a +/-:* 2.07 ± 0.10, U = 18.0, p = 0.41; *foxg1a -/-:* 2.13 ± 0.13, U = 11.0, p = 0.42) and duration (*foxg1a +/+:* 2.47 ± 0.04*; foxg1a +/-:* 2.55 ± 0.11, U = 20.0, p = 0.47; *foxg1a -/-:* 2.68 ± 0.17, U = 10.0, p = 0.34) in foxg -/- fish compared with wildtype, although the change is not significant (Figure 3.17). There is also no change in avalanche slope following PTZ exposure for avalanche size (*foxg1a +/+:* 1.95 ± 0.05, *foxg1a +/-:* 1.99 ± 0.07, U = 15.0, p = 0.47; *foxg1a -/-:* 2.15 ± 0.16, U = 5.0, p = 0.14) and duration (*foxg1a +/+:* 2.43 ± 0.08*; foxg1a +/-:* 2.49 ± 0.11, U = 15.0, p = 0.47; *foxg1a -/-:* 2.80 ± 0.29, U = 5.0, p = 0.14) in foxg -/- fish compared with wildtype. Therefore this indicates, while single neuron firing properties exhibit reduced excitability due to EI imbalance, these do not manifest at the population scale with altered avalanche exponents or power law relationships.

Next, to robustly assess whether single neuron firing property changes in FOXG1 syndrome push the brain away from criticality, I measured critical statistics during spontaneous activity. Interestingly, I found marginal increases in DCC values for foxg1a -/- fish compared with wildtype, (*foxg1a +/+:* 0.06 ± 0.02*; foxg1a +/-:* 0.10 ± 0.03, U = 14.0, p = 0.0.21; *foxg1a -/-:* 0.16 ± 0.04, U = 3.0, p = 0.03), although this change does not survive multiple comparison corrections (Figure 3.18B). I also see a similar trend in PTZ data (*foxg1a +/+:* 0.07 ± 0.02*; foxg1a +/-:* 0.13 ± 0.03, U = 9.0, p = 0.13; *foxg1a -/-:* 0.25 ± 0.03, U = 2.0, p = 0.03). Therefore, this indicates that genetically induced EI imbalance might cause a loss of exponent relations away from critical values, however this loss is likely marginal and still lies within critical values reported in previous studies (Ma et al., 2019).

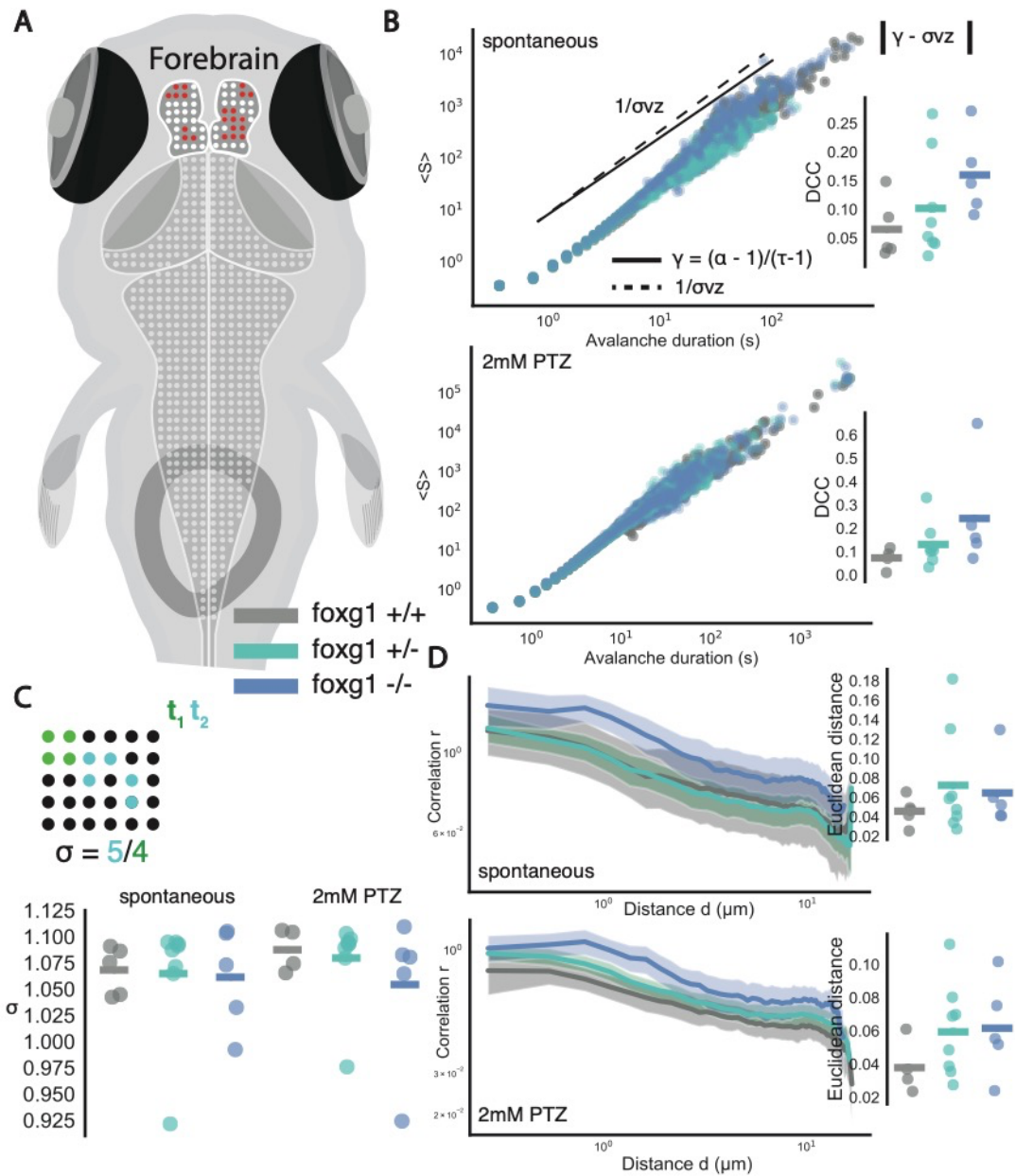

**Figure 3. 19 Genetically-induced EI imbalance effects on critical statistics.**

(A) Larval zebrafish brain schematic, with all foxg1a analyses performed across forebrain neurons. (B) Scaling exponent $1/\sigma\nu z$ captures the relationship between avalanche size and duration, here visualised as the mean size $\langle S \rangle$ for each duration in spontaneous (top) and 2mM PTZ (bottom) conditions. Each dot represents an $\langle S \rangle$ (T) relationship for a single fish, compared across foxg1+/+ (grey), foxg1+/- (teal), and foxg1-/- (blue) fish. (Outset) DCC values are plotted for each fish, comparing different genotypes. (C) Branching ratio relationship σ compared across foxg1 genotypes for spontaneous (left) and 2mM PTZ (right) conditions. (top) Avalanche schematic demonstrating calculation of branching ratio. Black neurons are off. Coloured neurons represent active neurons at time point $t_x$. Branching ratio σ is the number of descendants at a given time step ($t_2$) divided by the number of ancestors ($t_1$). (D) Correlation function r(d) compared across foxg1 genotypes for spontaneous (top) and 2mM PTZ (bottom) conditions. Shaded regions represents the standard deviation. (Outset) Euclidean distance from fitted power law compares power law fit compared across foxg1 genotypes.

Next, I assessed branching parameter σ which should be close to 1 at criticality. I found no significant changes to σ values over spontaneous (*foxg1a +/+:* 1.07 ± 0.01*; foxg1a +/-:* 1.06 ± 0.02, U = 12.0, p = 0.14; *foxg1a -/-:* 1.06 ± 0.02, U = 12.0, p = 0.5), or PTZ datasets (*foxg1a +/+:* 1.09 ± 0.01*; foxg1a +/-:* 1.08 ± 0.01, U = 14.0, p = 0.4; *foxg1a -/-:* 1.06 ± 0.03, U = 8.0, p = 0.36) (Figure 3.18C). This indicates that all foxg1a mutations do not change critical branching relationships exhibited in wildtype dynamics.

Finally, I assessed changes to the correlation function, in particular the presence of correlation function power law relationships. However, as above I found no changes in distances to power laws in foxg1a mutant fish compared with wildtype in spontaneous (*foxg1a +/+:* 0.05 ± 0.01*; foxg1a +/-:* 0.07 ± 0.02, U = 16.0, p = 0.30; *foxg1a -/-:* 0.06 ± 0.02, U = 10.0, p = 0.33), or PTZ datasets (*foxg1a +/+:* 0.04 ± 0.01*; foxg1a +/-:* 0.06 ± 0.01, U = 7.0, p = 0.07; *foxg1a -/-:* 0.06 ± 0.01, U = 5.0, p = 0.14) (Figure 3.18D). Furthermore I found no changes to the slope of the correlation functions as well in spontaneous (*foxg1a +/+:* 0.19 ± 0.05*; foxg1a +/-:* 0.18 ± 0.04, U = 20.0, p = 0.47; *foxg1a -/-:* 0.21 ± 0.05, U = 8.0, p = 0.20), or PTZ datasets (*foxg1a +/+:* 0.22 ± 0.05*; foxg1a +/-:* 0.23 ± 0.04, U = 16.0, p = 0.47; *foxg1a -/-:* 0.21 ± 0.04, U = 10.0, p = 0.45). This indicates that the correlation structure does not significantly change due to foxg1a mutations, indicating no changes to critical dynamics.

Taken together, I find that although genetically-induced EI imbalance through the foxg1a mutation gives rise to altered single neuron firing properties, this does not manifest as changes to avalanche dynamics or critical statistics. Therefore, EI imbalance disorders such as FOXG1 syndrome may not manifest as a loss of criticality or the emergence of supercritical states.

## 3.3   Discussion

In this chapter, I estimated the spatiotemporal propagation of neural activity to understand the role of EI balance in shaping criticality in whole-brain cellular-resolution networks *in vivo*. I aimed to answer three questions. First, are whole brain single cell networks organised around criticality? Second, does EI balance act as a control parameter for regulating criticality? Thirdly, do hyper-excitation EI imbalance disorders emerge as a supercritical state away from criticality? The statistics of whole brain dynamics indicate that single cell activity organises the brain to criticality in healthy networks. Furthermore, subtle changes to EI balance give rise to dramatic deviations from critical statistics supporting the role of EI balance in shaping critical dynamics. Importantly, I found that pharmacologically-induced EI imbalance giving rise to seizures caused a disruption of key critical statistics suggestive of a loss of criticality. Thus, these results suggest that EI balance may serve to shape brain dynamics within a functionally stable critical regime, and that the network instability of epileptic seizures emerges as a loss of criticality.

### 3.3.1   Identifying Mechanisms Supporting Self-Organising Criticality

Criticality describes a system in which dynamics are organised near to a phase transition, such that regimes of order and disorder can co-exist. An open question is how the brain can maintain dynamics perfectly at this balancing point, given the limited parameter range that defines a bifurcation region in a quasi-critical system (Bonachela et al., 2010). In the classical Ising model, the experimenter manually tunes the control parameter (temperature) to the exact value of the phase transition. Biological systems, however, must continuously adapt their network interactions in response to changing inputs to maintain dynamics at a bifurcation (Bornholdt & Rohlf, 2000). In this way spontaneous and robust critical dynamics require the self-tuning of a control parameter to criticality, a phenomenon known as self-organising criticality (SOC). SOC was popularised through the famous sandpile avalanche model (Bak et al., 1987b, 1988), in which a sandpile spontaneously organises its slope to the point at which scale-invariant avalanches occur, due to the feedback between local forces exerted by new grains falling on the pile and the slope angle. Maintaining critical dynamics in the brain would also require appropriate feedback mechanisms between system state and control parameter values (Chialvo et al., 2020; Sornette et al., 1995), a difficult feat in a system that must process a diverse range of inputs and engage in a diverse range of states. The requirement of adaptive feedback points towards plasticity mechanisms (Levina et al., 2007; Zeraati et al., 2021) in particular homeostatic plasticity, which acts to regulate firing rates around a fixed point in response to excitability changes (Abbott & Nelson, 2000; Keck et al., 2017; Tetzlaff et al., 2011). In fact, firing rate regulation follows the same time course as the establishment of criticality following monocular deprivation (Ma et al., 2019), indicating that criticality may be an endpoint for homeostatic plasticity in the brain. Key to the regulation of firing rates is the constant maintenance of appropriate excitation and inhibition (Turrigiano & Nelson, 2004), through activity-dependent changes to GABA (Kilman et al., 2002), NMDA and AMPA receptor expression (Lissin et al., 1998). Given that homeostatic plasticity serves to regulate EI balance, the separation of time scales that such slow acting plasticity mechanisms would support, and the known functional consequences related to EI imbalance (Fritschy, 2008; R. Gao & Penzes, 2015; Lee et al., 2017; Žiburkus et al., 2013), it is highly plausible that the regulation of EI balance could serve as a control mechanism for SOC.

Our data point towards a causal link between EI balance and critical dynamics *in vivo*. This is supported by work from *in vitro* studies which have shown that inhibitory blockade leads to supercritical behaviour, while reduced excitation causes subcritical dynamics (Beggs & Plenz, 2003; Bellay et al., 2015; Pasquale et al., 2008; Shew et al., 2009). Our work builds on these studies by, for the first time, investigating the link between EI balance and criticality using whole brain activity *in vivo* in up to ten thousand neurons compared to hundreds of neurons *in vitro*. Given that critical statistics deviate dramatically following graded changes to EI balance, this indicates that brain dynamics may be tightly poised close to a phase transition as previously reported (Beggs & Plenz, 2003; Meisel et al., 2012; Ponce-Alvarez et al., 2018; Tagliazucchi et al., 2012). Importantly, critical statistics measured across the whole brain were highly sensitive to changes to EI balance, as expected for a control parameter. Thus, I demonstrate the sufficiency of EI balance for criticality in systems with sensory input and intact network connectivity, as the presence of EI balance is sufficient for the presence of critical dynamics. Furthermore, I demonstrate that synaptic EI balance shapes avalanche power law exponents in whole brain networks, confirming previous *in silico* work which demonstrated the regulation of avalanche power law (Poil et al., 2012) and 1/f power spectra exponents by tuning EI balance (Lombardi et al., 2017). This confirms the role of synaptic EI balance in shaping population dynamics through avalanche cascades.

I also find that our data show variable avalanche exponents which on average are higher than other studies (Beggs & Plenz, 2003; Ponce-Alvarez et al., 2018), which supports the notion that brain state influences avalanche exponents (Fontenele et al., 2019). That our avalanche exponents differ from theoretically derived values indicate that neuronal avalanche dynamics may not belong to a known universality class, with different microscopic properties to other phase transitions (Perković et al., 1995). Importantly, brain dynamics can be poised to a phase transition, without showing universality (Iliass et al., 2016).

Our data builds on previous work with several novel insights. Firstly, we explore the link between EI balance and scaling and exponent relationships. That I demonstrate a loss of critical exponent relationships due to EI imbalance is a strong indicator of a deviation from criticality, as non-critical systems which exhibit power law scaling, critical exponents and universal scaling fail to show critical exponent relations (Sethna et al., 2001; Touboul & Destexhe, 2017). Similarly, that our scaling relationships deviate from values reported in purportedly critical systems *in vitro* and *in vivo* indicates a role of EI balance in driving dynamics around criticality. These changes are driven by an inability to predict avalanche size from duration, due to the emergence of large avalanches which grow quickly in time such that a single exponent fails to predict avalanche size across all scales, indicating a loss of self similarity. Furthermore, I demonstrated the emergence of characteristic scale in the correlation function due to EI imbalance, as expected for a supercritical system. The fact that this power-law relationship is lost due to EI imbalance indicates that a loss of scale invariance in avalanche dynamics occurs due to excessive correlation between distant units such that large scale events occur frequently. This further demonstrates the sufficiency of EI balance for the presence of criticality.

Taken together our data strongly indicate a causal link between EI balance and criticality. However, it is vital to note that the critical statistics I have measured are purely correlative and can appear in non-critical systems (Destexhe & Touboul, 2021). Furthermore, it is unlikely that brains operate at a singular critical point due to finite-size effects and metabolic costs of wiring (Amaral et al., 2000; Clauset et al., 2009). Thus the measurement of critical statistics will differ from idealised models and therefore any claims about the universality of living systems or extrapolations from theoretical models to the brain should be made with caution (Priesemann et al., 2009). Importantly, while the tuning of EI balance alters critical statistics as expected for a control parameter, further evidence is required to demonstrate the active role of homeostatic EI balance mechanisms in organising the system to criticality. To support this claim, one would need to track synaptic EI balance changes *in vivo* alongside critical dynamics. Furthermore, I cannot eliminate the possibility of downstream mechanisms being influenced by changes to EI balance that may instead act as control parameters for criticality. For example, network topology likely plays a role in SOC (Bornholdt & Rohlf, 2000), and effective and structural network topology will be influenced by broad changes to EI balance. Similarly, regulating gap junctions also gives rise to a deviation from criticality (Ponce-Alvarez et al., 2018). Therefore, while EI balance clearly shapes criticality, it is likely that multiple network mechanisms converge to maintain or deviate the brain from a phase transition. In this way, I cannot unequivocally claim that a direct relationship between EI balance and criticality demonstrates its regulating role for organising dynamics within a functional, critical range. Nonetheless, the data indeed support the notion that EI balance is sufficient for *in vivo* neuronal networks to give rise to criticality. This opens up avenues for exploring the role of the dysregulation of EI balance in causing network instability in epilepsy.

### 3.3.2 Understanding EI Imbalance Through the Lens of Critical Systems

A key feature of critical systems is that a multitude of computational properties are optimised at the phase transition, including information transmission, adaptability, network mediated separation and dynamic range (Bertschinger & Natschläger, 2004; Langton, 1990; Packard, 1988). Thus, if criticality is an optimal homeostatic endpoint for EI balanced networks, a loss of SOC due to EI imbalance would impair brain function. Early theoretical work postulated the link between epileptic seizures, which can emerge due to EI imbalance, and the emergence of a supercritical state (Beggs & Plenz, 2003; Haldeman & Beggs, 2005; Hsu et al., 2008). Such a state is characterised by an increase in connection strengths between units, giving rise to exponentially growing avalanches that saturate the network and should impair its optimal response properties. However, whether epileptic seizures truly emerge as a supercritical state *in vivo* is unclear. For example, theoretical work predicts the emergence of $\sigma > 1$ in supercritical networks (Haldeman & Beggs, 2005; Shew et al., 2009), which has been reported in inter-epileptiform activity in epilepsy patients (Arviv et al., 2016). However, human epileptogenic tissue has shown reductions in $\sigma$, which would be expected in a subcritical state (Hobbs et al., 2010; Hsu et al., 2008). Importantly, our data exhibit $\sigma > 1$ in generalised seizure transitions which matches theoretical predictions for $\sigma$ values in supercritical systems. I note that $\sigma \sim 1$ may also occur in supercritical systems (Hesse & Gross, 2014), and so exact values of $\sigma$ are relatively weak assessments for criticality. Nonetheless, given that our study records from vastly greater numbers of neurons than previous epilepsy studies, $\sigma$ will be more accurately assessed due to less subsampling and averaging effects (Priesemann et al., 2009). Having on average greater than 1 descendant for each ancestor would lead to exponentially growing avalanches, preventing scale invariance and reducing the range of inputs the network can process, as expected for a supercritical state (Harris, 1963).

Several differences arise in our data, compared with previous studies looking at supercriticality in epilepsy. In particular, *in vitro* studies exhibit an increase in avalanche slope, with the emergence of a bimodal distribution with increased probability for large avalanches but decreased probability for medium avalanches (Beggs & Plenz, 2003; Bellay et al., 2015; Pasquale et al., 2008). Notably, our avalanche distributions exhibit an increase in probability for both medium and large sized events, giving rise to a shallower power law slope as found previously in epilepsy patients (Arviv et al., 2016). The lack of a bimodal appearance is likely due to the scale difference in recordings, as our data can capture the full extent of supercritical cascades giving rise to both medium and large sized events, while supercritical cascades are likely to recruit the entire array in small population recordings. Thus, the appearance of a bimodal distribution in supercritical systems may be an artefact of the small recordings size, and seizures may instead emerge as a deviation in power law exponent. Taken together this suggests that epileptic seizures may well emerge as a supercritical state away from criticality, which can help to drive our understanding of abnormal brain dynamics in epilepsy and seizures.

Interestingly, some 20mM PTZ recordings do exhibit a slope change in the tail ends of the complementary cumulative distribution, which are suggestive of multiple scaling regimes in the data rather than a universal scaling function. This could indicate the existence of multiple avalanche regimes in the 20mM condition, consisting of frequent medium sized avalanches and unusually frequent large avalanches. This observation might suggest the presence of different brain states that co-occur in the generalised seizure condition, which could include pre-ictal, ictal and post-ictal states. In fact, PTZ is known to induce a diversity of brain dynamics and states (Reichert et al., 2019). In this scenario, multiple avalanche dynamics may occur during the recording period giving rise to multiple scaling regimes. Future work should aim to characterise these distinct scaling regimes in seizure data.

I also looked at EI imbalanced dynamics in a genetic model of foxg1 syndrome, in which patients abnormal inter-ictal dynamics and seizures (Cellini et al., 2016; Seltzer et al., 2014). Interestingly, while the foxg1a mutation reduces the number of interneurons in the network, this reduces single neuron excitability. The reason for this surprising effect is difficult to ascertain from my data alone, but could be caused by homeostatic compensations to synaptic wiring over development due to a loss of interneurons, which could render the network as a whole hypo-excitable. Importantly, foxg1a homozygous fish die early in development (Bruce, 2022) and therefore reduced excitability may instead be a downstream effect of severely disrupted homeostatic mechanisms. Importantly, reduced excitability does not influence critical statistics, causing no significant changes to avalanche exponents, branching parameters, or correlation functions. As such, while whole brain pathology during generalised seizures may emerge as a supercritical state, there is no evidence to suggest that EI imbalanced dynamics in foxg1 syndrome also emerge as a supercritical state.

It is worth noting that the accurate estimation of neuronal excitability, from spike rates, and of critical statistics, from avalanche dynamics, relies on the robust inference of calcium transients from fluorescence data. Here I applied a hidden Markov model developed previously by the lab (Diana et al., 2019). Importantly, all spike inference methods come with their caveats and state-of-the-art methods can only achieve moderate accuracy – an investigation by Theis et al. into 8 different methods showed that the best model only achieved ~0.5-0.6 correlation between true and inferred spike rates (2016). With this in mind, one should use spike inference methods with caution as they will always lead to false positives and negatives. In particular, the HMM used in this study requires the selection of several parameters, including the event probability and decay time. Here, I chose to use the same parameters for baseline, 5mM PTZ and 20mM PTZ conditions to avoid adding in systematic bias across conditions. Nonetheless, given that each condition likely has different system parameters, one cannot avoid the conclusion that the HMM may be inferring spikes sub-optimally for each condition. With this in mind, any inferences which rely on spike counts should be viewed with caution. Nonetheless, given that alternative metrics relying on raw fluorescence data (e.g. calcium amplitude measures) show conserved alterations in excitability as observed in the HMM, this suggests that the HMM can identify key excitability changes present in our data. Future work should aim to develop alternative methods that are not reliant on specific parameterisations.

One should also note that changes in GCaMP signal, either through protein translation or calcium influxes (e.g. during a seizure), might influence the cell segmentation steps outlined in Section 2.2. Cell segmentation relies on the identification of correlated, contiguous pixel clusters – higher GCaMP signal (such as during a seizure) could therefore lead to a greater number of cells being identified, which could bias the density of network sampling towards seizure datasets. This has implications for comparisons across these different groups, as seizure datasets might be less sparsely sampled than spontaneous data. Future work should aim to characterise the relationship between image brightness and cell number, and develop segmentation algorithms which are not reliant on image brightness.

To summarise, I find that i) whole brain networks at single cell resolution reside at criticality in healthy networks, ii) EI balance plays a role in shaping critical dynamics and is a sufficient component of criticality, and iii) epileptic seizures, but not genetically-induced EI imbalance cause a transition from critical to supercritical dynamics. Thus, criticality can serve as a useful framework for understanding pathological whole brain dynamics and emergent network instability.

# Chapter 4

## Neuronal Network Mechanisms Driving Dynamics and Functional Impairments in Seizures

This chapter contains text, data and figures from some of my work currently in preprint (Burrows et al., 2021).

## 4.1 Introduction

The dynamics of the brain are constrained by the connectivity patterns of the underlying network (Ju et al., 2020). This network structure is defined by synaptic connections between neurons, the strengths of which may be dynamically regulated by neuromodulation and plasticity mechanisms (Abbott & Nelson, 2000; Dembrow et al., 2010; Turrigiano & Nelson, 2004). Importantly, neuronal subtypes exhibit diverse and highly complex synaptic connectivity patterns both locally and distally (Gerfen et al., 2018; Lim et al., 2018), producing complex network topologies (Gal et al., 2017). Such complex and dynamic network structure ultimately gives rise to the neuronal dynamics that have been described at different scales (Bellay et al., 2015b; F. Lopes da Silva et al., 2003; Stringer, Pachitariu, Steinmetz, Reddy, et al., 2019), and linked to a variety of different brain functions (Buzsáki, 2002; de Arcangelis et al., 2006). Interestingly, a diversity of disorders of brain development implicate synaptic connectivity genes in aberrant brain dynamics and function, such as in epilepsy (R. Rosch et al., 2019). Therefore, central to understanding how functional impairment in brain disorders occurs, is understanding how changes in synaptic connectivity gives rise to emergent aberrant brain dynamics.

Epilepsy is characterised primarily by the recurrence of seizures, which emerge as aberrant brain dynamics punctuated by excessive network synchronisation and increased firing rates (Weiss et al., 2013). Such abnormal dynamics can give rise to impaired brain function – epileptic seizures often cause a loss of awareness and impaired responsiveness to stimuli (Inoue & Mihara, 1998), while those diagnosed with epilepsy and associated excitation-inhibition (EI) imbalance conditions exhibit wide ranging cognitive impairments (Butler & Zeman, 2008; Thompson & Duncan, 2005; Wong et al., 2019, p. 1). Identifying the microscale synaptic changes that give rise to such functional network instability in epilepsy is challenging (see Section 1.3.1), given that seizure dynamics often engage distributed brain-wide networks and having access to synaptic connectivity in such large networks in not currently feasible (see Section 1.1). Importantly, the larval zebrafish provides access to single neuron dynamics across the whole brain – this can help to bridge the gap between microscale neuronal activity and emergent global dynamics and function (see Section 1.1.4). However, the synaptic connections giving rise to these global dynamics still cannot be measured directly in such large, whole brain networks. A pragmatic solution is the use of data–driven network models in which underlying connectivity patterns are inferred from empirically derived whole brain data at single cell resolution (see Section 1.2). For example, data-driven network modelling of cellular resolution larval zebrafish recordings have uncovered certain effective neuronal connectivity motifs that render a network more susceptible to seizure transitions, indicating roles for feedforward hubs with highly connected neighbours (Hadjiabadi et al., 2021). Thus, the identification of network motifs in single neuron networks using data–constrained models can identify key neuronal mechanisms that drive global dynamics into seizure states. Such approaches may be further extended to understand how neuronal changes might give rise to brain dysfunction – for example, large scale brain models at neuronal resolution and constrained by diffusion imaging data, have shown that inhibitory synaptic feedback alters network response properties to sensory input (Deco et al., 2014). In this way data-driven, cellular resolution network models may be used to understand how microscale connectivity changes give rise to impaired global dynamics and function, such as in epilepsy.

A useful framework for relating microscale neuronal properties with global network dynamics is criticality (see Section 1.3.3) – here the interaction strengths of single neurons are tuned such that global network dynamics reside at a transition point between order and disorder (Ising, 1925). Interestingly, synaptic plasticity may enable the brain to homeostatically self-organise to criticality (Zeraati et al., 2021), while seizures could emerge due to a failure of homeostatic critical dynamics (see Chapter 3) (Meisel et al., 2012). Given that various network models have demonstrated a diversity of synaptic mechanisms supporting self-organising criticality, these approaches may provide clues into the synaptic pathways which, when dysregulated in epilepsy, give rise to a loss of criticality and ultimately seizures. For example, network models of self-organising criticality have demonstrated roles for the formation and retraction of neuronal connections (Bornholdt & Roehl, 2003; Bornholdt & Rohlf, 2000; Tetzlaff et al., 2010; van Ooyen & Butz-Ostendorf, 2019), alterations in synaptic strengths (de Arcangelis et al., 2006; Rubinov et al., 2011) and intrinsic neuronal excitability changes (Levina et al., 2007) in stabilising brain dynamics to criticality *in silico*. Interestingly, these changes have all been implicated in the emergence of seizures, with excessive synaptic connections due to impaired synapse refinement (Chu et al., 2010), alterations to synaptic strengths through shifting ratios of continuous to discontinuous postsynaptic densities (Geinisman et al., 1988), and increased intrinsic neuronal excitability through more vesicles in the readily releasable pool (Upreti et al., 2012), all reported in animal models of seizures. Clearly then changes to neuronal network connectivity, synaptic strengths and intrinsic neuronal excitability may be central for destabilising global critical dynamics during seizures. An investigation comparing the roles of these microscale neuronal changes using data-driven network models, will disentangle the differential contribution of neuronal parameters in driving global brain dynamics away from criticality into seizure states.

Critical phenomena are also a useful framework for studying brain dysfunction, as systems at criticality maximise various computational capacities (Bertschinger & Natschläger, 2004; Haldeman & Beggs, 2005; Kinouchi & Copelli, 2006; Legenstein & Maass, 2007; Shew et al., 2011). Given that epileptic seizure and EI imbalance-associated dynamics may emerge as a loss of criticality (Zimmern, 2020), emergent brain dysfunction could occur due to the suboptimal computational capacities found in networks away from a phase transition. In particular, systems at criticality are believed to: i) operate at the edge of chaos, resulting in neutral dynamics that support network-mediated separation and maximal memory capacity (Bertschinger & Natschläger, 2004; Legenstein & Maass, 2007; Maass et al., 2002), ii) maximise their dynamic range, a measure of the range of inputs the network can represent (Gautam et al., 2015; Kinouchi & Copelli, 2006; Shew et al., 2011), and iii) maximise their metastability (Deco et al., 2017; Haldeman & Beggs, 2005; Wildie & Shanahan, 2012), a tendency of the brain to transiently explore a wide range of semi-stable states supporting flexible dynamics (Bak & Chialvo, 2001; Fingelkurts & Fingelkurts, 2001; W. J. Freeman & Holmes, 2005; Friston, 1997). Critical systems theory therefore predicts that neuronal alterations driving brain dynamics away from criticality such as in epilepsy, should disrupt neutral and flexible dynamics, and impair network response properties. Therefore, given the identification of synaptic parameters driving aberrant brain dynamics away from criticality *in vivo* or *in silico*, critical systems theory may be used to relate such microscale alterations with emergent brain dysfunction. However, a direct investigation into the link between neuronal alterations driving aberrant brain dynamics in epilepsy and the functional properties of critical systems has yet to be carried out. Such an investigation would help to bridge the gap between microscale synaptic changes driving abnormal brain dynamics and emergent global network dysfunction in epilepsy.

In this chapter, I construct data-driven elementary network models from brain–wide cellular resolution recordings from the larval zebrafish to infer the neuronal connectivity alterations driving epileptic seizures. Furthermore, I extend network modelling and dynamical systems approaches to aberrant brain dynamics in seizures and EI imbalance disorders, to relate microscale network alterations with emergent brain dysfunction. I test 2 key hypotheses: i) seizure dynamics are a convergent endpoint emerging from non-specific changes to excitability (see Section 1.1.1), through changes to network connectivity, synaptic strengths or intrinsic excitability, and ii) microscale neuronal network alterations cause brain dysfunction in epileptic seizures and associated EI imbalance disorders, by disrupting the optimal properties of critical networks. I demonstrate that while focal seizures emerge out of non-specific changes, generalised seizures can only emerge through increases in neuronal connections in the network, with changes to intrinsic excitability and synaptic strengths both failing to explain empirical seizure dynamics. Furthermore, such increases in neuronal connectivity give rise to a loss of neutral and flexible dynamics and impaired network response properties in seizures, directly linking brain dysfunction in epilepsy with a loss of criticality.

## 4.2   Results

### 4.2.1   The Construction of a Cellular-Resolution Network Model of Brain-Wide Seizure Dynamics

To probe the microscale neuronal network changes driving seizures in whole brain networks, I performed 2-photon imaging of the larval zebrafish during pharmacologically-induced seizures (see Section 2.2.2). I was particularly interested in the network changes that separate out functionally stable, locally unstable and globally unstable dynamics (see Section 1.3.1). To clearly delineate locally from globally unstable networks, I took advantage of the characteristic time course of brain dynamics following 20mM PTZ administration – here, dynamics sharply transition from aberrant, yet locally synchronous activity to generalised, globally synchronous seizures (see Figure 2.2) (Diaz Verdugo et al., 2019a). To study functionally stable and unstable regimes, I isolated the following recording periods for each fish (n=9): i) *baseline* periods – 400 frames of randomly selected spontaneous activity recordings which capture functionally stable dynamics, ii) *pre-ictal* periods – 400 frames immediately preceding the generalised seizure which capture locally synchronous events that do not generalise, and iii) *ictal-onset* periods – the first 400 frames of the generalised seizure, where global cascades spread through the entire network (see Section 2.2.10). To describe dynamics in these 3 distinct regimes, I required a method of capturing the spatiotemporal progression of neuronal activity through the network. The neuronal avalanche framework is particularly useful in this context as it describes the statistics of propagation events through brain networks (see Section 3.2.1) and can distinguish between healthy and seizure regimes (see Section 3.2.7). Therefore, I use the avalanche framework to describe *baseline*, *pre-ictal* and *ictal-onset* regimes in my empirical data.

In order to understand the network mechanisms giving rise to empirical avalanche dynamics across each regime, I needed to construct a network model capable of capturing observed avalanche dynamics. Key to the emergence of brain network dynamics is the embedding of nodes (neurons) in 3-dimensional space, which is defined by constraints imposed on neuron location and density by brain structure and anatomy. To accurately capture this relationship in the larval zebrafish brain, I embedded my network in a 3-dimensional brain space that recapitulates the structural boundaries, anatomically-defined neuron densities, and the neuron-neuron distance distributions found in the larval zebrafish brain. Specifically, I registered all neuron coordinates across all brains (10 fish, 3 conditions) to a standard space (Tabor et al., 2019). Next, I performed k-means clustering on the spatial locations of all neurons (k = 8990, the mean number of cells across all datasets), with resulting cluster centroids used as network node locations (see Figure 4.1A). This enabled the construction of an 'average' larval zebrafish brain, which respected the structural boundaries and the spatial neuronal distributions of the brain, at the scale of my empirical recordings.

In order to accurately capture empirically observed avalanche dynamics, it was important to model the network connectivity which gave rise to such activity patterns. Although the network topology defining the larval zebrafish brain is unknown, here I assumed that the network follows a scale free topology (see Figure 4.1B), which has been reported in brain data (Bassett & Bullmore, 2017; Eguíluz et al., 2005). Network connectivity was defined by the growth and preferential-attachment algorithm (Barabási & Albert, 1999). The algorithm begins with $m$ initial connected nodes and progresses by sequentially adding a new node that forms $m$ connections to existing nodes, where the probability $p_i$ that each new node is connected to node $i$ is proportional to the degree of node $i$ defined as

$$p_i = \frac{k_i}{\sum_j k_j} \ , \quad (4.1)$$

where $k_i$ is the degree of node $i$ and $k_j$ is summed over all pre-existing nodes. This procedure allowed me to tune the number of connections in the network, while maintaining the same scale free topology across all models. It is also important to note that, while neuronal systems show clearly defined local connectivity patterns, the growth and preferential attachment algorithm used here does not account for distance. Therefore this model also assumes equally likely connection formation to nearby and distant neurons – this assumption is partially mitigated by the use of exponentially decaying synaptic weights (see Section 4.2.2).

Also important in driving network activity is the evolution of activity on each node, which can be defined using differential equations. I wanted a model which captured the non-linear input-output functions which shape neuronal network signal propagation, while remaining computationally tractable for large networks. To this end, I used leaky-integrate-and-fire dynamics for single neurons, which have been extensively applied to modelling neuronal avalanche dynamics (Levina et al., 2007). I considered a network of leaky integrate-and-fire neurons with $N$ excitatory neurons ($N$ = 8990), and $E$ external excitatory inputs ($E$ = 1000) onto each neuron. For the sake of simplicity, I chose to ignore inhibitory neurons. While this is of course an over-simplification, I reasoned that I could capture the effect of inhibitory changes induced by PTZ, through network parameter changes to connectivity, synaptic weights and intrinsic excitability in a purely excitatory network. This enabled me to reduce the parameter space for my model. Therefore, this model assumes a degree of homogeneity across recorded neuron types, while subsuming inhibitory neuronal subtype effects onto network parameter changes.

The membrane voltage for neuron $i$ is defined at subthreshold voltages by the differential equation

$$\tau_m \frac{dV_i}{dt} = \frac{-(V_i - V_r)}{R_m} + I_i \ , \quad (4.2)$$

where $V_i$ is the membrane voltage, $\tau_m$ is the membrane time constant ($\tau_m$ = 20), $V_r$ is the resting membrane potential ($V_r$ = 0), $R_m$ is the resistance, and $I_i$ is the input current. For voltages beyond the voltage threshold $V_{th}$ the neuron fires a spike, at which point $V_i$ is reset to $V_r$.

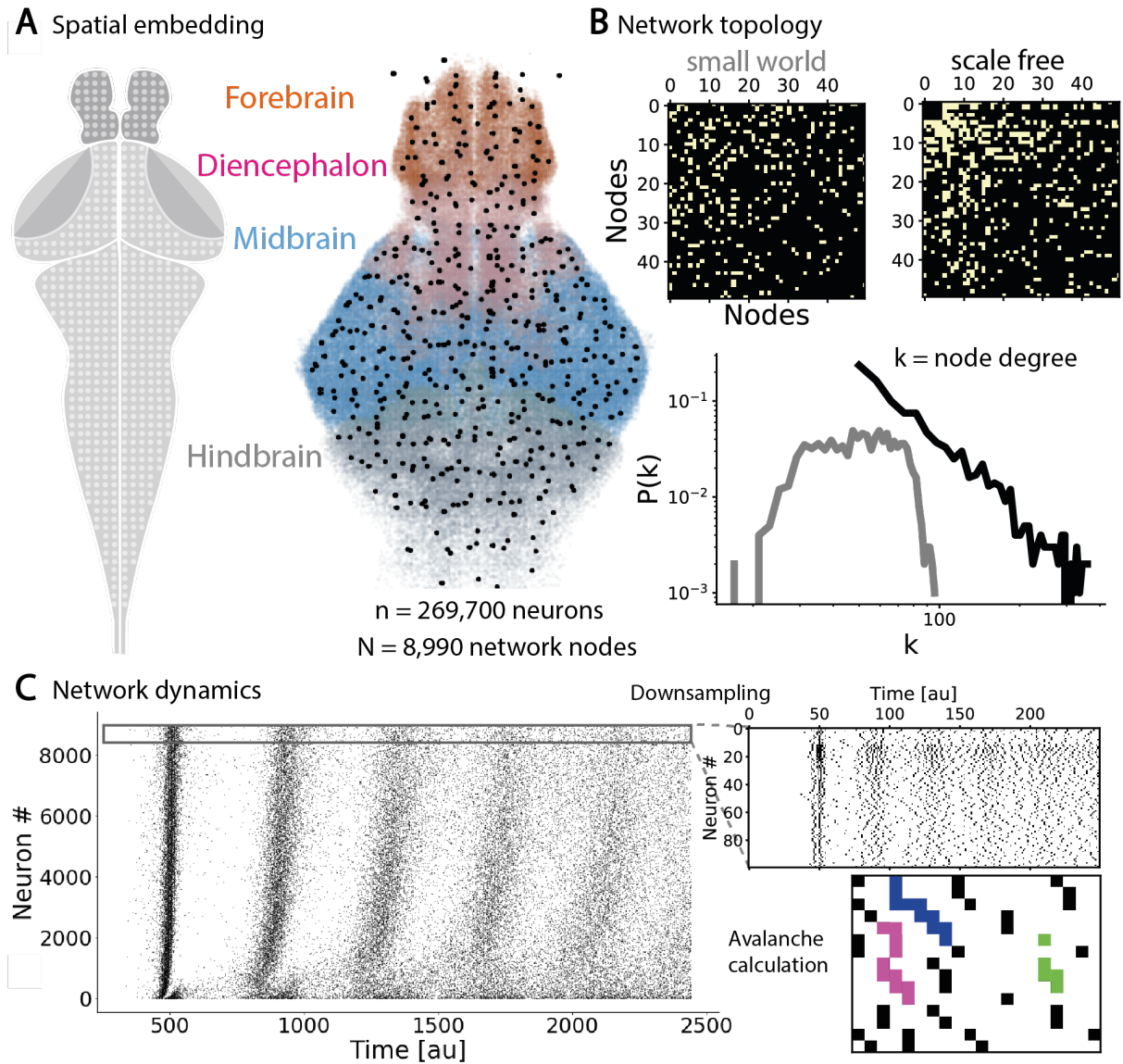

**Figure 4.1    Network model construction from larval zebrafish brain data.**

(A) Schematic showing spatial embedding procedure of model brain network. A rudimentary larval zebrafish brain structure is shown (left). The k-means algorithm was used to build an average larval zebrafish brain, where N clusters were constructed from n segmented neurons across 30 fish brains (right). (B) Network topologies were constructed using both the small world (left) and scale free (right) network construction procedures, with adjacency matrices shown for both , using networks of 50 neurons (small world network parameters: p=0.9, m=5; scale free network parameters: m=5). Empirical distributions for the degree *k* shown for scale free (black, p=0.9, m=50) and small world (grey, m=50) networks of 500 neurons – note how scale free networks exhibit power law distributions. (C) Network dynamics were constructed for scale free networks and simulated for 2500 time steps, with each node defined as a leaky integrate-and-fire neuron. Each time series was downsampled to allow for neuron spikes to be grouped into bins for avalanche calculation. Avalanches were calculated using the same procedure in Section 3.2.1, with avalanche events shown in non-black colours using a hypothetical spike matrix for illustrative purposes.

The input to neuron $i$ at time $t$ is defined by

$$I_i(t) = \sum J_{ij}\Delta(t - \tau_d) + \sum U_{ie}\Delta(Poisson, \psi). \quad (4.3)$$

The first term describes the input from other neurons in the network, where $J_{ij}$ is the weighted, directed adjacency matrix, between the presynaptic neuron $i$ and postsynaptic neuron $j$. A spike from neuron $j$ will affect neuron $i$ after a synaptic delay $\tau_d = 1$, through Dirac's delta function. The second term describes the input from the external current, where $U$ is the weighted adjacency matrix between the presynaptic neuron $i$ and the external neuron $e$. External input spikes are modelled as a Poisson point process with rate $\psi = 10$. $U_{ie}$ for all connections were set to 0.1. In this way, using spatially-constrained neurons connected in a scale free network with leaky-integrate-and-fire dynamics, I was able to simulate brain network activity (Figure 4.1C).

Finally, in order to calculate avalanches from my spiking network, I needed to capture the spatiotemporal propagation of activity. In my model, unlike in my empirical recordings, I have access to the exact connections between neurons and could track the transmission of signal from one neuron to the next. However, to ensure comparability between my empirically-derived and model-derived avalanche data, I applied the same avalanche algorithm outlined in Section 3.2.1 to my spiking network. Spiking activity was binned into 10 timestep windows for avalanche estimation (see Figure 4.1C).

*4.2.2 The Neuronal Network Mechanisms Driving Seizure State Transitions*

Once I had defined the spatial embedding, topology, dynamics and avalanche estimation procedures in my model network, I needed to define candidate seizure-driving parameters. I set out to understand the key microscale neuronal changes which drive the emergence of locally unstable and globally unstable dynamics following bath application of 20mM PTZ. As outlined in Section 4.1, changes in the number of connections (Chu et al., 2010), the strengths of synapses (Geinisman et al., 1988) and the intrinsic excitability of neurons (Upreti et al., 2012) are all associated with seizures. To distinguish between the differential involvement of these changes in local and global seizure dynamics, I defined 3 key parameters in my neuronal network: 1) *network connectivity* – the number of connections between neurons defined by parameter $m$, 2) *synaptic weights* – the strength of synapses between neurons defined by parameter $r$, and 3) *intrinsic excitability* – the propensity of individual neurons to spike defined by parameter $V_{th}$ (see Figure 4.6).

*Network connectivity* was varied by altering $m$ in the growth and preferential-attachment algorithm, thus changing the number of binary edges between neurons. Given that the formation and elimination of new synapses can take hours to days (Le Bé & Markram, 2006; Ozcan, 2017), I supposed that the formation of new structural synapses over the periods between *baseline, pre-ictal* and *ictal-onset* conditions (minutes-seconds) may be limited. However, evidence from animal seizure models supports roles for extrasynaptic mechanisms in driving seizures, for example through astrocyte-astrocyte coupling via gap junctions and rhythmic network synchronisation via volume-transmitted GABA (Diaz Verdugo et al., 2019a; Magloire et al., 2022). Such extrasynaptic pathways could help to synchronise non-synaptically coupled neurons, giving rise to new effective connections between neurons. Given that the timescales of these changes can occur over seconds to minutes, I supposed that effective connectivity changes may be able to drive unstable seizure dynamics. Therefore, here I use changes to network connectivity to relate to alterations in both effective and structural connectivity. In particular, I wanted to test the hypothesis that altering the number of connections in the network was sufficient to drive local and globally unstable seizure states.

*Synaptic weights* defines the distributions of edge weights across the network. Due to the short timescales over which synaptic strengths may be altered, for example through increased dendritic spine size (Matsuzaki et al., 2004) or AMPA receptor trafficking (Matsuzaki et al., 2004; Park, 2018; Y. Yang et al., 2008), I supposed that weight alterations could plausibly occur over the recording periods. In order to define the strength of synaptic weights across the network, I took into account the metabolic constraints imposed on the formation of synapses over long distances (Ahn et al., 2006) – I modelled synaptic weights throughout the network as an exponentially decaying function showing reduced weights as distance increases (Figure 4.2). To vary *synaptic weights* smoothly I created a weight function *w(d)* which defines synaptic weights as a function of distance, where *w(d)* is defined

$$w(d) = i + e^{\left(\frac{-s}{e^r}\right)d} , \quad (4.4)$$

where *i* is the initial non-scaled synaptic weight (*i* = 1.2), *s* is the softening parameter that dictates the magnitude of exponential decay for the synaptic weight over distance (*s* = 0.1), *d* is the neuron-neuron distance and *r* defines the strength of the synaptic weights across the network (weak = low *r*, strong = high *r*). This allowed me to smoothly vary the weight-distance relationship across all neurons, from a regime in which synaptic weights were weak for most neuron pairs except neighbouring ones, to one in which synaptic weights were strong across all neuron pairs (Figure 4.2). Here, I wanted to test the hypothesis that changing the synaptic weight distributions throughout the network was sufficient to cause unstable seizure states.

Finally, *intrinsic excitability* was represented as the spike threshold of each neuron in the network. This was controlled by varying $V_{th}$ for each neuron in the leaky-integrate-and-fire model. Here, I supposed that changes to intrinsic neuronal excitability, for example through alterations in the size of the readily releasable pool in the pre-synapse (Kaeser & Regehr, 2017), or metabolic-induced changes to membrane potential regulation, could plausibly occur over the recording periods. I tested the hypothesis that changes to intrinsic excitability alone was sufficient to drive the brain into unstable seizure states.

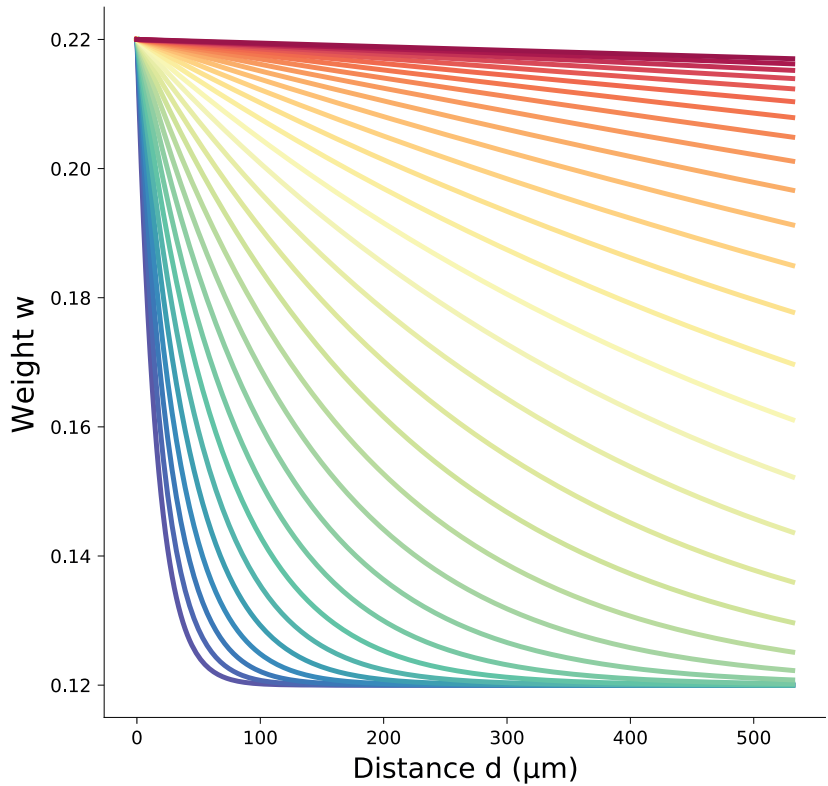

**Figure 4. 2   Synaptic weights parameter *r*.**

Weight function w(d) describes synaptic weights as a function of distance. Parameter r defines the distribution of edge weights between nearby vs distant neurons. Low values of r indicate low synaptic weight values for all neurons except nearby neighbours (blue), while increasing r increases synaptic weights across all neurons regardless of distance (red). Therefore, r dictates the strength of synaptic weights across the network.

In order to fit network model parameters to my empirical avalanche data, I simulated network dynamics for 4000 time steps. Model avalanche distributions were fit to empirical data generated from concatenated distributions for avalanche size of *baseline, pre-ictal* and *ictal-onset* datasets (n=9) (Figure 4.3). Avalanche simulations were run 9 times, and concatenated together to generate fits to empirical data.

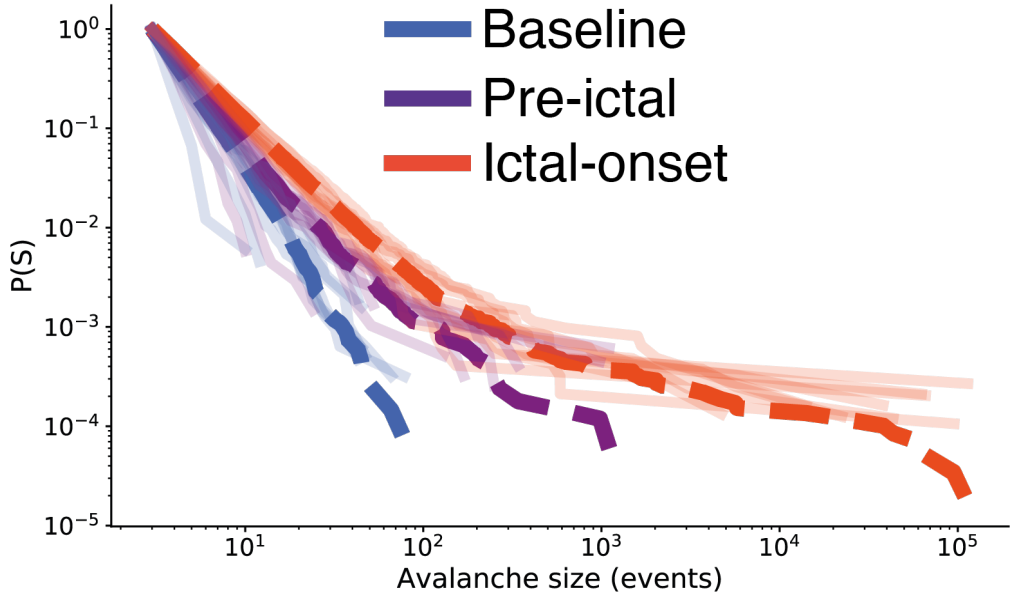

**Figure 4. 3  Concatenated avalanche distributions for baseline, pre-ictal and ictal-onset periods.**

Empirical avalanche distributions from n=9 fish were used to generate concatenated avalanche distributions for baseline (blue), pre-ictal (purple) and ictal-onset (crimson) periods. Concatenated avalanche distributions (dotted lines) were generated as a concatenated vector of all avalanches across all fish for a given condition. Empirical avalanche distributions for each dataset (n=9) and condition are also visualised (opaque solid lines).

In order to assess the closeness of fit between model and empirical data, I created a cost function which quantified the distance between empirical and model avalanche distributions. While the mean squared error function would usually suffice, I had to adapt it with a regularisation term in order to account for short-tailed distributions that match the candidate distribution over short-medium avalanche sizes, but have empty bins over the longer tail of the distribution. As such I designed a cost function as a regularised mean squared error term defined as

$$Cost = MSE\ R$$

$$MSE = \frac{\sum_{i=1}^{n}(y_i - \hat{y}_i)^2}{n - k}$$

$$R = \theta e^{\beta 10^{-5}}, \quad (4.5)$$

where $y_i$ = $P(S)$ for each avalanche bin, $n$ is the number of avalanche distribution bins, $k$ is the number of parameters, $\beta$ is the difference in non-empty bin numbers between model and empirical data, and $\theta$ scales the effect of the regularisation ($\alpha$ = 0.09). To decide on the method for model fitting, I first performed a coarse-grained sweep over my parameters to assess the landscape of the cost function. As is clear from Figure 4.4, the landscape is highly non-convex with multiple local minima and no clear global minima. As such I concluded that optimisation procedures such as gradient descent would get stuck in local minima, and so instead I opted for a grid-search approach.

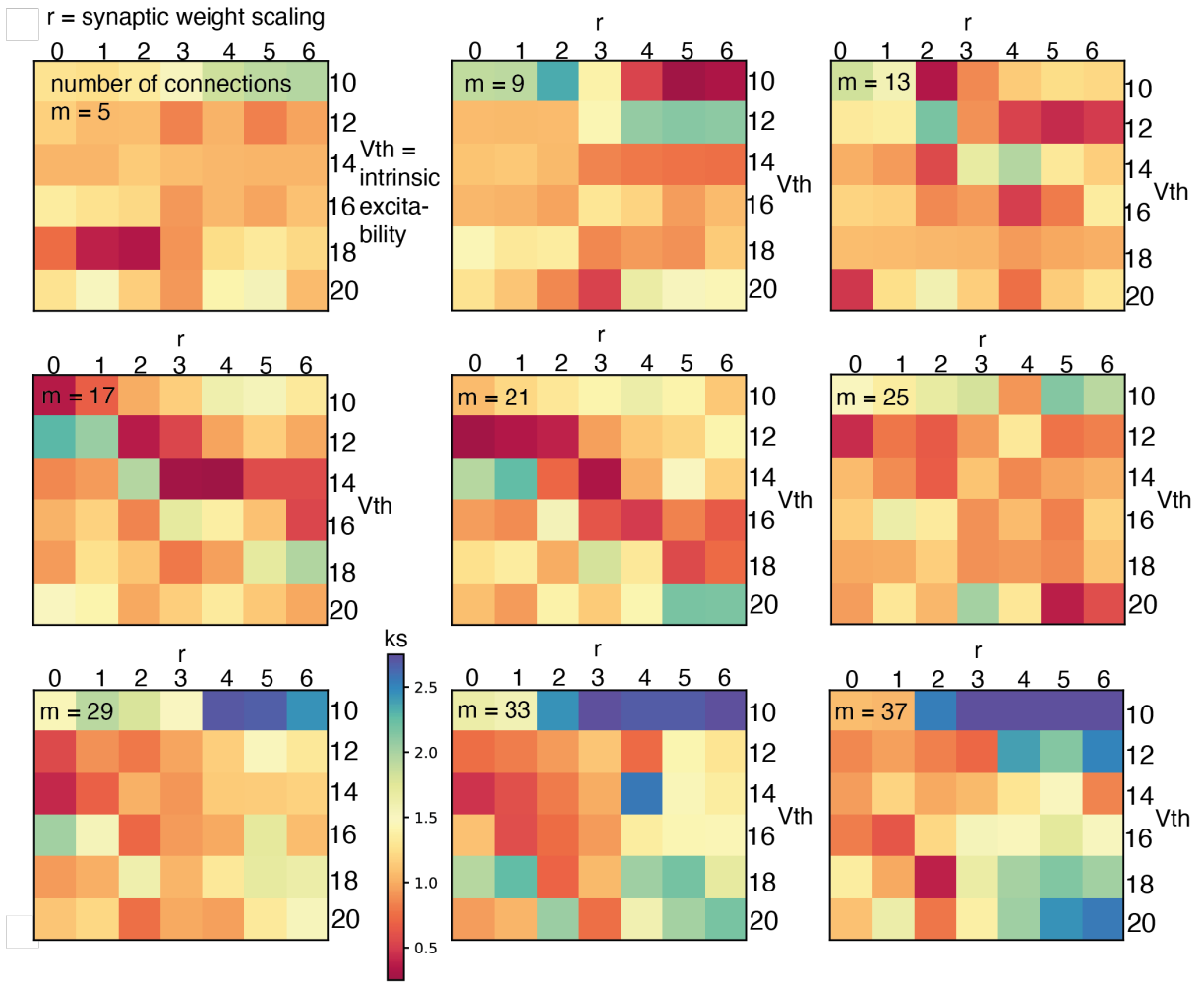

**Figure 4. 4   Network model fits to baseline avalanche data over a range of parameter values.**

Model fits to baseline avalanche data for coarse network parameter ranges for the number of connections in the network (m), the synaptic weight scaling (r) and the intrinsic excitability (Vth). Each model fit is assessed using the maximum distance between the empirical and the network avalanche distribution, the KS distance. Note how for the majority of heatmaps (for a fixed m), the model landscape is highly non-convex, with multiple local minima and few observable global minima.

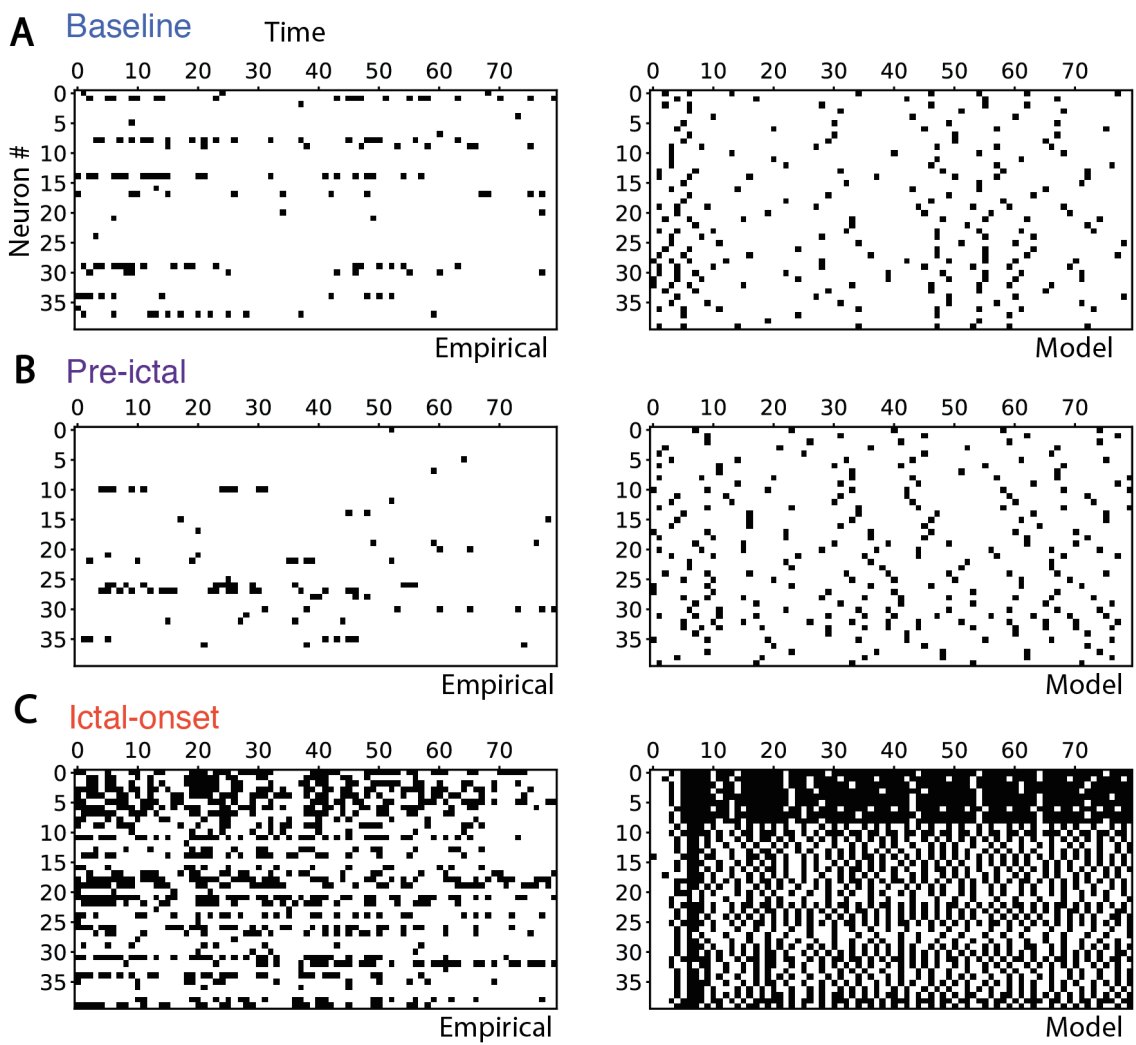

**Figure 4. 5   Spiking network model raster plot comparison.**

Spike raster plots comparing empirical (left) and model (right) spikes for the same example fish and network model, over the same 40 neurons for a representative period. Spike rasters are compared across baseline (A), pre-ictal (b) and ictal-onset (c) recording periods, with downsampled spike periods for spiking network models fit with all parameters.

I performed a grid-search of ~1,400 parameter combinations across my 3 parameters, across the range of parameter values which captured the full extent of empirical avalanche distribution shapes (*network connectivity (m)* = 5 - 33, *synaptic weights (r)* = 0 - 7, *intrinsic excitability ($V_{th}$) = 15 - 20*). This allowed me to identify the parameter combinations which had the lowest cost given the empirical data, and thus best captured *baseline, pre-ictal* and *ictal-onset* avalanche dynamics. Using all 3 parameters, I generated approximate model fits to *baseline* (network connectivity (m) = 7, synaptic weights (r) = 5, intrinsic excitability ($V_{th}$) = 20, cost = 0.113), *pre-ictal* (network connectivity (m) = 6, synaptic weights (r) = 0, intrinsic excitability ($V_{th}$) = 16, cost = 0.176), and *ictal-onset* data (network connectivity (m) = 31, synaptic weights (r) = 1, intrinsic excitability ($V_{th}$) = 17, cost = 0.120) (Figure 4.6A). On inspection of the spike data for the best models, I found that raster plots exhibited reasonably similar single neuron on/off dynamics as empirical datasets across conditions. Specifically, I found sparse firing in the *baseline* and *pre-ictal* states for model and empirical data, followed by a transition to clustered activity patterns in space and time during *ictal-onset* (Figure 4.5).

In order to compare the importance of each network parameter in the emergence of local and global seizure states, I explored a subset of parameters that were free to vary, while keeping others fixed. Specifically, to model the emergence of the locally unstable *pre-ictal* state, I fixed parameters to the best *baseline* model fit while allowing subsets of parameters to vary freely for model fitting. This was done because the locally unstable dynamics of the *pre-ictal* state were preceded by a baseline activity state before the addition of PTZ and thus empirical dynamics transition from stable to locally unstable dynamics. Allowing only single parameters to vary freely for fitting demonstrated that the *intrinsic excitability* parameter provided the best fit to *pre-ictal* data (*$V_{th}$* = 19, cost = 0.223) (Figure 4.6D), through increases in intrinsic excitability. *Network connectivity* changes alone gave rise to avalanche distributions with excessive heavy tails (*m* = 15, cost = 0.241) (Figure 4.6B), while *synaptic weights* changes alone were unable to generate sufficiently heavy tails (*r* = 6, cost = 1.06 ) (Figure 4.6C). However, I found that the model with all 3 parameters free to vary provides the best *pre-ictal* fit (Figure 4.6E), while models with any

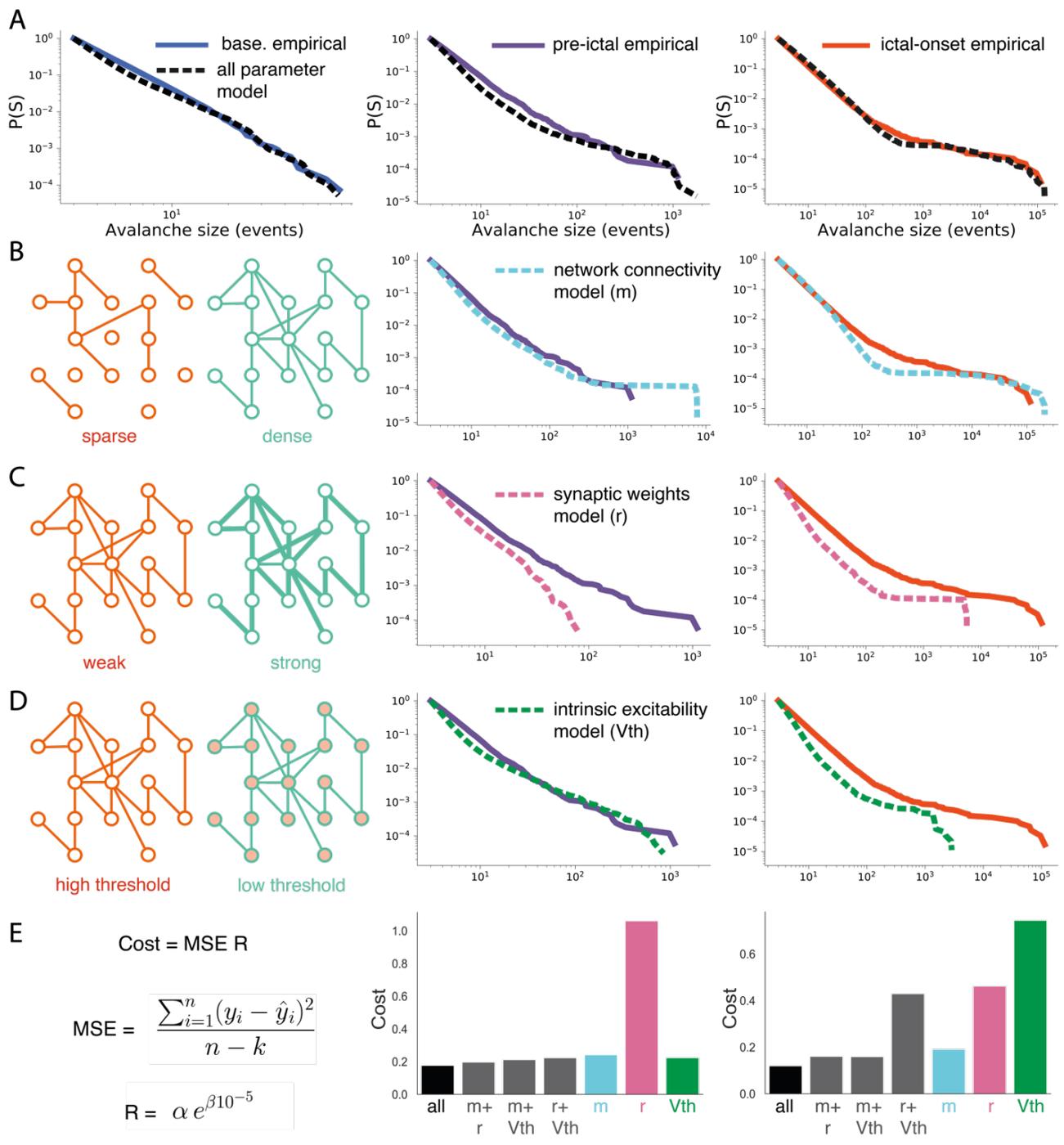

**Figure 4. 6   Spiking network model fits to pre-ictal and ictal data for single parameter models.**

(A) Avalanche distributions from state transition datasets for spontaneous (blue), pre-ictal (purple) and ictal-onset (red) periods, alongside best model fits using network connectivity, synaptic weights and intrinsic excitability parameters (black, dotted). (B) Network connectivity schematic for sparse and dense networks (left). Best model fits using network connectivity m parameter only (cyan, dotted) to pre-ictal (middle) and ictal-onset data (right). (C) Synaptic weights model schematic for uniform and local weight scaling (left, edge thickness = synaptic strength). Best model fits using synaptic weights r parameter only (magenta, dotted) to pre-ictal (middle) and ictal-onset data (right). (D) Intrinsic excitability schematic for high threshold to spike and low threshold to spike networks (left, red node = closer to spike). Best model fits using intrinsic excitability $V_{th}$ parameter only (green, dotted) to pre-ictal (middle) and ictal-onset data (right). (E) Model comparison for 3 parameter models (black), 2 parameter models (grey) and single parameter models (colours as above), for pre-ictal (middle) and ictal-onset data (right). The cost function used for model comparison was a mean squared error (MSE) term scaled by a regularisation parameter (R) to reduce overfitting of distributions with unequal number of non-empty bins (left). yi = P(S) for each avalanche bin, n is the number of avalanche distribution bins, k is the number of parameters, β = the difference in non-empty bin number between model and empirical data, α = scales the effect of the regularisation (0.09).

combination of 2 free parameters provide better fits than *intrinsic excitability* alone (Figure 4.7). Therefore, while increases in intrinsic excitability is the singular parameter change that best explains the pre-ictal avalanche dynamics captured by the model, non-specific combinations of all three parameters can more accurately describe such dynamics. This indicates that the *pre-ictal* state likely emerges from small, non-specific changes to *network connectivity*, *synaptic weights* and *intrinsic excitability*

To model the emergence of the *ictal-onset* state, I fixed parameters to the best *pre-ictal* model fit while allowing subsets of parameters to vary freely for model fitting. This was done because the globally unstable dynamics of the *ictal-onset* state were immediately preceded by a locally unstable state, which emerges a few minutes after 20mM PTZ administration. Allowing only single parameters to vary freely for fitting demonstrated that only the *network connectivity* parameter provided a reasonable fit to *ictal-onset* data (*m* = 30, cost = 0.190) (Figure 4.6B), emerging due to an increase in the number of connections from the *pre-ictal* state. Changing both *synaptic weights* (*r* = 7, cost = 0.465) and *intrinsic excitability* (*$V_{th}$* = 15, cost = 0.744) parameters alone failed to generate sufficiently heavy tails (Figure 4.6C & D). This suggests that increases in the number of effective or structural connections in the network, may

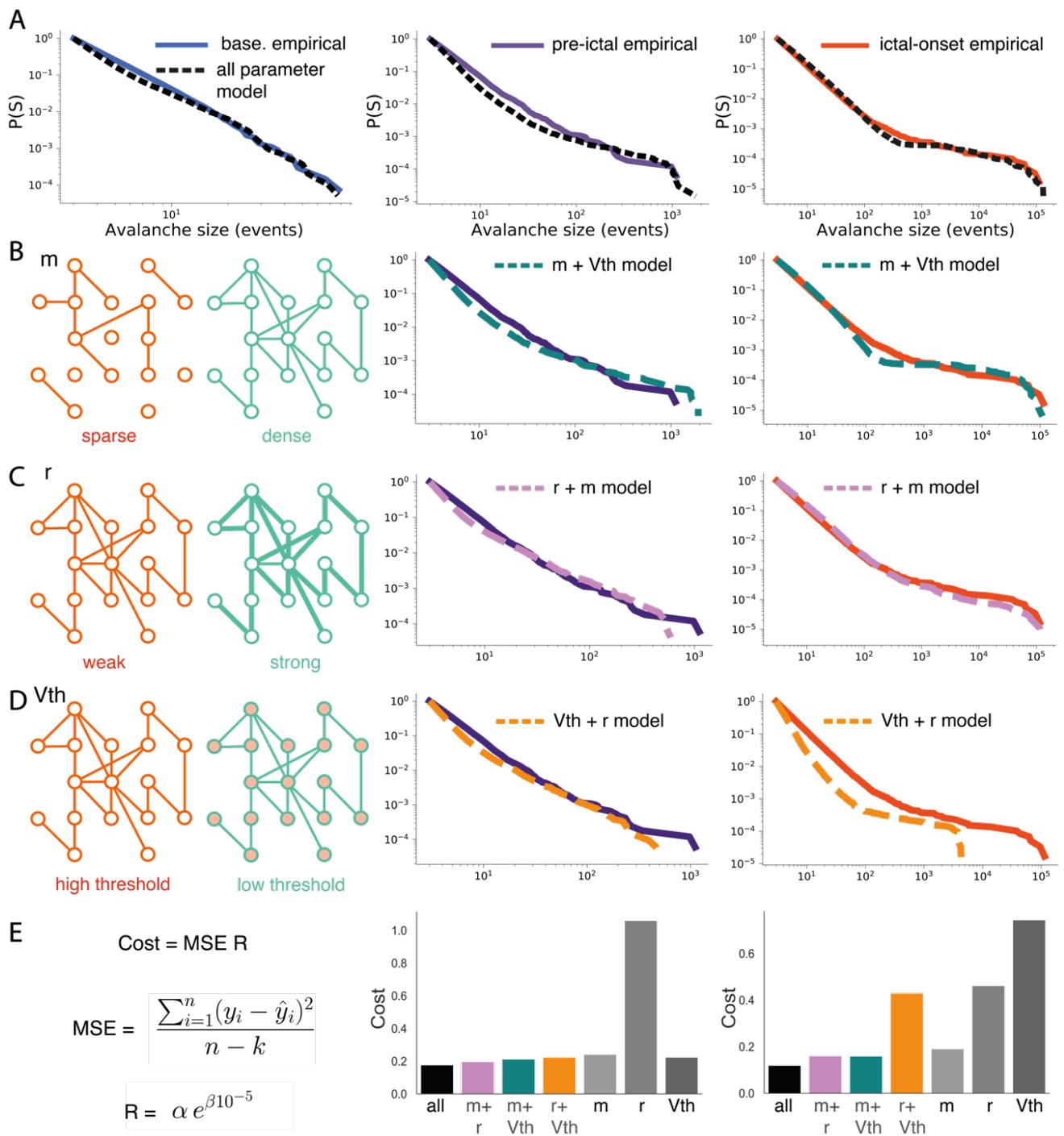

**Figure 4. 7 Spiking network model fits to pre-ictal and ictal data for 2 parameter models.**

(A) Avalanche distributions from state transition datasets for spontaneous (blue), pre-ictal (purple) and ictal-onset (red) periods, alongside best model fits using network connectivity, synaptic weights and intrinsic excitability parameters (black, dotted). (B) Network connectivity schematic for sparse and dense networks (left). Best model fits using network connectivity (m) and intrinsic excitability (Vth) (teal, dotted) to pre-ictal (middle) and ictal-onset data (right). (C) Synaptic weights model schematic for uniform and local weight scaling (left, edge thickness = synaptic strength). Best model fits using synaptic weight scaling (r) and network connectivity parameters (purple, dotted) to pre-ictal (middle) and ictal-onset data (right). (D) Intrinsic excitability schematic for high threshold to spike and low threshold to spike networks (left, red node = closer to spike). Best model fits using intrinsic excitability ($V_{th}$) and synaptic weight scaling (r) parameters (orange, dotted) to pre-ictal (middle) and ictal-onset data (right). (E) Model comparison for 3 parameter models (black), 1 parameter models (grey) and 2 parameter models (colours as above), for pre-ictal (middle) and ictal-onset data (right). The cost function used for model comparison was a mean squared error (MSE) term scaled by a regularisation parameter (R) to reduce overfitting of distributions with unequal number of non-empty bins (left). $y_i$ = P(S) for each avalanche bin, n is the number of avalanche distribution bins, k is the number of parameters, β = the difference in non-empty bin number between model and empirical data, α = scales the effect of the regularisation (0.09).

best explain the generalised seizure onset avalanche dynamics captured by the model. In fact, while model fits with all 3 parameters provided the best overall fit, the *network connectivity* model alone provided a better fit than the model with both *synaptic weights* and *intrinsic excitability* free to vary (Figure 4.6E & 4.7). Here both increasing *synaptic weights* and reducing *intrinsic excitability* still failed to generate sufficiently heavy tails to explain *ictal-onset* data (Figure 4.7D). Therefore, increasing the connectivity of the network alone is sufficient to drive heavy-tailed avalanche dynamics that co-occur with the onset of generalised seizures.

Taken together, I find that the emergence of locally unstable seizure dynamics can be driven by small, non-specific changes to combinations of neuronal connectivity, synapse strength and node excitability. Conversely, globally unstable generalised seizure dynamics require increases in neuronal connectivity.

### 4.2.3 The Effect of Neuronal Network Alterations on Dynamical Stability During Generalised Seizures

Having identified key neuronal parameters driving aberrant brain dynamics in epileptic seizures, I wanted to understand the functional implications of such microscale changes. In particular, generalised seizure dynamics are regularly associated with a complete loss of awareness (Blumenfeld, 2012). Due to the well characterised functional impairments present in generalised seizures, I refined further analysis to *20mM PTZ* recording periods in which generalised seizures occur (see Section 2.2.3). As discussed in Section 4.1, critical systems theory can provide a framework for linking aberrant brain dynamics with functional impairments in epilepsy. Given that generalised seizures emerge as a loss of criticality (see Section 3.2.7), I theorised that subsequent brain dysfunction would be caused by the suboptimal computational capacities of supercritical networks. Importantly, having identified specific synaptic mechanisms driving epileptic seizures in our data, I can test how such microscale changes disrupt the optimal network properties of critical systems *in vivo* and *in silico*.

One key prediction for critical networks is that the phase transition can separate out dynamically stable and unstable states, where the critical point exhibits neutral dynamics (Haldeman & Beggs, 2005). Such neutral dynamics mean that points in state space maintain their distance over time (Figure 4.8A), which gives rise to long autocorrelation and minimal information loss about inputs. Therefore, if critical brain dynamics exist at the boundary between dynamical stability and instability, known as the edge of chaos, then one would expect a loss of neutral dynamics during generalised seizures (see Section 1.3.3). Such a change would impair the ability of the network to retain information about inputs over time. Furthermore, if identified neuronal alterations driving generalised seizures also drive subsequent functional impairments, one would expect increases in network connectivity to directly disrupt neutral dynamics.

Firstly, I set out to demonstrate that empirically observed generalised seizures truly emerge as a disruption of neutral dynamics. Here, I considered 2 imaging conditions: i) *spontaneous* – 30 minute recording periods capturing baseline critical dynamics, ii) *20mM PTZ* – 30 minute recording periods capturing a frequent hypersynchronous,

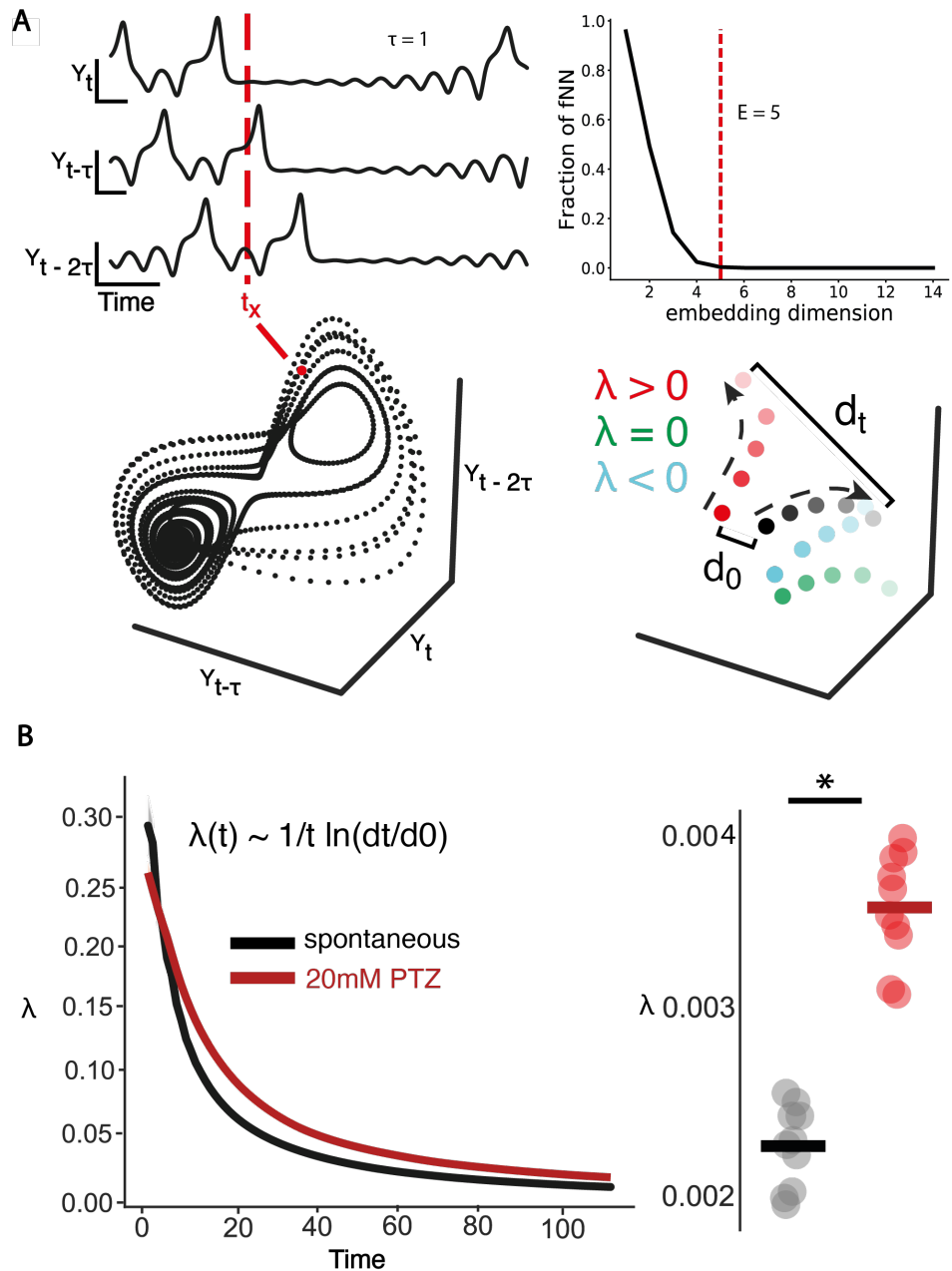

**Figure 4. 8    Lagged coordinate embedding approach to calculate dynamical stability of seizure dynamics.**

(A) Schematic demonstrating the lagged coordinate embedding approach for system reconstruction. (A, top) A single variable (Y) can be used to reconstruct an attractor that is topologically equivalent to the full system, by using a series of delayed variables (Y, $Y_{t-\tau}$ , $Y_{t-2\tau}$ …, $Y_{t-(m-1)\tau}$) of delay τ and dimension E. (A, bottom left) Embedding each lagged variable into state space provides the reconstructed attractor, where tx is the position in state space at time x corresponding to the dotted red line in the above panel. For all analyses τ was set to 1, while E was estimated as the embedding dimension at which the proportion of false nearest neighbours approaches 0 (A, top right). (A, bottom right) Schematic outlining the meaning of different values of the Lyapunov exponent (λ). Each colour represents the trajectory over time for a specific initial point along the attractor (high to low brightness represents movement in time). λ is calculated as the ratio between the difference between 2 points at the start (d0) and at timepoint t (dt). λ for each trajectory is calculated against the black trajectory. λ > 0 (blue trajectory): distances grow over time (chaos). λ < 0 (red trajectory): distances shrink over time (stability). λ = 0 (green trajectory): distances are constant over time (neutral). (B) Mean λ shown for spontaneous (black line) and 20mM PTZ conditions over time (red line). (B, right) λ compared across spontaneous (black bar = mean) and 20mM PTZ conditions over time (red bar = mean). * = p < 0.05.

high activity state consistent with generalised epileptic seizure activity (see Section 2.2.3). To assess the dynamical stability of the dynamics, I approximate the largest Lyapunov exponent (λ), which estimates the divergence of nearby trajectories in phase space, a defining feature of dynamical stability (Babloyantz & Destexhe, 1986; Rosenstein et al., 1993). To estimate λ from time series data, one needs to reconstruct the attractor of the underlying dynamical system (Figure 4.8A). This can be achieved using delay embedding theorem which states that a reconstructed manifold constructed from a variable $Y_0$, given by $\{Y_0(t), Y_0(t - \tau), ..., Y_0(t - (E - 1)\tau)\}$ is topologically equivalent to the full dynamical system $\{Y_0(t), Y_1(t), ..., Y_{E-1}(t)\}$, where $E$ is the dimension of the system and $\tau$ is the time lag (Takens, 1981) (Figure 4.8A).

I reconstructed the attractor using the first principal component of each dataset, as it captures the most variance in the system. In order to estimate $\tau$, one usually finds the $\tau$ that minimises the mutual information between time series. However, given the slow sampling rate of my recordings relative to the kinetics of GCaMP6s, I opted to use $\tau = 1$. Next, the embedding dimension $E$ was estimated using a false nearest neighbours approach which assumes that at the correct dimension, nearest neighbours should be retained at higher dimensions. Points in state space were nearest neighbours for a given $E$ if their distance was smaller than the standard deviation. False nearest neighbours were defined as

$$ R_i = \frac{|x_i - x_j|}{\|p_i - p_j\|} > R_{th} , \quad (4.6) $$

where $p_i$ and $p_j$ are nearest neighbours embedded in $E$ dimensions, $x_i$ and $x_j$ are the same points embedded in $E + 1$ dimensions and $R_{th}$ is the threshold ($R_{th} = 10$). The embedding dimension was defined for each dataset as the $E$ at which the fraction of false nearest neighbours approached 0. Using this approach I was able to generate embeddings for each dataset, which estimated attractors of the dynamics for *spontaneous* and *20mM PTZ* recording periods (Figure 4.9).

From these data, I was able to approximate the largest Lyapunov exponent ($\lambda$). Here, $\lambda$ captures changes in distance over time between two nearby points in state space: i) unstable, chaotic dynamics – $\lambda > 0$ which implies that points grow further apart over time, ii) stable dynamics – $\lambda < 0$ which implies that points get closer over time, iii) neutral dynamics – $\lambda \sim 0$ which implies that points retain their distances over time (Figure 4.8A).

To estimate $\lambda$ from embedded attractors I firstly locate the nearest neighbour for each point along the reconstructed attractor expressed as

$$d_i(0) = min\left\|X_i - X_j\right\|, \quad (4.7)$$

where $d_i(0)$ is the initial distance between $X_i$ and its nearest neighbour $X_j$. From this I calculate the change in distance over time for all nearest neighbours along the attractor which for each $t$ is

$$\lambda(t) = \frac{1}{t} \frac{1}{M-t} \sum_{t=1}^{M-t} \ln\frac{d_i(t)}{d_i(0)}, \quad (4.8)$$

where $d_i(t)$ is the distance between $X_i$ and $X_j$ at $t$, and $M$ is the number of points on the embedded attractor (Rosenstein et al., 1993; Sato et al., 1987). The largest Lyapunov exponent is thus the mean separation rate for nearest neighbour points over a given time period.

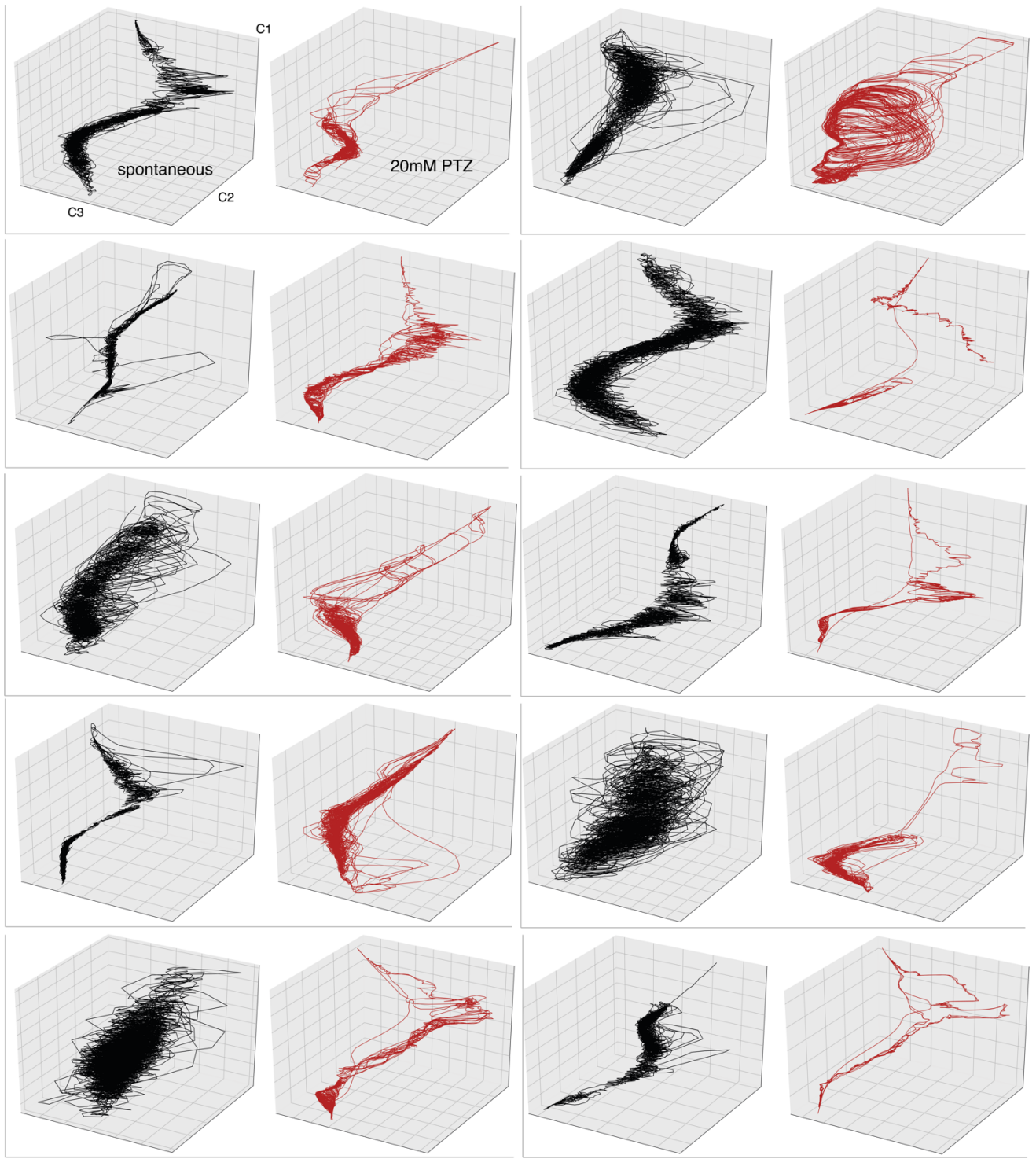

**Figure 4. 9   Reconstructed attractors from lagged coordinate embedding for spontaneous and generalised seizure dynamics.**

Reconstructed attractors using lagged coordinate embedding of the first principal component of the data across all fish (n=10). Isomap embedding was used to embed the attractor in 3 dimensional space and the embedding was gaussian filtered ($\sigma$ = 0.9) for visualisation.

I found that *spontaneous* dynamics exhibit $\lambda$ close to 0 ($\lambda$ = 0.0026 ± 0.0025), indicating relatively neutral dynamics with similar trajectories maintaining similar distances over time (Figure 4.8B). Interestingly, generalised seizures give rise to significantly more chaotic dynamics (*20mM PTZ:* $\lambda$ = 0.0034 ± 0.0001, w = 8.0, p <0.05). In this case, a more positive $\lambda$ suggests that close-by trajectories will grow further apart over time, suggesting a heightened sensitivity to initial conditions in the seizure state (Figure 4.8A,B). Therefore, this indicates a slight loss of the neutral dynamics expected in critical systems, and the emergence of a more chaotic state.

To confirm the presence of chaotic dynamics during generalised seizures, I also calculated $\lambda$ in my network model. Given that I have full control of my *in silico* system, $\lambda$ can be calculated by perturbing the network and following trajectories over time. To estimate $\lambda$ in my model I apply small perturbations to the network model once it has reached steady state (4000 time steps) and followed the trajectories over time, calculated as follows

$$\lambda(t) = \frac{1}{t} \ln \left| \frac{d_i(t)}{d_i(0)} \right| . \quad (4.9)$$

I calculated $\lambda$ in my *ictal-onset* network that was generated using all 3 parameters for model fitting. In line with my empirical data I find that *ictal-onset* networks are significantly more chaotic than *baseline* networks (*baseline:* $\lambda$ = 0.0002 ± 0.0001, *ictal-onset:* $\lambda$ = 0.0065 ± 0.0001,  t = -45.5, p < 0.0001), using models with all parameters free to vary (Figure 4.10A). Such chaotic dynamics during seizures would give rise to faster memory loss, as ongoing, noisy activity will disrupt network representations about a given input, such that initial conditions of a trajectory are lost. Given that seizure networks show an increase in chaoticity from relatively neutral dynamics, my data suggests that brain dynamics do indeed reside at the edge of chaos, separating regimes of order and disorder, as predicted *in silico* (Haldeman & Beggs, 2005).

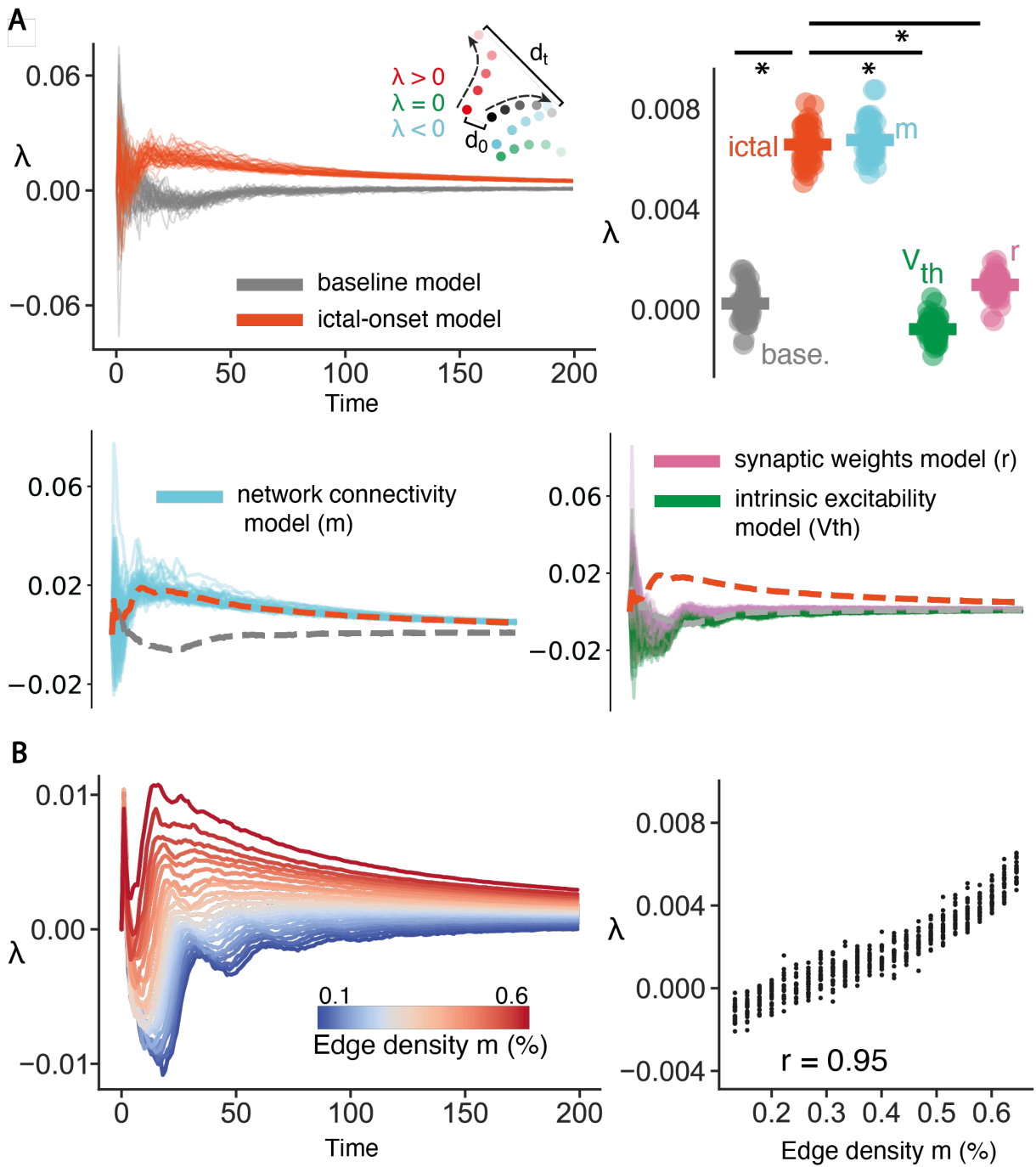

**Figure 4.10  Dynamical stability in spiking network model.**

(A, top left) Lyapunov exponent ($\lambda$) over time compared for spontaneous (grey) and ictal-onset models (crimson), using models with all parameters free to vary and showing all 50 simulations per model. Schematic outlining the meaning of different values of the Lyapunov exponent ($\lambda$) (outset). Mean $\lambda$ compared for spontaneous (grey) and ictal-onset models (crimson) using models with all parameters free to vary, and single parameter models for network connectivity (m, blue), synaptic weights (r, magenta) and intrinsic excitability ($V_{th}$, green) (top right). (A, bottom) $\lambda$ over time comparing average values over full parameter models for baseline (grey, dotted) and ictal-onset (crimson, dotted), with single parameter models for network connectivity (left, blue), synaptic weights (right, magenta) and intrinsic excitability (right, green), showing 50 simulations. (B) λ over time compared for increasing edge density m parameter values, ranging from pre-ictal values to ictal-onset values (left). Each line represents the mean λ over time for each model over 50 simulations. Correlation between edge density (m) and mean λ, ranging from pre-ictal values to ictal-onset values (bottom). * = p < 0.05.

Next, to relate observed chaotic dynamics in generalised seizures with underlying neuronal parameter changes, I tested the link between *λ* and *network connectivity* in my network model. To do this I compared *λ* across my single parameter models, with fixed parameters set to *pre-ictal* values and free parameters fitted to *ictal-onset* data. As expected, only the *network connectivity m* model produced λ that matched the full model (0.0067 ± 0.0001, U = 1108.0, p = 0.16), while *synaptic weights r* (0.0010 ± 0.0001, U = 0.0, p < 0.0001) and *intrinsic excitability* $V_{th}$ (-0.0010 ± 0.0001, U = 0.0, p < 0.0001) models failed to produce more chaotic dynamics (Figure 4.10A). This suggests that only the *network connectivity* parameter can drive the emergence of chaotic dynamics observed during generalised seizures. Importantly, I found a significant positive correlation between the number of connections *m* and *λ* (r = 0.95, p < 0.0001) (Figure 4.10B), with increasing *m* from *pre-ictal* to *ictal-onset* values resulting in a loss of neutral dynamics and the emergence of chaos. Therefore, the increased network connectivity which gives rise to generalised seizure avalanches, also results in more chaotic dynamics. This will impair the network's ability to maintain maximal memory about inputs, thus giving rise to network dysfunction during seizures (Haldeman & Beggs, 2005).

### 4.2.4  The Effect of Neuronal Network Alterations on Response Properties During Generalised Seizures

Another key property at criticality is that certain network response properties are optimised. In particular, evidence from *in vitro* and *in silico* studies have demonstrated that i) the network-mediated separation (NMS) – a network's ability to discriminate between inputs (Bertschinger & Natschläger, 2004), and ii) the dynamic

range – a network's ability to represent a wide range of inputs (Kinouchi & Copelli, 2006; Shew et al., 2009), should be optimised at the critical point. Therefore, if generalised seizures emerge as a loss of criticality, one would expect emergent brain dysfunction due to impaired network response properties. Furthermore, if the neuronal alterations driving generalised seizures are truly responsible for such dysfunction, one would expect increases in network connectivity to drive disrupted network response properties.

Firstly, I set out to demonstrate that the NMS is in fact disrupted during generalised seizures. Here, I assessed the NMS property in my network model, as identifying exact input-output relationships is not feasible *in vivo*. This is because one cannot disentangle a networks' responses to input from its ongoing activity. NMS is the property of a network to encode distinct inputs with distinct network states, enabling similar inputs to be separated and discriminated by a readout function (Schrauwen et al., 2008) (Figure 4.11A). To calculate NMS I provided pairs of inputs, *a* and *b* separated by distance *z* onto different network models. For each input, *x* nodes were randomly activated once the network had reached steady state at time *t* (4000 time steps). Here, network activity was binned into 10 time step windows to approximate network outputs. In order to distinguish between the separation of network activity due to the input responses, as opposed to the intrinsic variability in the network, I define the normalised NMS property as

$$\|q_a - g_b\| - \|q - g\|, \quad (4.10)$$

where $\| \|$ denotes the Euclidean distance. $q_a$ and $q$ are the binarised state vectors at time *t*, for identical instantiations of a network that received input *a* or received no input, respectively. $g_b$ and $g$ are the binarised state vectors at time *t*, for identical instantiations of a network that received input *b* or received no input, respectively. This allowed me to normalise the distance between two network states following similar inputs *a* and *b*, to the exact distance between those two states as expected in non-perturbed networks. To assess the slope of the NMS curve I performed linear regression on relationship between the mean NMS and the difference between the sizes of the inputs *a* and *b*.

Using network models with all parameters free to vary, I found that *baseline* networks assume higher NMS values across the range of input sizes compared with *ictal-onset* networks (*baseline*: 0.91 ± 0.01, *ictal-onset*: 0.18 ± 0.00,  U = 556991.0, p < 0.0001) (Figure 4.11A). Therefore, generalised seizures caused a reduction in the ability of the network to discriminate between inputs. Interestingly, *ictal-onset* networks exhibit a significantly more shallow NMS slope (*baseline*: m = 0.0034 ± 0.0000, *ictal-onset*: m = 0.0007 ± 0.0000,  t = 109.5, p < 0.0001), in line with edge of chaos computing predictions for chaotic networks (Figure 4.11A) (Bertschinger & Natschläger, 2004). This indicates a reduced sensitivity of the NMS property to input size differences, demonstrating an impaired ability to distinguish between similar input pairs and highly different input pairs. Therefore, generalised seizures impair the ability of the brain to not only discriminate between inputs, but also represent input pairs according to their similarity.

Next, to relate NMS reductions with underlying neuronal parameter changes, I tested the link between NMS and *network connectivity* in my network model. To do this I compared NMS across my single parameter models with fixed parameters set to *pre-ictal* values and free parameters fitted to *ictal-onset* data. I found that only the *network topology m* model produced NMS values matching the full model (0.18 ± 0.00, U = 3081443.4, p = 0.20), while *synaptic weights r* (0.49 ± 0.01, U = 1104945.0, p < 0.0001) and *intrinsic excitability V$_{th}$* (0.51 ± 0.01, U = 1546742.0, p <0.0001) failed to reduce NMS sufficiently to *ictal-onset* levels (Figure 4.11B) (A, top right) Schematic showing the calculation of network mediated separation (NMS). For 2 similar inputs ($i_1$ and $i_2$) onto the network, NMS is calculated as the euclidean distance between the two corresponding network states ($s_1$ and $s_2$). High NMS allows the brain to encode distinct inputs with distinct network states. (A, top left) NMS as a function of input difference compared for spontaneous (grey) and ictal-onset models (crimson), using models with all parameters free to vary and showing all 50 simulations per model (top). (A, bottom) NMS average values compared over full parameter models for baseline (grey, dotted) and ictal-onset (crimson, dotted), with single parameter models for network connectivity (left, blue), synaptic weights (right, magenta) and intrinsic excitability (right, green), showing 50 simulations. (B, left) Mean NMS compared for spontaneous (grey) and ictal-onset models (crimson) using models with all parameters free to vary, and single parameter models for network connectivity (m, blue), synaptic weights (r, magenta) and intrinsic excitability ($V_{th}$, green). (B, right) NMS slope compared for spontaneous (grey) and ictal-onset models (crimson) using models with all parameters free to vary. (C, left) NMS as a function of input difference compared for increasing edge density parameter m, ranging from pre-ictal values to ictal-onset values. (C, right) Correlation between edge density m and mean NMS, ranging from pre-ictal values to ictal-onset values (bottom). * = $p < 0.05$.

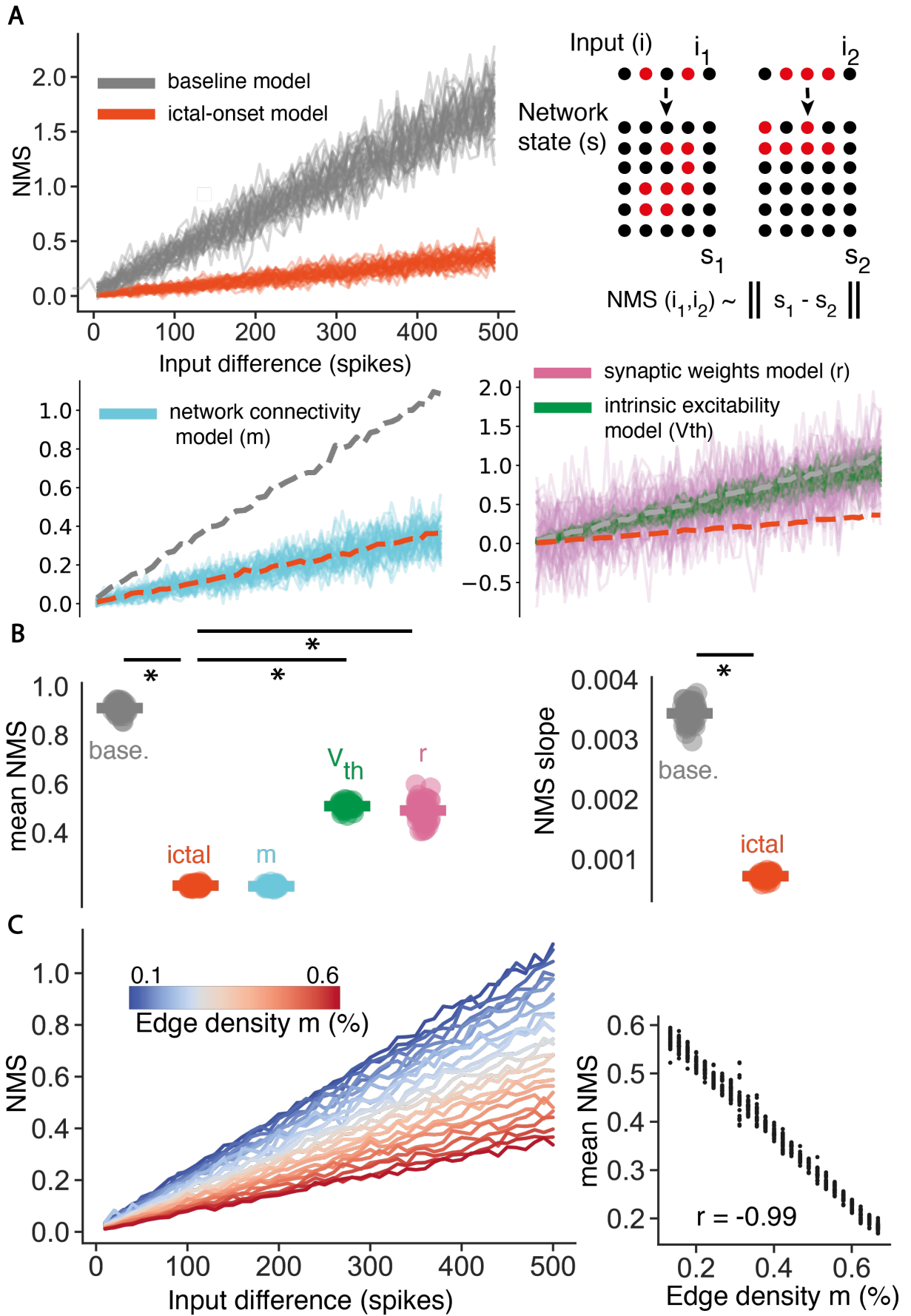

**Figure 4.11  Network mediated separation property in spiking network model.**

This demonstrates that of my 3 network parameter changes, only increasing the network connectivity can drive the loss of NMS observed in the seizure state. Interestingly, while increasing intrinsic excitability from *pre-ictal* to *ictal-onset* levels marginally reduced NMS, increasing synaptic weights hugely increased the variance of the NMS property (Figure 4.11A). This indicates that a high synaptic weights regime may be an unreliable mode for representing inputs. Importantly, I also found significant negative correlations between network connectivity *m* and NMS values (mean NMS: r =- 0.99, $p < 0.0001$, NMS slope: r = -0.99, $p < 0.0001$) (Figure 4.11C). Thus my data suggest an increase in the connectivity of the network which drives generalised seizures also causes a reduction in the network mediated separation property. Such microscale network alterations may thus disrupt brain function during seizures by impairing the optimised NMS property of critical networks.

Having demonstrated a link between generalised seizures and NMS, I set out to confirm that the dynamic range is also disrupted in seizure networks. The dynamic range $\delta$ is a measure of the range of input sizes that a system can represent, and therefore is a related but distinct measure of network response properties to NMS (Figure 4.12A). In order to assess the range of information that the network can hold I calculated $\delta$ by adapting previous measures (Shew et al., 2009). To calculate $\delta$ I provided input to the network by randomly activating $x$ neurons across a range of input sizes ($x$ range = 5 - 500, stepsize = 10), once the network had reached steady state (4000 time steps). As above network activity was binned into 10 time step windows to estimate the network response. In order to separate out the network response to a given input from the ongoing network activity, the corresponding output size was calculated as

$$\sum q_{ai} - \sum q_i, \quad (4.11)$$

where $q_a$ and $q$ are binarised state vectors for identical instantiations of a network that received input $a$ or did not receive input at time $t$, respectively. This enabled the separation of active neurons due to the input from active neurons due to the ongoing activity in the network.

$\delta$ is then defined across the range of input sizes as

$$\delta = 10 \, log_{10} \left( \frac{S_{max}}{S_{min}} \right), \quad (4.11)$$

where $S_{max}$ and $S_{min}$ are the input sizes leading to 90th and 10th percentile over the range of output sizes, respectively. Thus, a high $\delta$ network is one in which the ratio between the inputs causing its 90th and 10th percentile of output sizes is large, thus indicating a wide range of inputs distinctly represented in the network. Conversely, a small $\delta$ network is one in which this ratio is smaller, indicating a narrower range of input sizes that can be represented (Figure 4.12A). Using network models with all parameters free to vary, I found that *baseline* networks exhibit significantly higher $\delta$ compared with *ictal-onset* models (*baseline*: 8.22 ± 0.06, *ictal-onset*: 7.70 ± 0.07,  t = 5.83, p < 0.0001) (Figure 4.12A). Therefore, generalised seizures cause a reduction in the dynamic range, indicating a narrower range of inputs the network can represent.

Next, to relate observed reductions in $\delta$ in generalised seizures with underlying neuronal parameter changes, I tested the link between $\delta$ and *network connectivity* in my network model. To do this, I compared $\delta$ across my single parameter models with fixed parameters set to *pre-ictal* values and free parameters fitted to *ictal-onset* data. As above, I found that only the *network topology m* model produced $\delta$ values that matched the full *ictal-onset* model (*network topology m*: t = -0.95, p = 0.34, *synaptic weights r*: U = 593.0, p < 0.0001, *intrinsic excitability $V_{th}$*: t = 4.28, p <0.0001) (Figure 4.12A,B). This demonstrates that of all parameter changes driving seizures, the increase in network connectivity is central in driving a loss of dynamic range. Interestingly, while the number of connections *m* was significantly negatively correlated with $\delta$ (r = -0.71, p < 0.0001) on visual inspection I note a partially non-linear relationship. Here similar $\delta$ is maintained up to ~ 0.55% edge density followed by a drastic reduction (Figure 4.12C). This suggests that spontaneous networks may reside near to but not exactly at a phase transition point, as operating too close may easily spill over into suboptimal dynamics due to noise. Given that increasing network connectivity correlates with reduced dynamic range, my data suggest the synaptic alterations driving generalised seizures also impair the networks' response properties.

Taken together, I have shown that generalised seizures impair the optimal response properties of critical networks, resulting in reductions in NMS and dynamic range. Overall, this suggests that brain dysfunction emerging during generalised seizures can emerge due to a disruption of the optimal network response properties at criticality. Importantly, impairments to network responses are directly driven by the network connectivity changes that cause generalised seizures, linking microscale parameter changes with emergent global dysfunction in epilepsy.

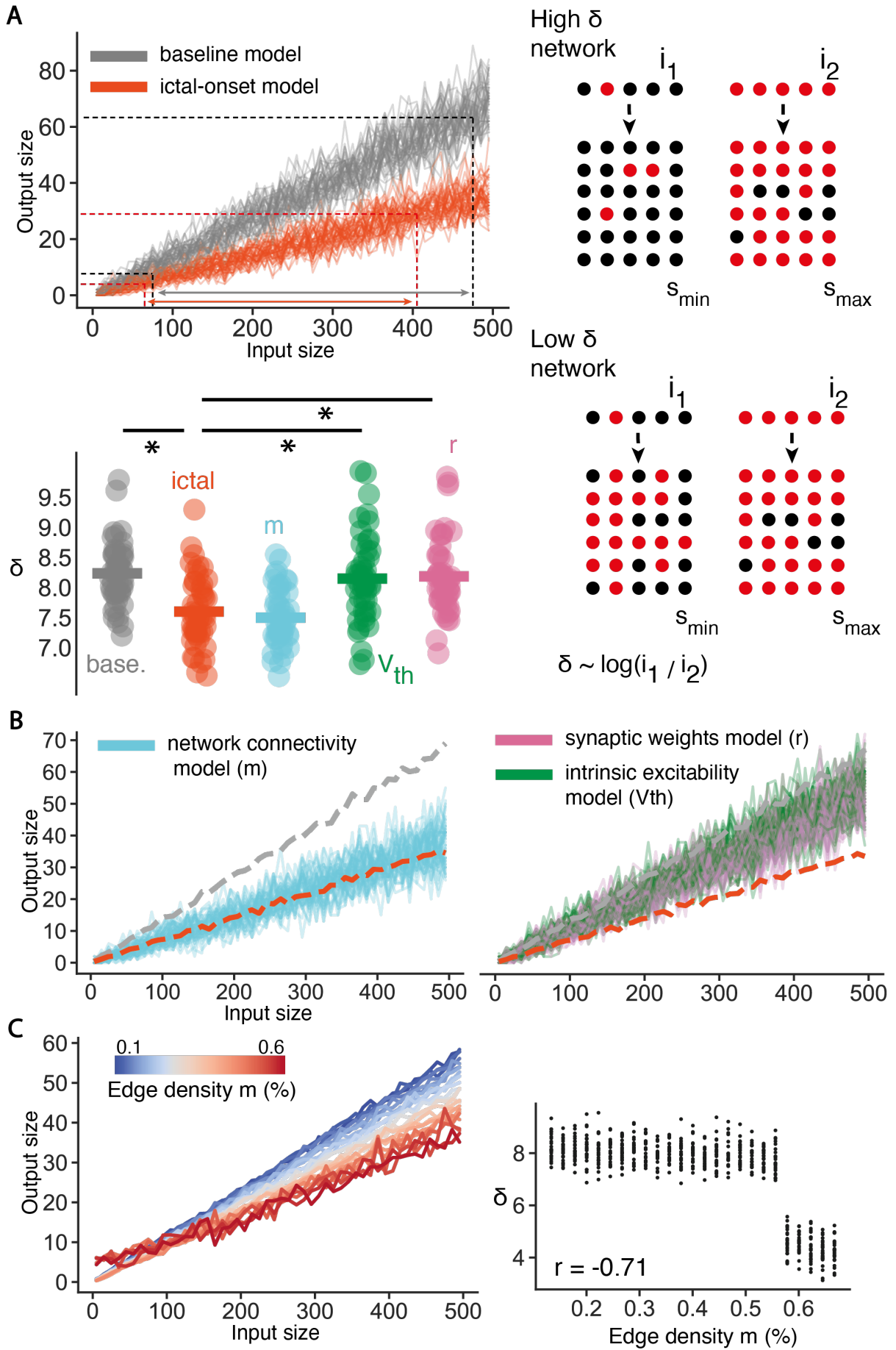

**A**

baseline model
ictal-onset model

High δ network

$i_1$   $i_2$

$s_{min}$   $s_{max}$

Low δ network

$i_1$   $i_2$

$s_{min}$   $s_{max}$

$\delta \sim \log(i_1 / i_2)$

base.   ictal   m   $V_{th}$   r

**B**

network connectivity model (m)

synaptic weights model (r)
intrinsic excitability model (Vth)

**C**

Edge density m (%)
0.1 — 0.6

r = -0.71

**Figure 4.12  Dynamic range property in spiking network model.**

(A, right) Schematic showing the calculation of dynamic range δ for two networks. For a range of different input size onto the network, δ is calculated as the log ratio of the inputs ($i_1$ and $i_2$) that give rise to the largest and smallest network responses ($s_{max}$ and $s_{min}$). High δ networks can represent a wide range of inputs, while low δ networks are easily saturated by inputs. (A, top left) Output size as a function of input size compared for spontaneous (grey) and ictal-onset models (crimson), using models with all parameters free to vary and showing all 50 simulations per model. Dotted lines show approximate $s_{max}$ and $s_{min}$ (top and bottom lines for each colour) for baseline (black, dotted) and ictal-onset models (crimson, dotted), where the length of the arrow is proportional to δ. (A, bottom left) δ compared for spontaneous (grey) and ictal-onset models (crimson) using models with all parameters free to vary, and single parameter models for network connectivity (m, blue), synaptic weights (r, magenta) and intrinsic excitability ($V_{th}$, green). (B) Output size as a function of input size compared over full parameter models for baseline (grey, dotted) and ictal-onset (crimson, dotted), with single parameter models for network connectivity (left, blue), synaptic weights (right, magenta) and intrinsic excitability (right, green), showing 50 simulations. (C, left) Output size as a function of input size compared for increasing edge density parameter m, ranging from pre-ictal values to ictal-onset values. (C, right) Correlation between edge density parameter m and δ, ranging from pre-ictal values to ictal-onset values (bottom). * = p<0.01.

*4.2.5  The Effect of Generalised Seizures on Flexible Dynamics*

Having confirmed that network impairments during generalised seizures can emerge from alterations to dynamical stability and network response properties, I wanted to probe the impact on state transition dynamics. In particular, a key property of systems at criticality is the ability to access a diversity of brain states (Figure 4.13A) – the number of states available to the system is maximal at the critical point (Haldeman & Beggs, 2005). Such *diverse* dynamics indicate the flexible transitioning into the full dynamic repertoire of the system (Figure 4.13A) (Cabral et al., 2011). Therefore, given that generalised seizures emerge as a loss of criticality, I theorised that functional impairments could arise through the disruption of diverse dynamics, specifically by reducing the number of available states.

To test the effect of generalised seizures on the diversity of the dynamics, I compared the number of available states during *spontaneous* and generalised seizure (*20mM PTZ*) periods in empirical recordings. To this end I adapt methods used in a previous study which estimates the number of metastable states from network activity (Haldeman & Beggs, 2005). Here, a metastable state is defined as a set of state vectors that are more similar to one another than expected in a random system. To identify metastable states in empirical data I performed affinity propagation across all state vectors for each condition. Affinity propagation is a clustering method which uses message passing to identify cluster exemplars, and which does not require cluster number to be defined *a priori*. Any clusters for which the number of state vectors belonging to that cluster was equal to 1 were removed. To calculate the similarity between state vectors that belong to a cluster, I define

$$Sim(\Omega^i, \Omega^j) = \frac{\langle \Omega^i, \Omega^j \rangle}{\langle \Omega^i, \Omega^i \rangle + \langle \Omega^j, \Omega^j \rangle - \langle \Omega^i, \Omega^j \rangle} \quad (4.12)$$

where $\Omega$ is a state vector and < , > denotes the dot product. For clusters to be identified as metastable, they needed to show greater similarity than expected by chance. I estimated the chance level similarity by performing affinity propagation clustering and similarity calculation on a null network using event-count matched shuffling. Any cluster with higher average similarity than the null network was declared a metastable state. One dataset failed to return any clusters for the *spontaneous* block and therefore I removed all conditions relating to this fish for further analyses.

As predicted, I found that generalised seizure dynamics exhibit significantly fewer metastable states than spontaneous dynamics (*spontaneous*: 20.9 ± 3.30, *PTZ 20mM*: 11.9 ± 2.31, w = 1.9, p < 0.05) (Figure 4.13A). This indicates that generalised seizures limit the diversity of states the brain can enter into, giving rise to more *homogeneous* dynamics. This likely impairs the flexible transitioning across the full dynamic repertoire of brain states, thus disrupting the diverse dynamics expected at criticality.

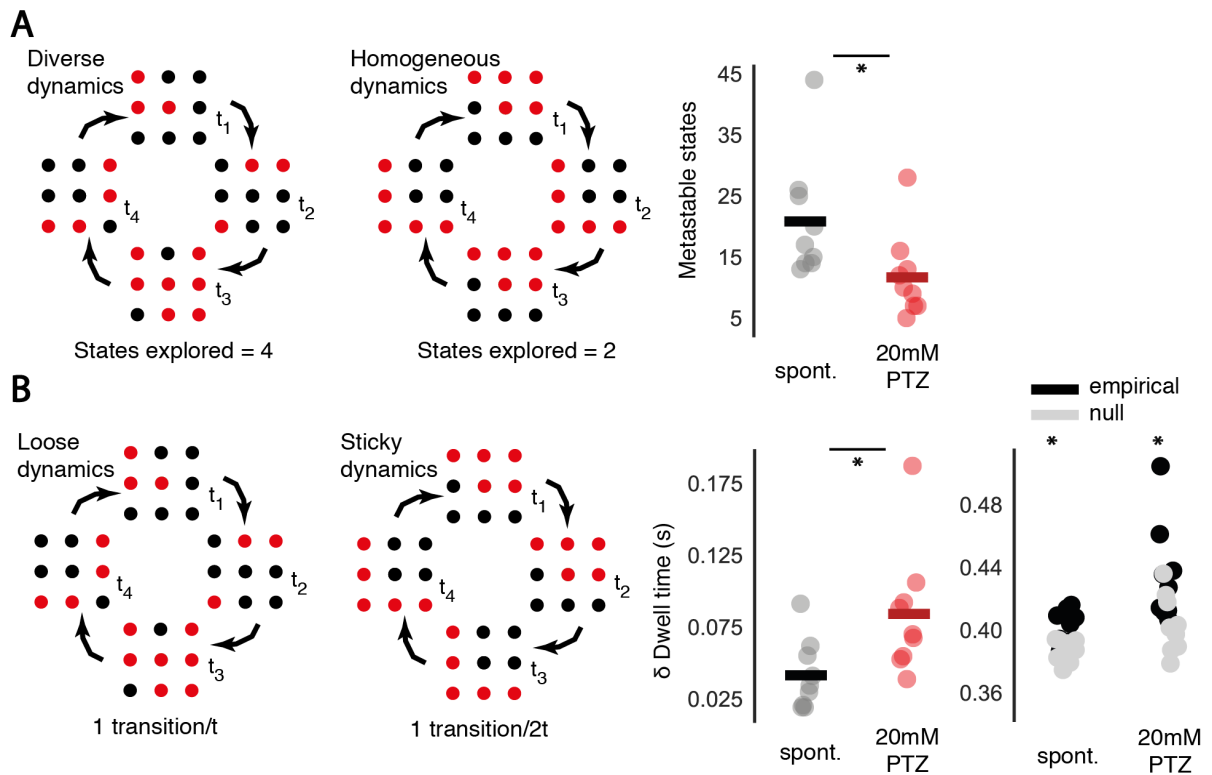

**Figure 4. 13 Flexible dynamics in spontaneous and seizure networks.**

(A, left) Schematic showing the exploration of brain states for flexible and inflexible systems. A system with diverse dynamics can explore a greater subset of its possible brain states (flexible, left), while a system with homogeneous dynamics may only explore a more limited subset (inflexible, right). (A, right) Number of metastable states compared across spontaneous (black) and 20mM PTZ conditions (red). (B, left) A system with loose dynamics can quickly transition between different states (flexible, left), while a system with sticky dynamics may reside for longer in a given state (inflexible, right). (B, right) Mean dwell time spent in each state compared across spontaneous and 20mM PTZ conditions, plotting both empirical (black) and null datapoints (grey). (B, middle) Change in dwell time from null models compared across spontaneous (black) and 20mM PTZ conditions (red). * = p < 0.01.

Next, I was interested in the effect of generalised seizures on the rate of state transitions. Here, I reasoned that the flexible dynamics expressed by critical networks might support rapid transitions across network states. Such *loose* dynamics could enable flexible transitions between states according to task requirements (Figure 4.13B). To assess the rate of state transitions in empirical data, I estimated the mean dwell time in each metastable state – this was calculated as the mean number of time frames for which the dynamics spent in each state. Interestingly, I found significantly increased dwell times during generalised seizures (*spontaneous*: states = 20.9 ± 3.30, *20mM PTZ*: *states* = 11.9 ± 2.31, w = 1.0, p < 0.05), suggesting a slower rate of state transitions – which I refer to as *sticky* dynamics (Figure 4.13B).

However, longer dwell times could occur by chance in systems with fewer states. Therefore, I calculated null models for each system, by randomising the dynamics and evaluating the dwell time expected by chance. In order to estimate the dwell time expected by chance for a system with *n* clusters, I randomly drew values between 1 and *n* from a uniform distribution, *t* times. The $\delta$ dwell time for each dataset was then calculated as the empirical dwell time minus the null dwell time. This allowed me to calculate the dwell time for a given fish normalised to the expected dwell time given a certain number of states in a random system. Interestingly, I found that the dynamics for spontaneous and generalised seizure periods are both non-random, with longer dwell times than expected by chance (*spontaneous*: t = 5.12, p < 0.001, *20mM PTZ*: w = 0.0, p < 0.01). Importantly, generalised seizures show significantly longer dwell times even when accounting for the fewer available states (*spontaneous*: $\delta$ dwell time = 0.04 ± 0.01s, *20mM PTZ*: $\delta$ dwell time = 0.08 ± 0.01s, w = 2.0, p < 0.05) (Figure 4.13B). Therefore, this indicates a slower rate of state transitions during generalised seizures, suggesting the emergence of sticky dynamics.

Taken together, I find that two proxy features of flexible dynamics, diverse states and loose dynamics, are disrupted during generalised seizures. These properties are in line with expectations for systems driven away from criticality. Sticky, homogeneous dynamics would likely prevent flexible responses to inputs and impair the optimal exploration of semi-stable states required for scale-invariant dynamics at criticality, giving rise to brain dysfunction in epilepsy.

### 4.2.6 Population Mechanisms Driving a Loss of Flexible Dynamics During Generalised Seizures

Next, I wanted to probe the population mechanisms driving the emergence of homogenous and sticky dynamics during generalised seizures. Having identified increased network connectivity as a driver of seizure dynamics, I reasoned that fewer brain states (i.e. homogeneous dynamics) might be a natural consequence of the emergent correlational structure of the dynamics in such densely connected networks. In particular, seizure networks may express large subnetworks of highly interconnected neurons – the activation of only a limited subset of constituent neurons could activate the entire subnetwork giving rise to strong multidimensional correlations. Correlations across a large number of neurons for any given subnetwork, would naturally limit the diversity of brain states by restricting the space of possible neuronal activation patterns to a more restricted subset, that is a lower dimensional manifold.

In order to test this hypothesis, I first wanted to demonstrate that seizure networks do indeed express higher multidimensional correlation. To do this I use a measure of the dimensionality defined in a previous study (Recanatesi et al., 2019), as

$$\frac{(\sum \omega_i)^2}{\sum \omega_i^2} \, , \qquad (4.13).$$

where $\omega_i$ captures the eigenvalue of the $i^{th}$ principal component. Here, the dimensionality captures the extent to which the variance over the principal components is shared equally or dominated by a few components. If one large, highly correlated subnetwork dominates the dynamics then the point cloud in state space would fall nearly onto a line – only a few eigenvectors would be needed to capture the variance of the data, while other orthogonal eigenvectors would be redundant (Figure 4.14A). Therefore, if the multidimensional correlation is high then the dimensionality is low, as defined by (4.13).

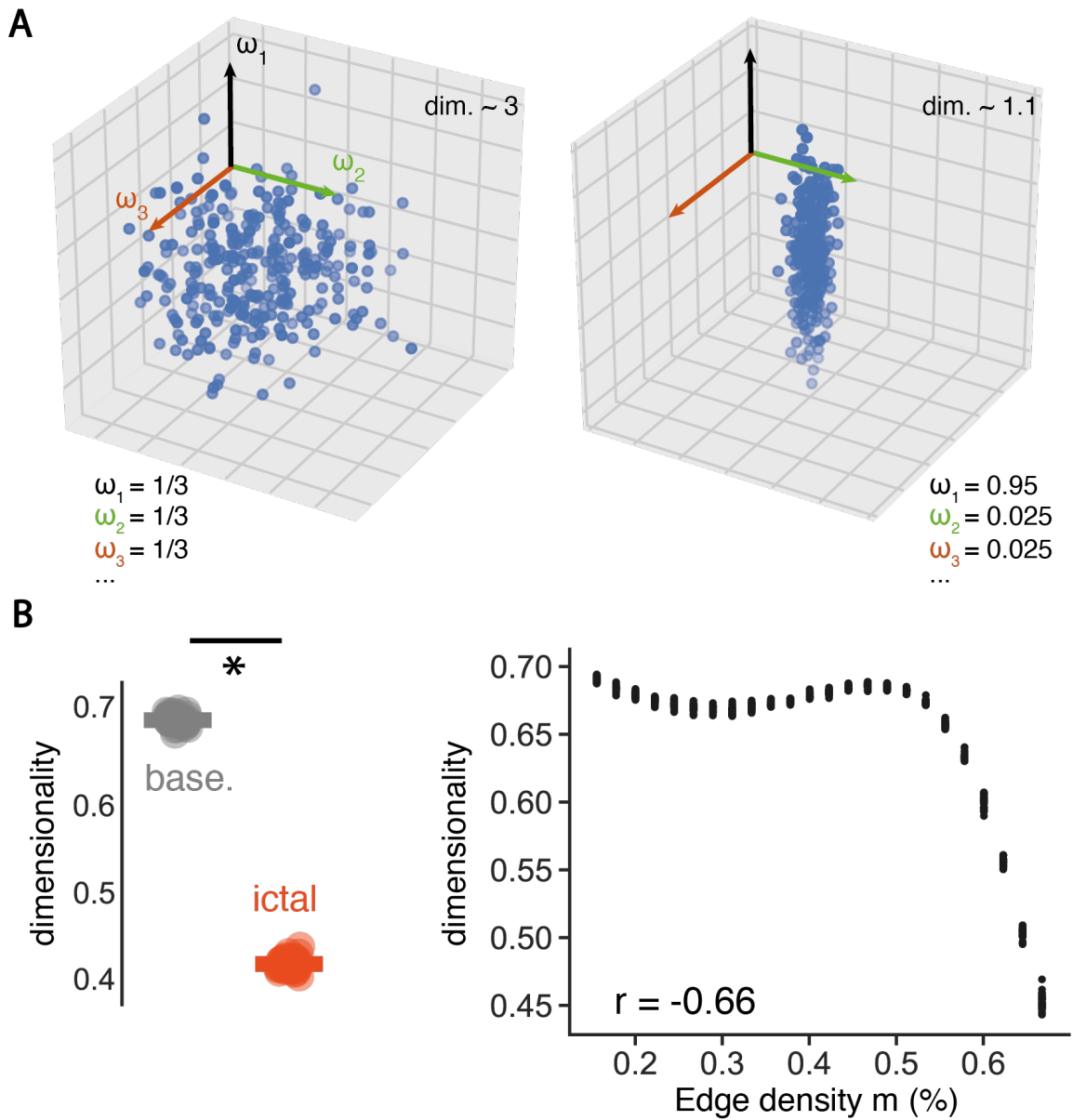

**Figure 4.14  Reduced dimensionality is driven by increased edge density in spiking network model.**

(A) Schematic explaining the dimensionality calculation method. Each point in the space represents a combination of weights across the first 3 principal components ($\lambda$) for a given timepoint. When the variance across each PC is relatively uniform across components the dimensionality is high (left), but when the variance is dominated by a few components the dimensionality is low (right). Figure adapted from Recanatesi et al., 2019. (B, left) Dimensionality compared across baseline (grey) and ictal-onset (crimson) models, using all parameters for model fitting. (B, right) The relationship between dimensionality and edge density parameter m.

I calculated the dimensionality as a proxy for the multidimensional correlation on my network model using all 3 parameters. As predicted, I found that the dimensionality is significantly lower in the *ictal-onset* network compared with the *baseline* network (Figure 4.14B), indicating higher multidimensional correlation during seizures. Next, I calculated the dimensionality for increasing numbers of connections *m*. Here, I found that *m* significantly negatively correlates with the dimensionality, with more edges causing a reduction in the dimensionality (r = -0.66, p < 0.0001) (Figure 4.15B). Therefore, increases in network connectivity give rise to a higher multidimensional correlation and lower dimensional dynamics. Such dynamics would naturally reduce the diversity of brain states available to the system, as the dynamics are dominated by a limited number of highly correlated modes restricting the space of possible neuronal activation patterns to a more restricted subset. As such, increased edge density drives the emergence of homogeneous, inflexible dynamics during seizures.

Next, I wanted to probe the population mechanisms that might drive sticky dynamics during seizures. Here, I build on previous work which has identified a link between multidimensional correlation and the smoothness of neural codes (Stringer, Pachitariu, Steinmetz, Carandini, et al., 2019). Given that the smoothness of population activity is defined by the specific trajectories of brain dynamics in state space, I reasoned that multidimensional correlation might also drive properties of state space trajectories relevant for state transitions. To examine the link between multidimensional correlation and state space trajectories, I use a different measure of multidimensional correlation to (4.13), which is employed in the above study. Specifically, I use the eigenspectrum function which defines the amount of variance $v(n)$ explained by the $n^{th}$ principal component. Here, the eigenspectrum slope $\phi$ dictates the degree of variance captured by early versus late components, with a high slope indicating high relative variance captured by early components and thus high multidimensional correlation and low dimensionality. I estimated the eigenspectrum function by performing principal components analysis (PCA) on a *k x m* matrix of *k* neurons and *m* time frames. Eigenspectra were calculated on raw fluorescence traces.

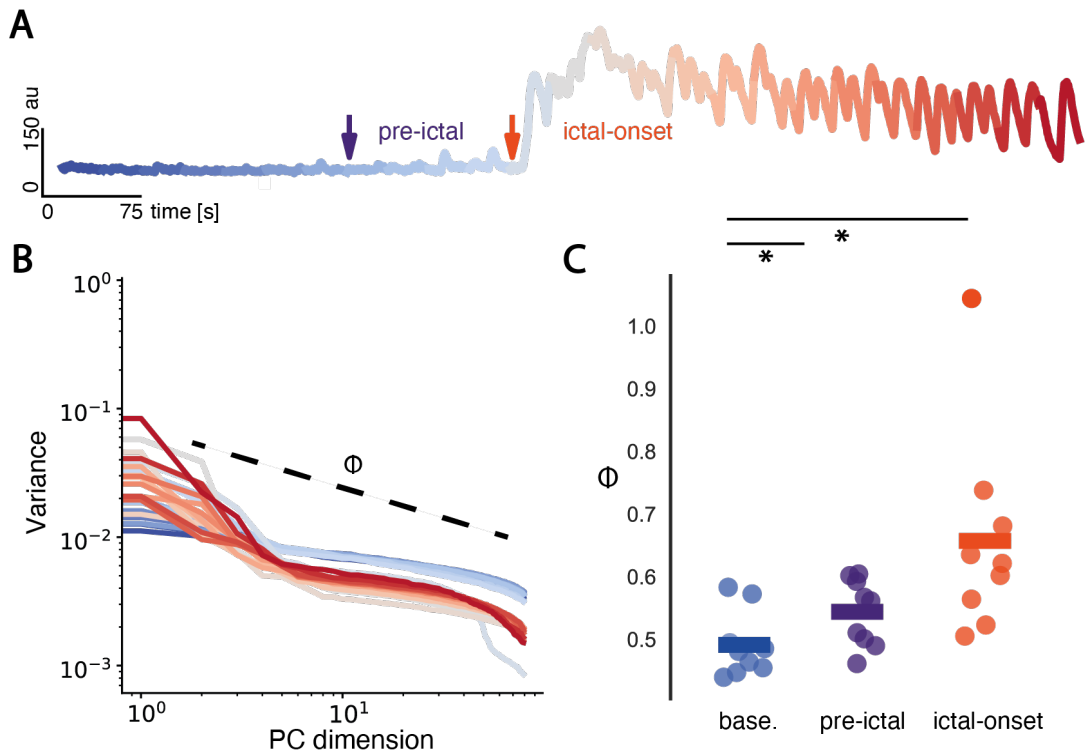

**Figure 4.15  Changes to multidimensional correlation during generalised seizure transitions.**

(A) Mean whole brain fluorescence coloured by time for an example fish following exposure to 20mM PTZ. Pre-ictal periods are classed as 400 frames immediately preceding the generalised state transition (from purple arrow, see Section 2.2.10), while ictal-onset periods are the first 400 frames immediately preceding the start of the transition (from crimson arrow). (B) Eigenspectrum function, which captures the variance that each subsequent principal component (PC) captures, is plotted over time for small segments of the recording period (colours as in A). Notice how the eigenspectrum slope $\Phi$ increases as the generalised seizure emerges suggesting an increase in population covariance. (C) Eigenspectrum slope $\Phi$ over entire 400 frame baseline (blue), pre-ictal (purple) and ictal-onset (crimson) periods.

Firstly, to examine multidimensional correlation changes that precede generalised seizures in empirical data, I calculated the eigenspectrum function in small bins in the lead up to the seizure. The eigenspectrum slope ($\phi$) dynamically reorganised in the lead up to the seizure and during the seizure itself, suggesting dynamic changes to multidimensional correlation. $\phi$ significantly increased in the *pre-ictal* period before the generalised seizure, and during the *ictal-onset* period of the seizure itself (Figure 4.15). This demonstrates that multidimensional correlation increases during and before generalised seizures.

Next, to directly study the relationship between the eigenspectrum slope and state trajectories, I use a 1-dimensional model introduced by Stringer, Pachitariu, Steinmetz, Carandini et al. I define a function *f(x)* which describes the variance of *n* components across *x* samples as

$$f(x)_{2n-1} = \frac{\cos(nx)}{n^{\frac{\phi}{2}}}$$

$$f(x)_{2n} = \frac{\sin(nx)}{n^{\frac{\phi}{2}}}, \quad (4.14)$$

where *n* is uniformly distributed between 0 and 2π. By construction the covariance eigenvalues across all components follow $n^{\phi}$. By increasing $\phi$ I was able to change the variance captured across each component, which allowed me to directly alter the eigenspectrum slope (Figure 4.16). Here, components represent different modes of the data, which can be combined to reconstruct the full data matrix.

Interestingly, in the model I note a direct relationship between $\phi$ and the velocity of the underlying dynamics. Here, I define the velocity as the normalised Euclidean distance travelled per unit time across the whole neuronal population. As $\phi$ increases, due to greater variance captured in earlier components, population dynamics exhibit slower transitions in state space (Figure 4.17A). To visualise this, I simulate eigenspectra and randomly project the generated data matrix *X* (*n* x *x*) through a 3 dimensional randomised weight matrix *W* (3 x *n*). As is clearly visualised, lower dimensionality in the model gives rise to slower dynamics in state space (Figure 4.18A).

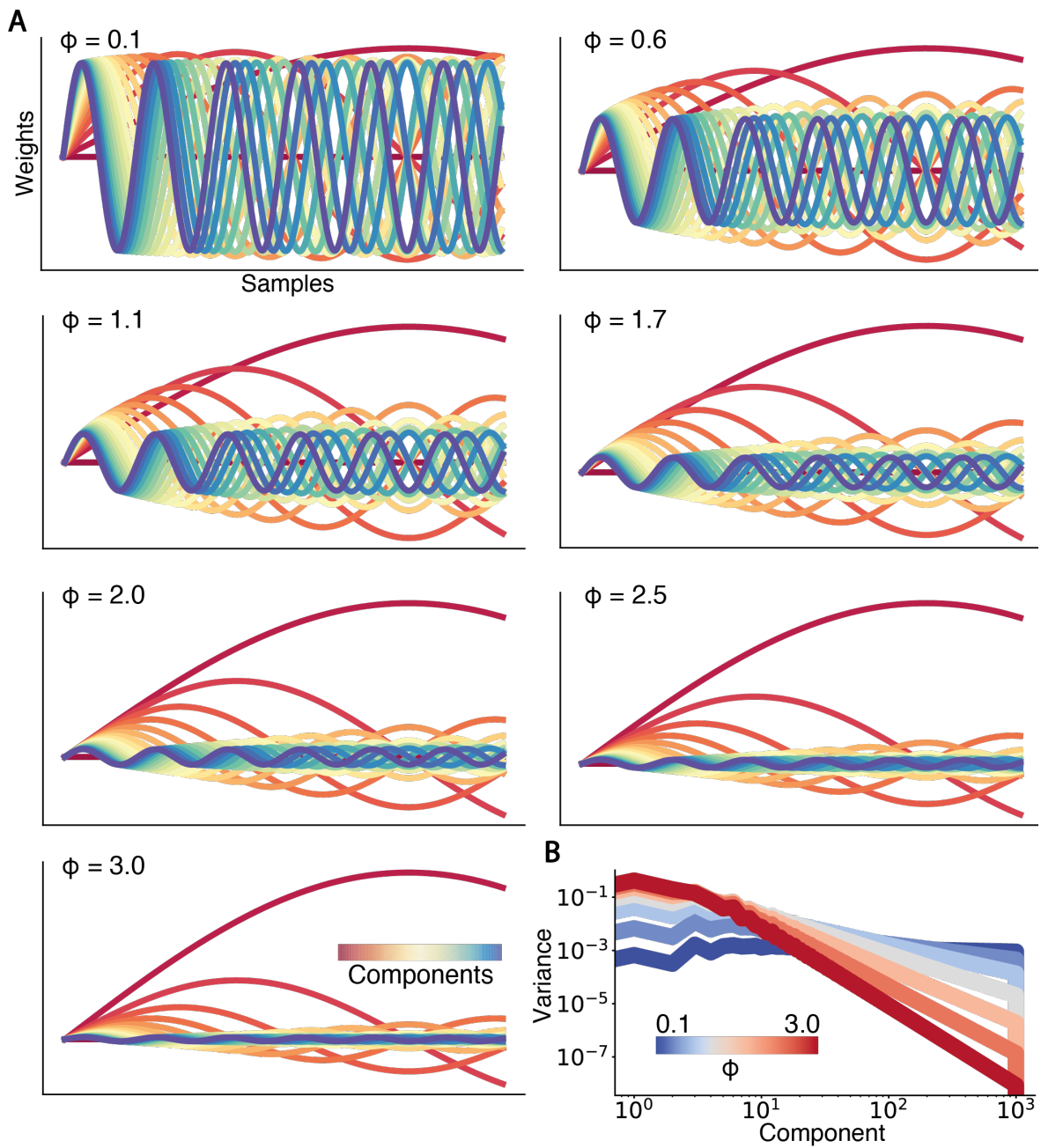

**Figure 4. 16 Simulating eigenspectra.**

(A) Simulated components of different variance shown for different eigenspectra slope parameters ϕ. 1000 simulated components were sampled over time, with ϕ defining the degree to which variance is shared across components (low ϕ) or dominated by early components (high ϕ). (B) For each ϕ value eigenspectra were calculated as the variance captured of each component, showing increased eigenspectral slope for higher ϕ values.

I wanted to confirm this identified relationship between eigenspectra slope and velocity from a simplified model, in my empirically derived data. Here, I calculate eigenspectra and velocity as above on *spontaneous* and generalised seizure (*20mM PTZ*) conditions. When looking across entire 30 minute periods, I found that the eigenspectrum slope significantly increases during generalised seizures (spontaneous: φ = 1.20 ± 0.04; 20mM PTZ: φ = 1.41 ± 0.03, t = -6.05, p < 0.001) (Figure 4.17B). This confirms greater variance captured in the first few components and increased multidimensional correlation during generalised seizures. Interestingly, I found significantly slower dynamics in state space during generalised seizures compared with spontaneous activity (spontaneous: velocity = 0.86 ± 0.01, 20mM PTZ: velocity = 0.70 ± 0.03, w = 0.0, p < 0.01) (Figure 4.17B, 4.18B). This confirms the prediction from the model, that higher multidimensional correlation causing lower dimensionality, is linked to the slowing down of dynamics in state space. Slower dynamics occur due to earlier components dominating the variance, such that the variance of reconstructed trajectories in state space is driven by only a few key modes. Such slow dynamics, are resemblant of sticky dynamics reported above in state transitions. Therefore reduced velocity from one point to the next in state space, could impair the network's ability to flexibly and quickly transition across different brain states. This corroborates predictions from critical networks that epileptic seizures should disrupt flexible dynamics, and provides a population mechanism through which it could occur.

Taken together, I find that densely connected networks drive increased multidimensional correlation. This can explain both homogeneous dynamics, driven through a restriction of neuronal activation patterns to a limited subset, and sticky dynamics, driven by reduced velocity in state space. As such, generalised seizures can give rise to impaired brain function through microscale synaptic alterations disrupting global brain dynamics, and impairing the optimal network response properties and dynamics of critical networks.

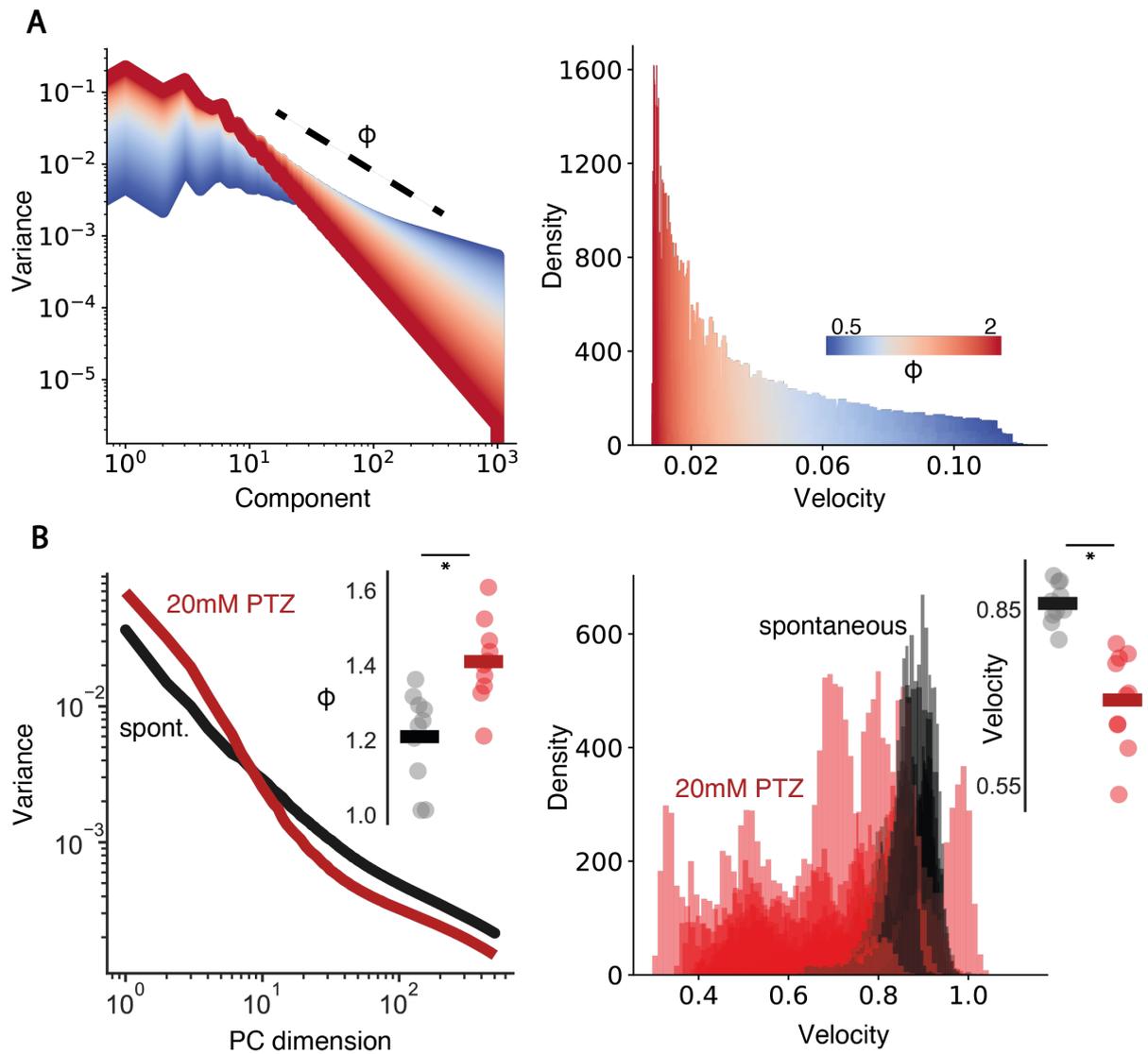

**Figure 4.17  The effect of population covariance changes on the velocity of brain dynamics.**

(A, left) Eigenspectrum function plotted for increasing slope ϕ. (A, right) State space velocity probability densities plotted as a function of ϕ, demonstrating slower dynamics with increasing ϕ. (B, left) Empirical mean eigenspectrum function plotted across spontaneous (black) and 20mM PTZ conditions (red). Eigenspectrum slope ϕ across spontaneous (black) and 20mM PTZ conditions (red) (outset). (B, right) State space velocity probability densities plotted for all fish, comparing spontaneous (black) and 20mM PTZ (red) conditions (bottom). Comparison of mean velocity for spontaneous (black) and 20mM PTZ conditions (red) (outset). * = p<0.01.

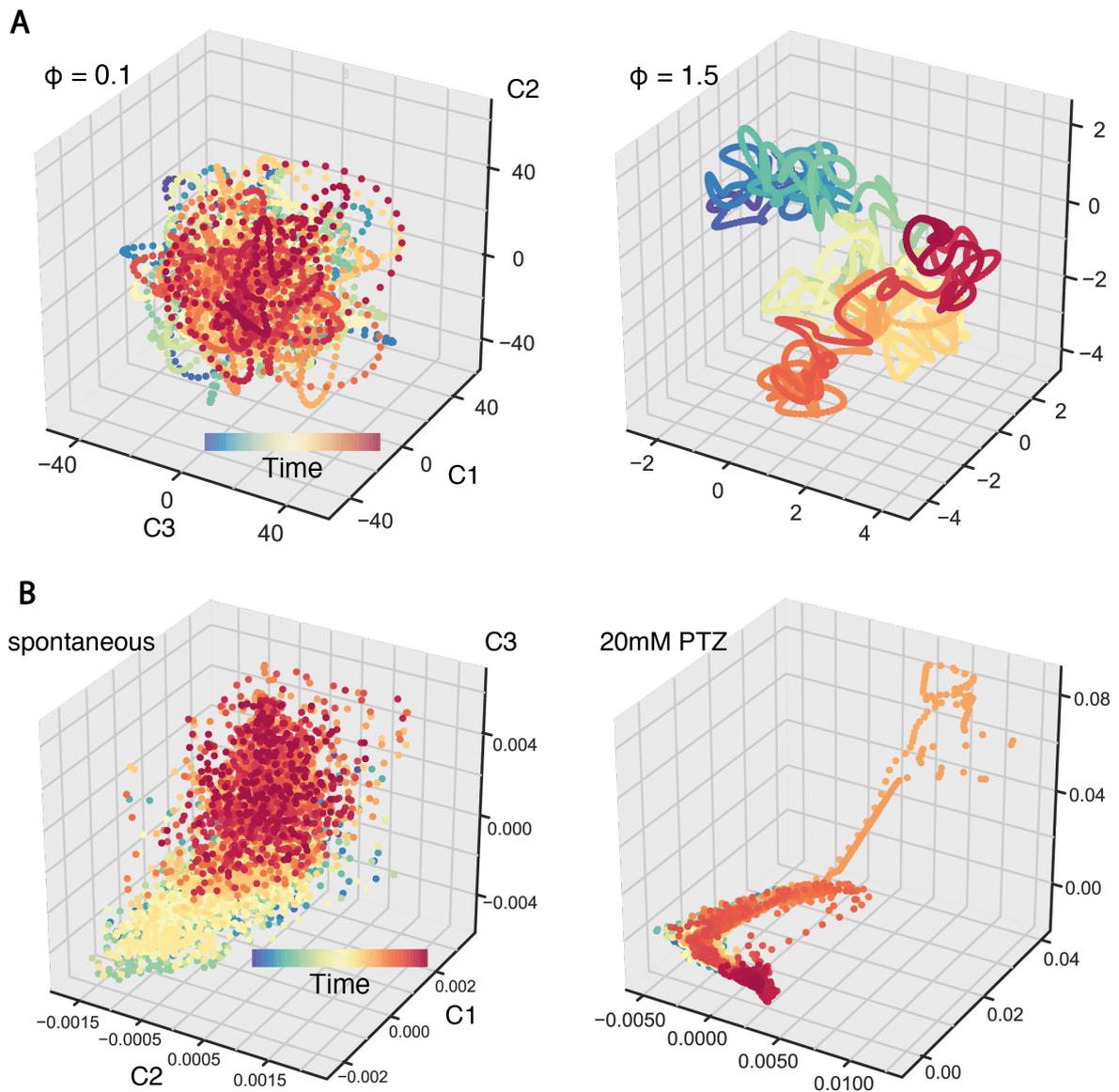

**Figure 4.18  The effect of increasing population covariance on velocity visualised in 3 dimensions.**

(A) Random projection of simulated eigenspectra into state space for different $\phi$. Note how higher $\phi$ (population covariance) causes slower transitions from one point in time to the next in state space. Each point is coloured by time. (B) Isomap embedding of reconstructed attractor for spontaneous (left) and 20mM PTZ (right) conditions for an example fish, demonstrating lower velocity dynamics in the seizure condition.

*4.2.7  The Effect of Foxg1a Mutations on Dynamical Stability and Flexible Dynamics*

Finally, having confirmed the link between optimal network properties at criticality and brain dysfunction during pharmacologically-induced seizures, I wanted to extend my analyses to genetically-induced EI imbalance (Sohal & Rubenstein, 2019). In particular, I wanted to ask whether the cognitive impairment that occurs in EI imbalance disorders may also emerge due to suboptimal network properties away from criticality. In particular, FOXG1 syndrome manifests as severe intellectual disability, lack of language and seizures, and which may be driven by a loss of interneuron number (Bruce, 2022).

To test whether genetic perturbations to EI balance disrupt critical properties in brain networks, I use foxg1a homozygous (-/-) and heterozygous (+/-) zebrafish mutants as models for genetically-induced EI imbalance (see Section 2.1.1). Here, I assess the dynamical stability and flexibility of brain dynamics in foxg1a mutant fish, supposing that observed network alterations in seizures might also occur in foxg1a mutants in the inter-ictal state. Given that observations of foxg1 mutant behaviour suggested no obvious seizures, I imaged both spontaneous and low concentration PTZ conditions (*2mM PTZ*), with the aim of capturing resting state activity and activity in seizure prone networks under conditions of reduced inhibition, to capture the full diversity of brain dynamics in genetically perturbed EI imbalanced networks (see Section 2.2.3). Finally, reductions in inhibitory neuron number are only known to occur in the telencephalon (Bruce, 2022) and therefore I restricted my analysis to telencephalic neurons only (see Section 2.2.7).

Firstly, I calculated the dynamical stability of brain dynamics, reasoning that foxg1a-induced EI imbalance might disrupt the neutral dynamics expected at criticality. I found that during spontaneous activity the largest Lyapunov exponent ($\lambda$) did not significantly change across genotypes (*foxg1a +/+:* 0.0013 ± 0.00*;  foxg1a +/-:* 0.0013 ± 0.00, U = 17.0, p = 0.36; *foxg1a -/-:* 0.0014 ± 0.00, U = 10.0, p = 0.34) (Figure 4.19B). Similarly, in the presence of *2mM PTZ*, I also found no changes to $\lambda$ (*foxg1a +/+:* 0.0014 ± 0.00*;  foxg1a +/-:* 0.0012 ± 0.00, U = 4.0, p = 0.025; *foxg1a -/-:* 0.0014 ± 0.00, U = 8.0, p = 0.36). This indicates that foxg1a mutations do not alter the dynamical stability of brain dynamics, with brain dynamics remaining relatively neutral.

Next, I evaluated the presence of inflexible dynamics, reasoning that foxg1a-induced EI imbalance might disrupt the flexible dynamics of critical systems. First, I calculated the number of states available to the system to investigate the presence of homogeneous dynamics. However I found no significant difference in the number of metastable states across genotypes in *spontaneous* (*foxg1a +/+:* 146.6 ± 26.9*; foxg1a +/-:* 136.8 ± 20.2, U = 18.0, p = 0.41; *foxg1a -/-:* 116.6 ± 15.1, U = 9.5, p = 0.30) and *2mM PTZ* conditions (*foxg1a +/+:* 105.5 ± 15.8*; foxg1a +/-:* 86.6 ± 9.2, U = 13.0, p = 0.34; *foxg1a -/-:* 73.6 ± 10.4, U = 2.0, p = 0.09) (Figure 4.19C). Similarly, I also reported no change in the $\delta$ dwell time spent in each state, across genotypes in *spontaneous* (*foxg1a +/+:* 0.09 ± 0.04*; foxg1a +/-:* 0.10 ± 0.05, U = 18.0, p = 0.41; *foxg1a -/-:* 0.05 ± 0.03, U = 9.0, p = 0.27) and *2mM PTZ* conditions (*foxg1a +/+:* 0.042 ± 0.01*; foxg1a +/-:* 0.06 ± 0.03, U = 16.0, p = 0.47; *foxg1a -/-:* 0.03 ± 0.02, U = 6.0, p = 0.20) (Figure 4.19C). Finally, I assessed the velocity of the dynamics, to investigate the presence of sticky dynamics as found in seizures. As above I found no significant difference in the velocity of the dynamics across genotypes in *spontaneous* (*foxg1a +/+:* 0.75 ± 0.03*; foxg1a +/-:* 0.76 ± 0.02, U = 16.0, p = 0.30; *foxg1a -/-:* 0.75 ± 0.04, U = 11.0, p = 0.42) and *2mM PTZ* conditions (*foxg1a +/+:* 0.75 ± 0.03*; foxg1a +/-:* 0.75 ± 0.02, U = 16.0, p = 0.47; *foxg1a -/-:* 0.76 ± 0.03, U = 9.0, p = 0.45) (Figure 4.19D). This indicates the foxg1a mutations do not give rise to more homogeneous or sticky dynamics than wildtype.

Taken together, the brain dynamics in foxg1 mutant fish do not exhibit the statistical properties of networks driven away from a phase transition. As such, emergent brain dysfunction that occurs in FOXG1 disorders is unlikely to emerge due to the suboptimal computational capacities found in networks away from criticality.

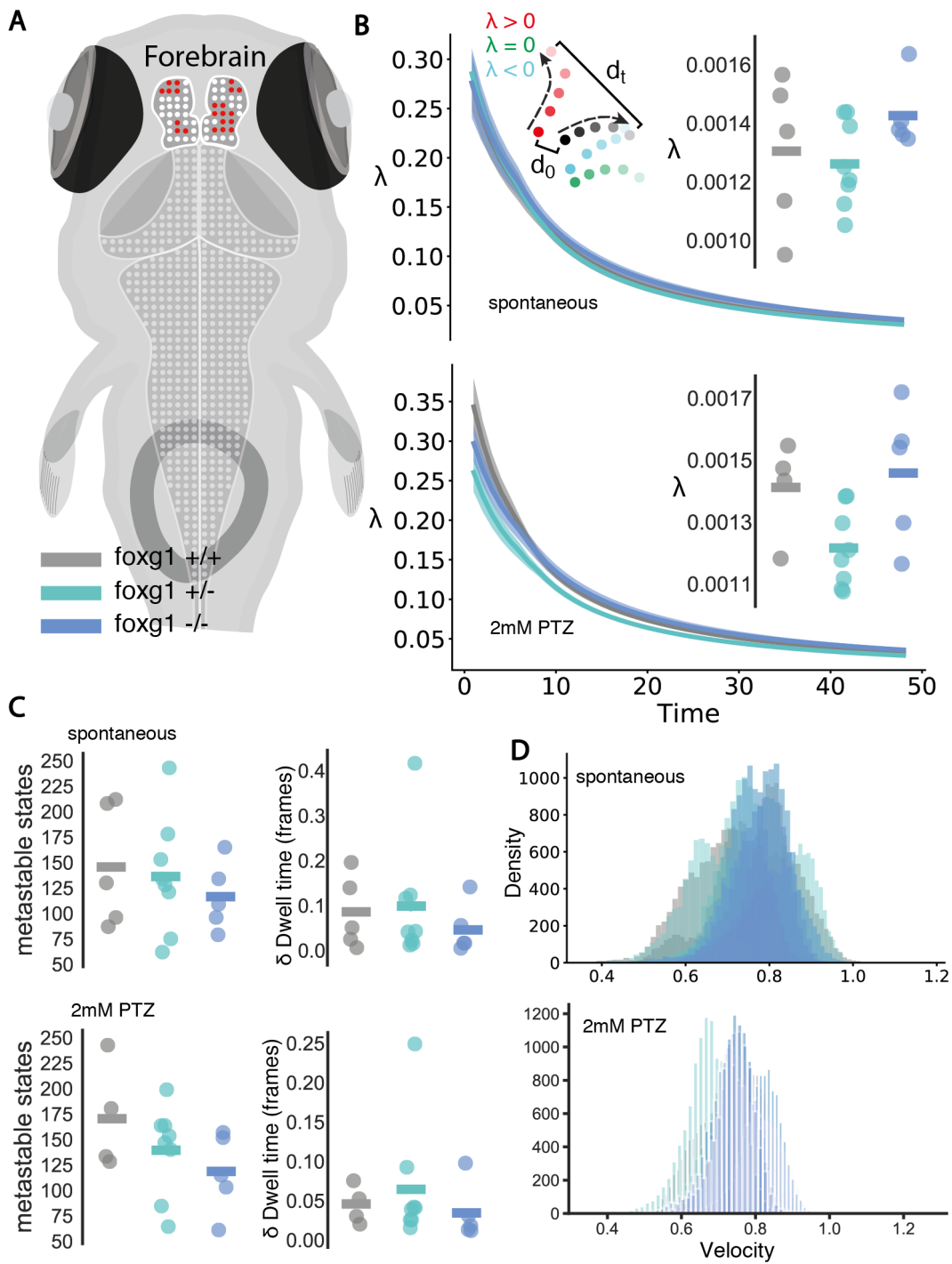

**Figure 4. 19 The effect of foxg1a mutations on dynamical stability, and flexible dynamics.**

(A) Larval zebrafish brain schematic, with all foxg1 analyses performed across forebrain neurons. (B) Mean λ shown over time across each genotype for spontaneous (top) and 2mM PTZ conditions (bottom). Shaded error is standard deviation. λ across all fish compared across homozygous, heterozygous and wildtype for spontaneous (top) and 2mM PTZ conditions (bottom). (B, top, inset) Schematic outlining the meaning of different values of the Lyapunov exponent (λ). Each colour represents the trajectory over time for a specific initial point along the attractor (high to low brightness represents movement in time). λ is calculated as the ratio between the difference between 2 points at the start (d0) and at timepoint t (dt). λ for each trajectory is calculated against the black trajectory. λ > 0 (blue trajectory) : distances grow over time (chaos). λ < 0 (red trajectory) : distances shrink over time (stability). λ = 0 (green trajectory) : distances are constant over time (neutral). (C) Number of metastable states (left) and dwell time (right) compared across all fish compared across homozygous, heterozygous and wildtype for spontaneous (top) and 2mM PTZ conditions (bottom). (D) State space velocity probability densities plotted for all fish, comparing across homozygous, heterozygous and wildtype fish for spontaneous (top) and 2mM PTZ (bottom) conditions.

## 4.3  Discussion

In this chapter, I built a single neuron network model of the larval zebrafish brain to understand the cellular alterations giving rise to epileptic seizures and subsequent functional impairments. I aimed to answer two key questions. First, are seizure dynamics a convergent endpoint emerging from non-specific changes to excitability, through changes to network connectivity, synaptic strengths or intrinsic excitability? Second, does functional impairment in epileptic seizures and associated EI imbalance disorders occur through disruption of the optimal network dynamics of critical networks? My findings revealed that local seizures emerge out of non-specific changes to excitability, yet generalised seizures can only emerge through increases in neuronal connections in the network. Furthermore, such increases in neuronal connectivity give rise to a loss of neutral and flexible dynamics and impaired network response properties in seizures, directly linking brain dysfunction in epilepsy with a loss of critical dynamics. Thus, these results identify exact microscale synaptic changes driving emergent brain dynamics and dysfunction in epilepsy.

### 4.3.1  Microscale Neuronal Mechanisms Driving Global Seizure Dynamics
This chapter demonstrates the microscale network alterations driving local and generalised seizure dynamics. A handful of studies have previously investigated the microscale neuronal properties driving global seizure dynamics in the larval zebrafish. Of particular interest, Diaz Verdugo et al. reported long range synchrony emerging across distant brain areas during epileptic seizures (2019). My work builds on these descriptive analyses, by demonstrating an exact neuronal mechanism that can give rise to such aberrant correlations through the formation of new network connections. The increased correlation between neurons (Diaz Verdugo et al., 2019a), and coarse brain areas (Gotman & Levtova, 1996; Schindler et al., 2007) that has been described during seizures, is thus a natural consequence of a more densely connected network. My work also corroborates previous whole brain modelling approaches which implicate feedforward hub neurons embedded in a densely connected neighbourhood in the emergence of state transitions (Hadjiabadi et al., 2021). More connections in my model network would both increase the propensity for highly connected neighbourhoods and drive the emergence of hub neurons due to the growth and preferential attachment algorithm used to construct my networks (see Equation 4.1). Therefore, previously identified feedforward hubs may naturally emerge in my model as a function of connectivity. However, a direct investigation into the role of such motifs is beyond the scope of this work. Nonetheless, my model supports the previously supposed notion of a pathologically interconnected state responsible for driving generalised seizure dynamics. That an increase in network connectivity drives the transition from a pre-ictal to a generalised seizure state in my model, suggests that increased network connectivity may be causal in such state transitions.

This study expands on previous work by comparing the role of specific network parameter changes in different seizure types. In particular, I tested the specificity of synaptic excitability changes to locally and globally synchronous seizure dynamics. Interestingly, seizures can emerge in brains throughout the animal kingdom (Buckmaster et al., 2014; Jirsa et al., 2014a; Podell, 1996), and may be triggered by a diversity of conditions such as sleep deprivation, stress, hypoxia, physical trauma, genetic mutations, and chemical and electrical stimulation (Jett, 2012; Luttges & McGaugh, 1967; Nakken et al., 2005). The common link between such seizure triggers is disruptions in the homeostatic regulation of neuronal processes, which can lead to increased excitability through impaired regulation of membrane potential (R.

Rosch et al., 2019). Therefore, seizures may well be a common endpoint for brain networks due to increased excitability. In fact, studies of the microscale changes driving seizures have identified a diversity of neuronal alterations that would cause increased excitability. In particular, impaired synapse refinement (Chu et al., 2010), alterations to synaptic strengths through shifting ratios of continuous to discontinuous postsynaptic densities (Geinisman et al., 1988), and increased intrinsic neuronal excitability through more vesicles in the readily releasable pool (Upreti et al., 2012), would all increase neuronal excitability and are linked to seizures. Using my network model, I was able to test the hypothesis that seizures emerge from non-specific changes to excitability, and thus may be a common end-point for disrupted brain networks. That local seizure dynamics could be well fit by 2 parameter models capturing any of intrinsic excitability, synaptic strength and connectivity changes suggests that seizures that do not generalise may emerge out of non-specific neuronal changes. This indicates that locally restricted seizure states, such as focal seizures, may be a convergent end point following non-specific increases in excitability. Interestingly, such locally synchronous seizures do not require alterations in the network topology, but can rely on the existing network structure for local propagation of activity.

However, in stark contrast to local seizures, I found that only increases in network connectivity in the model were able to accurately capture globally synchronous dynamics. The fact that the network topology model alone provided a much better fit than the model using all other parameters, suggests that increased connectivity may be sufficient to explain generalised seizure dynamics. This indicates that generalised seizures, are not a convergent endpoint to non-specific excitability changes, but in fact require specific increases in network connectivity. This finding has implications for our understanding of the network changes that separate out local and generalised seizures. Specifically, unlike in local seizures, generalised seizures cannot spread sufficiently using the existing network topology but require the addition of new connections. This suggests that healthy network topologies are favoured towards limiting the propensity for excessive global synchrony, and generalised seizures may emerge due to the maladaptive regulation of network topology.

While my network model has identified the differential contribution of simplified network alterations in driving seizures, mapping relatively ambiguous parameter shifts onto specific biological changes is non-trivial. Given the relatively short timescales that separate out baseline and locally unstable dynamics following PTZ administration (30-60 minutes), the non-specific changes driving local seizures in empirical data must occur over minutes. Increases in synaptic strengths, could occur due to early-long term potentiation (LTP) changes which can increase excitatory postsynaptic potentials within seconds-minutes due to the insertion of AMPARs into dendritic spines (Barco et al., 2002; Penn et al., 2017). In fact, partial GABAa receptor blockade would likely induce Hebbian learning by releasing excitatory neurons from their inhibitory restraint, and subsequent fast AMPA insertion might destabilise the network too quickly for homeostatic scaling to take effect (Hsu et al., 2008). As such, increased synaptic weights can drive a locally unstable seizure state through early-LTP. Similarly, increased intrinsic excitability could also occur over such short timescales. For example, partial GABAa blockade leads to excessive depolarisation of excitatory neurons which could cause the accumulation of extracellular $K+$ (Librizzi et al., 2017), the depolarisation of the $K+$ reversal potential thus ultimately pushing neuronal membrane potential closer to spike threshold. In fact, $K+$ dynamics during ictal and inter-ictal periods change substantially over seconds to minutes (Jensen & Yaari, 1997), and thus could feasibly drive intrinsic excitability changes over timescales relevant for the emergence of local seizures. In this way, increases in intrinsic excitability through a breakdown of neuronal membrane potential homeostasis might also drive local seizures in our data. As such, alterations to synaptic weights and intrinsic excitability driving local seizures in our model can be reduced down to a more limited subset of biological parameter changes that might occur over seconds to minutes, following the addition of PTZ.

Changes to network connectivity are more challenging to directly map onto specific network changes in our model. The formation of new structural synapses is likely limited during the short timescales between locally synchronous and globally synchronous seizures following PTZ administration (2-30 minutes) (Le Bé & Markram, 2006; Ozcan, 2017). Therefore, increased network connectivity may instead be explained by the formation of new effective connections between neurons, giving rise to dense effective connectivity. Changes to effective connectivity might emerge through a diversity of non-synaptic pathways. Interestingly, evidence from larval zebrafish recordings suggests that synchronous neuronal activity may be driven by glia-glia coupling (Diaz Verdugo et al., 2019a). Glia coupled non-synaptically via gap junctions have been shown to synchronise calcium transients over long distances during generalised seizures. Given that such synchronised calcium transients within glia give rise to glutamate release and excitation of neighbouring neurons, such glial-glial coupling can provide a means of synchronising non-synaptically coupled neurons over long distances. This provides a possible mechanism for increased effective connections that emerge over short timescales during generalised seizures to drive globally synchronous cascades. An alternative mechanism for the emergence of effective connections over short timescales, is volume-transmitted extracellular GABA release acting extrasynaptically via GABAa receptors. Interestingly, evidence suggests that interneuron discharge can lead to high extracellular GABA, which can actually entrain interictal discharges (Magloire et al., 2022). Here, the emergence of extracellular GABA waves and tonic GABA conductance dynamics through extrasynaptic receptors on interneurons, could entrain inhibitory network rhythms during seizures over short timescales. In this way, effective network connections might emerge between nonsynaptically coupled neurons through the volume-transmitted effects of extracellular GABA through the network. Taken together, the emergence of new connections in my network model over short timescales indicates non-synaptic mechanisms driving generalised seizures. Nonetheless, direct synaptic mechanisms may also play key roles in increasing functional connectivity – in particular a failure of feed-forward inhibition or long range inhibition onto local, connected excitatory neurons would increase the functional coherence of neighbouring excitatory neurons as well.

My modelling approach has several key limitations. In particular, I make several assumptions that will influence the performance of the model and should be taken into account. Firstly, I assumed that the topology of the network was scale free. However, I did not compare whether other network topologies such as small world, or random networks would provide better fits to my empirical data. Given that the underlying topology of the larval zebrafish brain is unknown, a broader comparison of network regimes driving seizure states across different network topologies would be useful to demonstrate whether such parameter changes are invariant to network type. Additionally, the growth and preferential attachment algorithm does not take into account the distances between neurons when forming connections. While, I used the synaptic weights exponential decay function to ensure a local preference for synaptic connections, the specific topology of the network would fail to account for metabolic constraints to wiring. Similarly, I assumed that the probability of forming connections was uniform across and between brain areas. Future work could include a distance constraint for the formation of connections, and include region specific wiring probabilities. I also note that, while I used distances to avalanche distributions to fit my models, there are likely many different network patterns that could give rise to a given distribution. A more specific, but computationally expensive approach would be to use exact neuronal spike patterns for model fitting, using approaches such as FORCE learning (Sussillo & Abbott, 2009). For the sake of simplicity, I ignored inhibitory neurons in my model – future work should characterise the specific inhibitory and excitatory synaptic parameters driving seizure dynamics, to justify the inclusion or exclusion of inhibitory neurons. Nonetheless, this work provides novel insight into the general principles driving locally and globally unstable dynamics during epileptic seizures.

### 4.3.2  Understanding Emergent Brain Dysfunction in Seizures and EI-Imbalance Disorders

I extended my network models of seizures, to understand how microscale changes can give rise to brain dysfunction. In particular, given that epileptic seizures cause a loss of criticality (see Chapter 3), I theorised that subsequent brain dysfunction might emerge due to the suboptimal computational capacities of brain networks driven away from a phase transition.

Key theoretical predictions relating to supercritical networks suggest a loss of neutral dynamics and the emergence of chaos (Haldeman & Beggs, 2005). However, whether such predictions from simplified models would translate to brain dynamics is unclear. Furthermore, the exact type of bifurcation at which brain dynamics may reside is up for dispute, as transitioning from critical to supercritical avalanche dynamics *in silico* does not necessarily alter the chaotic nature of the underlying dynamics (Kanders et al., 2017). Here, I demonstrate that seizures give rise to more positive Lyapunov exponents suggesting more chaotic dynamics in both empirical data and network models. This supports the theoretical prediction that brain dynamics exist at the edge of chaos, separating out states of order and disorder (Beggs, 2008; Haldeman & Beggs, 2005; Magnasco et al., 2009). Therefore, our data support the notion that spontaneous dynamics lie on the phase transition between order and disorder (λ~0), with seizures emerging as a chaotic, disordered state (λ > 0). However, one should note that comparing the largest Lyapunov exponents across different systems can be problematic, as λ also depends on the time scale of the system. Nonetheless, given that λ increases are invariant across empirical and model data and different estimation methods, this provides reasonable evidence for the emergence of chaos.

A loss of neutral dynamics and the emergence of chaos during seizures would reduce the network's memory about recent inputs, due to high reverberating activity in the network. This could help to explain impaired memory of seizure manifestation in patients with epilepsy, with between 44-63% of seizures going unnoticed (Blum et al., 1996; Hoppe et al., 2007; Kerling et al., 2006). For example, chaotic dynamics due to the emergence of dense connectivity in the brain, would disrupt the ability of the network to represent sensory input during the seizure, such that downstream readout functions might fail to accurately decode the network state and store it into memory.

The result of more chaotic dynamics during seizures, is contrasted by *in silico* modelling of avalanche dynamics generated from phase locking of EEG channels, which indicate more ordered dynamics during seizures (Meisel et al., 2012). The differences in our data likely lies in the type of phase transition studied – the avalanche construction used by Meisel et al. is based on phase synchrony between units, and thus seizures may emerge as a more ordered (synchronous) state, with respect to a synchrony-desynchrony transition (Kitzbichler et al., 2009; Linkenkaer-Hansen et al., 2001). Interestingly, the brain may lie on multiple phase transitions at once (Meisel et al., 2013; H. Yang et al., 2012). As such seizure dynamics might give rise to chaotic dynamics with regard to an activity-inactivity phase transition, while also emerging as a more ordered state regarding synchrony-desynchrony.

Some evidence also points to the role of criticality in shaping network response properties, specifically by maximising the dynamic range (Gautam et al., 2015; Kinouchi & Copelli, 2006; Shew et al., 2011) and network-mediated separation (NMS) (Bertschinger & Natschläger, 2004; Legenstein & Maass, 2007; Maass et al., 2002). My modelling data extends these findings into seizure networks, as I demonstrate reduced dynamic range and NMS in seizure networks in line. This finding naturally corroborates predictions for critical branching networks in which the branching ratio increases above 1(see Section 3.2.7) – such supercritical seizure networks should give rise to exponentially growing avalanches regardless of input size, both limiting the range of inputs the brain can represent but also the ability of the brain to separate out distinct inputs (Harris, 1963). These network features are relevant aspects of brain function, as they will dictate the ability of brain networks to represent and provide distinguishable readouts to downstream outputs (Bertschinger & Natschläger, 2004). Thus, my data provides novel insight into how epileptic seizures may give rise to impaired network properties, in line with a critical system driven away from the phase transition into disorder. The impaired ability to represent inputs would likely disrupt the brain's ability to respond appropriately to them. This could therefore could help to explain the impaired responsiveness that is present in many seizures – many epilepsy patients show the impaired ability to respond to stimuli during seizures (Inoue & Mihara, 1998).

Finally, the link between flexible spontaneous dynamics in critical systems and epileptic seizures has not been explored *in vivo* or *in silico*. Theoretical work indicates that critical systems should engage in metastable dynamics manifesting as the transient formation and dissipation of diverse semi-stable states (Haldeman & Beggs, 2005; Wildie & Shanahan, 2012). This feature has been shown to support a variety of flexible behaviours including, flexible learning (Bak & Chialvo, 2001), rapid state transitions (Fingelkurts & Fingelkurts, 2001), integrative and segregative activity across distributed brain regions (W. J. Freeman & Holmes, 2005), information storage (Hopfield, 1982), and flexible coupling (Friston, 1997). Furthermore, *in silico* and *in vivo* data point towards the presence of maximal metastability in spontaneous dynamics (Baker et al., 2014; Deco et al., 2017; Fingelkurts & Fingelkurts, 2001). My data demonstrate that epileptic seizures give rise to a reduction in the number of metastable states, suggesting a loss of the flexible state transitions that define metastability. The emergence of such homogeneous dynamics during generalised seizures, is supported by studies demonstrating a reduction in the diversity of brain states during a loss of consciousness induced by general anaesthesia (Schartner et al., 2015; Wenzel, Han, et al., 2019). Therefore, the loss of metastable dynamics and the emergence of homogeneity, could explain the loss of consciousness that occurs during absence and tonic-clonic seizures (Blumenfeld, 2012). However, I note that while criticality is one of multiple optimal regimes that the brain can engage in, non-critical and even chaotic networks can optimise computation as well (Farrell et al., 2019; Touboul & Destexhe, 2017). Therefore, it is plausible that epileptic seizures might disrupt network function via alternative mechanisms that do not involve critical to supercritical transitions.

I also uncover a novel link between multidimensional correlation and the speed of state transitions, suggesting that epileptic seizures give rise to sticky dynamics. This further supports the notion that seizures disrupt the flexible dynamics characteristic of metastability, which exhibits rapid state transitions (Fingelkurts & Fingelkurts, 2001). Interestingly, adapting an eigenspectrum model, I demonstrate a direct link between increased multidimensional correlation and the velocity of the dynamics. This suggests a novel mechanism through which microscale correlations can alter global dynamics and function. I note that this highly simplified model assumes that global dynamics may be reconstructed from a combination of model components, without the additional constraint that components are orthogonal or capture maximal variance, as in principal components analysis. As such comparisons from my eigenspectrum model and empirical data which relies on principal components analysis, should be interpreted with caution.

Finally, I also investigated whether the abnormal inter-ictal dynamics that is present in EI imbalance disorders linked to epilepsy, might give rise to impaired critical network properties. Using a model of FOXG1 syndrome which has known alterations to neuronal firing properties due to a loss of EI balance, I extended my analytical approaches to this data. Interestingly I found that FOXG1 brain dynamics do not influence key network properties and dynamics that are optimised in critical systems. In fact, homozygous and heterozygous dynamics exhibit optimal network properties as seen in wildtype animals. Therefore the cognitive impairments that occur in FOXG1 syndrome are unlikely to be caused by alterations to critical dynamics. This is in keeping with findings from Chapter 3 (see Section 3.2.8) which showed no changes to avalanche dynamics or critical statistics in foxg1a mutant zebrafish.

To summarise, I find that i) generalised seizures emerge in densely connected networks, while local seizures can emerge from non-specific changes to excitability, while ii) increases to network connectivity during generalised seizures disrupt the network's ability to represent inputs faithfully and engage in flexible dynamics. Therefore, network modelling of the larval zebrafish brain can help to link microscale neuronal alterations with emergent global dynamics and dysfunction in epileptic seizures.

# Chapter 5

## The Role of Non-Linear Network Interactions in Seizure Dynamics

### 5.1 <u>Introduction</u>

The International League Against Epilepsy define the cause of epileptic seizures as "abnormal excessive or synchronous neuronal activity" (Fisher et al., 2005), which is believed to result in a hyper-synchronous brain state (see Section 1.1) (Penfield & Jasper, 1954). Indeed, evidence across all scales of recordings report excessive neuronal synchrony during seizures, from single neuron activity (see Section 1.1.3) (J. Liu & Baraban, 2019), to large neuronal populations (Weiss et al., 2013), and entire brain areas (see Section 1.1.4) (Gotman & Levtova, 1996; Jiruska et al., 2013; Schindler et al., 2007). The view that epileptic seizures are generically synchronisation phenomena has shaped how we understand and attempt to prevent seizures. For example, highly synchronised dynamics suggest that epileptic neural activity will covary linearly and thus the network interdependencies driving seizures may be captured using linear techniques – many techniques used to describe seizure dynamics assume linear interactions (Arbib, 1995; Niedermeyer & Lopes da Silva, 2005). In fact, owing to the simplicity of such linear techniques and the coarse-grained nature of human brain recordings, some have claimed that linear models are sufficient to describe many observable brain dynamics (Nozari et al., 2020; Walter & Ross Adey, 1968). As such, linear techniques may be appropriate to describe and predict the synchronous neuronal dynamics of epileptic seizures.

However, this notion lies in stark contrast to the fact that neuronal dynamics at the microscale are highly non-linear (Izhikevich, 2000). This is because a neuron's membrane potential scales as a non-linear function of its inputs, due to the presence of voltage gated ion channels, intrinsic fluctuations in membrane potential and dynamic changes to synaptic weights through neuromodulation (Lehnertz et al., 2001; Nadim & Bucher, 2014). In fact, system reconstructions of epileptic electroencephalography (EEG) recordings (Takens, 1981), suggest the presence of non-linear dynamics in both ictal and inter-ictal data (Casdagli et al., 1997; Schreiber, 2000; Weber et al., 1998). Furthermore, analytical techniques which measure the extent of non-linear interdependencies across time series (Schiff et al., 1996) have shown key non-linear interactions between brain areas during seizures (Le Van Quyen et al., 1998). Therefore, seizure dynamics may emerge out of the non-linear interactions between neurons in the network. How then does this square with the notion of seizures as a highly correlated and thus linear state? Interestingly, evidence from microscale recordings challenge the hypothesis that neuronal interactions during seizures are purely hyper-synchronous (Jiruska et al., 2013) – large proportions of neurons can engage in highly heterogeneous firing patterns (Muldoon et al., 2013; Schevon et al., 2012; Truccolo et al., 2011). For example, some subpopulations can decouple from the network while others become tightly correlated during seizure onset (Meyer et al., 2018). Such diverse dynamics during seizures indicate complex, state-dependent neuronal interdependencies at the microscale which are characteristic of non-linear dynamics. In fact, cellular-resolution whole brain seizure networks exhibit chaotic dynamics indicating non-linear interdependencies between variables at the microscale (see Section 4.2.3). Taken together, this suggests that linear models will only partially capture microscale neuronal network dependencies in epilepsy, as neurons can engage in both linear and non-linear behaviours during seizures. However, the nature of such linear and non-linear interactions in seizures, in particular, the specific neuronal dynamics encompassing both linear and non-linear behaviours, and the cell subtypes differentially engaging in such behaviours, remains poorly understood. Uncovering the differential contribution of both linear and non-linear interactions in seizure networks, will help to identify novel cellular mechanisms that give rise to seizure dynamics.

Although at the microscale, non-linear interactions are key for driving neuronal dynamics, it remains uncertain whether such detailed, microscopic descriptions can be used for seizure intervention and prediction (Jiruska et al., 2013). Macroscale brain recordings coarse grain the underlying neuronal dynamics, filtering out non-linear input-output functions – as a consequence EEG dynamics can be accurately captured using linear models (Nozari et al., 2020). Given that current interventions to re-stabilise epileptic networks target coarse brain areas, such as neuromodulation or surgical intervention (Ryvlin et al., 2021), linear models may provide predictive power regarding the epileptogenicity of broad areas at the appropriate scale for seizure intervention. However, this does not preclude the possibility that targeted microscale interventions guided by non-linear techniques could offer higher precision and efficacy in re-stabilising network dynamics. In fact, microscale interventions on small groups of nodes guided by non-linear dynamical systems theory, have shown striking efficacy in reducing seizure propensity (Hadjiabadi et al., 2021) and restoring normal network dynamics *in silico* (Sanhedrai et al., 2022). Given that even single cell perturbations can alter brain state *in vivo* (Tanke et al., 2018), having access to the non-linear dependencies driving such microscale dynamics may aid in identifying specific neuron-neuron interactions that are key for the emergence of global instability, providing novel targets for network intervention. Furthermore, tracking microscale, non-linear dynamics during seizures, for example using phase space embedding procedures (Martinerie et al., 1998), could provide improved seizure forecasting above and beyond macroscale, linear approaches. In this way, non-linear microscale dynamics may provide improved avenues for seizure prediction, over conventional approaches applied to macroscale dynamics. To assess the relative utility of non-linear, microscale dynamics for seizure prediction and intervention, it is necessary to compare non-linear analytical techniques applied to microscale recordings with macroscale recordings such as EEG. This will help to identify the appropriate scale of recordings to most effectively predict and intervene in seizures.

Due to the computational cost of performing non-linear analyses, most previous approaches have been applied to low dimensional datasets – either to coarse grained recordings of the whole brain or single neuron recordings over a small area (Lehnertz et al., 2001). However, as discussed in Section 1.1, to accurately describe seizure dynamics, one needs single cell recordings across the entire brain. Here I take advantage of the larval zebrafish, a system which provides access to single neuron resolution across the entire brain, to perform 2-photon calcium imaging during induced seizures (see Section 1.1.4). Importantly, leveraging such microscale dynamics requires appropriate non-linear techniques. For example, neurons may exhibit strongly interdependent behaviour without exhibiting synchronous dynamics (Rulkov et al., 1995) – coupled deterministic dynamical systems can engage in transient correlations or show no correlation at all (Le Van Quyen et al., 1998; Sugihara et al., 2012). Linear approaches would fail to describe such state-dependent or non-synchronous, coupled behaviour (Chatfield, 1989). To study the linear and non-linear interdependencies driving seizures, I employ Convergent Cross Mapping (CCM), a computational technique which relies on lagged coordinate embedding to infer the causal relationships between time series (Sugihara et al., 2012). This approach not only accounts for non-linear interdependencies, but also infers the directionality of the relationship, neither of which are possible with linear techniques. Furthermore, I employ parallel distributed implementations of CCM to enable massively parallel inference across up to $4x10^8$ neuron-neuron comparisons per fish (Takahashi et al., 2021; Watanakeesuntorn et al., 2020).

Here, I test 2 key hypotheses: i) epileptic seizures emerge as both an increase in linear and non-linear behaviour, with increases in correlation and non-linear predictability, and ii) non-linear interdependencies are removed when spatially coarse graining, indicating that such information is lost at the macroscale. I find that during generalised seizures, alongside an increase in pairwise correlation, the amount of non-linear interactions between neuron pairs significantly increases. This suggests that seizure dynamics are both highly non-linear and linear at the single neuron scale. Furthermore, non-linear interdependencies are filtered out by spatially coarse graining which indicates that a large portion of information about node-node interdependence is lost in macroscale brain recordings. This opens up avenues for future research into the utility of such microscale non-linear dynamics in predicting and preventing seizures.

## 5.2    Results

I performed 2-photon imaging of the larval zebrafish during pharmacologically-induced seizures (see Section 2.2). This allowed me to record from single neuron activity across the whole brain during spontaneous and seizure dynamics. To capture spontaneous and induced-seizure activity periods, I recorded 3 x 30-60 minute imaging blocks for each fish: 1) *spontaneous* activity representing normal brain dynamics at rest, 2) *5mM PTZ* giving rise to focal seizures, and 3) *20mM PTZ* causing generalised seizures. *5mM PTZ* caused localised, bursts of synchronous activity while *20mM PTZ* caused long-lasting, high-amplitude, synchronous activity recruiting most of the brain consistent with generalised epileptic seizures (Figure 2.2). Datasets with low GCaMP expression or evidence of brain expansion during seizures were removed. Datasets were visually inspected for drift in the z-plane. Datasets where z-drift occurred, but a continuous period of at least 30 minutes with no drift was found, were included with drifting frames removed.

### 5.2.1  Convergent Cross Mapping

In order to estimate the causal, non-linear interdependencies between time series I made use of Convergent Cross Mapping (CCM), an algorithm that performs cross-manifold prediction to ascertain causal interdependence (Sugihara et al., 2012). CCM is able to account for both linear and non-linear directional dependencies between time series, where linear methods would fail. For example, given two time series X and Y that are perfectly synchronous, linear methods can capture their dependence as the variables linearly covary (Figure 5.1A). However, if X and Y are coupled via non-linear dynamics, the relationship between X and Y can form complex patterns which will not be captured with linear methods (Figure 5.1B). In this instance, methods which can reconstruct the non-linear structure of the underlying dynamical system can be used to demonstrate non-linear dependence (Schiff et al., 1996; Takens, 1981).

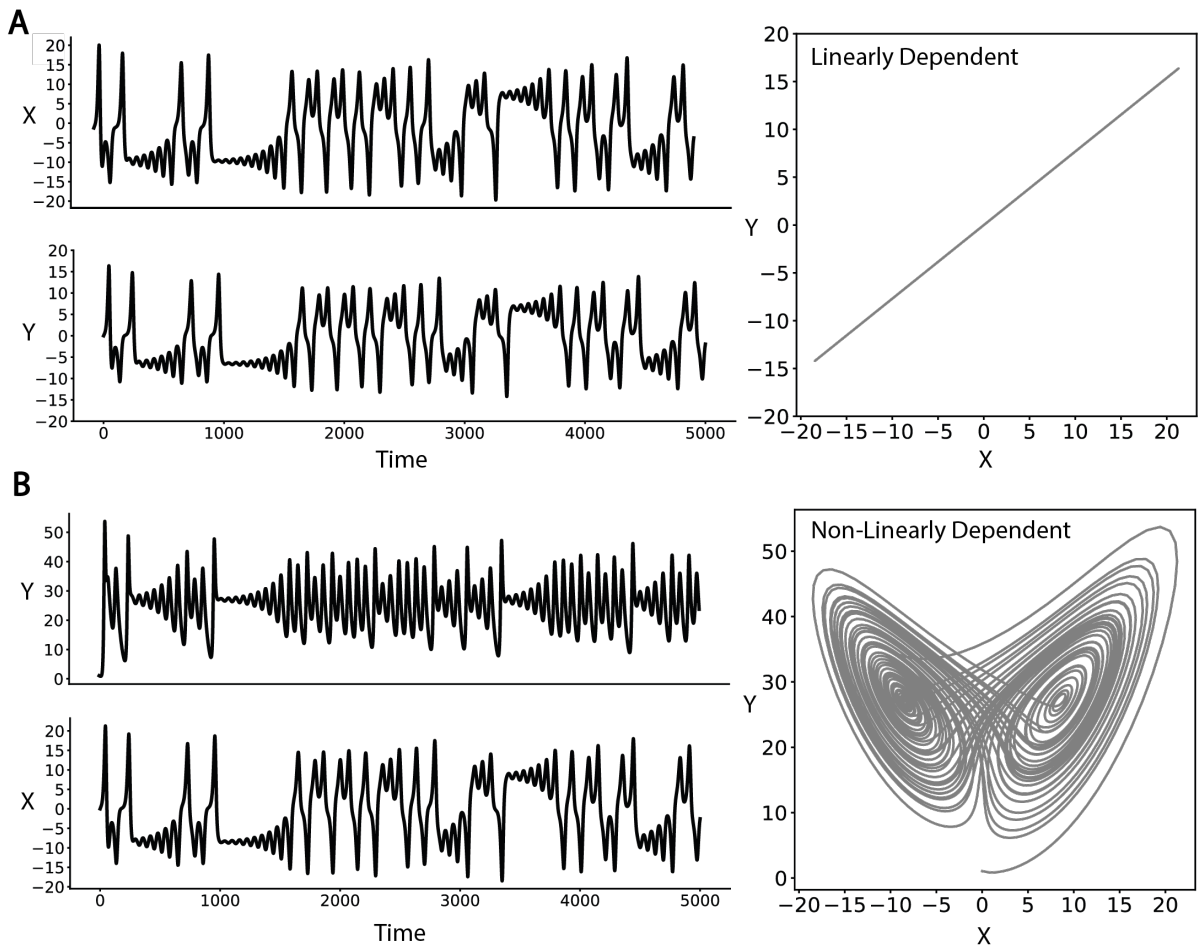

**Figure 5.1    Linear and non-linear dependence.**

(A) Two time series X and Y are correlated in time, and therefore their dependence may be described using correlation which measures the extent to which $X_i$ and $Y_i$ fall onto a line (right). (B) Two time series X and Y representing variables of the Lorenz system are non-linearly interdependent. Therefore the relationship between $X_i$ and $Y_i$ forms a highly complex shape, which requires non-linear methods to describe. Figure is adapted from Le van Quyen et al., 1998.

In order to perform CCM, the first step is to infer the manifold of the system, using a single observable variable of that system. A manifold is any geometric object that is *x* dimensional, but resembles *n* dimensional Euclidean space locally around each point. To infer the manifold of the system, I take advantage of both the Whitney and Takens embedding theorems (Takens, 1981; Whitney, 1936).

The Whitney embedding theorem states that any n-dimensional manifold may be embedded in a 2n+1 dimensional Euclidean space. Here, an embedding of a manifold $M$ onto a manifold $N$ defines a continuous map $f: M \rightarrow N$, where $f$ maps $M$ homeomorphically onto $N$. A homeomorphism is a continuous function between topological spaces, with a continuous inverse – this means that $M$ is continuously deformed to make $N$ such that the map $f$ has no discontinuities (e.g. gaps). Whitney's theorem proves that each point in the n-manifold $M$ maps uniquely to a distinct point in 2n+1 Euclidean space resulting in a completely unfolded manifold $N$ – this property is known as bijection. This implies that given 2n+1 observable variables of a system, it is possible to uniquely identify the state of the original n-manifold $M$, enabling reconstructing of the systems dynamics in state space.

The Takens embedding theorem went one step further to demonstrate that instead of using 2n+1 measurements, one can use time delayed values of a single observable variable to reconstruct the manifold $M$. Takens' theorem states that for a given time series $x = \{x(1), \ldots, x(n)\}$, one can use $E$ time delayed values of $x$ as dimensions in which to embed $x$ in an $E$ dimensional state space whose points construct a shadow manifold $M_x$ – a reconstruction of the original manifold M. Each point in $M_x$ is an $E$ dimensional vector $M_x(t)$, which contains $M_x(t) = \{x(t), x(t - \tau), x(t - 2\tau), \ldots, x(t - (E - 1)\tau)\}$, where $\tau$ is a positive time delay (Figure 5.2A). Here, $x$ is an observation function that maps points in $M$ to a real-valued scalar, and can be thought of as our measurement. Takens showed that any points $M_x(t)$ on the shadow manifold $M_x$ will map one to one with points $M(t)$ on the original manifold $M$ as a bijection, such that $M_x$ is a diffeomorphism of $M$ (formed by both continuous and differentiable deformations). Therefore, the reconstructed shadow manifold $M_x$ is topologically equivalent to the original manifold $M$ which contains all fundamental variables and trajectories of the full system. For example, delay embedding of a single variable of the Lorenz system constructs a shadow manifold that is locally equivalent to the original 3 dimensional system (Figure 5.2A).

In order to ensure a valid embedding of the shadow manifold $M_x$ one needs to estimate the correct embedding dimension $E$. To do this, one can use simplex projection which tests that the embedding is sufficiently unfolded such that attractor trajectories do not cross over (Sugihara & May, 1990). Briefly, simplex projection works by using nearest neighbours to a given point in the manifold, to predict where that point will end up in the future (Figure 5.2B). Here the logic is that an unfolded manifold will have higher predictability into the future, as the relative position of nearest neighbours to a given point should be conserved for longer.

Simplex projection works by the following steps:

(1) *Data split:* The time series of interest *x,* is split into library and prediction segments. Here *x* is split into equal segments.

*(2) Embedding:* Library and prediction segments are embedded using Takens' delay embedding approach, for a given *E* and *τ* (Figure 5.2A).

*(3) K-nearest neighbour search:* For every point *Y(t)* in the prediction manifold *Y*, containing *Y(t) = {x(t), x(t - τ), x(t - 2τ), ..., x(t - (E - 1)τ)}*, locate the *E* + 1 nearest neighbours in the library manifold *X*. These neighbour points in *X* form a simplex which encloses the point *Y(t)* in the state space. These neighbours are referred to as *X(n₁), X(n₂), …, X(nₑ₊₁)*, where *n* is the set of time indices defining the nearest neighbours (Figure 5.2B).

*(4) Prediction:* The prediction of a point *Y(t₀), k* time steps into the future, *Y(tₖ)*, is given by *Z(tₖ)* which is calculated as a linear combination of the exponentially weighted distances between nearest neighbours *X(n₁), X(n₂), …, X(nₑ₊₁)* and *Y(t)* at *t₀* , multiplied by the positions of the neighbours in the future *X(n₁+k), X(n₂+k), …, X(nₑ₊₁+k)* (Figure 5.2B). Specifically

$$Z(t_k) = \sum_{i=1}^{E+1} \frac{w_i}{\sum_{i=1}^{E+1} w_i} \ X(n_i + k), \quad (5.1)$$

where

$$w_i = \exp\left(-\frac{\|Y(t_0) - X(n_i)\|}{\min_{1 \le i \le E} \|Y(t_0) - X(n_i)\|}\right). \quad (5.2)$$

(5) *Assessment:* For a given time step *k*, step 4 is repeated for all points in the prediction manifold *Y*. At this point, the vectors *Z(t)* and *Y(t)* are projected into 1 dimensional space, to make the time series *z* and *y*, which represent the predicted and the observed time series (Figure 5.2C). To assess the performance of the prediction, the Pearson's correlation coefficient is calculated between *z* and *y*.

Steps 1-5 are repeated over a range of embedding dimensions *E* (Figure 5.2C). The *E* for which the correlation is greatest is used as the embedding dimension for Takens' delay embedding (Figure 5.2C). Given the relatively slow sampling rate of my imaging, compared with the underlying calcium dynamics, I assumed that each timestep was maximally independent. Therefore, *τ* was set to 1 for all embedding procedures. For simplex projection, I predicted 1 timestep into the future (k=1).

Once the manifold for each time series is constructed, one can perform CCM. Given two variables *x* and *y*, if they belong to the same dynamical system then they are causally linked – this implies that they will share the same underlying manifold *M* (Sugihara et al., 2012). Therefore, delay embedded attractors *X* and *Y*, generated from time series *x* and *y* respectively, will be topologically equivalent if they are causally linked. This means that points in *X* can be used to predict points in *Y*, and vice versa. Importantly, if variable *x* causes *y*, but the converse is not true, then one can use points in *Y* to predict *X*, but not vice versa. This is because in a deterministic dynamical system, only *x* causing *y* will mean that information about the states of *x* are present in the time series for *y*, but not vice versa (Sugihara et al., 2012). This enables the inference of directional causality across variables. CCM thus works by cross mapping from one manifold to another, to test the extent to which information about one variable is found within another. CCM uses a similar approach to simplex projection, by using nearest neighbours on one manifold to predict points on another, and thus measures how well local neighbourhoods are conserved across manifolds.

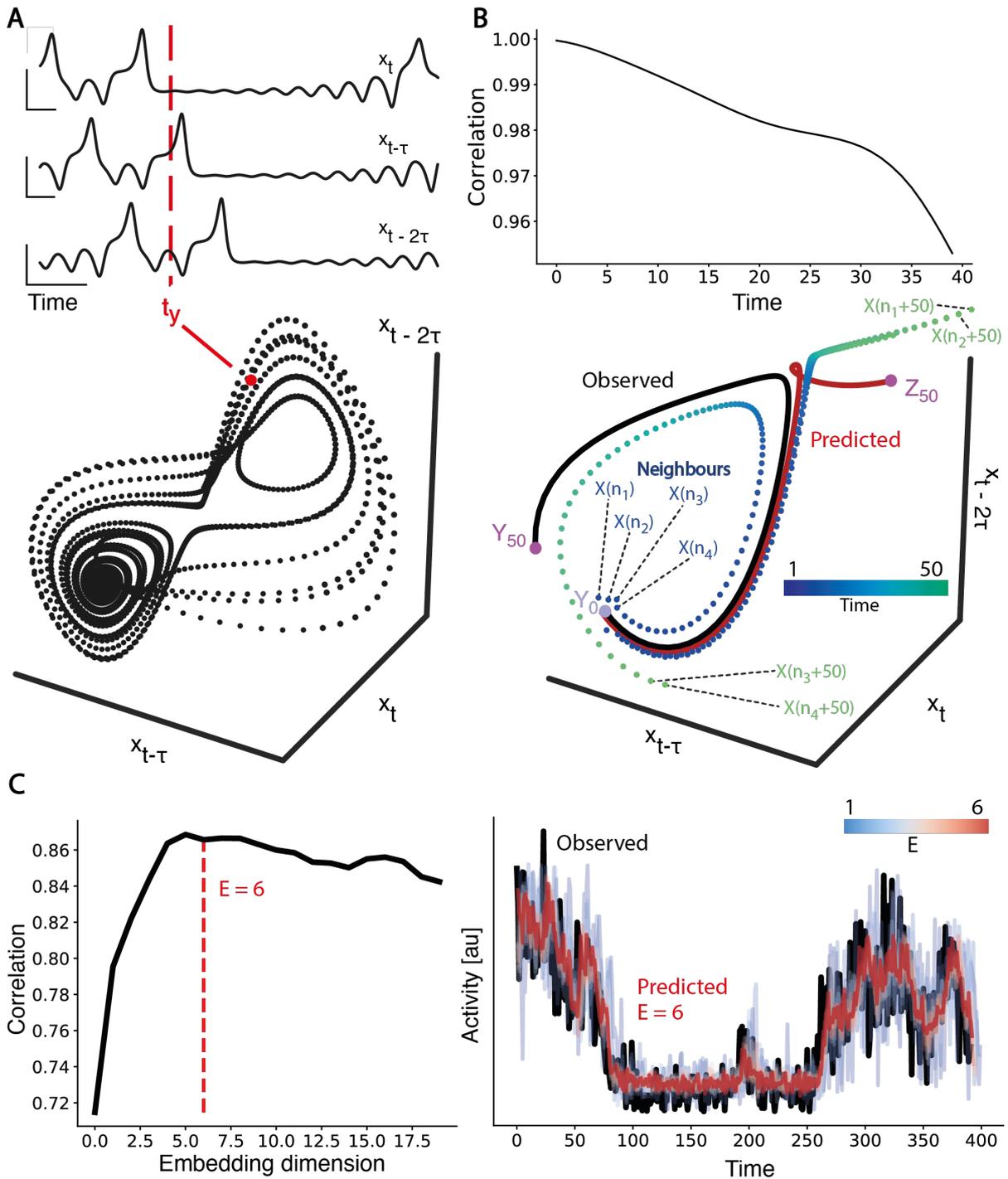

**Figure 5. 2  Lagged coordinate embedding and simplex projection.**

(A) Schematic demonstrating the lagged coordinate embedding approach for system reconstruction. (A, top) A single variable (x) can be used to reconstruct a shadow manifold that is topologically equivalent to the original manifold, by using a series of delayed variables (x, $x_{t-\tau}$, $x_{t-2\tau}$ ..., $x_{t-(E-1)\tau}$) of delay $\tau$ and dimension E. (A, bottom left) Embedding each lagged variable into state space provides the reconstructed manifold, where $t_y$ is the position in state space at time y corresponding to the dotted red line in the above panel. Using a single lagged variable reconstructs the local structure of the original Lorenz attractor. (B, bottom) Simplex projection predicts the location of a future point Y(k) on the manifold Y, by using the distance between Y(0) and its nearest neighbours $X(n_{1-4})$, and the positions of the same points k time steps into the future, given by $X(n_{1-4}+k)$. As I predict further into the future the prediction gets worse as the local structure of the manifold changes such that nearest neighbours at $t_0$ have changed their positions relative to our point of interest Y(k). This means that the correlation between observed and predicted trajectories decreases further into the future (B, top). (C) Manifolds that have been unfolded should have improved predictions into the future. This property is used to find the correct E. As E increases, the correlation reaches a maximum point around 6 (left). This is visualised as the reconstructed time series for a given neuron, which converges closer to the observed values around E = 6 (red).

CCM works by the following steps:

(1) *Embedding:* The time series *x* and *y* are embedded using Takens' delay embedding approach, for a given *E* and *τ* (Figure 5.3A). Note that *E* and *τ* may be different across each time series, depending on the dimensionality of each system. This process creates a prediction manifold *Y*, which refers to the potentially causative variable, and a library manifold *X*, referring to the caused variable (Figure 5.3A).

(2) *K-nearest neighbour search:* For every point on the library manifold *X(t)*, containing *X(t) = {x(t), x(t - τ), x(t - 2τ), ..., x(t - (E - 1)τ)}*, locate its *E* + 1 nearest neighbours. These neighbour points in *X* form a simplex which encloses the point *X(t)* in the state space. These neighbours are referred to as $X(n_1), X(n_2), ..., X(n_{E+1})$, where *n* is the set of time indices defining the nearest neighbours (Figure 5.3A).

(3) *Prediction:* The prediction of a point *Y(t),* given by *Z(t)*, is calculated as a linear combination of the exponentially weighted distances between nearest neighbours $X(n_1), X(n_2), ..., X(n_{E+1})$ and *X(t)*, multiplied by the positions of the corresponding neighbour points in Y: Y$(n_1), Y(n_2), ..., Y(n_{E+1})$ (Figure 5.3A). Specifically

$$Z(t) = \sum_{i=1}^{E+1} \frac{w_i}{\sum_{i=1}^{E+1} w_i} \; Y(n_i), \quad (5.3)$$

where

$$w_i = \exp\left(-\frac{\|X(t) - X(n_i)\|}{\min_{1 \le i \le E} \|X(t) - X(n_i)\|}\right). \quad (5.4)$$

(4) *Assessment:* Step 3 is repeated for all points in the prediction manifold *Y*. Vectors *Z(t)* and *Y(t)* are projected into 1 dimensional space, to make the time series *z* and *y*, which represent the predicted and the observed values. To assess the performance of the prediction, the Pearson's correlation coefficient is calculated between *z* and *y*.

If y causes x, then one should be able to use the neighbouring points on the manifold X to predict Y. Importantly, as the length of the library gets longer, and the manifold fills in with more neighbouring points, the prediction performance should monotonically increase (Sugihara et al., 2012) (Figure 5.3B). Therefore, a key property of causality is not only a high predictability but also a convergence of prediction performance as more neighbouring points are added.

### 5.2.2  Causal Seizure Networks

Firstly, I investigated the network interactions that give rise to focal and generalised seizure dynamics. While graph-based representations have been constructed of the larval zebrafish brain during seizures (Diaz Verdugo et al., 2019b; J. Liu & Baraban, 2019), these have all relied on correlation and thus are unidirectional and linear. Therefore, I supposed that using CCM to infer the causal, directional and non-linear interdependencies driving seizure dynamics, might provide novel insight into the specific regional interactions driving network instability.

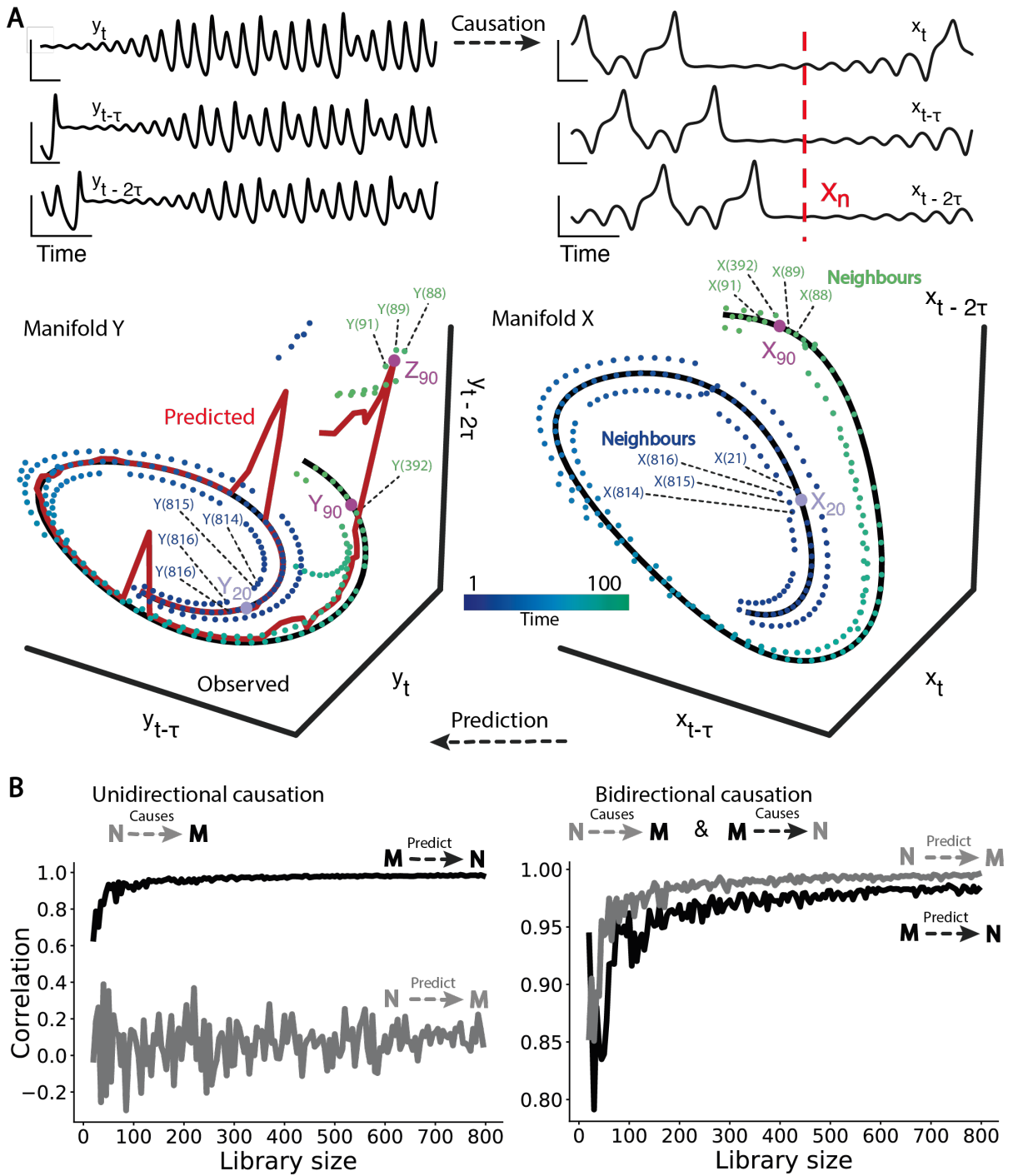

**Figure 5. 3   Convergent cross mapping.**

(A, top) The time series x (right) and y (left) are embedded using Takens' delay embedding approach, for a given E and τ. (A, bottom) Embedding each lagged variable into state space embeds the manifolds X (right) and Y (left), where Xn is the position in state space at time n for manifold X. To test if variable y causes x, I find the nearest neighbours to each point X(n) on the library manifold X (right), and locate the corresponding points by their time indeces on the prediction manifold Y (left). The prediction of Y(n), given by Z(n) is given by a linear combination of the exponentially weighted distances between X(n) and its nearest neighbours, multiplied by the positions of the corresponding neighbours in Y. At timepoint 20, the relative positions of the nearest neighbours to X(20): {X(21), X(814), X(815), X(816)}, and the relative positions of the set {Y(21), Y(814), Y(815), Y(816)} and the prediction point Y(20) are relatively conserved, meaning that the prediction Z(20) (point not shown) is very close to the observed Y(20) (left).  However, at timepoint 90, the relative positions of the nearest neighbours to X(90): {X(88), X(89), X(91), X(392)}, and the relative positions of the set {Y(88), Y(89), Y(91), Y(392)} and the prediction point Y(20) are poorly conserved, meaning that the prediction Z(90) is far from the observed Y(90) (left). If y causes x, then the prediction (red line) and observed (black line) values for the manifold Y should closely overlap, indicating information about y is found in the manifold for x. (B) To confirm causation, one needs to demonstrate that the correlation between predicted and observed values increases as more points are added into the state space. In the case of unidirectional causation, one can see that using variable m to predict n shows increased correlation as the library size increases, meaning that n causes m (left). However, the correlation does not increase when using variable n to predict m, meaning that m does not cause.

To this end, I calculated the average strength of causal interactions across neuron pairs, between each brain area. Specifically, I calculated the mean causal influence between two brain regions *i and j* for a given fish as

$$\frac{1}{n} \sum_{i=1}^{n} E_{ij}, \quad (5.5)$$

where *n* is the number of neuron pairs between region *i and j*, and $E_{ij}$ is the CCM predictability calculated when cross mapping from *i* to *j*. $E_{ij}$ is calculated as the Pearson's correlation coefficient between CCM predicted (*p*) and empirically observed (*o*) time series for *i*, as

$$E_{ij} = \frac{\sum (p_t - \bar{p})(o_t - \bar{o})}{\sigma p \ \sigma o}. \quad (5.6)$$

Thus equation 5.5 gives the mean effect of brain region $j$ to region $i$.

Using this approach, I first calculated the causal interactions across coarse brain areas to understand the macroscale network changes giving rise to focal and generalised seizures. For this analysis, I segmented the brain into 4 bilateral macroscale regions (forebrain, diencephalon, midbrain and hindbrain) using automated registration and atlas labelling techniques (see Section 2.2.7). During spontaneous activity, macroscale interactions involving the forebrain were particularly strong – connections within the forebrain, recurrent connections across forebrain and diencephalon, and unidirectional inputs onto the forebrain from the midbrain and hindbrain, showed particularly high interdependencies (CCM predictability > 0.018) (Figure 5.4A). This indicates a key role for the forebrain as an input area from neighbouring and distal regions. During focal seizures the mean CCM predictability increased across all regions, although none survived Bonferroni corrections for multiple comparisons (Figure 5.4A). However, generalised seizure networks exhibited significantly increased CCM predictability across multiple regional interactions. Specifically, within connections in the forebrain, diencephalon and midbrain, alongside recurrent interactions between the forebrain, diencephalon and midbrain all survived Bonferroni corrections ($p < 0.003$) (Figure 5.4A,B). This suggests that generalised seizures occur due to the emergence of recurrent interactions between forebrain, diencephalon and midbrain areas.

Next, I hypothesised that different brain areas might play specialised roles in sending outputs or receiving inputs during seizures. Given that only generalised seizures exhibited significant changes in network structure, I refined my analysis to generalised seizure dynamics. Here, I parcellated the brain into 111 granular structures to ascertain a more detailed view of the network changes driving such activity (see Section 2.2.7). Using equation 5.5, I was able to calculate the region-region interactions that significantly changed during the generalised seizures. I defined brain areas that showed a large number of significantly increased outgoing interactions as *out hubs* (>58), and those that that exhibited a large number of incoming interactions as *in hubs* (>58) (Figure 5.5). Given the large number of comparisons (12,321), I used the Benjamini Hochberg false discovery rate to test for

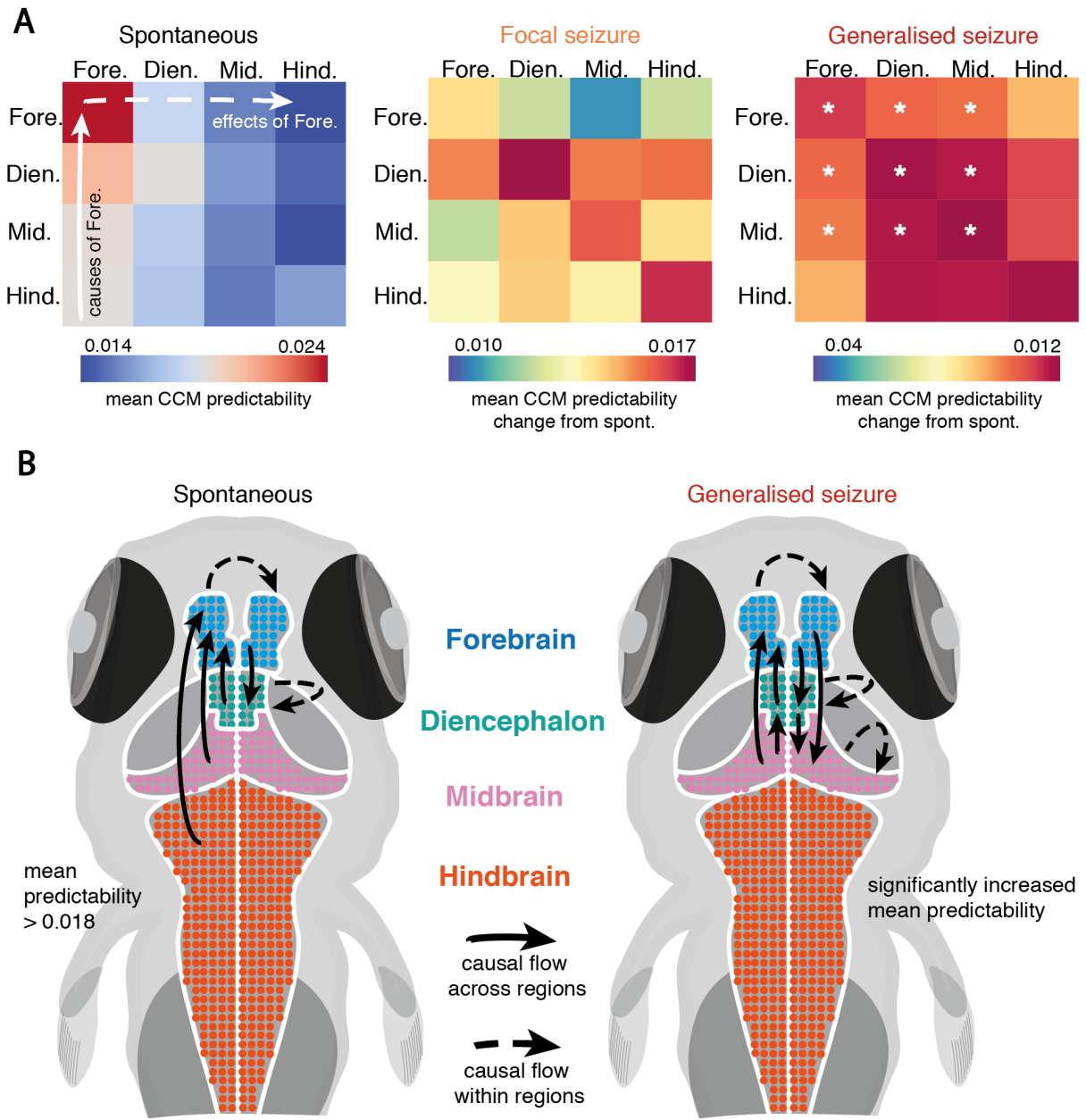

**Figure 5.4  Causal seizure networks.**

(A) Heatmap representing the mean CCM predictability across coarse brain areas for spontaneous activity (left). Cells in column x represent the strength of the cause that a given region has onto region x. Cells in row x represent the strength of the cause that the region x has on each region. Mean CCM predictability changes from spontaneous shows non-significant increases across network dependencies in focal seizures (middle), and significant, recurrent dependencies across forebrain, diencephalon and midbrain regions (right). (B) Schematic representing the flow of causal influence across the network in the spontaneous case (left), and in the seizure case (right) for significant connections. * = p < 0.003.

significance and correct for multiple comparisons. Interestingly, both in and out hubs were widely distributed throughout forebrain, diencephalon, midbrain and hindbrain regions (Figure 5.5A,B). This indicates that PTZ-induced generalised seizures emerge from widespread mesoscopic connectivity changes throughout much of the brain, with most in hubs also acting out hubs. However, by subtracting the number of significantly increased outgoing versus incoming connections, I was able to identify brain areas that acted preferentially as in or out hubs (Figure 5.5C). This identified several key structures as selective in hubs, including the inferior raphe, rhombomere 3,4 and statoacoustic ganglion, and selective out hubs, such as the mesencephalon, rhombomere 3,4, superior raphe and vestibulospinal neurons. Therefore, several diencephalic, midbrain and hindbrain structures exhibit distinct roles as in and out hubs during generalised seizures.

Finally, I was interested in understanding the variability of brain network interactions driving seizure dynamics. To calculate brain network variability, I calculated the mean strength of causal interactions between all brain regions using 111 mesoscopic brain areas. This resulted in a network dependency vector for each fish, which was 12,321 elements in length, with each element corresponding to a mean interaction strength between two areas. In order to calculate network variability for a given condition, I calculated the mean within group Euclidean distance for a given fish as

$$\frac{1}{n-1} \sum_{j=1}^{n-1} \|i - j\|, \quad (5.7)$$

where $i$ and $j$ are the network dependency vectors for fish $i$ and $j$ respectively, and $n$ is the number of datapoints in that condition. Interestingly, I found that both focal (within group Euclidean distance: 5.31 ± 0.37, U = 0.0, p < 0.001) and generalised seizure networks (within group Euclidean distance: 29.5 ± 1.76, U = 0.0, p < 0.01) exhibited significantly more variable network structures than spontaneous networks (within group Euclidean distance: 3.62 ± 0.17) (Figure 5.6A). This indicates that seizure networks, in particular generalised seizure networks, are highly diverse and engage distinct networks across different fish (Figure 5.6C).

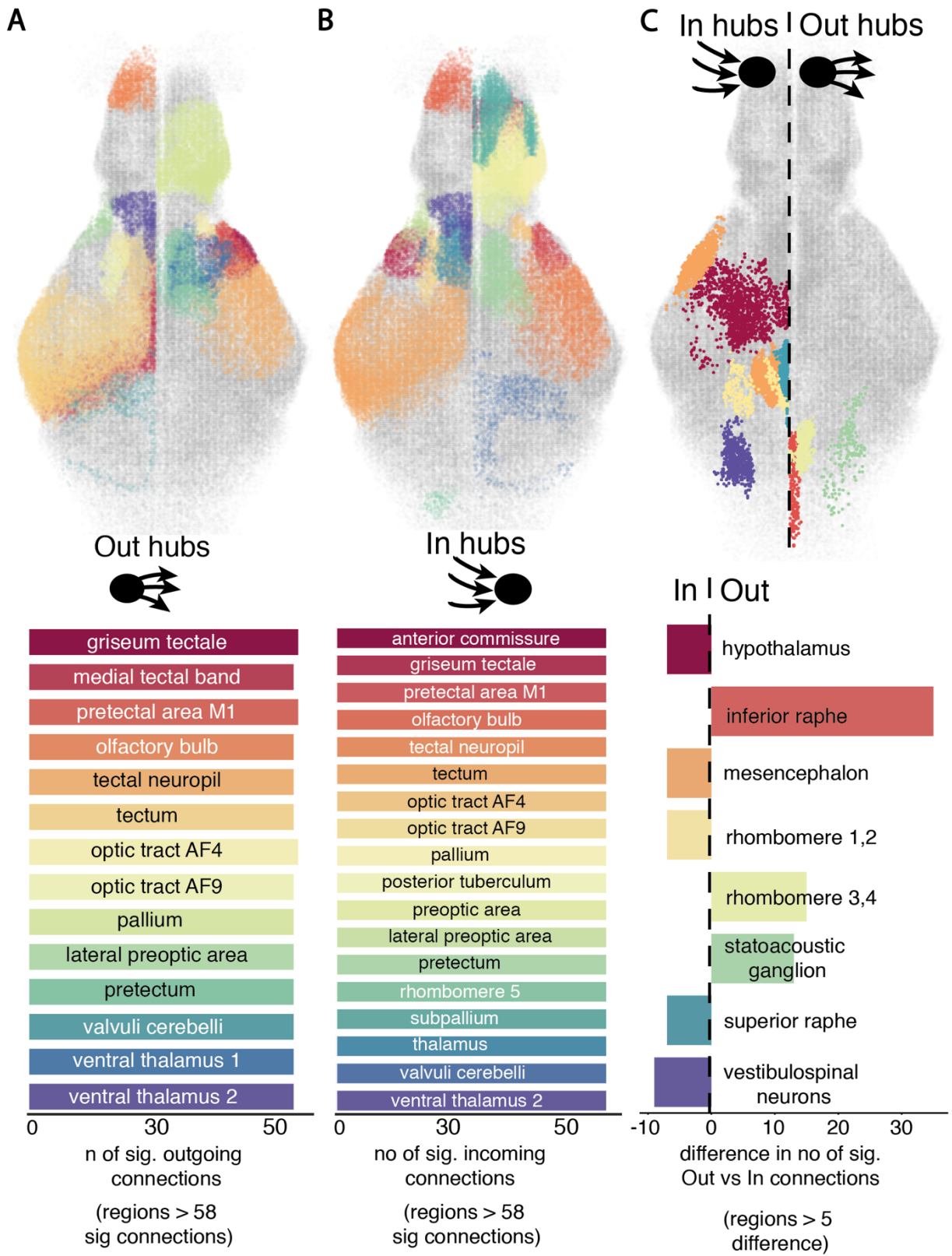

**Figure 5. 5  In and out hubs driving generalised seizures.**

Brain maps visualising the regions with significantly increased CCM predictability for outputs onto other brain regions (out hubs, A) and inputs from other regions (in hubs, B), with corresponding labels below. Significant regions are shown dispersed across right and left hemisphere for visual clarity – regions are separated randomly. (C) Brain map showing regions that have more significant incoming vs outgoing connections.

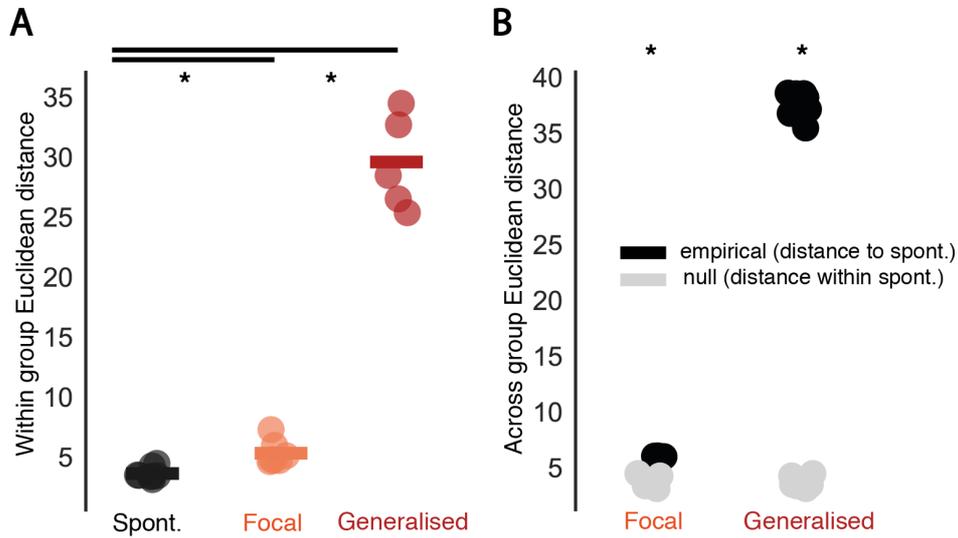

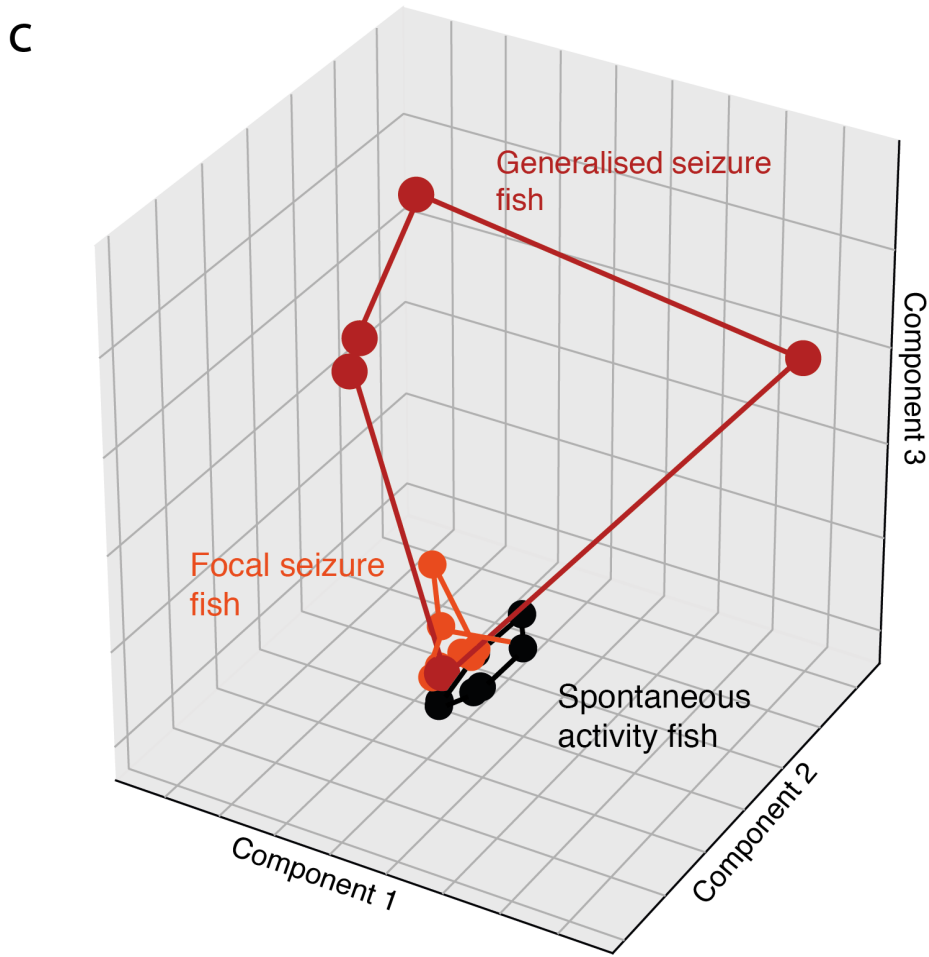

**Figure 5. 6   Seizure networks are highly diverse.**

(A) The Euclidean distance of mean regional network dependency vectors from each fish within a group, to show the diversity of network structures across spontaneous (black), focal (orange) and generalised seizure (red) conditions. (B) The Euclidean distance of mearegional network dependency vectors from each fish for a given seizure condition (focal or generalised) to each spontaneous fish vector (empirical black). This is compared with a null condition, which captures the within group distance for spontaneous fish. * =  p < 0.05. (C) Isomap embedding in 3 dimensions of mean regional network dependency vectors for each condition, illustrating high variance for generalised seizure networks.

*5.2.3  Non-Linear Seizure Neuron Identity & Dynamics*

Next, I wanted to investigate the extent to which seizure dynamics may be described as linear or non-linear. To calculate the linearity of brain dynamics for a given fish, I calculated the mean pairwise correlation between all neuron pairs as

$$\frac{1}{\text{n}} \sum_{i=1}^{n} R_i, \quad (5.7)$$

where $n$ is the number of neurons and $R_i$ is the mean pairwise Pearson's correlation between neuron $i$ and all other neurons in the network. $R_i$ is then calculated as

$$R_i = \frac{1}{\text{n}-1} \sum_{j=1}^{n-1} r_{ij}, \quad (5.8)$$

where $j$ is the number of other neurons in the population, and $r_{ij}$ is the Pearson's correlation between neuron $i$ and $j$ given by

$$r_{ij} = \frac{\sum (i_t - \bar{\imath})(j_t - \bar{\jmath})}{\sigma i \; \sigma j}. \quad (5.9)$$

I found that during generalised seizures, the mean correlation significantly increased compared with spontaneous activity (*Spontaneous:* 0.01 ± 0.00, *Generalised:* 0.32 ± 0.08, U = 1.0, p < 0.01), indicating increased linearity during seizures (Figure 5.7A).

To calculate the non-linearity of brain dynamics, I calculated the mean improved predictability across a given neuron and its neighbours, when using CCM compared with correlation. Specifically, the mean non-linearity for a given fish was defined as

$$\frac{1}{n^2} \sum_{i=1}^{n} \sum_{j=1}^{n} E_{ij} - R_{ij}, \qquad (5.10)$$

where *n* is the number of neurons, $E_{ij}$ is the correlation between observed and predicted time series from neuron *i* and *j* using cross map procedures, and $R_{ij}$ is the correlation between the raw time series for neuron *i* and *j*. The Fisher transformation was applied to both $E_{ij}$ *and* $R_{ij}$ before subtraction in order to normalise the underlying Pearson's r sampling distributions before comparison. Thus equation 5.10 captures the improved predictability across two time series, when using the non-linear structure of the manifold, over linear approaches, which is taken as an average over all neurons pairs for that fish.

Interestingly, the mean non-linearity also significantly increased during generalised seizures (*Spontaneous:* 0.020 ± 0.006, *Generalised:* 0.129 ± 0.015, U = 0.0, p < 0.01) (Figure 5.7B). Using Cohen's d to estimate the effect size, I found that the increase in non-linearity during generalised seizures was ~4.5 standard deviations suggesting a large effect size (d = 4.51). Therefore, seizures at the single cell level emerge through large increases in both linear and non-linear interdependencies.

Next, I wanted to probe whether the same neuronal populations driving linear behaviours were also driving non-linear behaviours. Here, I split up my data into 3 groups: *Linear neurons*: the top 5% most correlated neurons, *Non-linear integrators:*

the neurons with the top 5% mean non-linear inputs*, Non-linear senders:* the neurons with the top 5% mean non-linear outputs. This allowed me to identify the % overlap that each non-linear population had with the highly linear neurons. Interestingly,

during spontaneous activity non-linear integrators highly overlap with linear neurons

(*% overlap*: 51.23 ± 6.69), while non-linear senders showed little overlap (*% overlap*:

5.23 ± 2.25) (Figure 5.7C). This indicates that the neurons receiving the most non-

linear inputs are also engaging in highly linear behaviour. Remarkably, during

generalised seizures, the non-linear integrators became a separate population to

linear neurons (*% overlap*: 8.48 ± 3.66), while non-linear senders remained separate

(*% overlap*: 0.0 ± 0.0) (Figure 5.7C).  Therefore, highly linear and non-linear dynamics during generalised seizures are driven by separate neuronal subpopulations.

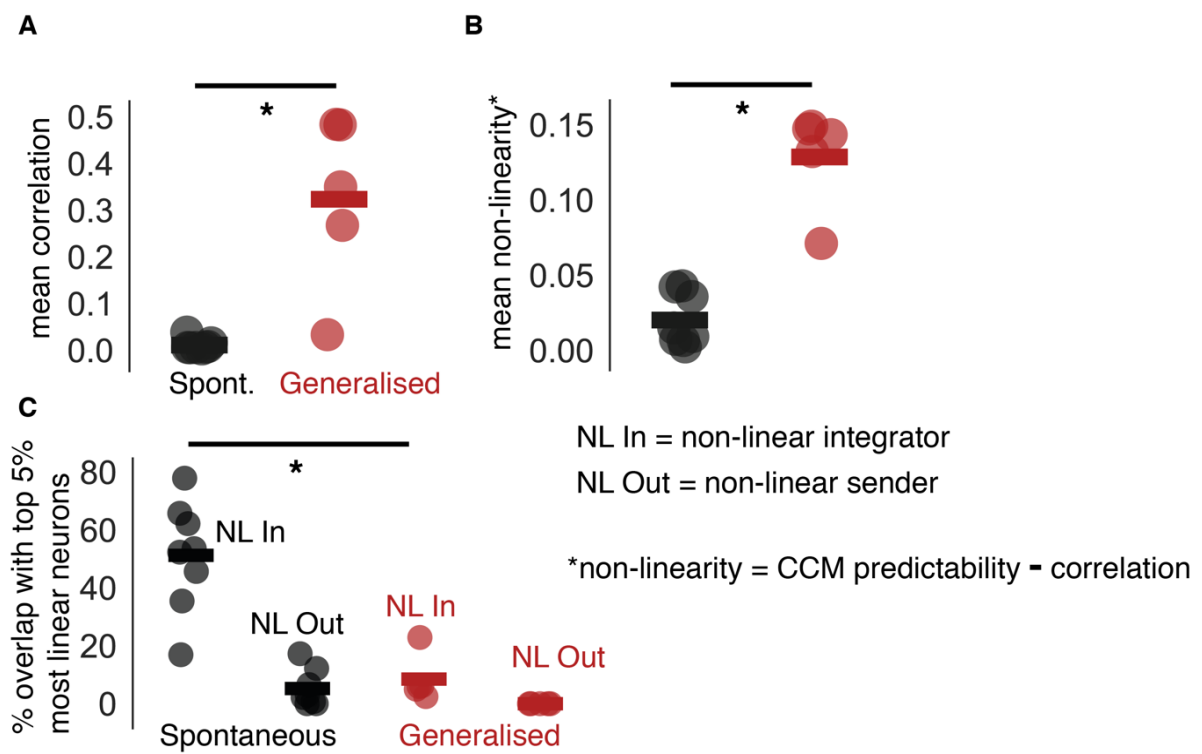

**Figure 5. 7   Seizure dynamics are characterised by increases in both linear
and non-linear dependencies.**

(A) Mean pairwise correlation across all neurons for a given fish, shows increased correlation

and thus linear behaviour during generalised seizures. (B) Mean non-linearity across all neurons for a given fish, calculated for each neuron-neuron connection as the difference between the CCM predictability and the correlation, shows increased non-linear behaviour during generalised seizures. (C) The % overlap for the top 5% non-linear neurons and the top 5% most linear neurons for a given fish, shows that in spontaneous dynamics non-linear integrators drive the linear behaviour, while during seizures non-linear neurons are separate populations to linear neurons. Non-linear integrator = those that are highly non-linear when considering the incoming inputs that neuron receives. Non-linear senders = those that are highly non-linear when considering the outgoing signals a neuron sends. * = $p < 0.05$.

Given that non-linear dynamics are highly prevalent during seizures, I wanted to further characterise the spatial and dynamical components of non-linear integrators and non-linear senders. I was particularly interested in non-linear integrators and senders as these two populations represent the neurons that integrate signals or send out signals yet do not correlate with their partners – therefore they are key network components that play a potentially novel role in signal propagation during seizures. Firstly, I calculated the percentage overlap of non-linear integrators and senders across datasets. Interestingly, in spontaneous and generalised seizure networks, non-linear integrators and senders showed very little overlap, while focal seizure networks showed partial overlap (*Spontaneous,* % overlap: 3.60 ± 1.17, *Focal,* % overlap*:* 14.60 ± 1.94, *Generalised,* % overlap*:* 2.20 ± 0.77). This suggests that across seizure and healthy brain dynamics, neurons that send non-linear outputs and receive non-linear inputs represent distinct subpopulations (Figure 5.8A).

Next, I tested to what extent these separate subpopulations engaging in opposing directional non-linear behaviour, exhibited distinct neuronal dynamics. Here, I supposed that identifying differential activity patterns that separate out non-linear senders and integrators may help to distinguish the type of non-linear dynamics that each population engages in. To this end, I performed k means clustering on the generalised seizures traces for a single example dataset (Figure 5.8B). Within these two clustered populations, I calculated the differential preference for non-linear sending and integrating behaviour as the ratio of the mean CCM predictability for outgoing connections divided by the mean CCM predictability for incoming connections for a given neuron. Interestingly, cluster 1 showed a significantly higher propensity for engaging in non-linear sender dynamics than cluster 2 (Cluster 1:

*mean non-linear sender/integrator:* 4.43 ± 3.53, Cluster 2: *mean non-linear sender/integrator:* 3.63 ± 0.52, U = 2724352.0, p < 0.001) (Figure 5.8B). Therefore, there are key features in the activity traces of seizure neurons that partially separate out non-linear integrators and senders, suggesting different dynamics defining each subpopulation. However, I note that both clusters have a mixture of non-linear integrators and senders within them (Figure 5.8B), suggesting that non-linear integrators/senders only partially capture the variance in activity from these two clusters. Nonetheless, the fact that clustering on the dynamics of the population partially segregates non-linear senders and integrators suggests information about these types of non-linear dynamics is present in neuronal traces. Therefore further investigation into the specific differential dynamics of non-linear integrators and senders may uncover the roles of each subpopulation in driving seizure dynamics.

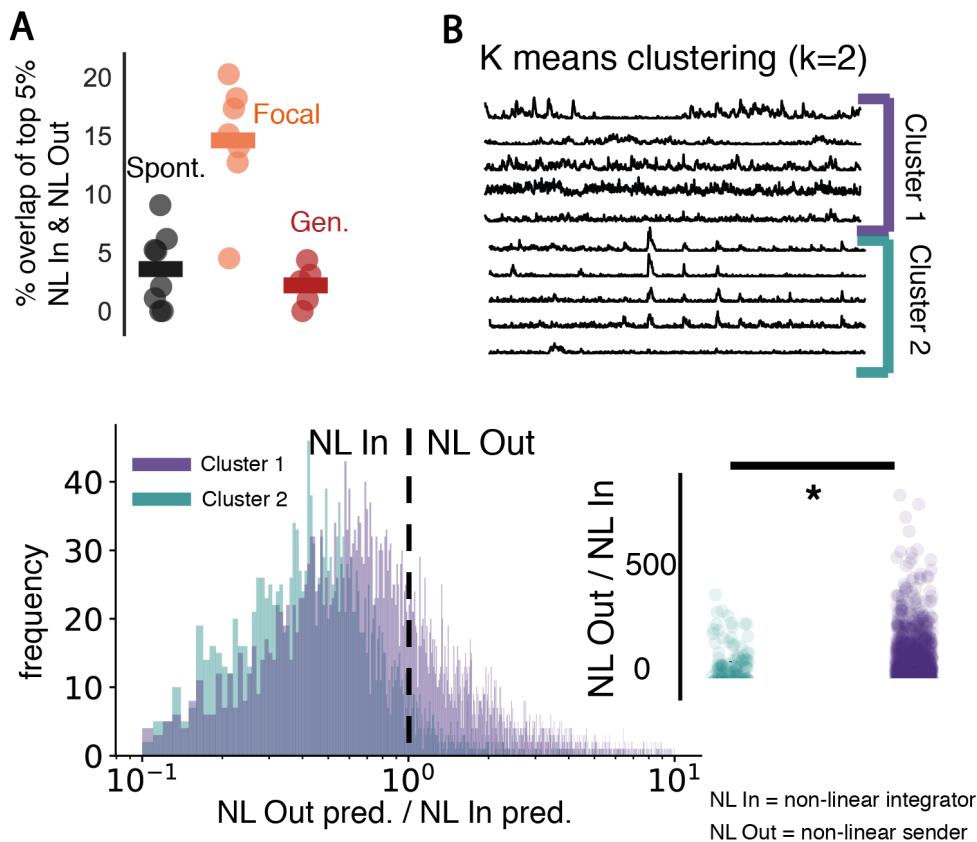

**Figure 5. 8  Non-linear integrators and senders are different populations and exhibit partially non-overlapping dynamics.**

(A) The % overlap of the top 5% non-linear integrators (NL In) and non-linear senders (NL Out) across spontaneous, focal and generalised seizure conditions. In the generalised seizure case, NL In and NL Out populations are distinct. (B) By clustering all traces for an example fish into 2 clusters, one notices that each cluster exhibits a differential preference for NL In and NL Out behaviour as quantified by the ratio of the mean CCM predictability for outgoing connections divided by the mean CCM predictability for incoming connections for a given neuron. Cluster 1 shows a significantly higher propensity for engaging in non-linear sender dynamics than Cluster 2.

### 5.2.4  The Effect of Coarse Graining on Non-Linear Network Interactions

Finally, I investigated whether the prevalence of non-linear interactions at the microscale during seizures might prove useful for seizure prediction and intervention. I supposed that conventional macroscale approaches for seizure detection, which coarse grain the underlying dynamics, would filter out such non-linear information, making it inaccessible. To this end, I spatially coarse grained my brain data for each fish. Specifically, I performed k-means clustering on the spatial locations of all neurons, with k decreasing in a logarithmic manner (Figure 5.9). Within each cluster I then averaged over the neuronal activity of all constituent neurons, before calculating

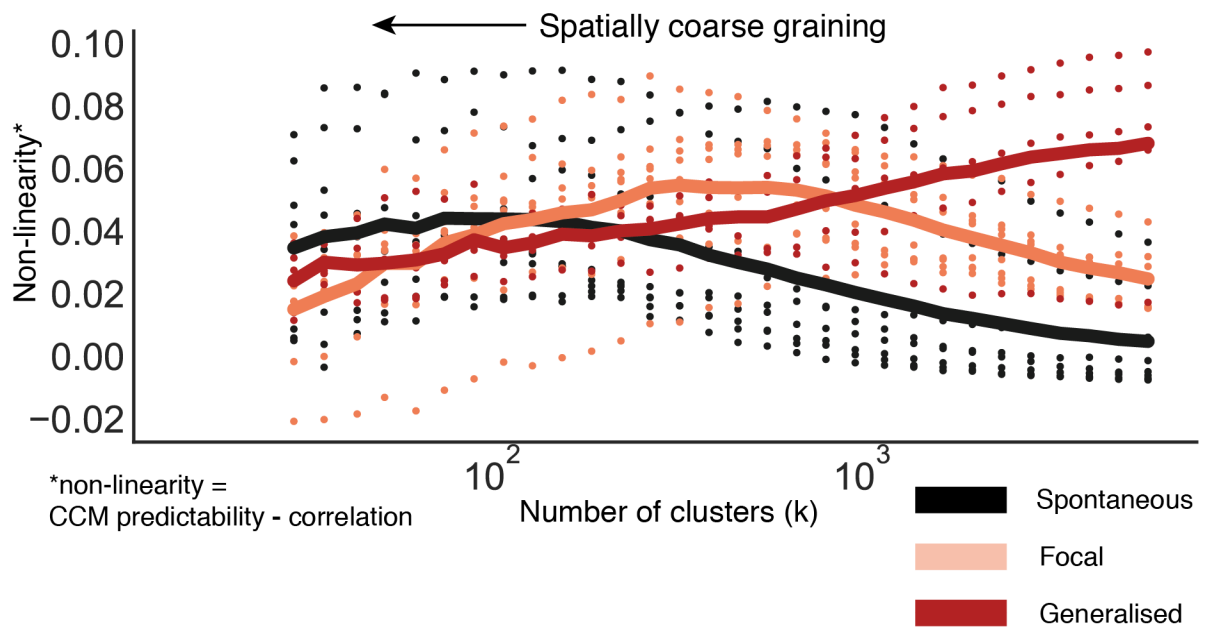

**Figure 5. 9   Spatially coarse graining removes non-linear dependencies between variables in generalised seizures.**

The non-linearity measured at different levels of spatial coarse graining across spontaneous, focal and generalised seizure conditions. To coarse grain, k-means clustering was performed on the spatial locations of neurons in the brain, with k selected in decreasing size. Each dot represents a fish.

the mean non-linear dependencies across cluster traces for each dataset using equation 5.10. This allowed me to assess the effect of spatially coarse graining on the non-linear interdepencies, across spontaneous, focal and generalised seizure dynamics. Interestingly, during generalised seizures spatially coarse graining caused a monotonic reduction in non-linearity (Figure 5.9). This demonstrates that macroscale recordings would cause a loss of information regarding the specific non-linear dependencies between variables. This has implications for applying non-linear techniques to macroscale recordings to identify ictogenic nodes or predicting seizure dynamics, as at this scale much non-linear information is lost. Interestingly, spatially coarse graining focal seizure dynamics caused an initial increase in non-linearity followed by a subsequent reduction. Conversely, spontaneous dynamics exhibited increases in non-linearity due to spatial coarse graining. Taken together, this indicates that seizure dynamics at the microscale hold fundamentally different information about network dependencies than macroscale recordings. Such information could be useful for seizure prediction or the identification of key node dependencies driving seizure dynamics – this finding opens up avenues for the investigation of the utility of such microscale information in predicting seizure before they happen and identifying ictogenic nodes.

## 5.3   Discussion

In this chapter, I performed non-linear causal inference on cellular-resolution whole brain activity data, to understand the microscale non-linear interactions that drive epileptic seizures. I set out to answer two key questions. First, do epileptic seizures emerge as an increase in linear or non-linear dynamics? And secondly, to what extent are non-linear dependencies driving seizures lost at the macroscale? I found that seizure dynamics occur through an increase in both linear and non-linear interactions, demonstrating the heterogeneity of the neuronal dynamics that underlie seizures. Importantly, these non-linear interactions are filtered out at the macroscale, suggesting that such dependencies would not be accessible with conventional macroscale recordings such as EEG. This chapter helps to uncover neuronal interactions that may be central for driving seizure emergence and could be used to inform seizure intervention.

### 5.3.1 Interpreting Non-Linear Dynamics During Epileptic Seizures

This chapter illustrates the importance of both linear and non-linear behaviours in driving epileptic seizures, using the non-linear causal inference technique convergent cross mapping (CCM). Interestingly, while seizures often appear as synchronised, and linear in macroscale recordings (Gotman & Levtova, 1996; Schindler et al., 2007), at the microscale neuronal activity is highly heterogeneous, with synchrony and asynchrony often co-occuring (Truccolo et al., 2011). Explaining such heterogeneity is challenging, but it has been linked to the engagement of spatially diverse neuronal assemblies (Muldoon et al., 2013), the differential engagement of neurons in the epileptogenic zone versus the surrounding area (Weiss et al., 2013), and differential engagement of cell subpopulations within the epileptogenic zone itself (Meyer et al., 2018). One possible explanation for such heterogeneity, is the prevalence of non-linear dependencies in neuronal networks which would give rise to non-trivial, complex dynamics during seizures. However, the nature of such non-linearities, and the relationship between linear and non-linear behaviours during seizures is poorly understood. I build on previous work, by demonstrating that epileptic seizures can be both highly linear and highly non-linear – neuronal networks can engage in both high synchrony and high non-linearity as quantified with CCM.

How do neurons within the network express this synchrony and non-linearity? – for example, it is possible that neurons engaging in high synchrony during seizure onset, might also engage in non-linear interactions at different parts of the seizure. Interestingly, I found that in spontaneous activity the neurons engaging in high synchrony also engaged in non-linear behaviours, by integrating non-linear inputs.

Conversely, during the seizure linear and non-linear neurons became separate subpopulations. This suggests that in the healthy network, synchronous activity is linked to the integration of non-linear inputs, while during seizures this relationship is decoupled. However, one should note that using the top 5% of neurons to define linear vs non-linear neurons in my network is a somewhat arbitrary choice. Future work should look at the overlap between linear and non-linear populations as a function of the % inclusion, to demonstrate such changes are invariant to population size. One potential explanation for a decoupling of linear and non-linear neurons during the seizure, could be that in healthy networks non-linear integrators receive few, strong common inputs, resulting in synchronous behaviour. However, following massively increased inputs from diverse sources onto a given neuron during the seizure (see Section 4.2.2), non-linear integrators may instead engage in noisy, uncorrelated behaviours. Further work is required to understand the exact neuronal source of such linear and non-linear behaviours during epileptic seizures.

There are several key limitations to my approach. Of course, CCM can identify non-linear dynamics, but cannot identify the specific nature of such non-linearities. In the words of Stanislaw Ulam, calling something non-linear is like "referring to the bulk of zoology as the study of non-elephant animals" (Campbell, 2004). Clearly, 'non-linear' is an ambiguous label and the majority of interactions in complex systems are non-linear. More work is therefore required to ascertain more precisely the type of interactions and dynamics that non-linear neurons are engaging in. One approach is separating neurons into non-linear senders and integrators as was done in this study, which can identify the directionality of the non-linear interaction. Another approach could be the use of simple network models, in which specific network parameters are altered to study the link between synaptic architecture and intrinsic neuronal dynamics and emergent non-linear interactions. This could provide a more solid biological foundation to help explain such non-linear alterations. CCM also assumes a constant relationship between different time series' during the recording period, and therefore assumes stationarity. Of course, spontaneous and seizure dynamics can change over short timescales and therefore one should note that neuronal dependencies that change quickly might be missed by the CCM reconstruction technique. Importantly, a key aspect to demonstrate CCM causality is the property of convergence, in which increasing the library size results in improved predictions (Sugihara et al., 2012). This property can separate out causally linked variables from random systems (see Secion 5.2.1). Due to constraints on time in the writing of this thesis, I was unable to test for the convergence property in my data. An important next step therefore is to confirm that non-linear changes and causal network interactions during seizures survive the filtering of non-convergent neuronal interactions.

*5.3.2   Utilising Non-Linear Microscale Dynamics for Seizure Intervention*
This chapter also identifies key network nodes that drive seizure emergence at coarse and granular scales. Importantly, the identification of ictogenic network nodes is important for targeted interventions and restoring normal network function (Goodfellow et al., 2016). At the macroscale, I found that generalised seizures emerged from connections within the forebrain, diencephalon and midbrain, alongside recurrent interactions between the forebrain, diencephalon and midbrain. This suggests a recurrently connected network predominating within more rostral brain areas. This is partially supported by previous work in larval zebrafish PTZ recordings which report midbrain and diencephalic regions as key seizure initiation zones (Niemeyer et al., 2022). Similarly, Rosch et al. used an alternative causal modelling approach, dynamic causal modelling, to show that seizure dynamics emerged from strong recurrent coupling within and onto the midbrain (Rosch et al., 2018). That alternative causal modelling approaches both identify a recurrently connected state during epileptic seizures supports the notion of generalised seizures as a dense, recurrently connected brain network at the macroscale. Importantly, I build on previous work by identifying granular populations of neurons that act as hubs for sending output and receiving input during seizures. For example, the inferior raphe, rhombomere 3,4 and statoacoustic ganglion act selectively as in hubs, while the mesencephalon, rhombomere 3,4, superior raphe and vestibulospinal neurons act selectively as out hubs. This suggests specialised roles for certain brain regions in sending or receiving inputs during generalised seizures that should be further investigated. Importantly, this in and out hub approach could be used to identify targets for network intervention in the larval zebrafish brain, which could be resected through laser ablation or opto/chemo-genetic silencing during or before seizures. This approach could offer a proof of principle for the utility of using non-linear, microscale recordings for identifying ictogenic nodes.

My work shows several key differences to previous studies. In particular, I found that the hindbrain does not significantly change its connection strengths across any brain areas. However, Niemeyer et al. identified the cerebellum as a key seizure initiation zone, while Rosch et al. also found that the within connectivity in the rostral hindbrain could greatly influence seizure dynamics. The reason for this difference could be methodological – given that significance tests in my approach were applied through average changes in CCM predictability across coarse brain areas, such hindbrain changes may not survive averaging. In fact, more granular parcellations indicate that hindbrain connectivity does significantly increase, specifically through the rhombomere 5 and the cerebellum. Therefore, hindbrain alterations during seizures may be spatially localised to granular areas. Furthermore, the above studies captured more of the caudal hindbrain than was possible using my volumetric imaging approach. Therefore, my imaging approach could be missing several key caudal hindbrain areas that could drive seizure emergence.

Finally, I showed that spatially coarse graining removes non-linear dependencies between variables. A similar result has been reported using *in silico* data (Nozari et al., 2020), but has not been applied to empirical recordings *in vivo* before. This indicates that information regarding the ictogenicity of neuronal network nodes at the microscale, would be lost in macroscale recordings such as EEG. This is unsurprising given that spatial coarse graining averages over neurons in space such that macroscopic dynamics represent averaged activity from heterogeneous sources. Given that microscale targeted interventions in human patients are still not currently possible, one could argue that such microscale information is unimportant. But with the advent of network modelling tools to investigate targeted, microscale interventions (Sanhedrai et al., 2022), it is important to consider the utility of such approaches in theoretical seizure networks. Demonstrating that microscale non-linear dynamics could be used to identify ictogenic network nodes with higher precision and efficacy than coarse brain approaches would illustrate the importance of developing such microscale interventional strategies. Similarly, non-linear dynamics can be used to identify phase transitions, such as seizures, before they happen (Scheffer et al., 2009). Given that microscale dynamics precede EEG abnormalities during seizures (Wenzel, Hamm, et al., 2019), such non-linear information could provide early warning signs to detect seizure before they are visible on macroscale recordings. A next step could therefore be to illustrate the predictive power of such microscale recordings above and beyond macroscale approaches, to demonstrate the utility of recording from microscale activity for seizure prediction.

To summarise, I find that i) seizure dynamics occur through an increase in both linear and non-linear interactions, and ii) these non-linear interactions are filtered out at the macroscale. Such information could be useful for seizure prediction or the identification of key node dependencies driving seizure dynamics – this finding opens up avenues for the investigation of the utility of such microscale information in seizure intervention.

# Chapter 6

## Conclusions and Perspective

This chapter contains some text from some of my previously published work (R. Rosch et al., 2019).

### 6.1  Summary

This thesis represents a comprehensive characterisation of the single cell network mechanisms driving whole brain dynamical states in epileptic seizures. The main findings of this work are:

1. Excitation-inhibition (EI) balance shapes critical avalanche dynamics in whole brain networks. Perturbing EI balance *in vivo* causes a loss of critical statistics as quantified through avalanche dynamics, demonstrating that EI balance is sufficient for the presence of criticality in the brain. This suggests that the developmental dysregulation of EI balance could disrupt critical dynamics (Chapter 3).

2. EI imbalance-induced epileptic seizures emerge as a loss of criticality across the whole brain. Using single cell dynamics, I was able to demonstrate that the collective behaviour of single neuron activity causes a loss of whole brain critical statistics. This provides a population mechanism for the emergence of epileptic seizures (Chapter 3).

3. Using neuronal network models fitted to empirical data, I found that locally synchronous seizure dynamics can be caused by non-specific changes to excitability through increases in synaptic weight strength, intrinsic excitability and neuronal connectivity. Conversely, globally synchronous seizures can only emerge in densely connected networks (Chapter 4).

4. Epileptic seizure networks *in vivo* and *in silico* disrupt key network response properties and dynamics expected to be optimal in critical systems. Specifically, seizures caused the emergence of chaotic dynamics which would impair the memory of system inputs over time. Seizures also disrupted the ability of the network to represent a wide range of inputs and discriminate between similar inputs. Finally, seizures caused a disruption in flexible dynamics, which emerges due to increased multidimensional correlation (Chapter 4).

5. Using non-linear causal inference, I showed that single cell dynamics during seizures emerge as an increase in synchrony and an increase in non-linear behaviours. Interestingly, linear and non-linear neurons represent distinct subpopulations during seizures. Importantly, this non-linear behaviour is filtered out by spatially coarse graining, suggesting that conventional macroscale recordings would miss such non-linear interdependencies (Chapter 5).

## 6.2   Bridging Scales

A major utility of the larval zebrafish in the study of seizure dynamics, is the ability to concurrently record from neuronal activity at multiple scales – from single neurons, to neuronal populations, to entire brain regions (Ahrens et al., 2013). However, making use of such multi-scale dynamics is non-trivial. In this work, I have used several approaches. Firstly, I made use of statistical physics methods which probabilistically describe the collective behaviour of individual units, to estimate the macroscopic behaviour of the brain. Such methods can relate the statistical behaviour of microscopic components with emergent properties at higher scales – in this case neuronal activity with avalanche dynamics and phase transitions. Here, I chose to make use of the avalanches framework because of evidence supporting critical brain dynamics in healthy networks (Beggs & Plenz, 2003), and a loss of critical dynamics in pathological networks (Hobbs et al., 2010). Demonstrating that epileptic seizures emerge as a loss of critical statistics provides a multi-scale description of the collective behaviour of individual neurons during seizures – one characterised by transitions from critical to supercritical dynamics. Importantly, given the potential to perturb any network node in the larval zebrafish brain *in vivo* using optogenetics, future work can unpick the exact relationship between synaptic plasticity mechanisms and the phase transition. For example, one could directly perturb single neurons, or neuronal subpopulations, or artificially entrain rhythmic activity to induce Hebbian plasticity, to study how microscale properties alter avalanche dynamics in real time. Such approaches would be useful to understand the vicinity that healthy dynamics reside to the critical point – some have theorised that safe brain dynamics are slightly subcritical (Wilting et al., 2018). This would suggest that healthy networks can actually be pushed closer to criticality, before transitioning to supercritical dynamics, while dysregulated networks might actually sit closer to criticality. Future investigations should work to directly modulate critical dynamics with targeted microscale perturbations, to study the link between microscale activity and emergent avalanche dynamics in healthy and pathological brain networks.

The second approach I used to make sense of multi-scale dynamics in seizures was network models. Here, I used both generative models of single neuron seizure dynamics, and data-driven causal inference models to infer network connectivity between neurons. Generative models are particularly useful for testing the role of specific network mechanisms in generating seizure data. Conversely, data-driven models can provide more empirically-constrained descriptions of seizure networks which can be useful for identifying exact ictogenic nodes or regions. While, epilepsy and epileptic seizures may emerge from neuronal networks, the intrinsic excitability, and synaptic connectivity of the network is itself an emergent property of intracellular protein and RNA expression. Therefore, truly multi-scale approaches to study seizure dynamics should not only rely on the study of network architectures giving rise to epileptic seizures, but the transcriptomic and proteomic dynamics that shape those network architectures. Epilepsy is clearly shaped by genetic alterations to a diversity of molecular processes (Rosch et al., 2019), while seizures themselves induce a plethora of immediate early gene expression changes (Kiessling & Gass, 1993). Such molecular alterations will dynamically change according to developmental programmes, and in response to environmental input, and may therefore provide insight into the molecular cause of network properties that are associated with epileptic seizures.

## 6.3  Translating Findings to Humans

Describing epileptic seizures using the framework of criticality can provide useful predictions about epileptic seizures. If the pre-ictal state is near to criticality, while seizures are supercritical, then seizures could emerge as brain networks get closer to the phase transition point, before bifurcating into supercriticality. If this is the case, then seizure transitions should be preceded by markers of critical transitions such as critical slowing – increasing autocorrelation and variance (Scheffer et al., 2009). In fact, evidence from EEG recordings suggest that signatures of an approaching critical transition occur before the seizure onset (Maturana et al., 2020). Although subsequent work of the pre-ictal state suggests no evidence for signs of critical transitions immediately preceding the seizure (Hagemann et al., 2021), Maturana et al. show evidence for critical transitions over longer timescales than reported by Hagemann et al. In this way, viewing epilepsy through the lens of criticality can help to translate theoretical predictions to patients. Furthermore, viewing seizures as the emergence of supercriticality can help to identify interventional strategies – for example, approaches which render the network more subcritical as a safety mechanism for preventing unwanted state transitions might prove effective. In fact, anti-seizure medication gives rise to subcritical avalanche dynamics in EEG (Meisel, 2020) – this could help to explain the cognitive side effects of anti-seizure medication, alongside their seizure-protective properties. In this way, multi-scale imaging approaches applied to seizure dynamics, can provide strong evidence regarding the dynamical mechanisms driving seizures about a phase transition – this can then be used to inform macroscale recording approaches in clinical settings.

The use of non-linear techniques applied to microscale recordings could be used to identify novel ictogenic network nodes. While having full access to microscale dynamics in epilepsy patients is not possible at present, illustrating the utility of such methods in preventing seizures above and beyond what is possible using coarse grained approaches, such as large tissue resection, can be an important proof of principle for the utility of gaining access to microscale data for network stabilisation. Moving forward, future work should aim to directly test the efficacy of non-linear inference techniques such as CCM in identifying key ictogenic nodes, through the use of optogenetic or laser ablation studies.

Finally, it is worth commenting on the generalisability of my findings to epilepsy. As outlined in Section 1.1, seizures may be a convergent endpoint for brain networks, and as such there may be general principles that define epilepsy as a whole. The hope is that the network and population mechanisms driving drug-induced seizures in larval zebrafish, are the same as those driving spontaneous seizures in human patients, to a certain coarse approximation. For example, this thesis suggests that seizures may generically emerge as a loss of criticality, which has been reported across brain scales (Meisel et al., 2012; Arviv et al., 2016). Identifying such general principles might then be used to help with seizure prediction and prevention, for example by identifying bio-markers of impending critical transitions across human seizures as discussed above. However, such generalisations will eventually break down at more fine-grained resolutions. In particular, given the known heterogeneity of seizure onset patterns (Perucca et al., 2014; Noachtar & Remi, 2009) and symptomatology (see Section 1.1), it is unlikely that different seizure types are driven by identical population mechanisms. It is therefore crucial to identify the point at which these generalisations fail, to gain a fine-grained view of distinct seizure types and their mechanisms. For example, do all seizure types emerge as a loss of criticality? If so, are avalanche exponents, correlational structures and dynamical stability measures also conserved across brain scales and seizure types? In this way quantifying such population dynamics can test the extent to which a departure from criticality may serve as a general principle for epilepsy as a whole, or whether different types of phase transitions might underlie different seizure types which would have implications for seizure prediction and prevention. Furthermore, CCM should also be applied to a range of different seizure networks, to identify whether identified changes to single cell connectivity and non-linearity are a general feature of all seizures, or a pathway specific to larval zebrafish brain networks.

## 6.4  Beyond Computation

Overall, the larval zebrafish is an incredibly useful model system for the study of network instability in brain pathology such as epilepsy. Moving forward, techniques to record brain activity are likely to become more sophisticated, and brain datasets will become richer and more complex. An important challenge will therefore be, how to make sense of such invariably complex patterns. Computational approaches are of course key, but as more sophisticated mathematical tools are inevitably called upon, the danger of losing oneself in computational theory and straying further from the biology is very real. Future computational approaches should strive to remain grounded in the biological basis of observed phenomena. Key to this is using experiments to inform models, and models to inform experiments.